\documentclass[paper]{JHEP3} % 10pt is ignored!
\JHEPspecialurl{http://jhep.sissa.it/JOURNAL/JHEP3.tar.gz}
\usepackage{epsfig,multicol}
\usepackage{cite}
\usepackage{amsmath}
\usepackage{amssymb}
\usepackage{subfigure}
\usepackage{here}
\preprint{KEK-TH-1482}
\title{A practical solution to the sign problem
in a matrix model for dynamical compactification
}
\author{Konstantinos~N.~Anagnostopoulos$^a$, 
Takehiro~Azuma$^{b}$ and Jun~Nishimura$^{c,d}$\\
\llap{$^a$}Physics Department, National Technical University, \\
Zografou Campus, GR-15780 Athens, Greece \\
\llap{$^b$}Institute for Fundamental Sciences, Setsunan University, \\
17-8 Ikeda Nakamachi, Neyagawa, Osaka, 572-8508, Japan \\
\llap{$^c$}KEK Theory Center, 
High Energy Accelerator Research Organization, \\
1-1 Oho, Tsukuba, Ibaraki, 305-0801, Japan \\
\llap{$^d$}Department of Particle and Nuclear Physics,\\
Graduate University for Advanced Studies (SOKENDAI),\\
1-1 Oho, Tsukuba, Ibaraki, 305-0801, Japan  \\
\email{konstant@mail.ntua.gr,
azuma@mpg.setsunan.ac.jp, 
jnishi@post.kek.jp}}
\abstract{%----------------------- ABSTRACT------------------------------
  The matrix model formulation of superstring theory
  offers the possibility
  to understand the appearance of 
  4d space-time from 10d
  as a consequence of spontaneous
  breaking of the SO(10) symmetry. 
Monte Carlo studies of this issue is technically difficult 
due to the so-called sign problem.
We present a practical solution to this problem
generalizing the factorization method proposed originally 
by two of the authors (K.N.A.~and J.N.).
Explicit Monte Carlo calculations and large-$N$ extrapolations
are performed 
in a simpler matrix model with similar properties,
and reproduce quantitative results obtained previously 
by the Gaussian expansion method.
Our results also confirm that 
the 
spontaneous symmetry breaking
indeed occurs due to the phase of the fermion determinant,
which vanishes
for collapsed configurations.
We clarify various generic features of this approach,
which would be useful in applying it
to other statistical systems with the sign problem.
}

\newcommand{\bel}{\begin{equation}\label}

\newcommand {\beq}{\begin{equation}}
\newcommand {\eeq}{\end{equation}}
\newcommand {\beqa}{\begin{eqnarray}}
\newcommand {\eeqa}{\end{eqnarray}}
\newcommand {\bc}{\begin{center}}
\newcommand {\ec}{\end{center}}
\newcommand {\tr}{{\rm tr\,}}
\newcommand {\Tr}{\mbox{Tr\,}}

\newcommand {\ee}{\mbox{e}}

\newcommand {\del}{\partial}

\newcommand {\vev}  [1]{\ensuremath{\langle #1 \rangle}}
\newcommand {\rf}   [1]{(\ref{#1})}
\newcommand {\eexp} [1]{\ensuremath{\mbox{e}^{#1}}}
\newcommand {\tl}      {{\tilde\lambda}}
\def\vs5{\vspace*{5mm}}
\def\vs1{\vspace*{1cm}}
\def\vs2{\vspace*{2cm}}
\def\hs5{\vspace*{5mm}}
\def\hs1{\hspace*{1cm}}
\def\hs2{\hspace*{2cm}}
\def\vs50{\vspace*{50mm}}
\def\vs20{\vspace*{20mm}}

\def\tr{\hbox{tr}}

\begin{document}
% %%%%%%%%%%%%%%%%%%%%%%%%%%%%%%%%%%%%%%%%%%%%%%%%%%%%%%%%%%%%%%%%%%%
% %%%%%%%%%%%%%%%%%%%%%%%%%%%%%%%%%%%%%%%%%%%%%%%%%%%%%%%%%%%%%%%%%%%
\section{Introduction}
\label{sec:intro}
% %%%%%%%%%%%%%%%%%%%%%%%%%%%%%%%%%%%%%%%%%%%%%%%%%%%%%%%%%%%%%%%%%%%
% %%%%%%%%%%%%%%%%%%%%%%%%%%%%%%%%%%%%%%%%%%%%%%%%%%%%%%%%%%%%%%%%%%%

One of the biggest puzzles in string theory is that
typical space-time dimensionality turns out to be higher than the
macroscopically observed four dimensions. 
As a possible approach to this problem,
one may think of space-time as an emergent notion,
which arises effectively from matrix degrees of freedom.
%a lower dimensional gauge theory.
This idea of ``emergent space-time'' is nicely realized 
in the gauge-gravity duality including the famous example of
the AdS/CFT correspondence\cite{Maldacena:1997re}.

Recently the gauge-gravity duality
has been demonstrated from first principles
by Monte Carlo simulations 
of supersymmetric gauge theory in one dimension 
with 16 supercharges \cite{Hanada:2007ti,Catterall:2007fp,%
Anagnostopoulos:2007fw,Catterall:2008yz,%
Hanada:2008gy,Hanada:2008ez,Catterall:2009xn,HNSY}.
(See ref.~\cite{Nishimura:2009xm} for a review.)
%% there was a considerable progress in this direction
%% based on Monte Carlo studies
%% of supersymmetric gauge theory in one dimension 
%% with 16 supercharges \cite{Hanada:2007ti,Catterall:2007fp,%
%% Anagnostopoulos:2007fw,Catterall:2008yz,%
%% Hanada:2008gy,Hanada:2008ez,Catterall:2009xn,HNSY}.
%% (See ref.\ \cite{Nishimura:2009xm} for a review.)
This theory can be obtained formally by
the one-dimensional reduction of
ten-dimensional $\mathcal{N}=1$ $\textrm{U}(N)$ super Yang-Mills theory,
and it provides a low energy description of
a stack of $N$ D0 branes in type IIA superstring theory.
In particular, in the 
%'t Hooft 
$N\rightarrow\infty$ limit 
with large 't Hooft coupling,
the one-dimensional gauge theory was conjectured to be dual 
to the 0-brane solution
in type IIA supergravity \cite{Itzhaki:1998dd}.
Monte Carlo calculations 
%in the gauge theory 
have been performed 
on the gauge theory side 
for various quantities such as the 
internal energy \cite{Anagnostopoulos:2007fw,Catterall:2008yz,
Hanada:2008ez,Catterall:2009xn},
%thermodynamic relation, 
the Wilson loop \cite{Hanada:2008gy} and
the two-point correlation functions \cite{HNSY},
% of BPS operators \cite{HNSY},
and the results provided first-principle confirmation
of corresponding predictions based on the gauge-gravity duality.
One of the most important conclusions \cite{Hanada:2008ez}
is that the black hole thermodynamics of the 0-brane solution
has been understood microscopically
in terms of open string degrees of freedom
attached to the D0 branes,
which are described by the gauge theory. 
%reproduced from the gauge theory 
%including the $\alpha '$ corrections \cite{Hanada:2008ez}.
In other words the 1d U($N$) supersymmetric gauge theory describes 
the interior structure of the black hole in ten dimensions.
%The black 0-brane geometry is a particular solution
%of type IIA superstring theory.

%% In a different large-$N$ limit,
%% the above one-dimensional gauge theory
%% is conjectured to be
%% a non-perturbative definition of M-theory 
%% in the infinite momentum frame \cite{Banks:1996vh}. 
%% %a special light-like frame \cite{Banks:1996vh}. 
%% M-theory is a 11D hypothetical theory \cite{Witten:1995ex}, 
%% which is described by 11D supergravity at low energy,
%% and it is believed to appear in the strong
%% coupling limit of 10D type IIA superstring theory.

As a closely related but more ambitious conjecture,
the IIB matrix model was proposed 
as a nonperturbative definition of type IIB superstring 
theory in ten dimensions \cite{Ishibashi:1996xs}.
This model can be formally obtained by the zero-dimensional reduction
of ten-dimensional $\mathcal{N}=1$ $\textrm{U}(N)$ 
super Yang-Mills theory, and it has manifest SO(10) symmetry.
Here the space-time is represented by the eigenvalue distribution of the
10 bosonic matrices \cite{9802085}.
Therefore, 
%it has 
the model offers
the possibility to
%can naturally 
realize
% if the eigenvalue distribution collapses to some 
% four-dimensional manifold, 
{\it dynamical} compactification,\footnote{See
ref.\ \cite{Kostelecky:1988zi} for an earlier work suggesting 
this possibility in the framework of bosonic string field theory.
}  
%of extra dimensions,
where the extra dimensions become small due to
the spontaneous symmetry breaking (SSB) of $\textrm{SO}(10)$.
%
%% This raises the possibility
%% of {\it dynamical} compactification of the extra dimensions,\footnote{See
%% ref.\ \cite{Kostelecky:1988zi} for an earlier work suggesting 
%% this possibility in the framework of bosonic string field theory.
%% } where the extra dimensions become small due to
%% the spontaneous symmetry breaking (SSB) of $\textrm{SO}(10)$.
%
%which due to nonperturbative effects. 
This scenario has been supported by explicit calculations based on
the Gaussian expansion method (GEM).
%provided some evidence that such a scenario is realized.
By comparing the free energy of 
the SO($d$) symmetric vacua
%space-time 
with $d=2,4,6,7$,
% dimensions,
it was shown that $d=4$ is favored at the 3rd order of the
expansion \cite{Nishimura:2001sx}.
Higher order calculations confirmed this conclusion \cite{higher}.
% compared with the other dimensions.
%higher dimensional ans\"atze.
(See refs.~\cite{Nishimura:2011xy,Kim:2011cr}, however.)

Just as Monte Carlo simulations 
have been useful in studying the gauge-gravity duality from
first principles, 
they are expected to be useful also in addressing
%such issues as 
the issue of dynamical compactification in the IIB matrix model
and many others.
Indeed early works on zero-dimensional matrix
models \cite{Krauth:1998xh,9811220,Ambjorn:2000bf,%
Ambjorn:2000dx,Bialas:2000gf} have
provided us with a wealth of information on
the large-$N$ limit of the models
and the nonperturbative dynamics of their degrees of freedom. 
When one applies such an approach to the IIB matrix model,
one encounters a serious technical problem
called the sign problem
%the most serious technical problem
since the determinant (or Pfaffian, strictly speaking)
one obtains from integrating out the fermionic matrices
is complex.
%which causes the so-called sign problem.
%degrees of freedom.
Due to the fluctuation of the phase,
huge cancellations occur in Monte Carlo integration
over the bosonic matrices.
%degrees of freedom
%and one needs to generate exponentially many configurations
%to achieve given accuracy as the system size increases.
To overcome this problem, 
a promising method termed the factorization method
has been proposed \cite{0108041},
tested on simple models \cite{Ambjorn:2003rr,Azcoiti:2002vk} and used
in simulations of finite density QCD \cite{Fodor:2007vv}. 
(See refs.~\cite{07063549,fdQCD} for related works on
the QCD phase diagram
and refs.~\cite{complex-action-prob} for
other approaches to the complex action problem.)
% of QCD
% at finite density 
It resembles previously proposed density of states methods \cite{dos} 
but suggests more general and efficient ways of slicing the configuration
space. With the proposed technique it is possible to compute 
the chosen distribution function
%the chosen density of states function 
%in extremely suppressed regions of configuration space. 
in the region which is enhanced due to relatively small fluctuations of 
the phase.
%the effects of the phase.
Expectation values can be computed as the easily
located minima of the free energy thereby drastically reducing the
propagation of errors.
% in finite size scaling. 
It also has the merit of
being in principle applicable to a wide range of models and its
possible success can have a great impact on 
%a wide range of 
various physical problems. 
%% This motivates the application of this method in models with
%% a strong complex action problem.  
%% Related works 
%% in the study of the
%% %on the
%% QCD phase diagram
%% % of QCD
%% % at finite density 
%% can be found in refs.~\cite{07063549,fdQCD}.

In fact the fluctuations of the phase of the fermion determinant is
%partition function is
expected to play a crucial role
in the 
%SSB mechanism of 
dynamical compactification 
in the IIB matrix model \cite{Nishimura:2000ds}.\footnote{See
refs.~\cite{9802085,Vernizzi:2002mu} for discussions on
other possible mechanisms.}
It was shown in ref.\ \cite{Nishimura:2000ds} that the phase
%becomes stationary 
vanishes
for collapsed configurations,
%It becomes more stationary for lower dimensions,
and that the enhancement due to this property
can compensate the entropic suppression for such configurations.
%which can compensate the entropical suppression for such configurations.
In ref.~\cite{0108041}
it was shown that the competition between the two effects
can make the dominant
%the effect of the phase is to favor 
configurations have very different length scales 
than in the phase-quenched model 
suggesting a mechanism for dynamically generating 
small and large dimensions. 
Models \emph{without} the phase 
factor have also been studied by Monte Carlo simulations,
which provided strong evidence that the SSB does not occur in 
such models \cite{9811220,Ambjorn:2000bf,Ambjorn:2000dx}.
These results stress the importance of the role played by the 
%strong cancellations induced by 
phase factor.
%fluctuations.

In this paper we study a simple 
zero-dimensional matrix model \cite{0108070},
% that has been proposed in ref.\ \cite{0108070} that 
which realizes the dynamical compactification.
% of the dimensions of space. 
%
%defined by the bosonic degrees of freedom. 
The (non-supersymmetric) model has $\textrm{SO}(4)$
rotational symmetry and consists of four $N\times N$ bosonic
matrices and $N_f$ flavors of Weyl fermions in the fundamental
representation of $\textrm{SU}(N)$. 
The integration over fermions yields a complex determinant.
%effective action. 
In ref.\ \cite{0108070}, the large-$N$ limit is taken by
keeping the ratio $r=N_f/N$ fixed, and 
it is shown analytically for infinitesimal $r$
that the $\textrm{SO}(4)$ symmetry is broken down to $\textrm{SO}(3)$. 
If the phase is quenched, the SSB is shown not to occur.
For finite $r$, the model has been studied using 
the GEM \cite{0412194}. 
While the exact results for infinitesimal $r$ are consistently reproduced,
%The results are consistent with
%the remaining $\textrm{SO}(3)$ symmetry at infinitesimal $r$, 
%but as $r$ increases,
it is shown that the symmetry actually breaks down to
$\textrm{SO}(2)$ at finite $r$
due to the stronger effect of the phase.
%and the effect of the phase becomes stronger, 

Motivated by these results, we perform
Monte Carlo simulation of this simplified model,
% using the factorization method.
%in order to study it from first principles. 
%This is 
which is expected to serve as a testing ground 
for the ideas discussed above concerning the SSB in the IIB matrix model 
as well as for the method of simulating systems
with a complex action.
% proposed in ref.\ \cite{0108041}. 
%We point out that the original version of the method actually does not
%solve remaining overlap
Besides being lower dimensional, 
the model is much easier 
to simulate than the IIB matrix model with the computational effort 
increasing as $N^3$ instead of $N^6$,
% for the IIB matrix model, 
since the fermions are in the fundamental representation instead of 
the adjoint representation.
%where the fermions are in the adjoint representation. 
%On the technical side, 
We find that the method with appropriate generalization
samples efficiently the configuration space, heavily suppressed
in the phase-quenched model but important for the full model,
overcoming the so-called overlap problem. 
%We also find that the model behaves
%similarly to the matrix model studied in ref.\ \cite{0108070}. 
%
%% We also find that the model behaves
%% similarly to the observations made
%% in ref.\ \cite{0108041}. 
%
%% The length scale of the largest (smallest)
%% dimension turns out to be 
%% larger (smaller) than the one in the phase-quenched model, 
%% which provides a clear indication of SSB.
%% For each intermediate dimension,
%% two different length scales appear,
%% and one of them dominates in the large-$N$ limit.
The distribution function 
%The density of states 
and the phase of the fermion determinant 
exhibit nice scaling properties
%for collapsed configurations,
%at large and small length scales,
which are important for the extrapolation to large $N$ 
and to the regions of configuration space
%parameter space,
in which the phase fluctuation obscures the Monte Carlo measurements. 
We present detailed analyses of this scaling,
and provide its clear understanding 
based on simple theoretical arguments.
Thus we have confirmed that the SSB indeed occurs in this model
due to the effect of the phase.

We first apply the factorization method to 
a single observable
as originally formulated in ref.~\cite{0108041}.
% for simplicity.
While the results show that the method works to some extent,
we also realize certain discrepancies from the GEM results 
%and some 
as well as some puzzles,
which we attribute to the remaining overlap problem.
In order to solve this problem,
we generalize the method to multiple observables,
and show that the GEM results can be consistently reproduced.
%% We argue that these puzzles are due to some remaining
%% overlap problem. In order to remove the remaining
%% overlap problem, we need to control a set of observables
%% that are chosen appropriately.
%
%We show that the factorization method indeed 
%confirms the results of the Gaussian expansion method.
%%which we hope mimics the situation that occurs 
%in the IIB matrix model, 
%and find that it is consistent with simple theoretical arguments.
The importance of controlling multiple observables in the factorization
method is 
%discussed in a general modoe
reported briefly 
in our previous publication 
\cite{Anagnostopoulos:2010ux}.

This paper is organized as follows.
In section \ref{section:model}
we review the basic properties of the model,
and discuss the sign problem that arises in its Monte Carlo studies.
%arising from the complex fermion determinant.
In section \ref{section:mc} we investigate the model
by the factorization method with a single observable.
% as originally formulated in ref.~\cite{0108041}.
%and describe how to overcome it.
%% While the results show that the method works to some extent,
%% we also realize discrepancies from the GEM results and some puzzles,
%% which we attribute to the remaining overlap problem.
%
%from the GEM results 
%due to the remaining overlap problem.
In section \ref{generalized_factorization} 
we generalize the factorization method to multiple observables.
%% in order to solve the problem with the single observable case.
%% We show, in particular, 
%% that the results of the GEM can be consistently reproduced.
%
%and show that the results of the GEM can be consistently reproduced.
Section \ref{letzte_ep} is devoted to a summary and discussions.
In appendix \ref{algorithm}
we describe the details of the algorithm used in our Monte Carlo simulation.
In appendix \ref{sec:pqm}
we discuss some properties of the functions that play an important
role in the method.
In appendices \ref{sec:largeNextrapol} and \ref{largeN-multi}
we present the details of the large-$N$ extrapolations made in our
analyses.

% %%%%%%%%%%%%%%%%%%%%%%%%%%%%%%%%%%%%%%%%%%%%%%%%%%%%%%%%%%%%%%%%%%%
% %%%%%%%%%%%%%%%%%%%%%%%%%%%%%%%%%%%%%%%%%%%%%%%%%%%%%%%%%%%%%%%%%%%
\section{The model and the sign problem}
\label{section:model}
% %%%%%%%%%%%%%%%%%%%%%%%%%%%%%%%%%%%%%%%%%%%%%%%%%%%%%%%%%%%%%%%%%%%
% %%%%%%%%%%%%%%%%%%%%%%%%%%%%%%%%%%%%%%%%%%%%%%%%%%%%%%%%%%%%%%%%%%%

The model we study in this paper is defined by the partition
function \cite{0108070}
\begin{eqnarray}
  Z &=& \int dA \, d \psi \, d {\bar \psi} \,
\ee^{-(S_{\rm b} + S_{\rm f})} \ ,
%  \textrm{ where } 
\label{total_pf} \\
  S_{\rm b} &=& \frac{1}{2} \, N \, \tr (A_{\mu})^{2} \ , 
\label{boson_action} \\
  S_{\rm f} &=& - {\bar \psi}^{f}_{\alpha}\, (\Gamma_{\mu})_{\alpha \beta}\,
  A_{\mu} \psi^{f}_{\beta} \ .
\label{ferm_action}
\end{eqnarray}
The bosonic degrees of freedom are represented by 
$N \times N$ Hermitian\footnote{In 
ref.\ \cite{0412194} the matrices
$A_{\mu}$ are assumed to be also traceless to simplify the calculations
in the Gaussian expansion. In our Monte Carlo studies, 
we do not impose the tracelessness condition to avoid unnecessary
complication, since it is irrelevant in the large-$N$ limit.} 
matrices $A_{\mu}$ $(\mu= 1,\cdots,D)$.
%transforming in the adjoint representation of
%$\textrm{SU}(N)$,
Hereafter we assume $D$ to be even.\footnote{For odd $D$,
the model can be defined using Dirac fermions instead \cite{0108070}. 
In that case, however, the fermion determinant is real,
% and it is real positive for even $N_{\rm f}$. 
and no SSB of $\textrm{SO}(D)$ is expected.}
The fermionic degrees of freedom are represented by
${\bar \psi}^{f}_{\alpha}$ and $\psi^{f}_{\alpha}$,
which have a spinor index $\alpha=1,\cdots,p$, where $p$ represents 
the number of components of a $D$-dimensional Weyl spinor
\beq
p=2^{D/2-1} \ .
\label{def-p}
\eeq
They also have a flavor index $f=1,\cdots,N_{\rm f}$, where
$N_{\rm f}$ represents the number of flavors.
The fermionic variables
${\bar \psi}^{f}_{\alpha}$ and $\psi^{f}_{\alpha}$
are $N$-dimensional row and column vectors, respectively,
so that the actions (\ref{boson_action}) and (\ref{ferm_action})
have an $\textrm{SU}(N)$ symmetry.
%Besides the $\textrm{SU}(N)$ vector index, which is suppressed,
The $p \times p$ matrices $\Gamma_\mu$ are $\textrm{SO}(D)$
gamma matrices after the Weyl projection.
Thus the actions (\ref{boson_action}) and (\ref{ferm_action})
%, transforming in
%the fundamental (anti-fundamental) representation of $\textrm{SU}(N)$.  
have an $\textrm{SO}(D)$ symmetry,
where the bosonic variables $A_\mu$
transform as a vector and 
the fermionic variables transform as Weyl spinors.
The fermionic part can
be regarded as the zero volume limit of Weyl fermions interacting
with a background gauge field via fundamental coupling.

Integrating out the fermions, we obtain
\begin{equation}
\label{z_int_out}
Z = \int dA \, \eexp{-S_{\rm b}} \, Z_{\rm f}[A] \ , 
\end{equation}
where 
$Z_{\rm f}[A]= (\det {\cal D})^{N_{\rm f}}$ and
${\cal D} = \Gamma_{\mu} A_{\mu}$ is a $pN \times pN$
matrix. Let us then discuss the properties
of the fermion determinant $\det {\cal D}$ for a single flavor.
First of all, it is complex in general. 
Under parity transformation $A_D\to -A_D$, $A_i\to A_i$ ($i
\ne D)$, it
%the fermion determinant 
transforms as 
$\det{\cal D}\to (\det{\cal D})^*$. 
This implies that $\det{\cal D}$ is real\footnote{In this case, the
fermion determinant $Z_{\rm f}[A]$ for $N_{\rm f}$ flavors
is actually
real positive for even $N_{\rm f}$, although it is 
not necessarily so for odd $N_{\rm f}$. 
There is some numerical evidence, however, that
configurations with positive determinant dominate statistically
in the large-$N$ limit. Therefore we consider that
there is no distinction between odd $N_{\rm f}$ and even $N_{\rm f}$
in the large-$N$ limit. In the present work, we always use 
even $N_{\rm f}$ to avoid this subtlety.
}
for configurations with $A_D=0$. 
{}From this fact alone, it follows that 
the phase of the determinant becomes stationary
for configurations with $A_D=A_{D-1}=0$
since one cannot have a phase fluctuation 
within a linear perturbation around such 
configurations.\footnote{Obviously
the same statements hold for configurations obtained by $\textrm{SO}(D)$ 
rotations.}
These properties of the fermion determinant
are analogous to those found in 
the IIB matrix model \cite{Nishimura:2000ds}.

As in the case of the IIB matrix model \cite{Nishimura:2000ds},
one can extend the above argument further to arrive at the following
statements, which actually play an important role in our analysis.
%will be used in understanding some important results of our simulation.
Let us define ``$d$-dimensional configurations'' ($d \ge 1$)
as such configurations $A_\mu$ that can be
transformed into $A_{d+1} = \cdots  = A_{D} =0$
by an SO($D$) transformation.
%% Let us define the sets $\Omega_{i}$ ($i=1,2,\cdots ,D-1$)
%% of ``$i$-dimensional configurations'' $A_\mu$, 
%which can be transformed into $A_{i+1} = \cdots  = A_{D} =0$
%by an SO($D$) transformation.
%% \begin{equation}
%%  \Omega_{i} = \Big\{ \, \{ A_\mu \} \, | \, 
%% n_{\mu}^{(k)} A_{\mu} = 0 
%%   \textrm{ for } ^{\exists} n^{(k)}_{\mu} (k=1,\cdots, 4-i) \ , 
%%   \textrm{ linearly independent} \Big\} \ . 
%%   \label{deg_conf}
%% \end{equation}
%% As we mentioned, all configurations in $\Omega_i$ for $i\le 3$ 
%% have real fermion determinant.
%% This should happen for every perturbation $\delta A_\mu$ of order
%% $(D-1)-i$ because it leaves the configuration within
%% $\Omega_{D-1}$, which implies
Then for general $d$-dimensional configurations, 
we have 
 \begin{equation}
 \frac
  {\partial^{k} \Gamma}
  {\partial A_{\mu_{1}} \cdots \partial A_{\mu_{k}} } 
    = 0 
%\ , 
%  \hspace{2mm} \textrm{ for } \hspace{2mm} 
  \hspace{4mm} \textrm{ for } \hspace{2mm} 
  k=1,\cdots, (D-1)-d \ ,
\label{pert_phase}
\end{equation}
where $\Gamma$ represents the phase of the
fermion determinant $Z_{\rm f}[A]$.

We take the large-$N$ limit with $r=N_{\rm f}/N$ fixed,
which corresponds to the Veneziano limit.\footnote{For $r=1$
  the fermionic variables can be written in terms of $N\times N$
  matrices $(\Psi_\alpha)_{if}$ and $(\bar\Psi_\alpha)_{fi}$. The
  fermionic part of the action becomes
  $S_{\rm f}=-(\Gamma_\mu)_{\alpha\beta}
\tr(\bar\Psi_\alpha A_\mu \Psi_\beta)$,
which can be compared with a term
  $S_{\rm f}=-(\Gamma_\mu)_{\alpha\beta}
\tr(\bar\Psi_\alpha [A_\mu , \Psi_\beta])$
in the IIB matrix model.} 
This is needed to make the fermionic degrees of freedom 
contribute to the partition function comparably to
the bosonic degrees of freedom in the large-$N$ limit.
Whether the SSB of $\textrm{SO}(D)$ occurs in that limit
is the issue we would like to address.
%in the large-$N$ limit.
For that purpose, we consider
the ``moment of inertia tensor'' \cite{9802085,9811220}
\begin{equation}
\label{inertia}
 T_{\mu \nu} = \frac{1}{N} \tr (A_{\mu} A_{\nu}) \ ,
\end{equation}
and its real positive eigenvalues $\lambda_{n}$ ($n=1,\cdots,D$) 
ordered as
\begin{equation}
\label{eq:lams}
 \lambda_{1} \geq \lambda_{2} \geq  \cdots  \geq \lambda_{D} \ . 
\end{equation}
The vacuum expectation values (VEVs) of these eigenvalues 
$\langle \lambda_{n} \rangle$ 
%taken with respect to the
%partition function (\ref{total_pf})
play the role of the order parameters.
If they turn out to be unequal in the large-$N$ limit,
it implies the SSB of $\textrm{SO}(D)$.
%% The same order parameter is used for the discussion of
%% SSB in the IIB matrix model.
%%  \cite{9802085}
%We define the $D$ dimensional Euclidean space from the eigenvalues of
%the bosonic matrices $A_\mu$ as in 
In this model, the sum of the VEV of all the eigenvalues
is given exactly as\footnote{This can be derived by 
rescaling 
$A_{\mu}' = \kappa^{\frac{1}{2}} A_{\mu}$.
Defining
$Z(\kappa) = \int dA \, Z_{\rm f}[A]
\exp \left[-  \frac{1}{2} N \tr (A'_{\mu})^{2} 
 \right]$ and using $Z_{\rm f}[A']=\kappa^{pr N^2/2} Z_{\rm f}[A]$,
we obtain 
$ Z(\kappa) = \kappa^{-N^{2}(D+pr)/2} Z(1)$. 
Finally $ \left\langle \frac{1}{N} \tr (A_{\mu})^{2} \right\rangle 
% \sum_{i=1}^{D} \langle \lambda_{i} \rangle 
= - \frac{2}{N^{2}}  \frac{\partial }
{\partial \kappa} \log Z(\kappa) \Big|_{\kappa=1}
=  D + pr$. The same derivation with the replacement
$Z_{\rm f}[A] \mapsto |Z_{\rm f}[A]|$
holds for the phase-quenched model.
\label{sum-rule-derive}
}
\begin{equation}
\sum_{n=1}^{D} \langle \lambda_{n} \rangle = 
\left\langle \frac{1}{N} \tr (A_{\mu})^{2} \right\rangle 
= D + p \, r \ . 
\label{a-tr_identity}
\end{equation}

From now on, let us consider the $D=4$ case, which implies
$p=2$ due to (\ref{def-p}).
The gamma matrices are given by
\begin{equation}
\label{gamma_matrices}
  \Gamma_{1} =  \sigma_{1} 
= \left( \begin{array}{cc} 0 & 1 \\ 1  & 0  \end{array} \right)\ , \hspace{2mm}
  \Gamma_{2} =  \sigma_{2} 
= \left( \begin{array}{cc} 0 & -i \\ i & 0  \end{array} \right)\ , \hspace{2mm}
  \Gamma_{3} =  \sigma_{3} 
= \left( \begin{array}{cc} 1 & 0 \\ 0  & -1 \end{array} \right)\ , \hspace{2mm}
  \Gamma_{4} = i   \sigma_{4} 
= \left( \begin{array}{cc} i & 0 \\ 0  & i  \end{array} \right)\ . \nonumber
\end{equation}
Using a simple technique familiar in random matrix theory, 
it was shown
in the large-$N$ limit and at infinitesimal $r$
% in ref.\ \cite{0108070} 
that 
%in the large-$N$ limit and at infinitesimal $r$,
the VEV $\vev{\lambda_n}$ is given as \cite{0108070}
\begin{equation}
\label{eq:toylam}
 \vev{\lambda_n} = 
\left\{ 
\begin{array}{ll} 
1 + r + {\rm o}(r)  & \mbox{~for~}n=1,2,3 \ , \\ 
1 -r + {\rm o}(r)  & \mbox{~for~}n=4 \ , 
\end{array} 
\right.
\end{equation} 
which implies that 
the $\textrm{SO}(4)$ symmetry is broken down to $\textrm{SO}(3)$.
This SSB is associated with the formation of a condensate
$\vev{\bar\psi_\alpha^f\psi_\alpha^f}$, 
which is invariant under $\textrm{SO}(3)$ only.
%but not under $\textrm{SO}(4)$. 

% %%%%%%%%%%%%%%%%%%%%%%%%%%%%%%%%%%%%%%%%%%%%%%%%%%%%%%%%%%%%%%%%%%%%%
\FIGURE{
    \epsfig{file=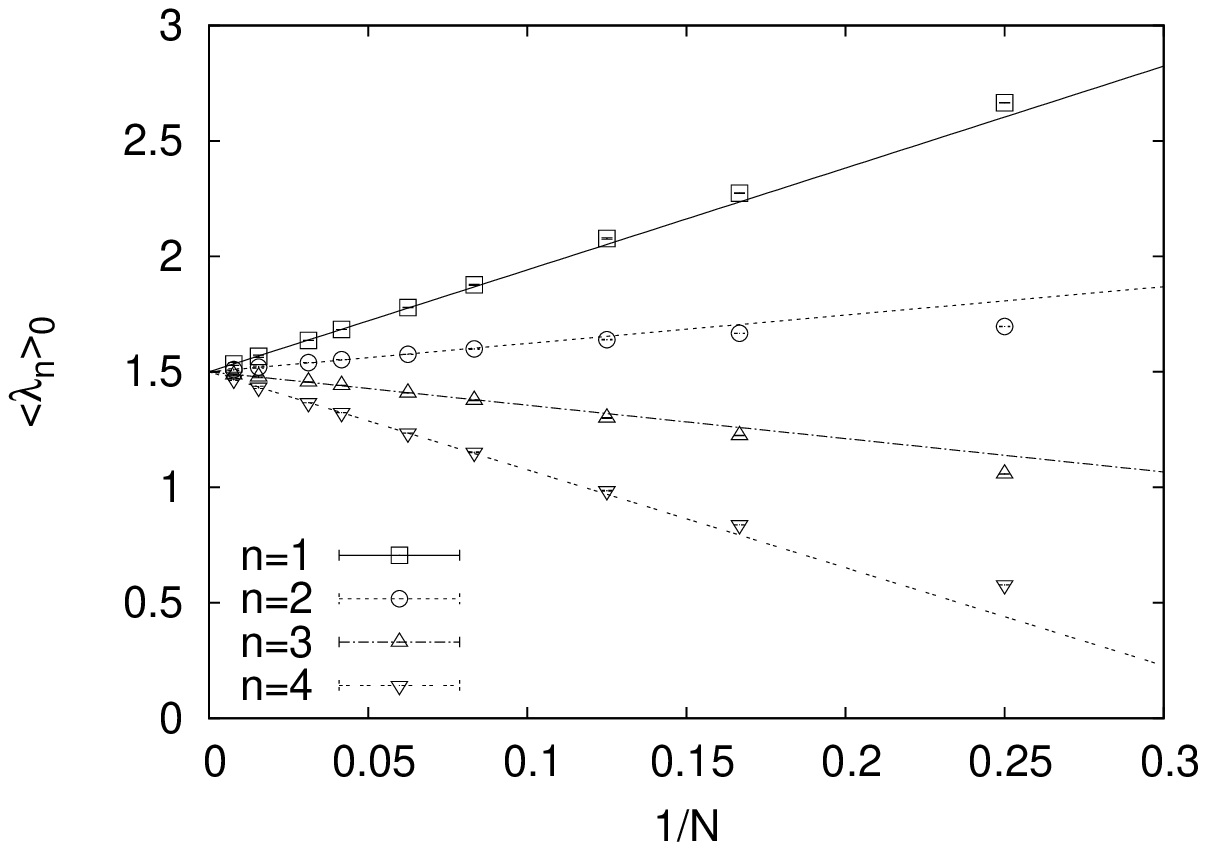,width=7.4cm}
    \epsfig{file=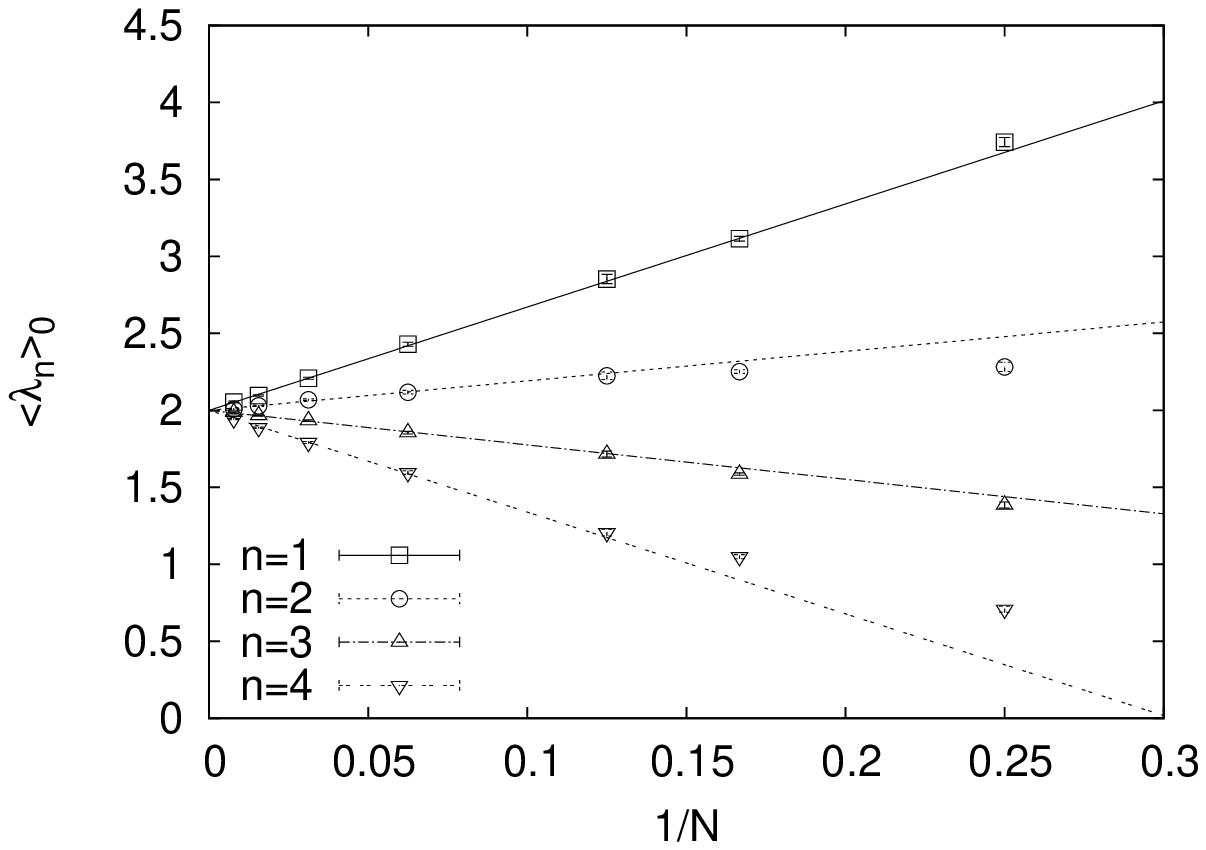,width=7.4cm}
    \caption{The VEVs $\langle \lambda_{n} \rangle_{0}$
      ($n=1,2, 3$ and $4$)
% of the moment of inertia tensor
%      $T_{\mu\nu}=\frac{1}{N}\tr (A_\mu A_\nu)$ 
in the phase-quenched model (\ref{eq:z0})
%$Z_0$ 
are plotted for $r=1$
%$\displaystyle r=1$
(Left) and $r=2$ (Right) against $1/N$.
%$\displaystyle \frac{1}{N}$. 
The data for $N\ge 8$ can be nicely fitted to
straight lines meeting at the same point $(1+r/2)$ at $N=\infty$,
which demonstrates the absence of SSB for each $r$.}
   \label{no-phase-li}
}
% %%%%%%%%%%%%%%%%%%%%%%%%%%%%%%%%%%%%%%%%%%%%%%%%%%%%%%%%%%%%%%%%%%%%%

Here the effect of the phase plays a crucial role.
To see that, let us consider the
``phase-quenched model''
\beqa
  \label{eq:z0}
%\label{real_part}
  Z_0 &=& \int dA \, e^{-S_{0}[A]} \ , \\
S_{0}[A] &=& 
% S_{\rm b}[A] -\Gamma_{\rm R}[A] 
S_{\rm b}[A] - N_{\rm f} \log |\det {\cal D}[A]| \ . 
\label{action-z0}
\eeqa
Eq.\ (\ref{a-tr_identity}) holds for this model as well 
(See footnote \ref{sum-rule-derive}.),
and therefore the absence of SSB would imply in the large-$N$ limit
\begin{equation}
  \label{eq:pqmlam}
\vev{\lambda_n}_0 =
%\vev{\lambda_i}=
1+\frac{r}{2} \qquad\mbox{for all}\quad n=1,2,3 , 4 \ ,
% \quad\mbox{and}\quad 4 \ ,
\end{equation}
where the VEV $\vev{\  \cdot \ }_0$ is taken 
with respect to $(\ref{eq:z0})$.
%In a phase-quenched model defined by setting $\Gamma=0$ in
%\rf{def_phase}, 
Indeed this can be confirmed at infinitesimal $r$ \cite{0108070}.
%and also at finite $r$ as shown in fig.~\ref{no-phase-li}.
Here 
%we see below.
we simulate the system $(\ref{eq:z0})$ for $r=1, 2$ and 
plot the eigenvalues $\vev{\lambda_n}_0$
against $1/N$
%, which are predicted as eq.\ \rf{eq:pqmlam} in the large-$N$ limit.
in fig.~\ref{no-phase-li}.
We find that the data can be nicely fitted to straight lines,
and the extrapolation to $N=\infty$ gives
the value $1+r/2$ for all $n=1,2,3$ and $4$ 
in accord with (\ref{eq:pqmlam})\footnote{Similar results have been
obtained for $r=0.25,0.5,1.5,4.0$.}
with accuracy better than $0.3\%$.
This implies, in particular, that 
the SSB of $\textrm{SO}(4)$ does not occur
in the phase-quenched model.

In what follows
%the following 
we will study 
the normalized 
eigenvalues
%principal moments of inertia
% $\tl_i$ 
\begin{equation}
{\tilde \lambda}_{n} \stackrel{\textrm{def}}{=} 
\frac{\lambda_{n}}{\langle \lambda_{n} \rangle_{0}} \ . 
\label{def_tilde}
\end{equation}
The deviation of $\langle {\tilde \lambda}_{n} \rangle$ from 1
represents the effect of the phase. The relevant question is whether
this deviation depends on $n$ in the large-$N$ limit.

% %%%%%%%%%%%%%%%%%%%%%%%%%%%%%%%%%%%%%%%%%%%%%%%%%%%%%%%%%%%%%%%%%%%%%
\TABULAR[h]{|c|c|c|c|c|}{
\hline%---------------------------------------------------------
    & \multicolumn{2}{|c|}{$r=1$} &\multicolumn{2}{|c|}{$r=2$}\\
\hline%---------------------------------------------------------
ansatz &  SO(3)  &  SO(2)  & 
%      $\vev{\tl_n}_{\mbox{\scriptsize{GEM}}}$  & 
      SO(3)  &  SO(2)  
% &  $\vev{\tl_n}_{\mbox{\scriptsize{GEM}}}$  
\\ 
\hline%---------------------------------------------------------
\vev{\tilde{\lambda}_1}   &  1.17   & 1.4 & 
%1.4&
  1.23  & 1.7
%& 1.7
\\
\vev{\tilde{\lambda}_2}   & 1.17 & 1.4
%& 1.4
& 1.23
& 1.7
%& 1.7
\\
% 3  & 0.67 & 1.13 & 0.7& 0.53 & 1.17 & 0.5\\
\vev{\tilde{\lambda}_3}   & 1.17 & 0.7
%& 0.7
& 1.23  & 0.5
%& 0.5
\\
\vev{\tilde{\lambda}_4}   & 0.5 &  0.5    
%& 0.5
& 0.31   &   0.1   
%& 0.1
\\
\hline%---------------------------------------------------------
free energy   & -1.5 &  -1.8
%& 0.5
& -1.9   &   -3.6
%& 0.1
\\
\hline%---------------------------------------------------------
}
%% % %%%%%%%%%%%%%%%%%%%%%%%%%%%%%%%%%%%%%%%%%%%%%%%%%%%%%%%%%%%%%%%%%%%%%
{The results for the normalized eigenvalues 
$\vev{\tilde{\lambda}_n}$
% defined by (\ref{def_tilde})
and the free energy obtained by the GEM at $N=\infty$
with the SO(3) and SO(2) ansatz \cite{0412194}.
\label{tab:GEMsummary}
}  
% %%%%%%%%%%%%%%%%%%%%%%%%%%%%%%%%%%%%%%%%%%%%%%%%%%%%%%%%%%%%%%%%%%%%%

At finite $r\le 2$, the large-$N$ limit of
%$N\rightarrow \infty$ limit of 
the full model (\ref{total_pf})
was studied by the GEM up to the 9-th order 
\cite{0412194} in the same way as it was used 
in the IIB matrix model \cite{Nishimura:2001sx,higher}.
The results for $r=1$ and $r=2$, which are the cases
we focus on in this paper,
are summarized in Table \ref{tab:GEMsummary}.
%Note that 
We have translated the results for the unnormalized
quantities $\langle \lambda_{n} \rangle$ 
%presented 
in ref.~\cite{0412194}
to the normalized ones $\langle {\tilde \lambda}_{n} \rangle$
using (\ref{eq:pqmlam}) and (\ref{def_tilde}).

The free energy obtained with the $\textrm{SO}(2)$ ansatz 
was found to be smaller than the one with the
$\textrm{SO}(3)$ ansatz,
which implies that the true vacuum is only $\textrm{SO}(2)$ symmetric.
%Using these ans\"atze, 
%% The expectation values $\vev{\lambda_n}$ were also calculated, and the
%% results suggest the large-$r$ behavior in the true vacuum to be roughly
%% \begin{equation}
%% \label{eq:toylam-large}
%%  \vev{\lambda_n} \sim
%% \left\{ 
%% \begin{array}{ll} 
%% r+ 1.5 - \frac{0.2}{r} + {\rm o}(r^{-1})  & \mbox{~for~}n=1,2 \ , \\ 
%% 1  + {\rm o}(r^{-1}) & \mbox{~for~}n=3 \ , \\ 
%% \frac{0.4}{r} + {\rm o}(r^{-1})  & \mbox{~for~}n=4 \ .
%% \end{array} 
%% \right.
%% \end{equation} 
%% Hence the configuration becomes essentially two-dimensional
%% in the large-$r$ limit.
%% This conclusion seems natural since
%% configurations with stationary phase are expected to dominate
%% in that limit.
%% As we mentioned below eq.\ (\ref{z_int_out}),
%% the phase indeed becomes stationary around two-dimensional
%% configurations.
At smaller $r$ \cite{0412194}, the free energy
and the VEV $\vev{\lambda_n}$ obtained with the $\textrm{SO}(2)$ ansatz
%in ref.\ \cite{Nishimura:2001sx}
asymptotes to the ones obtained with the $\textrm{SO}(3)$ ansatz,
which ensures the consistency with
%are in agreement with 
the analytic results (\ref{eq:toylam})
obtained at infinitesimal $r$.

%Hence, 
%in \rf{z_int_out} for a generic configuration $A$ is complex,
%let us define
%
%% \begin{equation}
%%   \label{eq:zf}
%%   Z_{\rm f}[A] = 
%% | Z_{\rm f}[A] | \, \ee ^{i \Gamma[A]} \ .
%% \end{equation}
%
%\exp\left( \Gamma_{\rm R}[A] + i \, \Gamma[A] \right) \ ,
%\end{equation}
%where $\Gamma_{\rm R}[A] = \log |Z_{\rm f}[A]| = N_{\rm f} 
%\log |\det {\cal D}|$.
%Plugging this into (\ref{z_int_out}),

Monte Carlo simulation of the full model
% (\ref{z_int_out})
is not straightforward
since the complex integrand of (\ref{z_int_out}) cannot be regarded
as the Boltzmann weight, and hence the idea of importance sampling
is not applicable.
One approach is to rewrite (\ref{z_int_out}) as
\beq
 Z = \int dA \, \eexp{-S_{0}[A]} \, \eexp{i \Gamma[A]} \ ,
\label{def_phase} \\
\eeq
and to calculate expectation values by reweighting
\begin{equation}
  \label{eq:ovev}
  \vev{{\cal O}} = \frac
 {\vev{{\cal O}\,\eexp{i\Gamma}}_0}
 {\vev{          \eexp{i\Gamma}}_0}\ ,
\end{equation}
where the VEVs on the right-hand side
are taken with respect to the phase-quenched model (\ref{eq:z0}),
and hence can be evaluated, in principle, 
by standard Monte Carlo techniques. 
The VEV $\vev{\eexp{i\Gamma}}_0$ is nothing but
the ratio of the partition functions $Z/Z_0$, 
and it decreases exponentially at large $N$ as 
$\eexp{-N^2 \Delta   F}$, where $\Delta F>0$ is 
the difference in free energy of the two systems. 
This can happen due to huge cancellations from the factor
$\eexp{i\Gamma}$. 
%% Indeed, for small changes $\delta A$ in the
%% configuration, $\delta\Gamma=\Gamma[A+\delta A]-\Gamma[A]$ grows
%% linearly with the fermionic degrees of freedom 
%% which are of ${\cal  O}(N^2)$ for finite $r$ 
%% and the corresponding fluctuations of the
%% phase (which all contribute terms of modulus one for each
%% configuration) are responsible for the cancellations. 
As a result, one needs 
% ${\cal O}(\eexp{\mbox{\scriptsize const.}N^2})$ 
${\rm O}(\eexp{\mbox{\scriptsize const.}N^2})$ 
configurations
to compute an observable 
with given accuracy by using the formula (\ref{eq:ovev}) directly.
This is called the sign problem.

Another closely related 
%but independent
%important 
problem is the so-called overlap problem. 
By using $Z_0$ in sampling configurations,
we usually sample the wrong region of
configuration space, which has little overlap with the dominant
configurations in the system given by $Z$. 
This problem becomes exponentially hard at large $N$ 
and makes importance sampling inefficient.

% %%%%%%%%%%%%%%%%%%%%%%%%%%%%%%%%%%%%%%%%%%%%%%%%%%%%%%%%%%%%%%%%%%%%%
% %%%%%%%%%%%%%%%%%%%%%%%%%%%%%%%%%%%%%%%%%%%%%%%%%%%%%%%%%%%%%%%%%%%%%
\section{Factorization method with a single observable}
\label{section:mc}
% %%%%%%%%%%%%%%%%%%%%%%%%%%%%%%%%%%%%%%%%%%%%%%%%%%%%%%%%%%%%%%%%%%%%%
% %%%%%%%%%%%%%%%%%%%%%%%%%%%%%%%%%%%%%%%%%%%%%%%%%%%%%%%%%%%%%%%%%%%%%

%In this paper we study the model described in the previous section
%by Monte Carlo simulation.
In this section we perform Monte Carlo studies of the model
(\ref{z_int_out})
by applying the factorization method to a single observable
as originally proposed in ref.\ \cite{0108041}.
While the analysis shows various encouraging behaviors,
we will see that the method is only partially successful
in reproducing the known results of the GEM.
We argue that this is due to the remaining overlap problem,
and discuss in section \ref{generalized_factorization}
how it can be solved by generalizing the method to multiple observables.

%the subsequent section.

\subsection{the basic idea} 
\label{single_factorization}

Let us consider calculating
the VEV $\langle {\tilde \lambda}_{n} \rangle$
by Monte Carlo simulation.
Instead of using the reweighting formula (\ref{eq:ovev})
% order to obtain the VEV
directly,
% to the VEV $\langle {\tilde \lambda}_{n} \rangle$, 
%we are going to use
we first rewrite the VEV as
\beq
\langle {\tilde \lambda}_{n} \rangle
= \int_0 ^{\infty}  dx \, x \, \rho_{n}(x)  
\label{vev-integral}
\eeq
in terms of the distribution function
%where we have defined 
\begin{equation}
\rho_{n}(x) = 
\Big\langle \delta (x-{\tilde \lambda}_{n}) \Big\rangle \ . 
\label{rho_full}
\end{equation}
%An important property of $\rho_{n}(x)$ is that it factorizes as
Applying now the reweighting (\ref{eq:ovev}) to the right-hand side,
one finds that it factorizes as
\begin{equation}
\label{eq:rhodef}
 \rho_{n}(x) = \frac{1}{C} \, \rho_{n}^{(0)}(x) \, w_{n}(x) \ .
\end{equation}
The real parameter $C$ is 
%an irrelevant 
a normalization constant given by\footnote{In the second equality,
we have used the fact that the phase $\Gamma$ flips its sign
%as $\Gamma \mapsto - \Gamma$ 
under the parity transformation $A_D \mapsto - A_D$,
which is a symmetry of the phase-quenched model (\ref{action-z0}).
A similar comment applies also to the second equality of (\ref{w_n}).
} 
\begin{equation}
\label{eq:cconst}  
C \stackrel{\textrm{def}}{=} 
  \langle e^{i \Gamma} \rangle_{0} = 
  \langle \cos \Gamma \rangle_{0} \ ,
\end{equation}
which need \emph{not} be calculated in the present method.
%turns out to be irrelevant in our analysis.
%This is in sharp contrast to the calculation
%of the VEV $\langle {\tilde \lambda}_{n} \rangle$
%by using the reweighting method (\ref{eq:ovev}) directly. 
The real function $\rho_{n}^{(0)}(x)$ is nothing
but the distribution function of $\tl_n$ in the phase-quenched
model defined by
\begin{equation}
\rho_{n}^{(0)}(x) \stackrel{\textrm{def}}{=} 
\Big \langle \delta (x- {\tilde \lambda}_{n}) \Big\rangle_{0} \ ,
\label{rho_naught}
\end{equation} 
which is peaked at $x=1$ due to the
chosen normalization \rf{def_tilde}.
The function $w_{n}(x)$ in \rf{eq:rhodef} is defined by
\begin{equation}
 w_{n}(x) 
\stackrel{\textrm{def}}{=} 
  \langle e^{i \Gamma} \rangle_{n,x} = 
  \langle \cos \Gamma  \rangle_{n,x} \ , 
\label{w_n}
\end{equation}
where $\langle \  \cdot  \ \rangle_{n,x}$ 
denotes a VEV with respect to
the partition function
\begin{equation}
 Z_{n,x} = \int dA \, \ee^{-S_{0}} \, \delta(x- {\tilde \lambda}_{n} ) \ . 
\label{ix_pf}
\end{equation}
It turns out that $w_n(x)>0$, which simplifies our analysis 
significantly.\footnote{This property does not always hold in
applications of the factorization method to a system with 
the complex action problem.
See ref.\ \cite{Ambjorn:2003rr} for a study in such a case.}

Using the saddle point approximation in 
the integral (\ref{vev-integral}),
the problem of determining $\vev{\tl_n}$ 
can be reduced to that of minimizing 
the ``free energy''
\begin{equation}
 {\cal F}_{n}(x) = - \log \rho_{n}(x) \ ,
\label{free_energy}
\end{equation}
which simply amounts to solving
\begin{equation}
 f^{(0)}_{n} (x) \stackrel{\textrm{def}}{=} 
 \frac{d}{dx} \log \rho_{n}^{(0)}(x) 
 = -
 \frac{d}{dx} \log w_n(x) \ .
% \Phi'(x)\, . 
\label{extrema}
\end{equation}
It turns out that both sides of this equation scales
as ${\rm O}(N^2)$.
Therefore, the saddle point approximation actually becomes
exact in the $N\rightarrow \infty$ limit.

\subsection{practical implementation of the method}
\label{prac-implement}

%Let us then discuss how to obtain
As we have discussed above, the calculation of the VEV $\vev{\tl_n}$
reduces to the determination of
the two functions $w_{n}(x)$ and $f_{n}^{(0)}(x)$.
%As was originally proposed in ref.\ \cite{0108041},
Here we discuss how we can measure them by Monte Carlo 
simulation following ref.\ \cite{0108041}.
%$w_{n}(x)$ and $f_{n}^{(0)}(x)$.
For that purpose we approximate
the delta function $\delta(x- {\tilde \lambda}_{n} )$ 
in \rf{ix_pf}
by the Gaussian function and simulate the system defined by 
\beq
Z_{n,V} = \int dA \, 
e^{- \{S_{0} + V(\lambda_{n}) \} } \ ,  \quad
%\label{ix_pf_gauss}  \\
% \textrm{ where } 
V(z) = \frac{1}{2} \gamma \, (z-\xi)^{2} 
\label{iV_partition} \ , 
\eeq
where $\gamma$ and $\xi$ are real parameters.
%% These parameters control the distribution of $\tl_n$,
%% which typically has a sharp peak for sufficiently large $\gamma$.
%% %% \footnote{The results turn out to be
%% %% independent of the choice of other functions of $\tl$.} 
%% %% $\eexp{-V(\tl_i)}$,
%% ($\gamma$ controls the width of the peak,
%% and $\xi$ controls the position of the peak.)
%
%We have to choose $\gamma$ large enough
%so that the results become independent of its value. 
%
% $\lambda_{i}$ does not fluctuate beyond the precision 
% of $x$ we require for the measurement of $w_i(x)$.
%
%% In our simulations we have used $\gamma$ in the range 
%% $10^3$---$10^7$.
%% Intermediate values of $x$ are easily simulated with
%% $\gamma=10^3$, while
%% we have to increase $\gamma$ as we 
%% try to drive the system to the small and large-$x$ regions.
%% We have checked that our results are independent of
%% $\gamma$ for a large range in $\gamma$.
%% The parameter $\xi$ is typically 
%% in the range $\sim [ -10\vev{\lambda_n}_0 , +100\vev{\lambda_n}_0]$. 
Let us then consider the distribution function of $\tl_n$
for the system (\ref{iV_partition}), which is given by
\beq
\rho_{n,V}(x) 
\stackrel{\textrm{def}}{=} 
\Big\langle \delta (x- {\tilde \lambda}_{n}) \Big\rangle_{n,V}  
 \propto 
 \rho_{n}^{(0)}(x) \, \exp
\Big\{ - V \Big(x \langle \lambda_{n} \rangle_{0}\Big) \Big\} \  , 
 \label{rho_v}
\eeq
%% \beqa
%% \rho_{n,V}(x) 
%% &\stackrel{\textrm{def}}{=} &
%% \Big\langle \delta (x- {\tilde \lambda}_{n}) \Big\rangle_{n,V}  \\
%% & \propto &
%%  \rho_{n}^{(0)}(x) \exp
%% \Big\{ - V \Big(x \langle \lambda_{n} \rangle_{0}\Big) \Big\} \  , 
%%  \label{rho_v}
%% \eeqa
where $\langle \ \cdot \ \rangle_{n,V}$ represents 
a VEV with respect to $Z_{n,V}$.
The position of the peak, which we denote by $x_{\rm p}$,
% of the distribution function $\rho_{n,V}(x)$
can be obtained by solving
\begin{equation}
0 = \frac{d}{dx} \log \rho_{n,V}(x) = 
f_{n}^{(0)}(x) - \langle \lambda_{n} \rangle_{0} \, 
V'\Big(x \langle \lambda_{n} \rangle_{0}\Big) \ .
\label{eq_xp}
\end{equation}
%% where we define $\displaystyle f^{(0)}_{n}(x)$ by
%% \begin{equation}
%%  f^{(0)}_{n} (x) \stackrel{\textrm{def}}{=} 
%%  \frac{d}{dx} \log \rho_{n}^{(0)}(x) \ . 
%% \label{f_naught_def}
%% \end{equation}
%We denote the solution by 

Since the distribution function $\rho_{n,V}(x)$
is sharply peaked at $x_{\rm p}$ for sufficiently large $\gamma$, 
we can use the VEV of ${\tilde \lambda}_{n}$ 
as an estimator of $x_{\rm p}$, \emph{i.e.},
\begin{equation}
x_{\rm p}  = \langle {\tilde \lambda}_{n} \rangle_{n,V}  \ .
\label{vev_nv}
\end{equation}
By varying the value of $\xi$ in (\ref{iV_partition}), we obtain
the functions $ w_{n} (x)$ and  $f_{n}^{(0)} (x)$ by\footnote{One might 
naively think that the function $\rho_{n}^{(0)}(x)$, and hence 
the function $f_{n}^{(0)}(x)$,
can be obtained by measuring the distribution of 
$\tilde{\lambda}_{n}$ in the phase-quenched model $Z_0$.
By such a direct method, however, accurate calculation
is possible only in the vicinity of the peak $x=1$.
Note that the solution of (\ref{extrema}) is 
typically not very close to the peak.}
\begin{eqnarray}
 w_{n} (x_{\rm p})
&=& \langle \cos \Gamma \rangle_{n,V} \ , 
\label{wi_vev} \\
 f_{n}^{(0)} (x_{\rm p})
%\Big(x=\langle {\tilde \lambda}_{n} \rangle_{n,V} \Big) 
&=&  \langle \lambda_{n} \rangle_{0}  \, 
V'\Big(\langle \lambda_{n} \rangle_{n,V} \Big) 
= \gamma \, \langle \lambda_{n} \rangle_{0}  \, 
\Big(\langle \lambda_{n} \rangle_{n,V} - \xi \Big) \ . 
\label{fi_vev}
\end{eqnarray}
The parameter $\gamma$ should be chosen large enough to
make the fluctuation of $\tilde{\lambda}_{n}$ smaller than the 
required resolution in $x$.
For the purpose of evaluating $f_{n}^{(0)}(x)$, however,
one should not use too large $\gamma$.
Theoretically, the right-hand side of (\ref{fi_vev}) should converge to 
the correct value in the $\gamma\rightarrow \infty$ limit,
which implies that
$\Bigl(\langle \lambda_{n} \rangle_{n,V} - \xi \Bigr) \propto 1/\gamma$.
Hence a small error in
% the evaluation of 
$\langle \lambda_{n} \rangle_{n,V}$ 
propagates to $f_{n}^{(0)}(x)$ 
%is magnified 
by the factor of $\gamma$.

%% If one uses too large $\gamma$, one obtains 
%% $\Bigl(\langle \lambda_{n} \rangle_{n,V} - \xi \Bigr) \propto 1/\gamma$,
%% and hence the small error in the evaluation of 
%% $\langle \lambda_{n} \rangle_{n,V}$ is magnified by the factor of $\gamma$.

%% By making a fit of the measured values of $f_{i}^{(0)}(x_k)$ to a
%% continuous function, we calculate
%% \begin{equation}
%% \rho_{i}^{(0)}(x) = 
%% \exp \left[ \int^{x}_{0} dz f_{i}^{(0)}(z) + \textrm{const} \right] \ . 
%% \label{r0_vev}
%% \end{equation}
%% Whenever similar values of $x_p$ are obtained
%% using different pairs of parameters $(\gamma,\xi)$, we check that our
%% measurements yield the same results. 

%In fact, from the same simulation of $Z_{n,V}$, it is possible

%% One might naively think 
%% that the function $\rho_{n}^{(0)}(x)$
%% can be obtained by measuring the distribution of 
%% $\lambda_{n}$ in the simulation of the phase-quenched model $Z_0$.
%% By such a direct method, however, one can precisely obtain 
%% the function only in the vicinity the peak.
%% %but not in the whole range of $x$.
%% This also explains why our method is different from the standard
%% reweighting method.

Note that the computation of $w_{n}(x)$ based on (\ref{wi_vev})
is hard due to the sign problem.
%the complex action problem. 
However, we only have to obtain $w_{n}(x)$ in the region
where the (possibly local) minima of the free energy 
(\ref{free_energy}) are likely to exist.
This typically implies the region where
$w_{n}(x)$ is not extremely small.
In such a region, 
one can
%hope to see the scaling of 
make a sensible large-$N$ extrapolation
\begin{equation}
  \label{eq:phidef}
  \Phi_n(x) = \lim_{N\rightarrow \infty}
\frac{1}{N^2} \log w_n(x) \ .
\end{equation}
%for various $N$.
Then one can use the equation
\begin{equation}
\frac{1}{N^2}  f_{n}^{(0)}(x) = -   
\frac{d}{dx}  \Phi_n  (x) 
% \Phi_i ' (x) 
% \Phi'(x)\, . 
\label{master-eq}
\end{equation}
to obtain the VEV $\vev{\tl_n}$.
Since there is no 
%complex action problem
sign problem
in the calculation of $f_{n}^{(0)}(x)$,
one may obtain the left-hand side of 
(\ref{master-eq}) at larger $N$ (if necessary)
than the values of $N$ used to obtain $\Phi_n(x)$ in
eq.\ (\ref{eq:phidef}).
Note also that the error on both sides of (\ref{master-eq})
due to statistics and finite $N$ does not propagate exponentially 
to $\vev{\tl_n}$
as a direct computation based on (\ref{vev-integral}) would imply.

%% In the following section we present the results of our numerical
%% simulations. 
%% %The model $Z_{i,V}$ is simulated 
%% %using the standard Hybrid Monte Carlo (HMC) algorithm, which 
%% The details of the algorithm are given in appendix \ref{algorithm}.
%% % we describe the details. 

\subsection{asymptotic behaviors of the functions 
$w_n(x)$}
\label{sec:asymp-w}

In our approach 
the effects of the phase $\Gamma$
%of the fermion determinant
are represented by the right-hand side of 
(\ref{master-eq}).
Therefore, the large-$N$ extrapolation (\ref{eq:phidef})
is the most crucial step for the success of the method.
In the left columns of  
figs.~\ref{res_r1-sol} and \ref{res_r2-sol},
we plot our results for $\frac{1}{N^2} \log w_n (x)$
for $r=1$ and $r=2$, respectively.
%in the linear scale 
%
%together with the scaling function $\Phi_n(x)$ obtained as follows.

As one can see from these plots,
the function $w_n(x)$ approaches 1 for $x\ll 1$ and/or $x\gg 1$.
%for $x\ll 1$, the function $w_n(x)$ ($n=2,3,4$) approaches unity,
%and for $x\gg 1$, the function $w_n(x)$ ($n=1,2,3$) approaches unity.
%
%This is clearly seen in the plots on the left in figures 
%\ref{res_r1-sol} and \ref{res_r2-sol}.
These are the regions, which are  
%Thus we can clearly identify the region which is 
favored by the effects of the phase.
We can understand these properties as follows.
Let us recall that $w_n(x)$ defined by (\ref{w_n})
is the VEV of $e^{i \Gamma}$ in the ensemble (\ref{ix_pf}).
The dominant configurations in the ensemble
%(\ref{ix_pf})
for $x \ll 1$ have $(5-n)$ shrunken directions due to the ordering
(\ref{eq:lams}), and hence they are approximately
configurations with dimension $d=(n-1)$.
Similarly, dominant configurations for $x \gg 1$
%the 1st through $n$-th directions extend, 
have $n$ extended directions, and hence they are approximately
configurations with dimension $d=n$.
%are close to $d=n$-dimensional.
Since the phase of the determinant vanishes for collapsed configurations
(i.e., with $1\le d \le 3$ dimensions)
as we explained below eq.~(\ref{z_int_out}),
the VEV of $e^{i \Gamma}$ 
%$w_n(x)$ 
approaches 1
when such configurations dominate in the corresponding ensemble.

%% for $x\ll 1$, the function $w_n(x)$ ($n=2,3,4$) approaches unity,
%% and for $x\gg 1$, the function $w_n(x)$ ($n=1,2,3$) approaches unity.
%% This is clearly seen in the plots on the left in figures 
%% \ref{res_r1-sol} and \ref{res_r2-sol}.
%% Thus we can clearly identify the region which is favored by the 
%% effects of the phase $\Gamma$.

%Let us discuss the effects of the phase of the fermion determinant.
%% We present our results for the $r=1$ and $r=2$ model 
%% of Eq.\ \rf{def_phase} 
%% by including the effect of the phase of the fermion
%% determinant. 
%% In order to calculate it,
%% we need to obtain the function $w_n(x)$ defined in eq.\ \rf{w_n}.
%% The value $w_n(x)$ at $x=x_p$ is estimated from \rf{wi_vev}. 
%% Using the measurements of $f_n^{(0)}(x)$ discussed in the previous 
%% subsection, we can obtain the VEV $\vev{\tilde\lambda_k}$
%% by solving eq.\ (\ref{master-eq}).
%% Due to the sign problem, it is difficult to obtain $w_n(x)$ 
%% %and hence the scaling function $\Phi_n(x)$ 
%% in the whole range of $x$.
%% However, there exist certain regions of $x$, in which
%% $w_n(x)$ is not extremely small and hence one can obtain it.
%% In fact those are the regions, which are favored by the
%% effect of the phase.

Moreover, we can deduce the asymptotic behaviors 
of the functions $w_n(x)$
% in the favored region 
as
%% At small and large values of $x$, the function $w_n(x)$ 
%% %$\Phi_n(x)$ in eq.\ \rf{eq:phidef} 
%% is expected to have the following asymptotic behavior 
%
\begin{equation}
\frac{1}{N^2} \log w_n(x) \simeq 
% \Phi_{n} (x) = 
  \left\{ 
  \begin{array}{ll} 
    -c_{n} x^{5-n} 
% +  \cdots 
& (x \ll 1, n=2,3,4)  \ , \\ 
    -d_{n} x^{-(4-n)} 
% +  \cdots 
& (x \gg 1, n=1,2,3)  \ .
  \end{array} 
  \right. 
\label{ii-fit}
\end{equation}
%The scaling behavior \rf{ii-fit} 
This can be derived from the property (\ref{pert_phase})
of the phase.
% \cite{Nishimura:2000ds}. 
%(Note that we are dealing with the $D=4$ model.)
It follows that
the fluctuation of the phase is of the order of 
$\delta\Gamma\sim (\delta A/|A|)^{4-d}$ 
around a collapsed configuration with $1\le d\le 3$ dimensions,
where $\delta A$ and $|A|$ represent the typical scale
% of fluctuations
of $A_\mu$ in the shrunken and extended directions, respectively.
%al configuration ($d=1,2,3$).
%
%For $x \ll 1$, the $n$-th through 4-th directions
%shrink, and the configuration becomes $(n-1)$-dimensional.
%
%% Then the saddle points of $\Gamma$
%% %$\eexp{i \Gamma}$ 
%% given by \rf{pert_phase} dominate and 
%% we expect the fluctuation of the phase to be
%% we expect the fluctuation of the phase to be
%% ${\rm O}((\delta A)^{4-(n-1)})={\rm O}((\delta A)^{5-n})$.
%
%Let us recall here that $\sqrt{x}$ has the same
%scaling dimension as $A_\mu$ due to the definition (\ref{inertia}). 
Due to the definition (\ref{inertia}), 
it is expected that $\delta A/|A|$ is proportional 
to $\sqrt{x}$ (and $1/\sqrt{x}$) for $x \ll 1$ (and $x \gg 1$), respectively.
When $x \ll 1$,
the distribution of the phase is therefore expected to
have a width $\sigma \propto (\sqrt{x})^{5-n}$.
Assuming that the distribution is 
Gaussian,\footnote{We have checked that this is indeed the case in
the region of $x$ where $w_n(x)$ is close to 1.}
we can evaluate the
expectation value of $e^{i \Gamma}$ by using the formula \cite{07063549}
\beq
\int d \Gamma \frac{1}{\sqrt{2\pi}\sigma} 
\exp\left(- \frac{1}{2 \sigma^2} \Gamma^2 \right) e^{i \Gamma}
= \exp \left( -\frac{1}{2} \sigma ^2\right)  \ .
\label{ejiri}
\eeq
Thus we obtain $ - \log w_n (x) = \frac{1}{2} \sigma^2
\propto  x^{5-n}$.
%from the argument of ref.\ 
Similarly, when $x \gg 1$, 
%% the 1st through $n$-th directions extend, and the configuration becomes
%% $n$-dimensional.
%% The 
the distribution
% of the phase 
is expected to 
have a width of the order of 
$(1/\sqrt{x})^{4-n}$,
and hence $ - \log w_n (x) \propto  x^{-(4-n)}$.

We fit our data to the
asymptotic behaviors (\ref{ii-fit}) and extract the coefficients
$c_n$ and $d_n$ as described in appendix \ref{sec:largeNextrapol}.
It turns out that the finite-$N$ effects in these coefficients are
of the order of $1/N$. Based on this observation, we make 
large-$N$ extrapolations to obtain (\ref{eq:phidef})
in the asymptotic regions,
which is also plotted in the left columns of  
figs.~\ref{res_r1-sol} and \ref{res_r2-sol}.
The two solid lines represent the margin of error
due to the uncertainty from the large-$N$ extrapolation.
%correspond to
%the upper bound and the lower bound representing 
%the uncertainty from the large-$N$ extrapolation.
%

%Next we study the density of states $\rho_n(x)$ of
%eq.\ \rf{rho_full}. 
%
%% In the left columns of  
%% Figures \ref{res_r1-sol} and \ref{res_r2-sol},
%% we plot the results for $\frac{1}{N^2} \log w_n (x)$
%% in the linear scale together with
%% the scaling function $\Phi_n(x)$ obtained as described above.
%% %The solid line represents the asymptotic behaviors obtained above.

The regions of $x$ in which $\Phi_n(x)$ becomes small 
represent the regions in which the sign problem becomes severe.
As we approach such regions, it becomes more difficult
to obtain data points,
% for $\frac{1}{N^2} \log w_n (x)$,
in particular, for larger $N$.
%the scaling in $N$ gets worse.
%
%In such regions we were able to obtain data for $\Phi_n(x)$ only 
%for $N=4$.
%
%finite $N$ effects in $\Phi_n(x)$ are expected to be large
%particularly in this regime, and the $N=4$ data cannot be used
%to discuss the large-$N$ limit.
%% On the other hand, we observe reasonably 
%% good large-$N$ scaling behaviors in the regions 
%% where we observe the asymptotic behaviors (\ref{ii-fit}).
A big virtue of the factorization method is 
that we can use the function $\Phi_n(x)$
% obtained above 
in the asymptotic regions to extrapolate towards
the region where the data points are not available due to
the sign problem.
%% Clearly one does not have this feature
%% when one uses the reweighting formula (\ref{eq:ovev}) 
%% directly to obtain $\vev{\tilde{\lambda}_n}$.
One should keep in mind, however, that
%It is, of course, likely that 
the true function $\Phi_n(x)$ may 
deviate from the asymptotic behavior as one approaches $x\sim 1$,
which causes certain systematic error.
%In principle, one may reduce it by improving the extrapolation
We will discuss this 
%important 
issue in sections \ref{sec:sys-error} and \ref{sec:calc-free}.

\subsection{results for $\vev{\tilde{\lambda}_n}$}
\label{lambda-results-single}

\begin{figure}[H]
%\begin{tabular}{cc}
\begin{center}
\begin{minipage}{.45\linewidth}
\includegraphics[width=1.0 \linewidth]{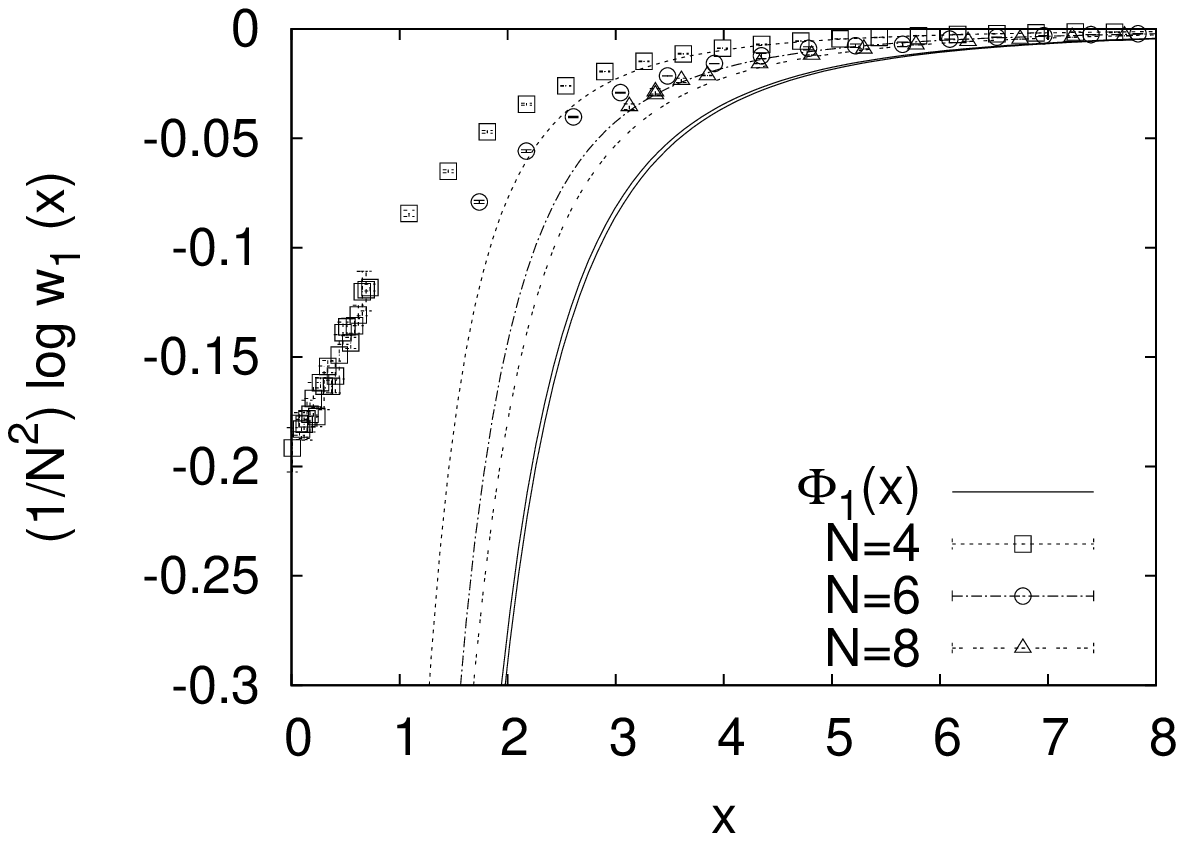}
  \end{minipage}
%  \end{center}
  \hspace{1.0pc}
%\begin{center}
\begin{minipage}{.45\linewidth}
\includegraphics[width=1.0 \linewidth]{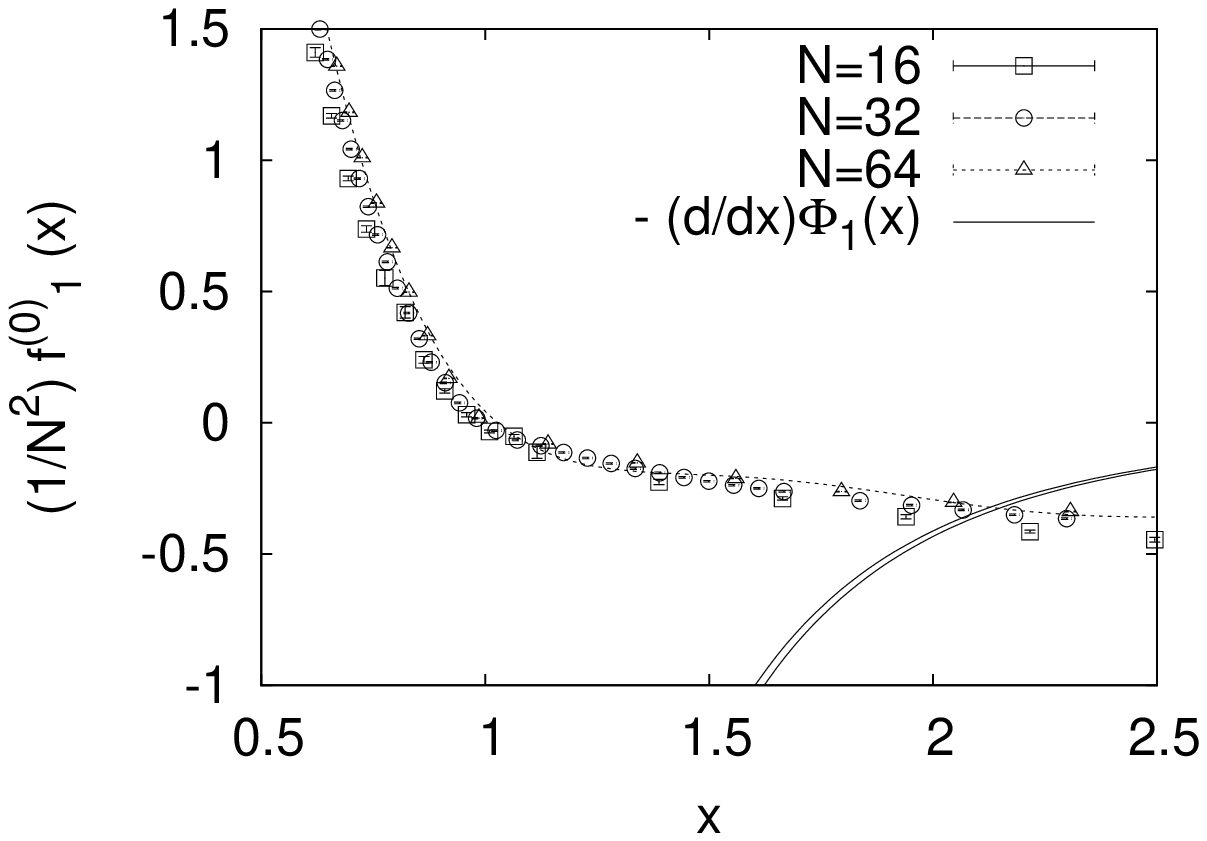}
%\caption{figure1}
  \end{minipage}
%\label{fig1}

\begin{minipage}{.45\linewidth}
\includegraphics[width=1.0 \linewidth]{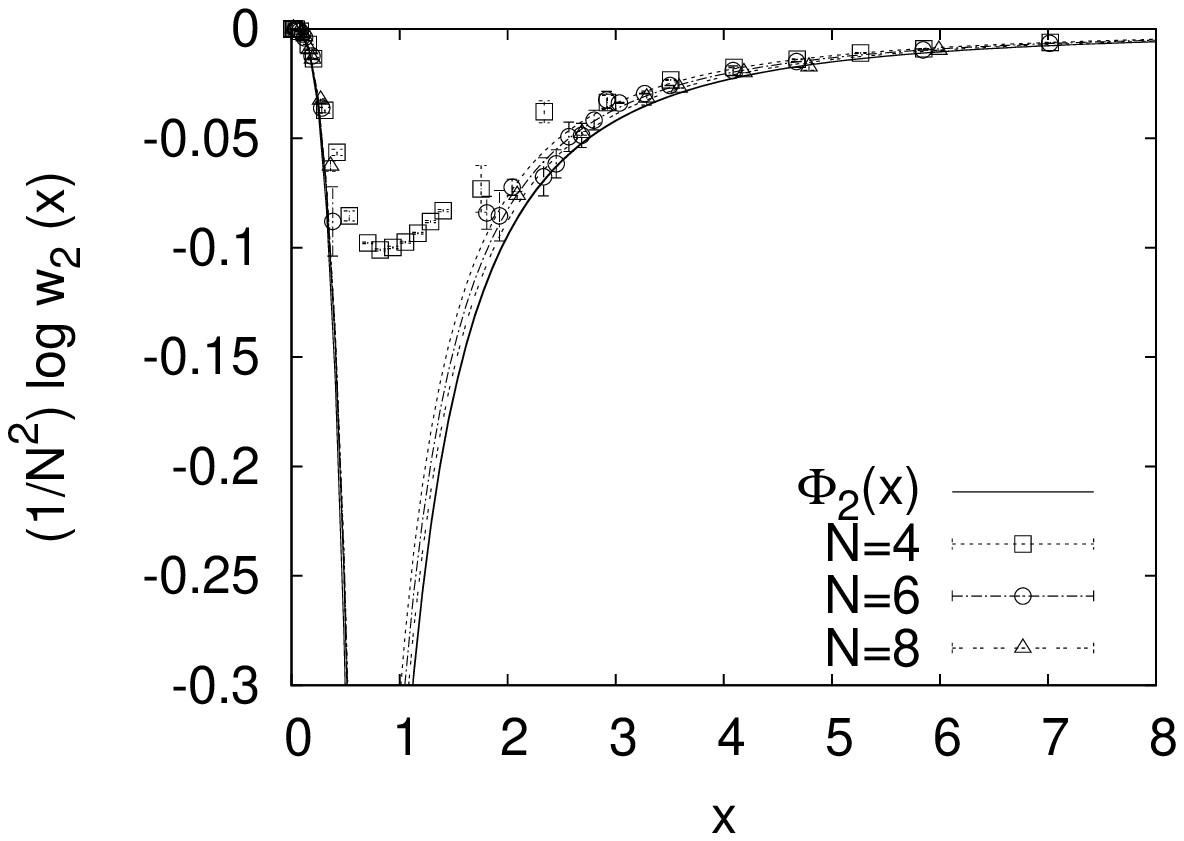}
  \end{minipage}
%  \end{center}
  \hspace{1.0pc}
%\begin{center}
\begin{minipage}{.45\linewidth}
\includegraphics[width=1.0 \linewidth]{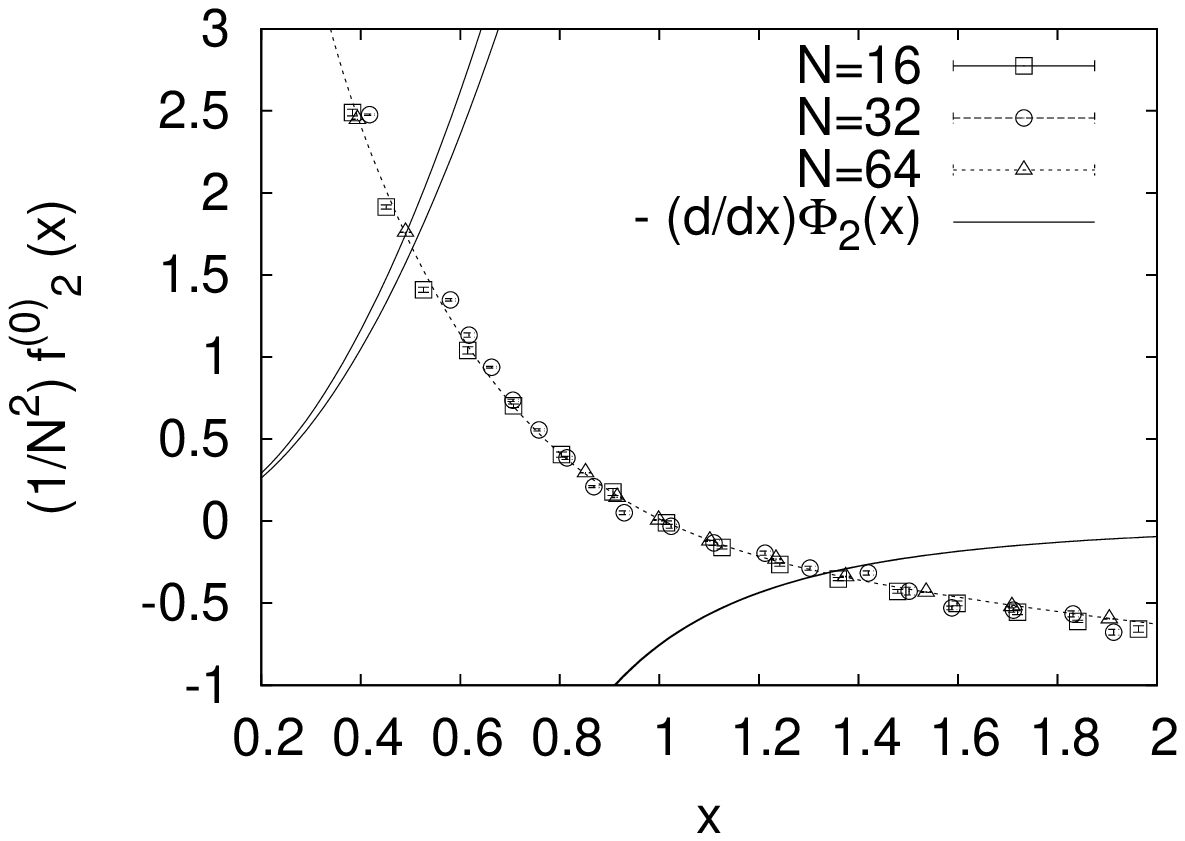}
%\caption{figure1}
  \end{minipage}
%\label{fig1}

\begin{minipage}{.45\linewidth}
\includegraphics[width=1.0 \linewidth]{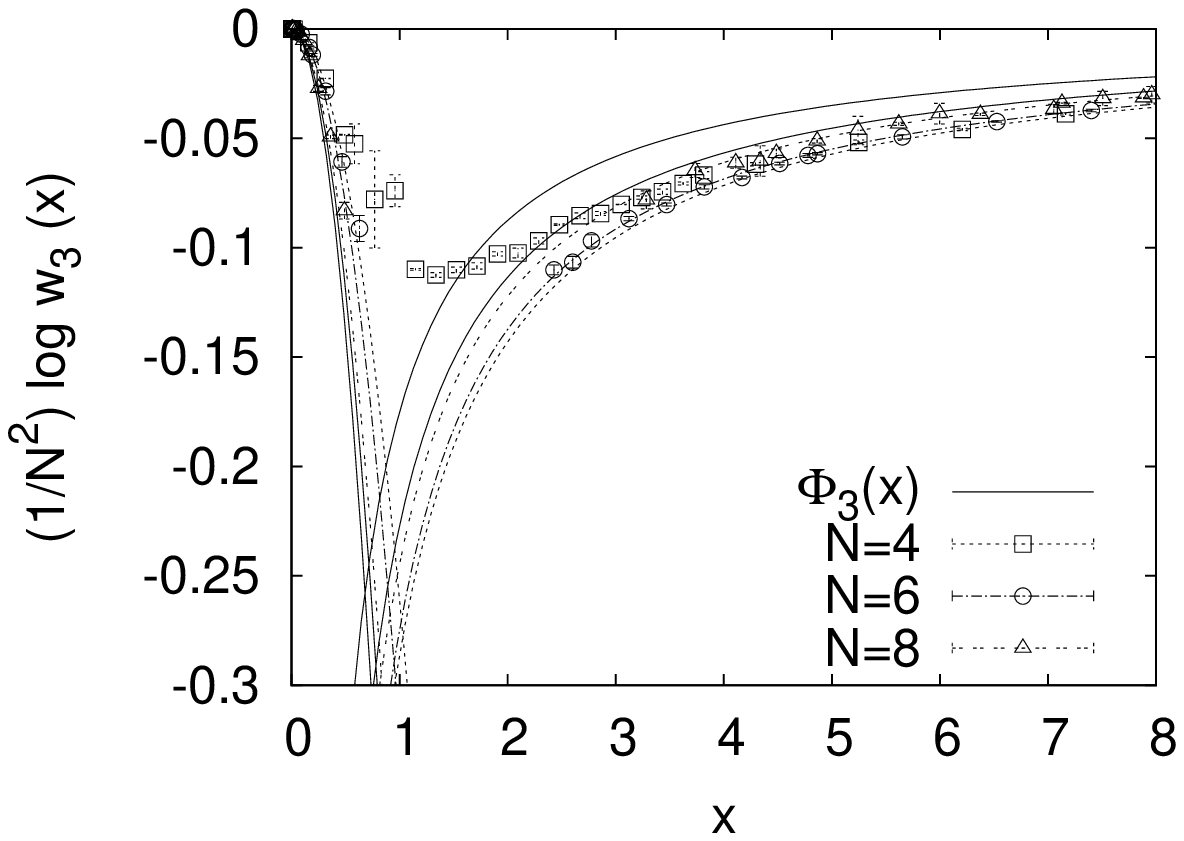}
  \end{minipage}
%  \end{center}
  \hspace{1.0pc}
%\begin{center}
\begin{minipage}{.45\linewidth}
\includegraphics[width=1.0 \linewidth]{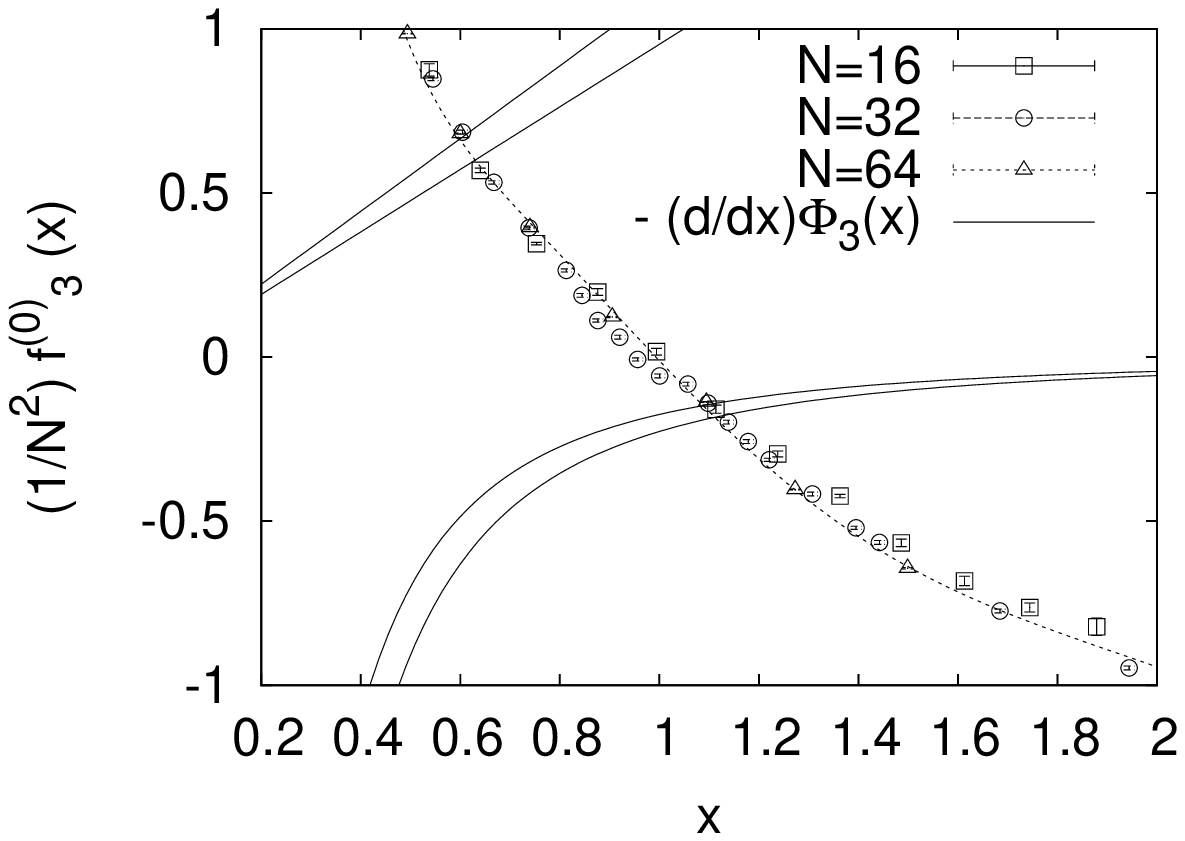}
%\caption{figure1}
  \end{minipage}
%\label{fig1}

\begin{minipage}{.45\linewidth}
\includegraphics[width=1.0 \linewidth]{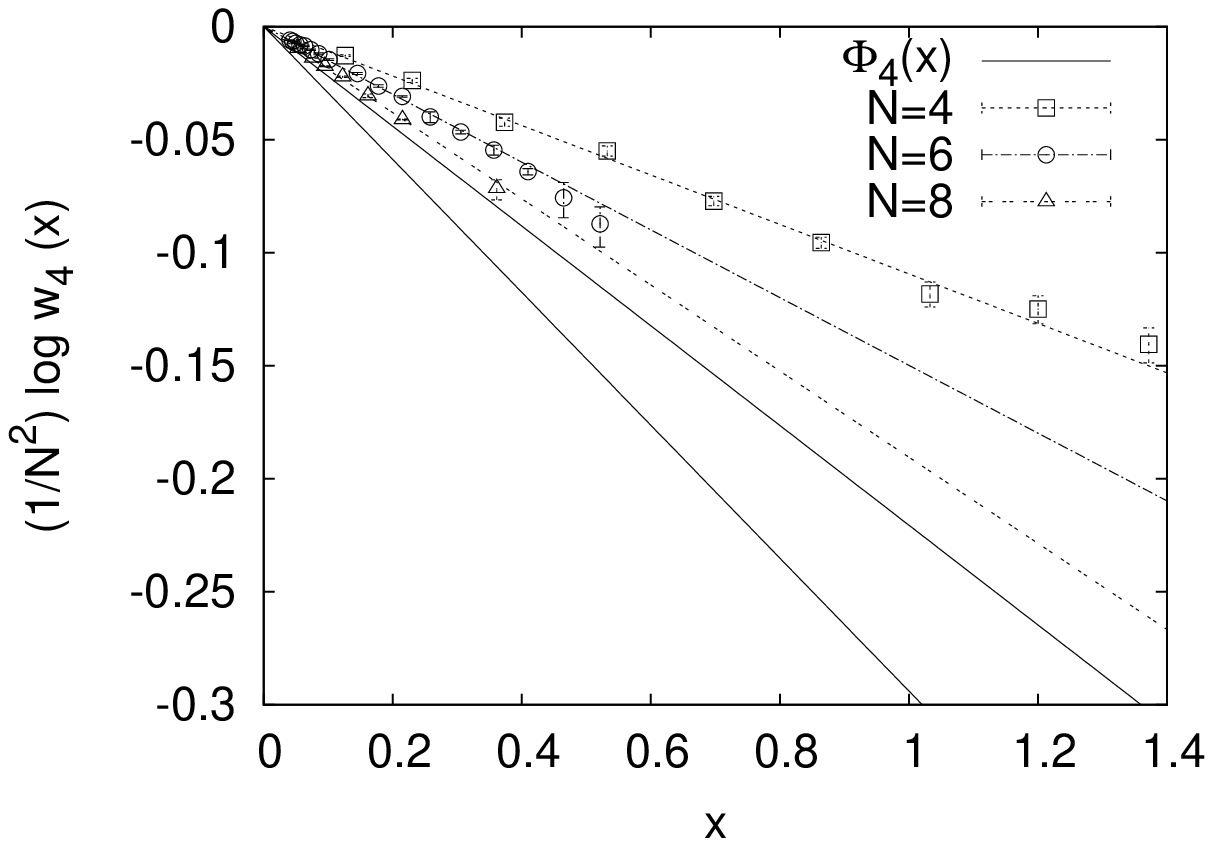}
  \end{minipage}
%  \end{center}
  \hspace{1.0pc}
%\begin{center}
\begin{minipage}{.45\linewidth}
\includegraphics[width=1.0 \linewidth]{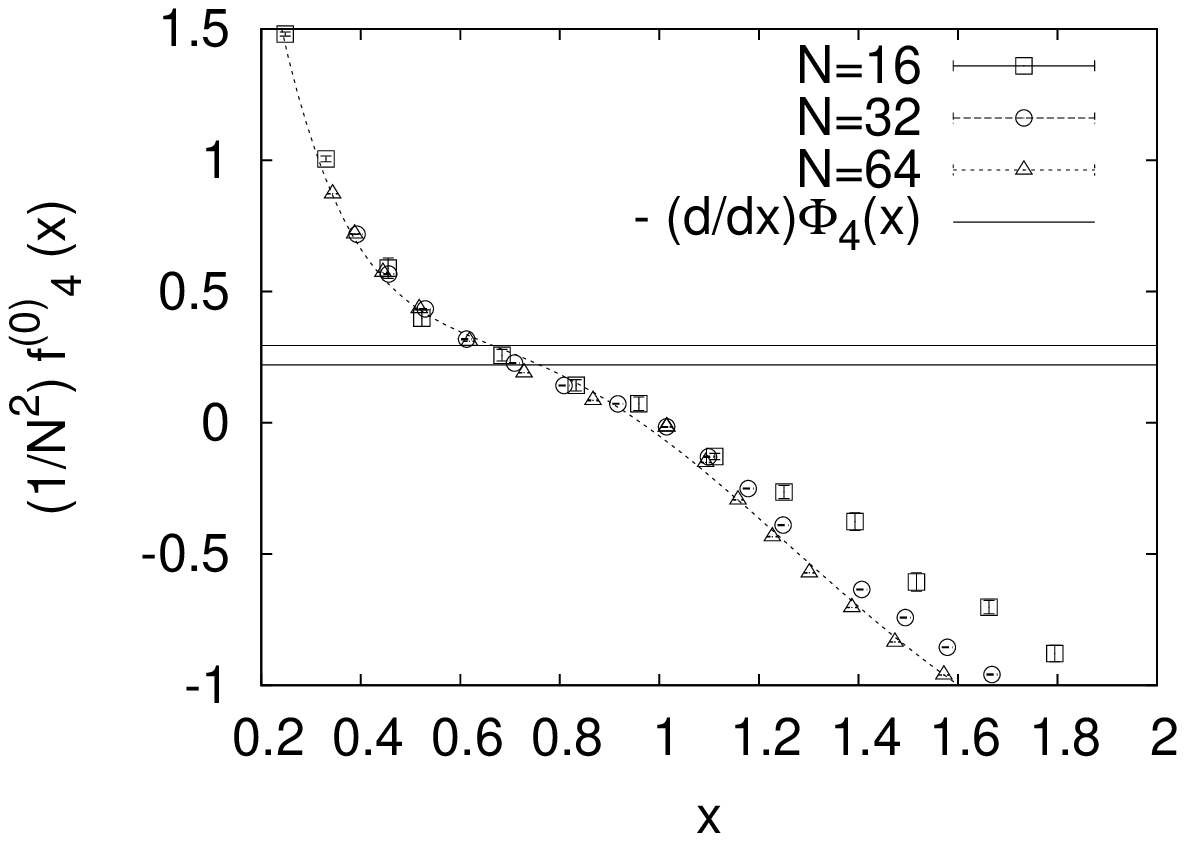}
%\caption{figure1}
  \end{minipage}
%\label{fig1}

\end{center}
\caption{
Results for the $r=1$ case. 
(Left) The function $\frac{1}{N^2} \log w_n (x)$
%effect of the phase $\Phi_{n}(x)$ 
is plotted for $N=4,6,8$
together with the scaling function $\Phi_{n}(x)$ 
obtained by large-$N$ extrapolations based on
%extracted from 
the asymptotic behavior (\ref{ii-fit}).
%by fitting $N=8$ data to the form predicted theoretically.
(Right) 
The function $\frac{1}{N^{2}} f^{(0)}_{n}(x)$ is plotted for 
$N=16,32,64$ together with $- \frac{d}{dx} \Phi_{n}(x)$.
The $x$ coordinate of the intersecting point gives
a solution to (\ref{master-eq}).
% is obtained
%by finding the intersections of 
%$y=\displaystyle \frac{1}{N^{2}} f^{(0)}_{n}(x)$ and
%$y= - \frac{d}{dx} \Phi_{n}(x)$.
%
%The position of the intersections is indicated by the arrows
%with the symbols $x_{\rm s}$  and $x_{\rm l}$
%for the regions $x<1$ and $x>1$, respectively. 
}
   \label{res_r1-sol}
  \end{figure}

\begin{figure}[H]
%\begin{tabular}{cc}
\begin{center}
\begin{minipage}{.45\linewidth}
\includegraphics[width=1.0 \linewidth]{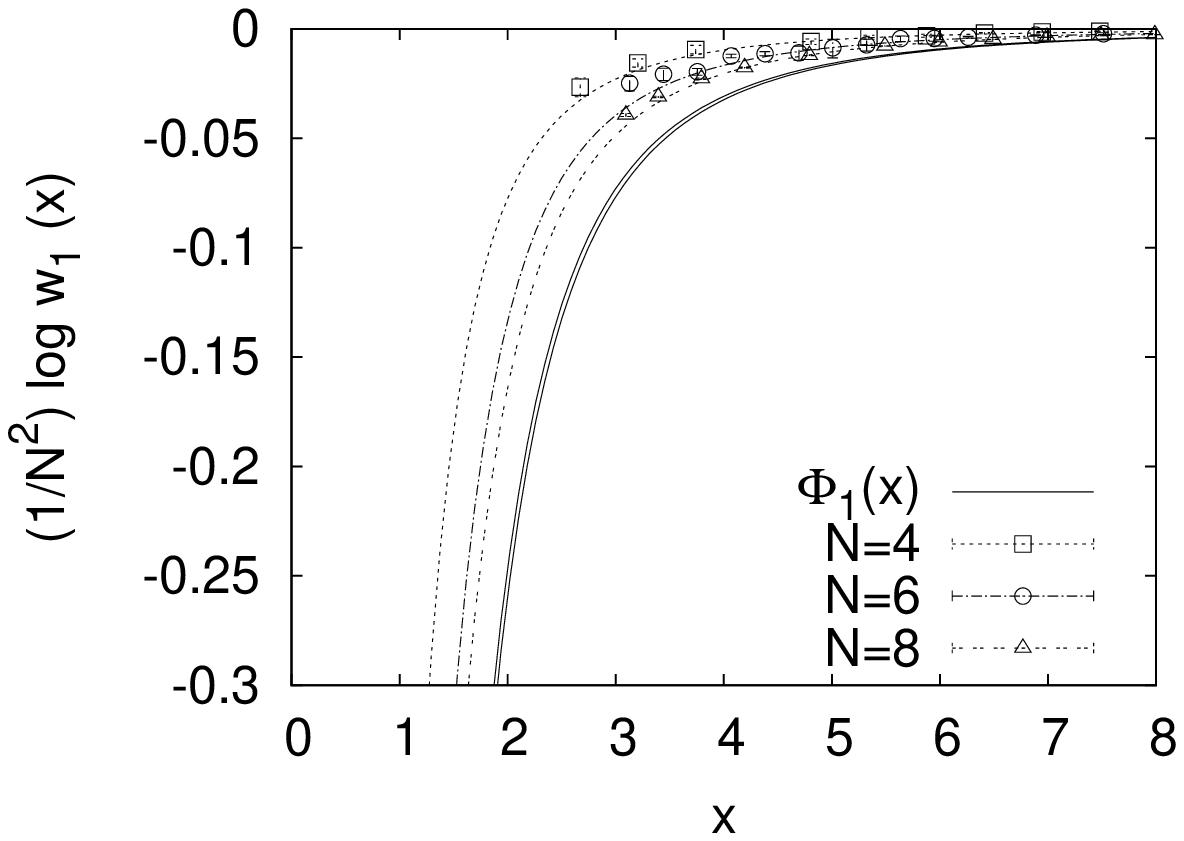}
  \end{minipage}
%  \end{center}
  \hspace{1.0pc}
%\begin{center}
\begin{minipage}{.45\linewidth}
\includegraphics[width=1.0 \linewidth]{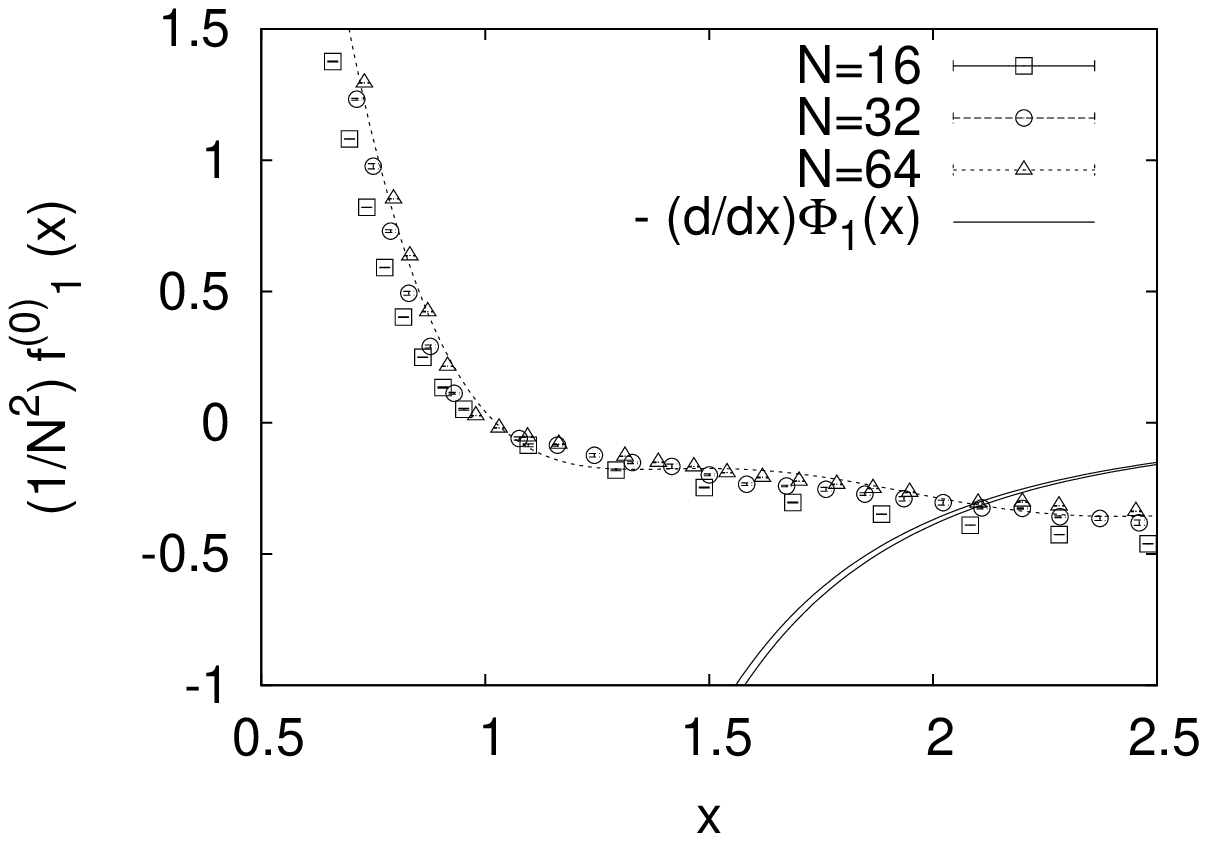}
%\caption{figure1}
  \end{minipage}
%\label{fig1}

\begin{minipage}{.45\linewidth}
\includegraphics[width=1.0 \linewidth]{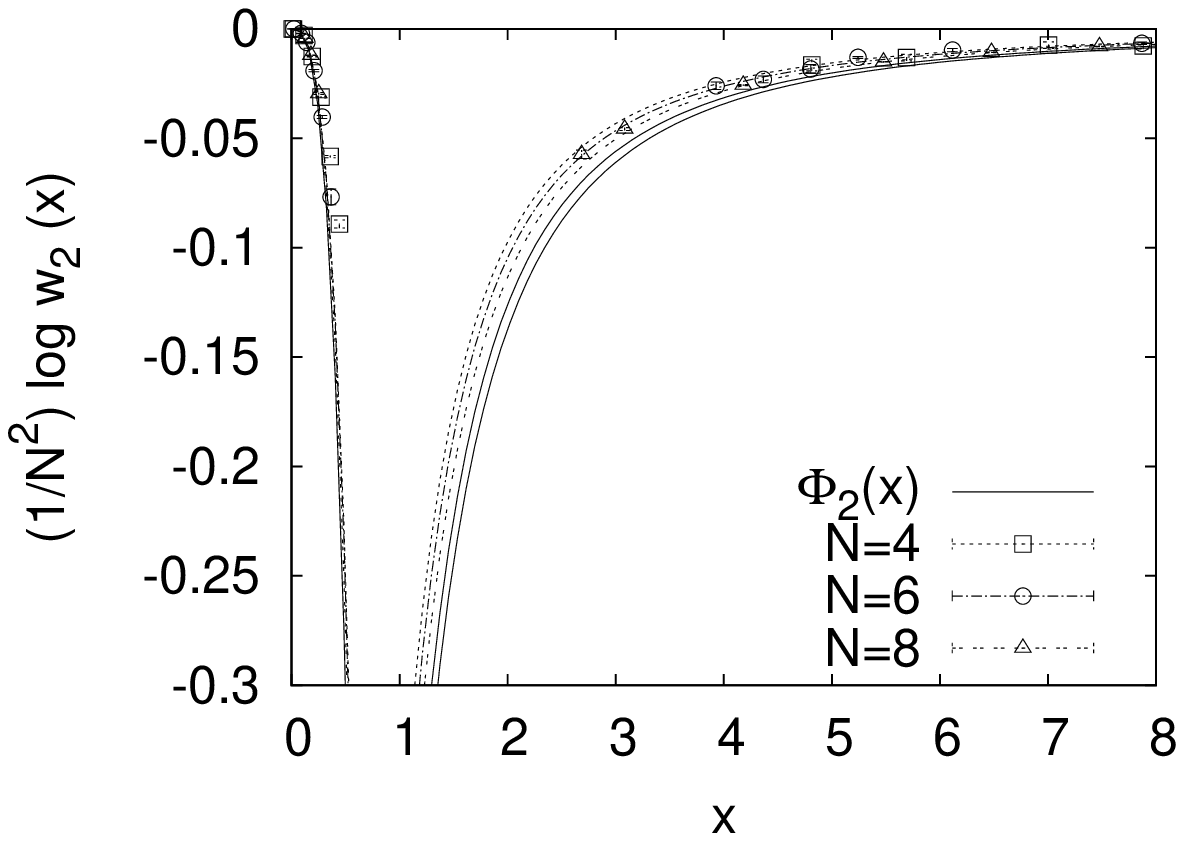}
  \end{minipage}
%  \end{center}
  \hspace{1.0pc}
%\begin{center}
\begin{minipage}{.45\linewidth}
\includegraphics[width=1.0 \linewidth]{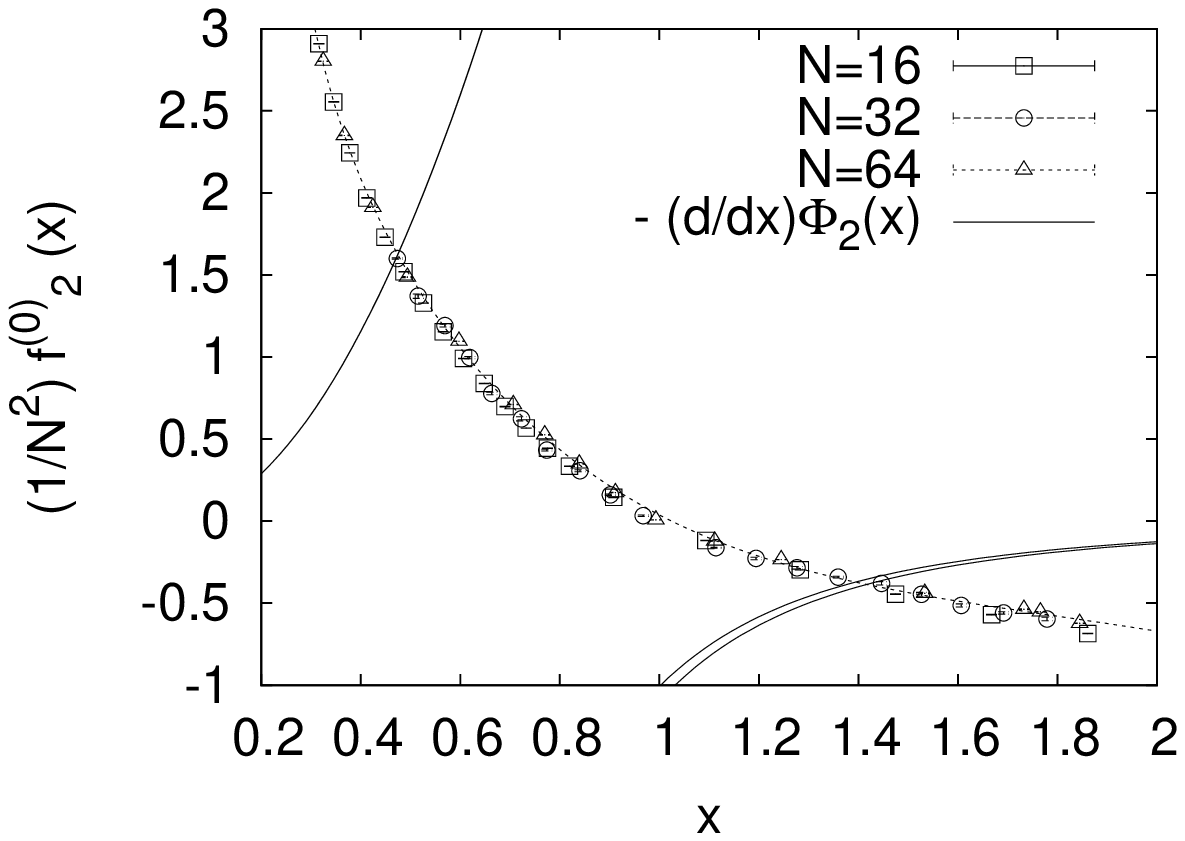}
%\caption{figure1}
  \end{minipage}
%\label{fig1}

\begin{minipage}{.45\linewidth}
\includegraphics[width=1.0 \linewidth]{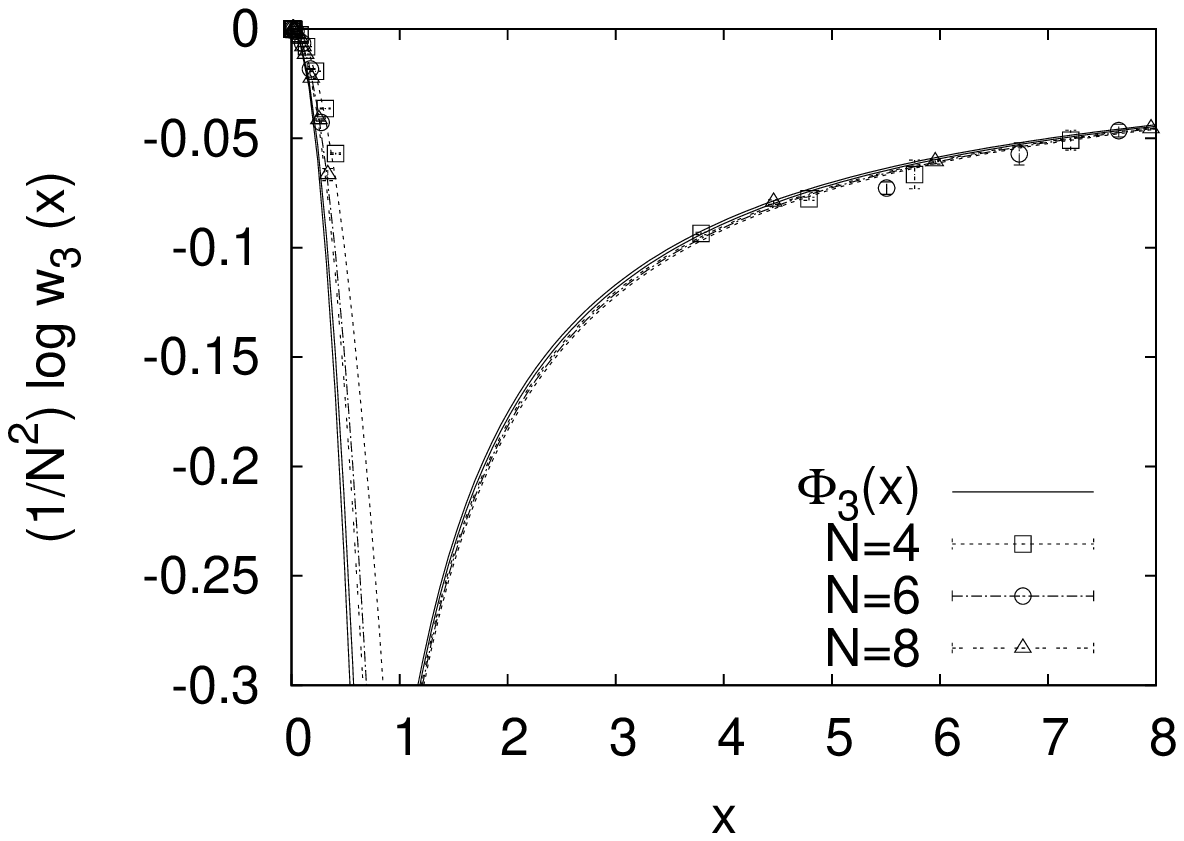}
  \end{minipage}
%  \end{center}
  \hspace{1.0pc}
%\begin{center}
\begin{minipage}{.45\linewidth}
\includegraphics[width=1.0 \linewidth]{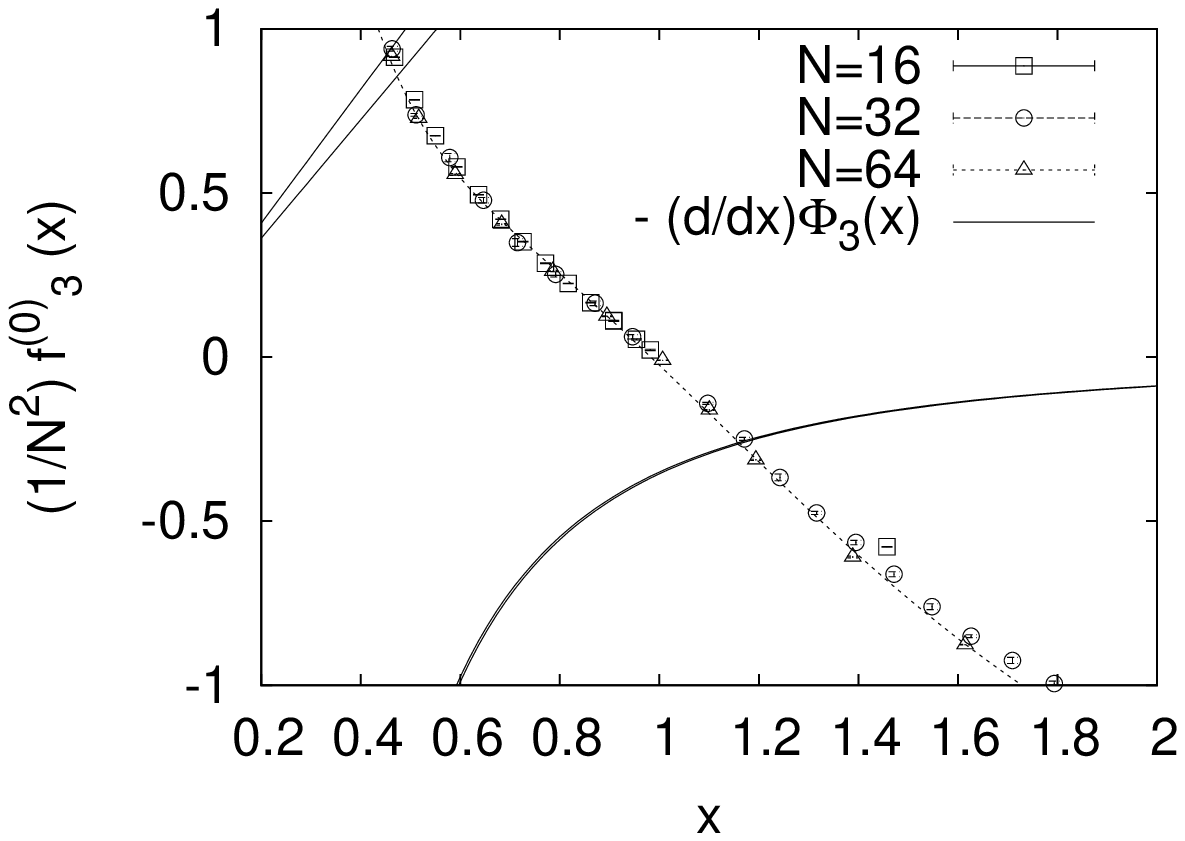}
%\caption{figure1}
  \end{minipage}
%\label{fig1}

\begin{minipage}{.45\linewidth}
\includegraphics[width=1.0 \linewidth]{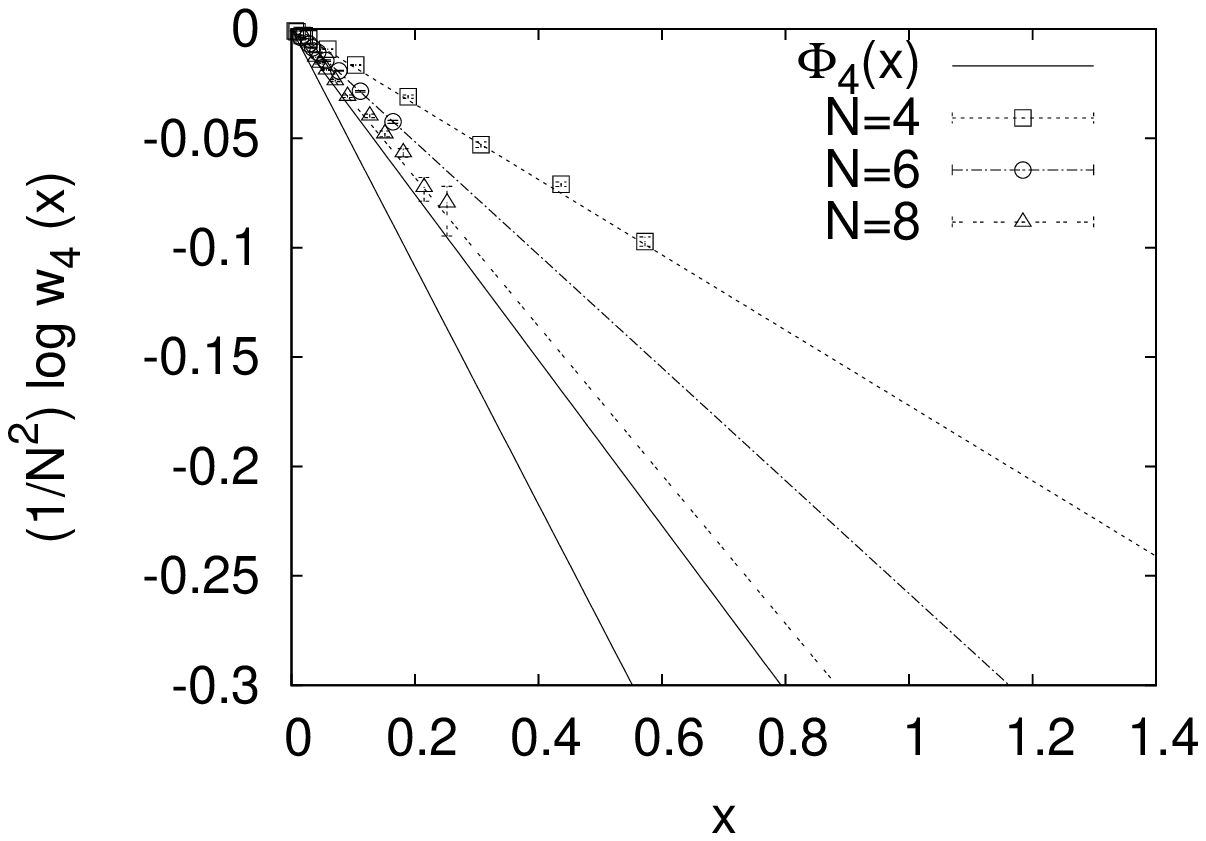}
  \end{minipage}
%  \end{center}
  \hspace{1.0pc}
%\begin{center}
\begin{minipage}{.45\linewidth}
\includegraphics[width=1.0 \linewidth]{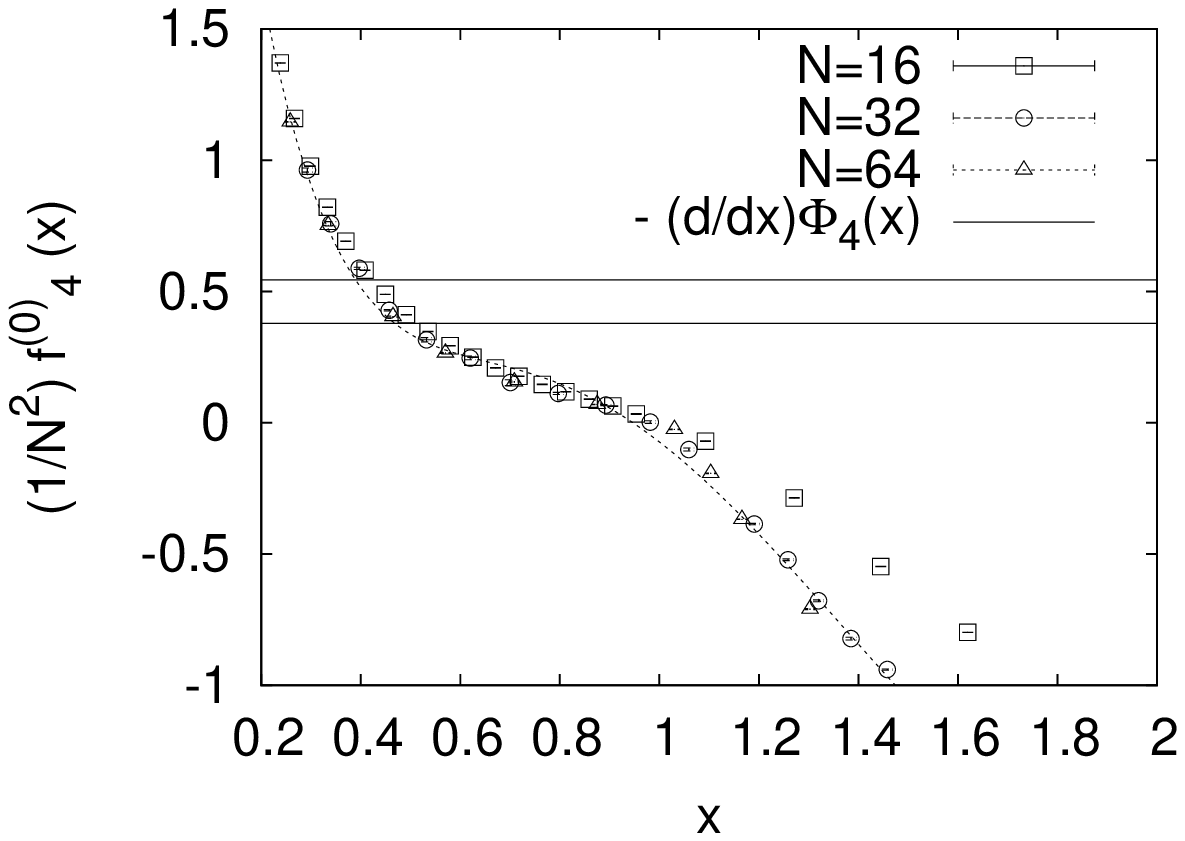}
%\caption{figure1}
  \end{minipage}
%\label{fig1}

\end{center}
\caption{
Results for the $r=2$ case. 
(Left) The function $\frac{1}{N^2} \log w_n (x)$
%effect of the phase $\Phi_{n}(x)$ 
is plotted for $N=4,6,8$
together with the scaling function $\Phi_{n}(x)$ 
obtained by large-$N$ extrapolations based on
%extracted from 
the asymptotic behavior (\ref{ii-fit}).
%by fitting $N=8$ data to the form predicted theoretically.
(Right) 
The function $\frac{1}{N^{2}} f^{(0)}_{n}(x)$ is plotted for 
$N=16,32,64$ together with $- \frac{d}{dx} \Phi_{n}(x)$.
The $x$ coordinate of the intersecting point gives
a solution to (\ref{master-eq}).
%The same as fig.~\ref{res_r1-sol} but for $r=2$.
%% (Left) The function $\frac{1}{N^2} \log w_n (x)$
%% %effect of the phase $\Phi_{n}(x)$ 
%% is plotted together with the scaling function $\Phi_{n}(x)$ 
%% extracted from the asymptotic behaviors (\ref{ii-fit}).
%% %by fitting $N=8$ data to the form predicted theoretically.
%% (Right) The solution to (\ref{master-eq}) is obtained
%% by finding the intersections of 
%% $y=\displaystyle \frac{1}{N^{2}} f^{(0)}_{n}(x)$ and
%% $y= - \frac{d}{dx} \Phi_{n}(x)$.
%% %The position of the intersections is indicated by the arrows
%% %with the symbols $x_{\rm s}$  and $x_{\rm l}$
%% %for the regions $x<1$ and $x>1$, respectively. 
}
   \label{res_r2-sol}
  \end{figure}

On the right columns of
figs.\ \ref{res_r1-sol} and \ref{res_r2-sol},
we plot our results for the function $\frac{1}{N^2}f_n^{(0)}(x)$
for $r=1$ and $r=2$, respectively.
Since there is no sign problem in this calculation, we
can obtain results for much larger $N$, which clearly show
the large-$N$ scaling behavior.
Note also that the function $\frac{1}{N^2}f_n^{(0)}(x)$
crosses zero at $x\sim 1$ corresponding to the fact
that $\rho_n^{(0)}(x)$ is peaked at $x\sim 1$.
The asymptotic behaviors of the 
functions $f_n^{(0)}(x)$ in the regions $x \ll 1$ and $x \gg 1$
%the favored region of $x$
are discussed in appendix \ref{sec:pqm}.
%% Using similar arguments as we gave below (\ref{ii-fit}),
%% we can deduce the asymtotic behaviors of the 
%% functions $f_n^{(0)}(x)$ in the regions $x \ll 1$ and $x \gg 1$
%% %the favored region of $x$
%% as described in appendix \ref{sec:pqm}.
In the same figures,
we also plot $-\frac{d}{dx}\Phi_n(x)$ using 
the scaling function $\Phi_n(x)$ obtained 
as described above.
%there are more than two 
%dynamically generated length scales, but the remaining ones are not
%seen in the present calculation.}
Then the solutions to (\ref{master-eq}) can be obtained
as the $x$ coordinates of the intersection points.\footnote{We will discuss
in section \ref{compare-single-multi} that in fact
there should be more intersecting points that are not seen
in these plots.
The statements given below ignore this subtle issue.}
%% Once $\Phi_n(x)$ is obtained as described above,
%% % as described above, 
%% we can use it in the equation (\ref{master-eq}) and solve it.
%% obtain $\vev{\tilde{\lambda}_n}$ by solving (\ref{master-eq}).
%% On the right column of figs.\ 
%% \ref{res_r1-sol} and \ref{res_r2-sol},
%% we plot our results for $\frac{1}{N^2}f_n^{(0)}$
%This is done graphically on the right columns of 
%% In figs.\ \ref{res_r1-sol} and \ref{res_r2-sol}
%% the solutions of (\ref{master-eq}) are obtained
%% graphically from the intersection points of 
%% the two curves $y=\frac{1}{N^2} f^{(0)}_n(x)$ and 
%% $y=-\frac{d}{dx}\Phi_n(x)$. 

For $n=1,4$ there is only one intersection.
% which we denote by $x_{\rm l}$ and $x_{\rm s}$, respectively. 
The effect of the phase is to
suppress the $x \sim 1$ region and to shift the corresponding scale to
larger/smaller scales, respectively. 
For $n=2,3$ there are two intersections, which 
correspond to the local maxima of $\rho_n(x)$.
%% \footnote{In fact 
%% if one uses the true function $\Phi_n(x)$,
%% there should be another intersecting point corresponding to 
%% the local minimum of $\rho_n(x)$ near $x=1$,
%% which is not of our interest.}
%
%, whose locations in $x$ are denoted as $x_{\rm s}$ and $x_{\rm l}$, 
% respectively ($x_{\rm s} < x_{\rm l}$).
%% These two solutions
%% % for our purpose.}
%% of (\ref{master-eq})
%% correspond to the local maxima of $\rho_n(x)$.
%This shows that the 
The effect of
the phase is to produce a double peak structure in $\rho_n(x)$, 
which corresponds to two {\it dynamically} generated length scales for the
corresponding dimensions.
In Table \ref{tab:xsxl} we present the solutions of (\ref{master-eq})
for each $n$, where $x_{\rm s}$ and $x_{\rm l}$ denotes
the solutions in the $x<1$ and $x>1$ regions, respectively.

%\begin{figure}[htbp]

%%%%%%%

%\begin{figure}[htbp]

% %%%%%%%%%%%%%%%%%%%%%%%%%%%%%%%%%%%%%%%%%%%%%%%%%%%%%%%%%%%%%%%%%%%%%
\TABULAR[h]{|c|c|c|c|c|c|c|c|c|}{
\hline%---------------------------------------------------------
    & \multicolumn{6}{|c|}{$r=1$} \\ \hline%---------------------------------------------------------
$n$ & $ x_{\rm s}$  & $ x_{\rm l}$ & $\Phi_n(x_{\rm s})$&$\Phi_n(x_{\rm l})$& 
$  \Xi_{n} $ & $\Delta_n$ \\
\hline%---------------------------------------------------------
% 1  &      & 2.1  &      &  & &        &      & 1.92 &      &   & &      \\
% 2  & 0.50 & 1.32 &      &  & & 0.33   & 0.48 & 1.36 &      &   & & 0.25 \\
% 3  & 0.66 & 1.12 &      &  & & -0.003 & 0.54 & 1.18 &      &   & & -0.02 \\
% 4  & 0.72 &      &      &  & &        & 0.50 &      &      &   & &       \\
%  r=1, i=2, i=3, i=1, i=4
%  r=2, i=3, i=2, i=1
% 3  &  -0.18 & -0.22 & 0.07 & 0.03 &  -0.19 & -0.33 
% & 0.12 & -0.02 \\
 1  &  & 2.14(1)  &      &  -0.231(3) &        &     \\
 2  &  0.49(1) & 1.317(1) & -0.274(3) & -0.218(1) &
0.27(1)  & 0.33(2)   \\
% 3  &  0.62(2)  & 1.129(1) &  -0.20(1) & -0.218(1) &  changed(2011.8.1)
 3  &  0.62(2)  & 1.11(2) &  -0.20(1) & -0.18(2) & 
%0.10(1) & 0.08(2)  \\  changed(2011.8.1)
0.10(2) & 0.11(4)  \\
 4  & 0.71(5) & & -0.20(2) &       &      &      \\
%%  1  &       & $-0.2334 \sim -0.22841$ &        &     \\
%%  2  &  $-0.2767 \sim -0.2707$ & $-0.2183 \sim -0.2170$ & 
%% $0.2558 \sim 0.2792$ & $0.3082 \sim 0.3390$  \\
%%  3  &  $-0.2035 \sim -0.1919$ & $-0.2194 \sim -0.2166$ & 
%% $0.0866 \sim 0.1054$ & $0.0591 \sim 0.0924$  \\
%%  4  &  $-0.1956 \sim -0.1673$ &       &      &      \\
\hline \hline
    & \multicolumn{6}{|c|}{$r=2$} \\ \hline%---------------------------------------------------------
$n$ &  $x_{\rm s}$  & $ x_{\rm l}$ & $\Phi_n(x_{\rm s})$&$\Phi_n(x_{\rm l})$& 
$  \Xi_{n} $ & $\Delta_n$ \\
\hline%---------------------------------------------------------
% 1  &      & 2.1  &      &  & &        &      & 1.92 &      &   & &      \\
% 2  & 0.50 & 1.32 &      &  & & 0.33   & 0.48 & 1.36 &      &   & & 0.25 \\
% 3  & 0.66 & 1.12 &      &  & & -0.003 & 0.54 & 1.18 &      &   & & -0.02 \\
% 4  & 0.72 &      &      &  & &        & 0.50 &      &      &   & &       \\
%  r=1, i=2, i=3, i=1, i=4
%  r=2, i=3, i=2, i=1
% 3  &  -0.18 & -0.22 & 0.07 & 0.03 &  -0.19 & -0.33 
% & 0.12 & -0.02 \\
 1  &  & 2.12(2) &      & -0.213(2) &        &     \\
 2  &  0.4718(1) & 1.39(1) & -0.2528(1) & -0.27(1) & 
 0.253(5)  & 0.23(1)  \\
 3  &  0.47(1) & 1.159(1)  & -0.208(3) & -0.305(2) & 
0.18(1) & 0.08(2)  \\
 4  & 0.44(4) & & -0.20(2) &       &      &      \\
%%  1  &       & $-0.2144 \sim -0.2108$ &        &     \\
%%  2  &  $-0.252808 \sim -0.252722$ & $-0.2794 \sim -0.2644$ & 
%% $0.2478 \sim 0.2565$ & $0.2211 \sim 0.2448$  \\
%%  3  &  $-0.2109 \sim -0.2047$ & $-0.3069 \sim -0.3028$ & 
%% $0.1654 \sim 0.1847$ & $0.0632 \sim 0.0929$  \\
%%  4  &  $-0.2171 \sim -0.1824$ &       &      &      \\
\hline
%---------------------------------------------------------
}
{The first two columns
show the solutions ($x_{\rm s}$ and $x_{\rm l}$)
to eq.~\rf{master-eq} that correspond to the (local)
maxima of $\rho_n(x)$.
%found by the single-observable analysis.
The other columns show the results for 
$\Phi_n(x_{\rm s})$, $\Phi_{n} (x_{\rm l})$,
$\Xi_n$ and $\Delta_n$ that appear 
in eq.\ (\ref{Delta-def}).
%is defined in eq.\ \rf{ab_diff} and 
%% A positive/negative value of $\Delta_n$ $(n=2,3)$ 
%% indicates the dominance of the $x_{\rm l}$/$x_{\rm s}$ peak,
%% respectively.
\label{tab:xsxl}
}
% %%%%%%%%%%%%%%%%%%%%%%%%%%%%%%%%%%%%%%%%%%%%%%%%%%%%%%%%%%%%%%%%%%%%%

%In order to compute $\vev{\tl_n}$ 
For $n=2,3$, 
%the final result for $\vev{\tl_n}$ is given by the peak
%$\rho_n(x)$ actually dominates in the $N\rightarrow \infty$ limit. 
we have to determine which of the two
peaks of $\rho_n(x)$ actually dominates in the $N\rightarrow \infty$ limit. 
For that purpose we calculate
%For that purpose we have to calculate 
%by calculating 
the difference of 
the free energy \cite{0108041}
\begin{eqnarray}
\Delta_{n} &\stackrel{\textrm{def}}{=}&  
   \frac{1}{N^{2}} 
\Bigl\{ \log \rho_{n} (x_{\rm l}) - \log \rho_{n}(x_{\rm s}) \Bigr\} 
%\nonumber \\  &=&  
=
\Bigl\{ \Phi_{n} (x_{\rm l}) - \Phi_n(x_{\rm s}) \Bigr\} +   \Xi_n    \ ,
\label{Delta-def} \\
\mbox{where~~~}\Xi_n &\stackrel{\textrm{def}}{=}&  
\int^{x_{\rm l}}_{x_{\rm s}} 
dx \, \left\{ \frac{1}{N^2} f^{(0)}_n(x)  \right\} \ .
  \label{ab_diff}
\end{eqnarray}
If we find that $\Delta_n$ is positive (negative), 
the peak at $x=x_{\rm l}$ ($x=x_{\rm s}$) 
dominates in the large-$N$ limit.
%which then gives the final result for $\vev{\tl_n}$.
%, respectively.
Note that $\Delta_n$
depends on the values of $\Phi_n(x)$ 
at $x_{\rm s}$ and $x_{\rm l}$, but
%which means that
%Therefore, 
there is no need to know the correct values of $\Phi_n(x)$ 
in 
%$x_{\rm s}$ and $x_{\rm l}$. in between corresponding to the 
the region where we have a severe sign problem.
The integral $\Xi_n$ can be computed by
fitting the data points for largest $N$
%$\frac{1}{N^2} f^{(0)}_n(x)$ 
to some known function with free parameters
in the region $x_{\rm s} \le x \le x_{\rm l}$, and then 
integrating that function over the region. 
In Table \ref{tab:xsxl} we also present
the results for $\Phi_n(x_{\rm s})$, $\Phi_{n} (x_{\rm l})$,
$\Xi_n$ and $\Delta_n$.
For $n=2$ we obtain a clearly positive value for $\Delta_n$,
%which provides evidence 
which means that the peak at $x=x_{\rm l}$ dominates.
%is the dominant one of $\rho_2(x)$ dominates. 
For $n=3$ the obtained value for $\Delta_n$
is too small to reliably determine the dominant peak 
considering the expected 
systematic errors (See section \ref{sec:calc-free}.) 
in
%uncertainties
%in each term of (\ref{Delta-def}).
%the estimates on 
$\Phi_n(x_{\rm s})$ and $\Phi_{n} (x_{\rm l})$.

The results presented above suggest
that $\vev{\tilde{\lambda}_n}$ indeed deviate
from 1 representing the strong effect of the phase,
and that we are likely to get
$\vev{\tilde{\lambda}_1} , \vev{\tilde{\lambda}_2} > 1
> \vev{\tilde{\lambda}_4}$, which is consistent
with the GEM results.
Thus our results support
the speculation that the SSB of SO(4) symmetry occurs 
due to the effect of the phase.
% of the fermion determinant.
Moreover, 
the value of $x_{\rm l}$ for $n=2$ 
and 
the value of $x_{\rm s}$ for $n=3$ 
are reasonably close to 
the GEM results for
$\vev{\tilde{\lambda}_2}$
%$\vev{\tilde{\lambda}_2} = 1.4, 1.7$ for $r=1$ and $r=2$, respectively
%$\vev{\tilde{\lambda}_2}$
and 
$\vev{\tilde{\lambda}_3}$, respectively,
%$\vev{\tilde{\lambda}_3} = 0.7, 0.5$ 
obtained with the SO(2) ansatz.
(See Table \ref{tab:GEMsummary}.)
%However, the comparison of the two peaks in the distribution 
%$\rho_3(x)$ is not clear.
%
%the value of $x_{\rm l}$ for $n=1$ and 
%the value of $x_{\rm s}$ for $n=4$ 
%

On the other hand, our results for $\vev{\tilde{\lambda}_1}$ 
and $\vev{\tilde{\lambda}_4}$ are not very 
close to the corresponding GEM results with the SO(2) ansatz.
%Let us point out that there is However, 
We also notice a puzzle.
In usual Monte Carlo calculations, one obtains various observables
using the same set of configurations.
% that are generated by Monte Carlo simulation.
However, in the present method, one has to perform independent
simulations to obtain $\vev{\tilde{\lambda}_n}$ for different $n$.
In fact the dominant configurations one obtains by simulating 
the constrained system (\ref{ix_pf})
with the parameter $x$ fixed at the GEM results for
%correct value of
$\vev{\tilde{\lambda}_n}$ can be quite different for different $n$.
While this is not a problem \emph{per se} at least theoretically,
it suggests the existence of a practical problem, which 
is closely related to 
%causes
the observed discrepancies.

For instance,
the dominant configurations one obtains from the constrained system
(\ref{ix_pf}) with $n=1$ and $x>1$ have 
$\tilde{\lambda}_1 (=x)
%(=\vev{\tilde{\lambda}_1}) 
> 1 > \tilde{\lambda}_2 ,
%\simeq 
\tilde{\lambda}_3 ,
%\simeq  
\tilde{\lambda}_4 $.
%which cannot be SO(2) symmetric.
Note that $w_1(x)$ is calculated as 
the VEV of $e^{i \Gamma}$ in such an ensemble.
When $x$ is close to $\vev{\tilde{\lambda}_1}$
obtained by the GEM with the SO(2) ansatz,
the dominant contribution to $w_1(x)$ 
is expected to come from 
SO(2) symmetric configurations
%which is possible 
due to the enhancement by $e^{i \Gamma}$.
% when $x$ is close to $\vev{\tilde{\lambda}_1}$.
This clearly suggests that the overlap problem still remains.
As a result, it is expected that
$w_1(x)$ is underestimated in that region of $x$,
which naturally explains the discrepancy 
%observed 
for $\vev{\tilde{\lambda}_1}$ observed above.
Due to this remaining overlap problem, we need to be careful in 
interpreting the results of the single-observable analysis.
We will come back to this point in section \ref{compare-single-multi}.

\section{Factorization method with multiple observables}
\label{generalized_factorization}

%The crucial point of the factorization method is that
%it enables us to explore the region of the configuration space
%that are not important in the phase-quenched model

%In the previous section we treated
%$\tilde{\lambda}_n$ for each $n$ separately.
The important point of the method described in the previous section
was to consider the distribution function
(\ref{rho_full}) instead of trying to calculate the VEV directly
by reweighting (\ref{eq:ovev}).
Then the effect of the phase was factorized and
represented nicely by the weight function
$w_n (x)$ defined by (\ref{w_n}).
This function suppresses the region near $x=1$, which is preferred
by the phase-quenched model, and enhances the regions 
$x \ll 1$ and/or $x\gg 1$. 
By solving the equation (\ref{master-eq}), we can obtain solutions
quite different from $x=1$.
If we simulate the phase-quenched model, 
the observables $\tilde{\lambda}_n$ fluctuate
around $\tilde{\lambda}_n=1$, and it would be difficult to sample
configurations having $\tilde{\lambda}_n$ different from 1,
which represents the overlap problem.
By simulating
%considering the distribution function and 
the constrained system (\ref{ix_pf}), 
we are able to estimate the effect of the phase
in the region strongly suppressed in the phase-quenched model,
and thus the overlap problem can be reduced.
% drastically.

However, 
%let us note here that
we also noticed that
the overlap problem can still remain when one evaluates $w_n (x)$
defined by eq.~(\ref{w_n})
since it is calculated 
as the VEV of $e^{i \Gamma}$ in the constrained system (\ref{ix_pf}).
There might be some region of the configuration space
which cannot be sampled efficiently by simulating
the constrained system (\ref{ix_pf}),
and yet has important contribution
to the VEV due to less fluctuation of $\Gamma$.
%as we discussed at the end of the previous section,
%
%% It could be that the region of configuration space
%% that is suppressed in (\ref{ix_pf}) is strongly enhanced 
%% by the effect of the phase, and one can miss it in
%% the simulation of (\ref{ix_pf}).
%% This will lead to underestimation of $w_n (x)$.
%
% That can explain the discrepancies from the GEM results
% and the puzzle discussed in the previous section.
%
%% can be naturally understood
%% by the overlap problem that remains when one evaluates $w_n (x)$
%% defined in eq.~(\ref{w_n}).
%% For instance, when we evaluate $w_1 (x)$
%% in the region of $x$ corresponding to the real solution 
%% $\vev{\tilde{\lambda}_1}$, it is expected that there is a region
%% in the configuration space in which $\tilde{\lambda}_n$ ($n=2,3,4$)
%% are close to $\vev{\tilde{\lambda}_n}$, respectively,
%% which gives the dominant contribution.
%% Such regions are suppressed in the (\ref{ix_pf}), and the overlap
%% problem can lead to underestimation of $w_1 (x)$.
%
%In what follows we first 
In this section we discuss
how we can solve this problem by generalizing
the factorization method to multiple observables.
Instead of considering the distribution function 
of $\tilde{\lambda}_n$ for each $n$ separately, 
we consider the distribution function 
%of $(\tilde{\lambda}_1 ,\tilde{\lambda}_2 ,
%\tilde{\lambda}_3 ,\tilde{\lambda}_4 )$.
%all of which are specified at the same time.
specifying
all the $\tilde{\lambda}_n$ ($n=1,2,3,4$) at the same time.
We discuss the possibility of including more observables 
in section \ref{more-obs}.

\subsection{the generalized formulation}
\label{sec:generalized-form}

We can actually generalize all the formulae in section 
%\ref{generalized_factorization} 
%\ref{section:mc}
\ref{single_factorization}
%can be
%generalized 
to the multi-observable case
in a straightforward manner.
The relevant functions for the single-observable case
can then be written in terms of those for the multi-observable case.
In this generalized formulation, one can clearly identify 
the overlap problem that still remains in the single-observable 
analysis.

%Since the phase of the determinant is most sensitive
Let us begin by defining the distribution functions for multiple
observables as 
\beqa
\label{rho_full_gen1}
\rho(x_1 , x_2 , x_3 , x_4) &=& 
\Big\langle \prod_{k}\delta (x_k -{\tilde \lambda}_{k}) \Big\rangle \ , \\
\rho^{(0)}(x_1 , x_2 , x_3 , x_4) &=& 
\Big\langle \prod_{k}\delta (x_k -{\tilde \lambda}_{k}) \Big\rangle_0 
\label{rho_full_gen}
\eeqa
for the full model and the phase-quenched model, respectively.
By definition (\ref{eq:lams}), 
these functions vanish unless
$x_1 \ge x_2 \ge x_3 \ge x_4$.
Applying the reweighting (\ref{eq:ovev}) to the right-hand side of
(\ref{rho_full_gen1}),
one finds that it factorizes as
%Following the arguments in section \ref{single_factorization},
%\ref{section:mc},
%the two functions are related to each other by
\beq
\label{eq:rhodef_gen}
\rho(x_1 , x_2 , x_3 , x_4)
 = \frac{1}{C} \, 
\rho^{(0)}(x_1 , x_2 , x_3 , x_4)
 \, w(x_1 , x_2 , x_3 , x_4) \ ,
\eeq
where $C$ is 
%an irrelevant 
a normalization constant given by eq.~(\ref{eq:cconst}).
The correction factor $w(x_1 , x_2 , x_3 , x_4)$ is defined by
\begin{equation}
w(x_1 , x_2 , x_3 , x_4) 
\stackrel{\textrm{def}}{=} 
  \langle e^{i \Gamma} \rangle_{x_1 , x_2 , x_3 , x_4} = 
  \langle \cos \Gamma  \rangle_{x_1 , x_2 , x_3 , x_4} \ , 
\label{w_n_gen}
\end{equation}
where $\langle \  \cdot  \ \rangle_{x_1 , x_2 , x_3 , x_4}$ 
denotes a VEV with respect to
the partition function
\begin{equation}
 Z_{x_1 , x_2 , x_3 , x_4} 
= \int dA \, \ee^{-S_{0}} \, 
\prod_{k} \delta (x_k -{\tilde \lambda}_{k})
%\delta(x- {\tilde \lambda}_{n} ) 
\ .
\label{ix_pf_gen}
\end{equation}
Note that 
all the four observables ${\tilde \lambda}_{n}$
($n=1,2,3,4$) are constrained here
in contrast to (\ref{ix_pf}).

%These are just straightforward generalization of the previous formulae
%to multiple observables.
We can rewrite the relevant functions 
in the single-observable case in terms of the functions defined above.
%% Let us then make connections to the functions 
%% we had in the case of 
%% a single observable.
For instance, we have
\beqa
\label{rho0-gen-full}
 \rho_{n}(x_n) 
&=& \int \prod_{k\neq n} d x_k \, \rho(x_1 , x_2 , x_3 , x_4) \ , \\
 \rho_{n}^{(0)}(x_n)  
&=& \int \prod_{k\neq n} d x_k \, \rho^{(0)}(x_1 , x_2 , x_3 , x_4) \ .
\label{rho0-gen}
\eeqa
Using (\ref{eq:rhodef}), we also obtain
\beq
 w_{n}(x_n) 
= \frac{C \rho_{n}(x_n) }{ \rho_{n}^{(0)}(x_n)}
= \frac{
\int \prod_{k\neq n} d x_k \, \rho^{(0)}(x_1 , x_2 , x_3 , x_4) 
\, w(x_1 , x_2 , x_3 , x_4) }{
\int \prod_{k\neq n} d x_k \, \rho^{(0)}(x_1 , x_2 , x_3 , x_4) 
} \ ,
\label{crucial-formulae}
\eeq 
where we have used (\ref{eq:rhodef_gen}), (\ref{rho0-gen-full}) and
(\ref{rho0-gen}) in the second equality.
%The denominator is nothing but $\rho_{n}^{(0)}(x_n)$
%as one can see from (\ref{rho0-gen}).
This formula reveals
%makes explicit 
the possibility 
%implies a possibility 
that the overlap problem can still
occur when one calculates
the function $w_n (x)$ defined by (\ref{w_n})
as the VEV of $e^{i \Gamma}$ in the system $Z_{n,x}$.
%(\ref{ix_pf}).
%and measured the expectation value of $e^{i \Gamma}$.
%$\cos \Gamma$.
The region that gives dominant contribution
in the numerator of (\ref{crucial-formulae})
may not be sampled efficiently by
simulating the system $Z_{n,x}$.
It is also clear that the overlap problem can be reduced further
by constraining all the four observables 
$\tilde{\lambda}_n$ ($n=1,2,3,4$) at the same time.
Now we have to maximize the
% generalized
distribution function $\rho(x_1 , x_2 , x_3 , x_4)$
with respect to $x_1$, $x_2$, $x_3$ and $x_4$.
This leads to the coupled equations
\beq
\frac{\del }{\del x_n} 
\log \rho^{(0)}(x_1 , x_2 , x_3 , x_4)  = 
- \frac{\del }{\del x_n}  \log w(x_1 , x_2 , x_3 , x_4) 
  \quad \mbox{for $n=1,2,3,4$}  \ .
\label{general-master-eq}
\eeq
The function on the left-hand side and 
$w(x_1 , x_2 , x_3 , x_4)$ defined by (\ref{w_n_gen})
can be obtained similarly 
to the single-observable case described in 
section \ref{prac-implement}.
%by simulating the system (\ref{ix_pf_gen}).
In fact it is a formidable task to search for solutions
in the 4d parameter space $(x_1 , x_2 , x_3 , x_4)$
in full generality.
%solve these equations as we did
%in the single variable case in the previous Section.
%However, in the present 
However, given the insights we obtained from the analysis
in the previous section,
we can restrict ourselves to the region of
$(x_1 , x_2 , x_3 , x_4)$ in which we expect to find
a solution.
In particular, it is expected that there exist solutions which satisfy
%% \begin{enumerate}
%% \item $x_1 =  x_2 = x_3  > 1 >  x_4$
%% \item $x_1 =  x_2 > 1 > x_3  > x_4$
%% \item $x_1 >  1 > x_2 > x_3  > x_4$ \\
%% \end{enumerate}
\beqa
\mbox{(a)} &~& x_1 =  x_2 = x_3  > 1 >  x_4  \ , 
\nonumber \\
% \label{case_a_gen} \\
\mbox{(b)} &~& x_1 =  x_2 > 1 > x_3  > x_4  \  ,
\nonumber \\
%\label{case_b_gen} \\
\mbox{(c)} &~& x_1 >  1 > x_2 > x_3  > x_4   \ .
\label{case_gen}
%\label{case_c_gen}
\eeqa 
The solutions one obtains for the 
cases (a) and (b) correspond to the SO(3) and SO(2) symmetric vacua,
respectively, and hence one can compare the results against
those obtained by the GEM 
with the SO(3) ansatz and the SO(2) ansatz.\footnote{The case (c) 
was not studied  by GEM in ref.\ \cite{0412194}
since it is technically more difficult than the
other cases due to the existence of more free parameters in the
Gaussian action. However, our result $\Delta_2>0$
discussed in section \ref{lambda-results-single} strongly suggests
that the solution for the case (c) has larger free energy
than the solution for the case (b).\label{footnote:gem-c}
}
That would still require solving coupled equations
for 2 variables in the case (a), and 
for 3 variables in the case (b).
In what follows,
we restrict ourselves to the $r=1$ case for simplicity,
and 
%In this work 
%we have therefore decided to 
%suffice ourselves by 
%we 
investigate whether the GEM results for $\vev{\tilde{\lambda}_n}$
with the SO(3) ansatz and the SO(2) ansatz
are indeed solutions to (\ref{general-master-eq}) 
in the cases (a) and (b), respectively.
We also discuss the calculation of free energy in 
section \ref{sec:calc-free}.

%% Using $\rho_{i}^{(0)}(x)$ and $w_{i}(x)$, the VEV
%% $\langle {\tilde \lambda}_{i} \rangle$ can be written as
%% \begin{equation}
%% \langle {\tilde \lambda}_{i} \rangle = 
%%     \int^{\infty}_{0} dx \, x \rho_{i}(x) = 
%%   \frac 
%%    {\int^{\infty}_{0} dx \, x \rho_{i}^{(0)} (x) w_{i} (x)} 
%%    {\int^{\infty}_{0} dx \,   \rho_{i}^{(0)} (x) w_{i} (x)} \ . 
%% \label{li_direct}
%% \end{equation}
%% It should be stressed that the above formula is not numerically 
%% the same as a direct evaluation based on eq.\ \rf{eq:ovev}. 
%% Using this method we are able to sample configurations 
%% where $|\rho_{i}^{(0)}(x) w_{i}(x)|$ is significant. 
%% These configurations are quite improbable in the ensemble 
%% defined by $Z_0$, but
%% we can force the system to sample the important region efficiently
%% by using $Z_{i,V}$. 

%The computation of $w_{i}(x)$ is still hard due to the complex
%action problem. 
%However, 

\subsection{analysis for the SO(3) symmetric vacuum}
\label{sec:res-SO3}

%% The result of GEM with the SO(3) ansatz is roughly
%% \begin{equation}
%% \label{eq:toylam-large-so3}
%%  \vev{\lambda_n} \sim
%% \left\{ 
%% \begin{array}{ll} 
%% 1  + \frac{3}{4} r  & \mbox{~for~}n=1,2,3 \ , \\ 
%% 1  - \frac{1}{4} r & \mbox{~for~}n=4 \ .
%% \end{array} 
%% \right.
%% \end{equation} 

%We study the case (a), which corresponds to the SO(3) ansatz of the GEM.

Let us consider the case (a) in (\ref{case_gen}).
We define the reduced functions
\beqa
\rho_{\rm SO(3)}^{(0)} (x, y) & 
%\equiv 
\stackrel{\textrm{def}}{=} 
&  \rho^{(0)} (x,x,x,y) \ , \\
w_{\rm SO(3)} (x, y) & 
%\equiv 
\stackrel{\textrm{def}}{=} 
&  w (x,x,x,y) \ . 
\eeqa
From the coupled equations (\ref{general-master-eq}),
we obtain the counterpart of (\ref{master-eq}) as
\beq
 \frac{1}{N^2} f^{(0)}_{\textrm{SO(3)},\zeta} (x,y) 
= - \frac{\partial}{\partial \zeta} \Phi_{\textrm{SO(3)}} (x,y) \ ,
\label{master_eq_so3}
\eeq
%% \beqa
%%  \frac{1}{N^2} f^{(0)}_{\textrm{SO(3)},x} (x,y) 
%% &=& - \frac{\partial}{\partial x} \Phi_{\textrm{SO(3)}} (x,y) 
%% \label{master_eq_so3} \\
%%  \frac{1}{N^2} f^{(0)}_{\textrm{SO(3)},y} (1.17,y) 
%% &=& - \frac{\partial}{\partial y} \Phi_{\textrm{SO(3)}} (1.17,y) \ , 
%% \label{master_eq_so3_2}
%% \eeqa
where $\zeta=x,y$,
%represents $x$ or $y$, 
and we have defined
\beqa
f_{{\rm SO(3)},\zeta} ^{(0)} (x, y)  
%&\equiv&
&\stackrel{\textrm{def}}{=} &
\frac{\del}{\del \zeta} \log \rho_{\rm SO(3)} ^{(0)} (x, y)  \ , \\
\Phi_{\textrm{SO(3)}}(x , y)
%&\equiv& 
&\stackrel{\textrm{def}}{=} &
\lim _{N \rightarrow \infty}
\frac{1}{N^2} \log w_{\textrm{SO(3)}} (x,y) \ .
\label{def-Phi-so3}
\eeqa
%% \beqa
%% f_{{\rm SO(3)},x} ^{(0)} (x, y)  
%% &\equiv&
%% \frac{\del}{\del x} \log \rho_{\rm SO(3)} ^{(0)} (x, y) \\
%% f_{{\rm SO(3)},y} ^{(0)} (x, y)  
%% &\equiv&
%% \frac{\del}{\del y} \log \rho_{\rm SO(3)} ^{(0)} (x, y) \\
%% \Phi_{\textrm{SO(3)}}(x , y)
%% &\equiv& \lim _{N \rightarrow \infty}
%% \frac{1}{N^2} \log w_{\textrm{SO(3)}} (x,y) \ .
%% \eeqa
%
%% % reduces to
%% \beqa
%% \label{del-x-rho-3d}
%% \frac{\del}{\del x} \log \rho_{\rm SO(3)} ^{(0)} (x, y)  &=& 
%% - \frac{\del}{\del x} \log w_{\rm SO(3)} (x, y)  \ , \\
%% \frac{\del}{\del y} \log \rho_{\rm SO(3)} ^{(0)} (x, y)  &=& 
%% - \frac{\del}{\del y} \log w_{\rm SO(3)} (x, y)  \ .
%% \label{del-x-rho-3d2}
%% \eeqa
The GEM results obtained for $r=1$ with 
the SO(3) ansatz are
$\vev{\tilde{\lambda}_n}_{\rm SO(3)}=1.17$ ($n=1,2,3$)
and $\vev{\tilde{\lambda}_4}_{\rm SO(3)}=0.5$
(See Table \ref{tab:GEMsummary}.)
We would like to check whether $(x,y)=(1.17,0.5)$ indeed
solves eq.~(\ref{master_eq_so3}).

First let us 
consider
%test whether the GEM results solves 
the $\zeta=x$ component of eq.~(\ref{master_eq_so3}).
%eq.~(\ref{del-x-rho-3d}).
Let us set $y$ to 0.5
%the value $\vev{\tilde{\lambda}_4}_{\rm SO(3)}=0.5$
and solve the equation
%the $\zeta =x$ component of eq.~(\ref{master_eq_so3}) 
%with $\zeta=x$ 
for $x$
to see if the solution agrees with the value 1.17.
%$\vev{\tilde{\lambda}_n}_{\rm SO(3)}=1.17$ ($n=1,2,3$) obtained by GEM.
We can calculate
$w_{\rm SO(3)}(x,0.5)$
and $f_{{\rm SO(3)},x} ^{(0)} (x, 0.5)$
using the method described in section \ref{prac-implement}.
%% We constrain $\lambda_3$ and $\lambda_4$
%% by the Gaussian potential\footnote{We decided not to constrain 
%% the larger eigenvalues $\lambda_1$ and $\lambda_2$, since they turn out to 
%% be close to $\lambda_3$ automatically in the parameter range that
%% has been explored. In fact we can use a larger coefficient $\gamma$ 
%% for constraining $\lambda_4$, since we do not need to calculate 
%% the gradient $f_{{\rm SO(3)},y} ^{(0)} (x, y)$ in the $y$ direction.
%% Similar comments apply to the other cases, too.
%% } 
%% as we did in (\ref{iV_partition}).
%
%% What we should do in the simulation is to add a Gaussian potential
%% for $\tilde{\lambda}_3$ and constrain it 
%% to the value obtained by GEM (1.17 for $r=1$).
%% Then $\tilde{\lambda}_1$ and $\tilde{\lambda}_2$ will be 
%% almost the same value as $\tilde{\lambda}_3$ automatically.
%% (I think we should not impose constraints on
%% all of $\tilde{\lambda}_1$, $\tilde{\lambda}_2$
%% and $\tilde{\lambda}_3$ since we will have much smaller acceptance
%% rate by doing so.)
In fig.~\ref{result_so3} (Left)
we plot the function $\frac{1}{N^2} \log w_{\textrm{SO(3)}} (x,0.5)$ 
for $N=8,12,16$.
%against $x$.
We make a large-$N$ extrapolation using the asymptotic behavior
at large $x$
to obtain $\Phi_{\textrm{SO(3)}}(x , 0.5)$
represented by the solid lines.
(See appendix \ref{SO3-lambda3} for the details.)
Figure \ref{result_so3} (Right) shows that the solution 
%to the $\zeta =x$ component of eq.~(\ref{master_eq_so3}) 
lies at $x=1.151(2)$,
%between $x=1.150359561004216$ and $x=1.154228782247431$, 
which is consistent with
%agrees well with 
the GEM value $1.17$.

  \FIGURE{
\epsfig{file=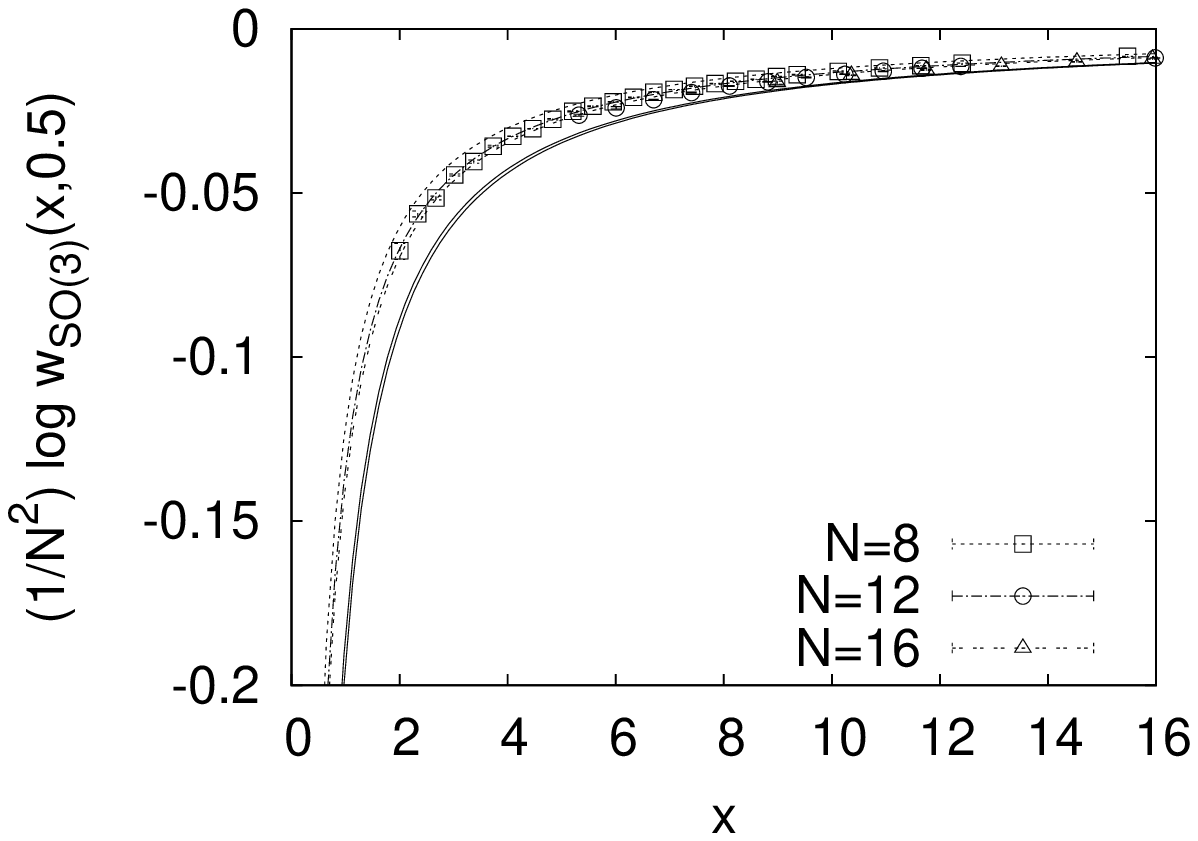,width=7.4cm}
\epsfig{file=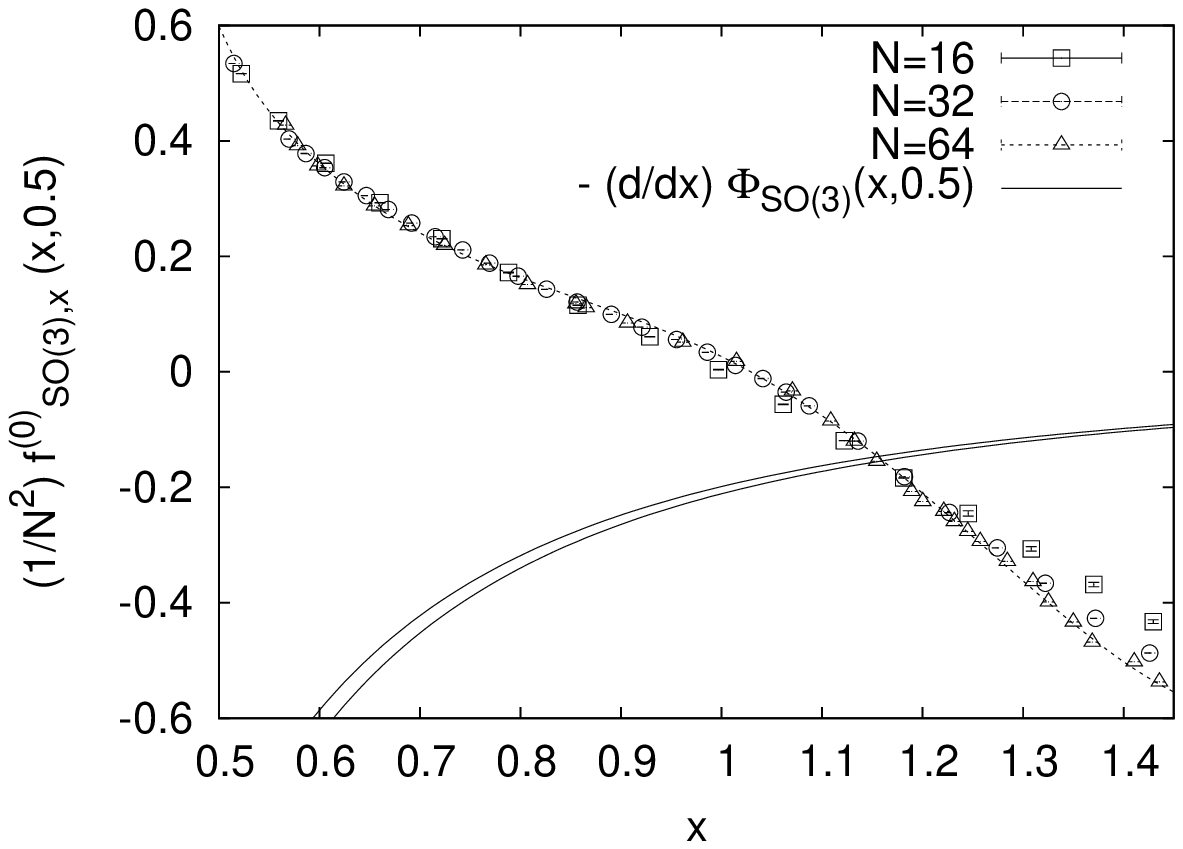,width=7.4cm}
   \caption{(Left) The function
$\frac{1}{N^2} \log w_{\textrm{SO(3)}} (x,0.5)$ 
is plotted 
against $x$
for $N=8,12,16$.
The solid lines represent
%, with margin of error, 
the function
$\Phi_{\textrm{SO(3)}} (x,0.5)$
obtained by extrapolation to
$N=\infty$ as described in appendix \ref{w_3d_x3}. 
(Right) The function
$\frac{1}{N^2} f^{(0)}_{\textrm{SO(3)},x} (x,0.5)$ 
is plotted 
against $x$ 
for $N=16,32,64$.
The solid lines represent
$- \frac{\partial}{\partial x} \Phi_{\textrm{SO(3)}} (x,0.5)$
obtained from the plot on the left.}
   \label{result_so3}}

Next let us consider 
%test that the GEM result solves 
the $\zeta=y$ component of eq.~(\ref{master_eq_so3}).
%eq.~(\ref{del-x-rho-3d2}).
%Note that this is an independent test.
%(The success of the previous test does not guarantee that of this one.)
%For that we 
We set $x$ to 1.17
%the value obtained by the GEM,
and solve the equation
%eq.~(\ref{del-x-rho-3d2}) 
for $y$.
In order to obtain $\Phi_{\textrm{SO(3)}} (1.17,y)$,
we have to make a large-$N$ extrapolation.
For that purpose we first tried to 
use the asymptotic behavior of
%$\log w_{\textrm{SO(3)}} (1.17,y)$ 
$\frac{1}{N^2}\log w_{\textrm{SO(3)}} (1.17,y)$ 
at small $y$.
However, it turned out that the finite-$N$ effects
in $\frac{1}{N^2}\log w_{\textrm{SO(3)}} (1.17,y)$ at small $y$
are much severer than in
$\frac{1}{N^2}\log w_{\textrm{SO(3)}} (x,0.5)$ at large $x$ 
investigated above.\footnote{The same trend can
also be seen in the single-observable analysis 
relevant to the SO(3) ansatz.
In fig.~\ref{res_r1-sol} (Left),
the finite-$N$ effects in $\log w_4(x)$ at $x<1$
are much severer than in $\log w_3(x)$ at $x>1$.}
We therefore use an ``orthogonal'' method here.

Note that we were able to obtain the function
$\Phi_{\textrm{SO(3)}} (x, y)$ at $y=0.5$ 
from fig.~\ref{result_so3} (Left).
We can do the same thing for different $y$ such as
$y=0.45,0.55$. (See appendix \ref{SO3-lambda3} for the details.)
In fig.~\ref{result_so3_2} (Left)
we plot $\Phi_{\textrm{SO(3)}} (1.17,y)$ for $y=0.45, 0.50, 0.55$.
By fitting the three points to a straight line, 
we obtain the derivative
$\frac{\partial}{\partial y} \Phi_{\textrm{SO(3)}} (1.17,y) 
= - 0.33(2)$ at $y=0.5$. 
%0.329064 \pm 0.02356$
Figure \ref{result_so3_2} (Right) shows that the solution 
%to the $\zeta=y$ component of eq.~(\ref{master_eq_so3})
lies 
at $y=0.59(2)$, 
which is reasonably close to the GEM value 0.5
given the uncertainties involved in the analysis.

%between $x=0.5692469049450339$ and $x=0.6065774876931346$.

  \FIGURE{
\epsfig{file=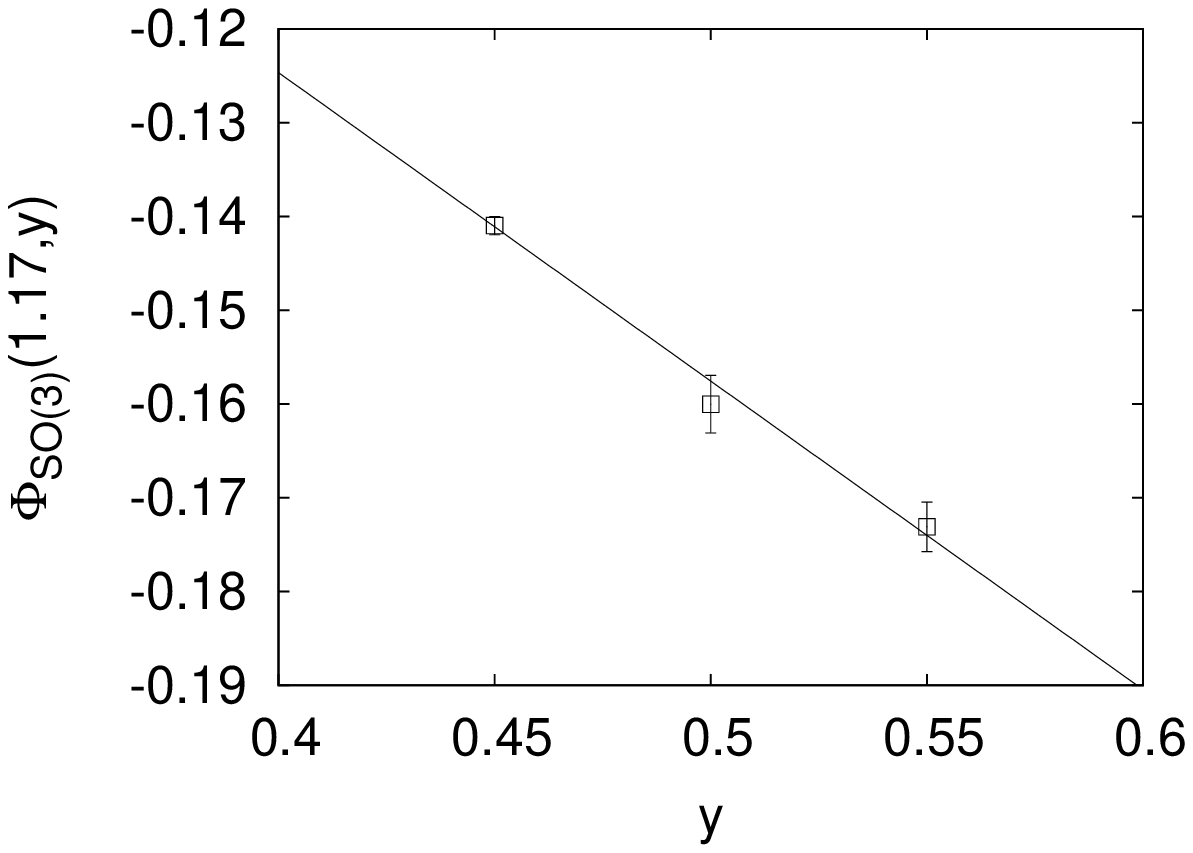,width=7.4cm}
\epsfig{file=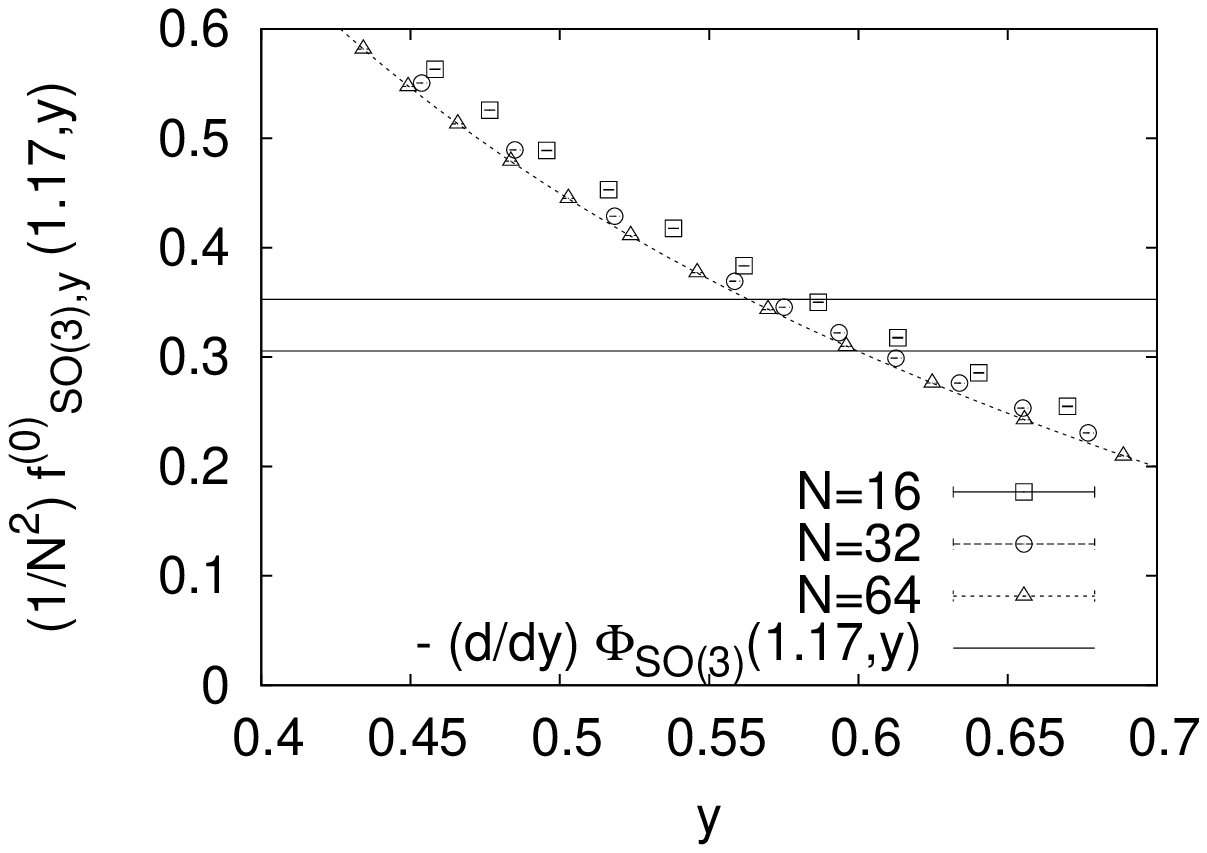,width=7.4cm}
   \caption{(Left) The values of 
the function $\Phi_{\textrm{SO(3)}} (1.17,y)$ 
at $y=0.45,0.5,0.55$
are plotted against $y$.
(Right) The function $\frac{1}{N^2} f^{(0)}_{\textrm{SO(3)},y} (1.17,y)$ 
is plotted 
against $y$ 
for $N=16,32,64$. The solid lines
represent
% the derivative
$- \frac{\partial}{\partial y} \Phi_{\textrm{SO(3)}} (1.17,y)$
at $y=0.5$ obtained from 
the plot on the left.
%the plot on the left.
}
\label{result_so3_2}}

%% We may compare the results of the present analysis with 
%% those of the single-observable analysis
%% with $n=3$ at $x>1$ and with $n=4$ at $x<1$.
%% There we obtained result for $\vev{\tilde{\lambda}_3}$ 
%% (1.129(1) for $r=1$) 
%% % and 1.159(1) for $r=2$
%% is already close 
%% to the GEM result (1.17 for $r=1$),
%% %and 1.23 for $r=2$),
%% and indeed we find above that constraining $\tilde{\lambda}_4$
%% to the GEM value does not alter the result much.
%% On the other hand, the obtained result for $\vev{\tilde{\lambda}_4}$
%% (0.71(5) for $r=1$)
%% % and 0.44(4) for $r=2$) 
%% still deviates
%% from the GEM result (0.5 for $r=1$).
%% % and 0.31 for $r=2$).
%% By constraining $\tilde{\lambda}_3$ to the GEM values,
%% we do find better consistency.

\subsection{analysis for the SO(2) symmetric vacuum}
\label{sec:res-SO2}

Let us consider the case (b) in (\ref{case_gen}).
We define the reduced functions
%Similarly to the SO(3) ansatz in the previous section, we define 
\begin{eqnarray}
 \rho^{(0)}_{\textrm{SO(2)}} (x,y,z) 
%&\equiv&
&\stackrel{\textrm{def}}{=} &
\rho^{(0)} (x,x,y,z) \ ,  \\
%\hspace{3mm}
 w_{\textrm{SO(2)}} (x,y,z) 
%&\equiv&
&\stackrel{\textrm{def}}{=} &
w(x,x,y,z) \ .
\label{so2_ansatz_obs}
\end{eqnarray}
From the coupled equations (\ref{general-master-eq}),
we obtain the counterpart of (\ref{master-eq}) as
\beq
 \frac{1}{N^2} f^{(0)}_{\textrm{SO(2)},\zeta} (x,y,z) 
= - \frac{\partial}{\partial \zeta} \Phi_{\textrm{SO(2)}} (x,y,z) \ ,
\label{master_eq_so2}
\eeq
%% \beqa
%%  \frac{1}{N^2} f^{(0)}_{\textrm{SO(3)},x} (x,y) 
%% &=& - \frac{\partial}{\partial x} \Phi_{\textrm{SO(3)}} (x,y) 
%% \label{master_eq_so3} \\
%%  \frac{1}{N^2} f^{(0)}_{\textrm{SO(3)},y} (1.17,y) 
%% &=& - \frac{\partial}{\partial y} \Phi_{\textrm{SO(3)}} (1.17,y) \ , 
%% \label{master_eq_so3_2}
%% \eeqa
where $\zeta=x,y,z$,
%represents $x$ or $y$, 
and we have defined
\beqa
f_{{\rm SO(2)},\zeta} ^{(0)} (x, y,z)  
%&\equiv&
&\stackrel{\textrm{def}}{=} &
\frac{\del}{\del \zeta} \log \rho_{\rm SO(2)} ^{(0)} (x, y, z)  \ , \\
\Phi_{\textrm{SO(2)}}(x , y, z)
%&\equiv& 
&\stackrel{\textrm{def}}{=} &
\lim _{N \rightarrow \infty}
\frac{1}{N^2} \log w_{\textrm{SO(2)}} (x,y,z) \ .
\label{def-Phi-so2}
\eeqa
The GEM results obtained for $r=1$ with 
the SO(2) ansatz are
$\vev{\tilde{\lambda}_n}_{\rm SO(2)}=1.4$ ($n=1,2$),
$\vev{\tilde{\lambda}_3}_{\rm SO(2)}=0.7$
and $\vev{\tilde{\lambda}_4}_{\rm SO(2)}=0.5$
(See Table \ref{tab:GEMsummary}.)
We would like to check whether $(x,y,z)=(1.4,0.7,0.5)$ indeed
solves eq.~(\ref{master_eq_so2}).

First let us 
consider
%test whether the GEM results solves 
the $\zeta=x$ component of eq.~(\ref{master_eq_so2}).
We set $y$ and $z$ to $0.7$ and $0.5$, respectively,
and solve the equation for $x$.
In fig.~\ref{result_so2_2} (Left)
we plot the function $\frac{1}{N^2} \log w_{\textrm{SO(2)}} (x,0.7,0.5)$ 
for $N=8,12,16$.
We make a large-$N$ extrapolation using the asymptotic behavior
at large $x$
to obtain $\Phi_{\textrm{SO(2)}}(x , 0.7, 0.5)$
represented by the solid line.
(See appendix \ref{SO2-1st} for the details.)
Figure \ref{result_so2_2} (Right) shows that the solution 
%to the $\zeta =x$ component of eq.~(\ref{master_eq_so2}) 
lies at $x=1.373(2)$, which is consistent with the GEM value 1.4.
%between $x=1.371437295790456$ and $x=1.374373932820009$.

  \FIGURE{
\epsfig{file=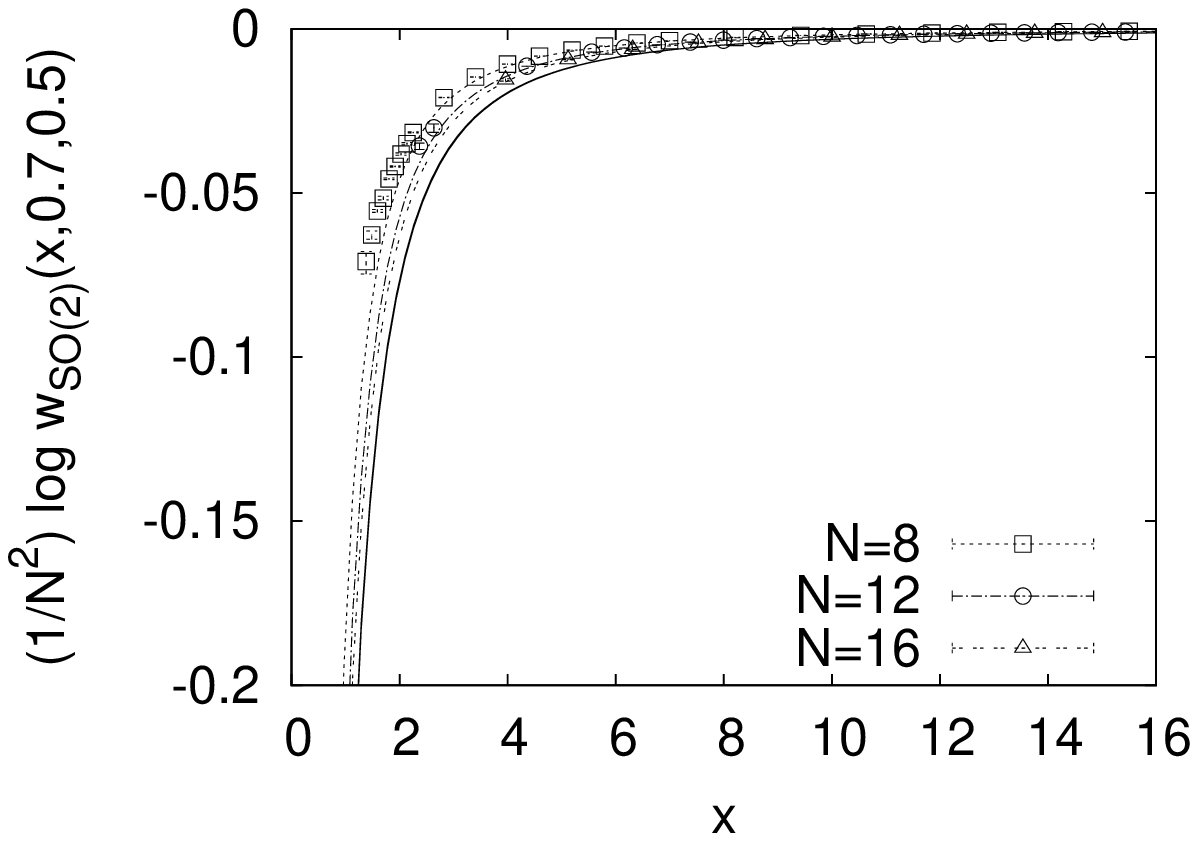,width=7.4cm}
\epsfig{file=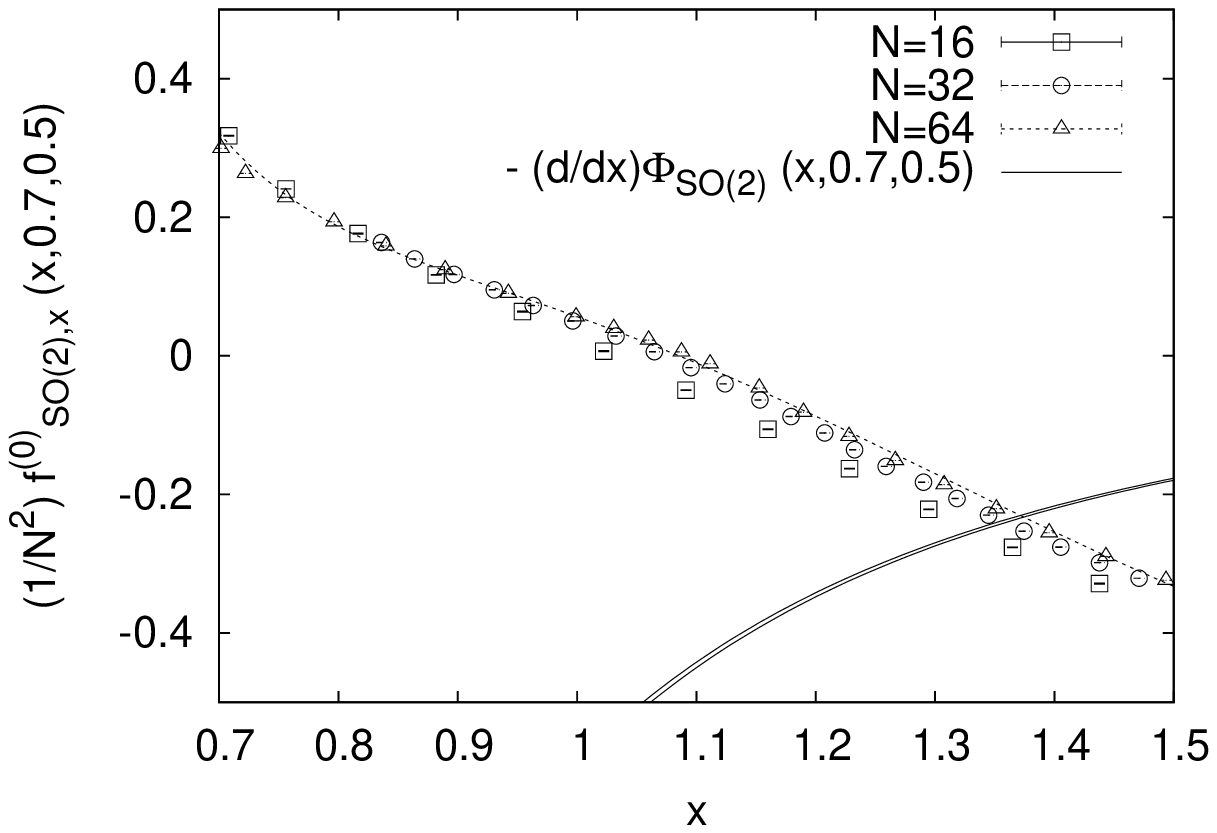,width=7.4cm}
   \caption{(Left) The function 
$\frac{1}{N^2} \log w_{\textrm{SO(2)}} (x,0.7,0.5)$ 
is plotted against $x$ for $N=8,12,16$. 
The solid line represents the function 
$\Phi_{\textrm{SO(2)}} (x,0.7,0.5)$
obtained by extrapolation to $N=\infty$ 
%using the asymptotic behavior (\ref{w_2d_x2})
as described in appendix \ref{SO2-1st}.
(Right) The function $\frac{1}{N^2} f^{(0)}_{\textrm{SO(2)},x} (x,0.7,0.5)$ 
is plotted against $x$ for $N=16,32,64$.
The solid lines represent
% the function
$- \frac{\partial}{\partial x} \Phi_{\textrm{SO(2)}} (x,0.7,0.5)$
obtained from the plot on the left. }
   \label{result_so2_2}}
   
  \FIGURE{
\epsfig{file=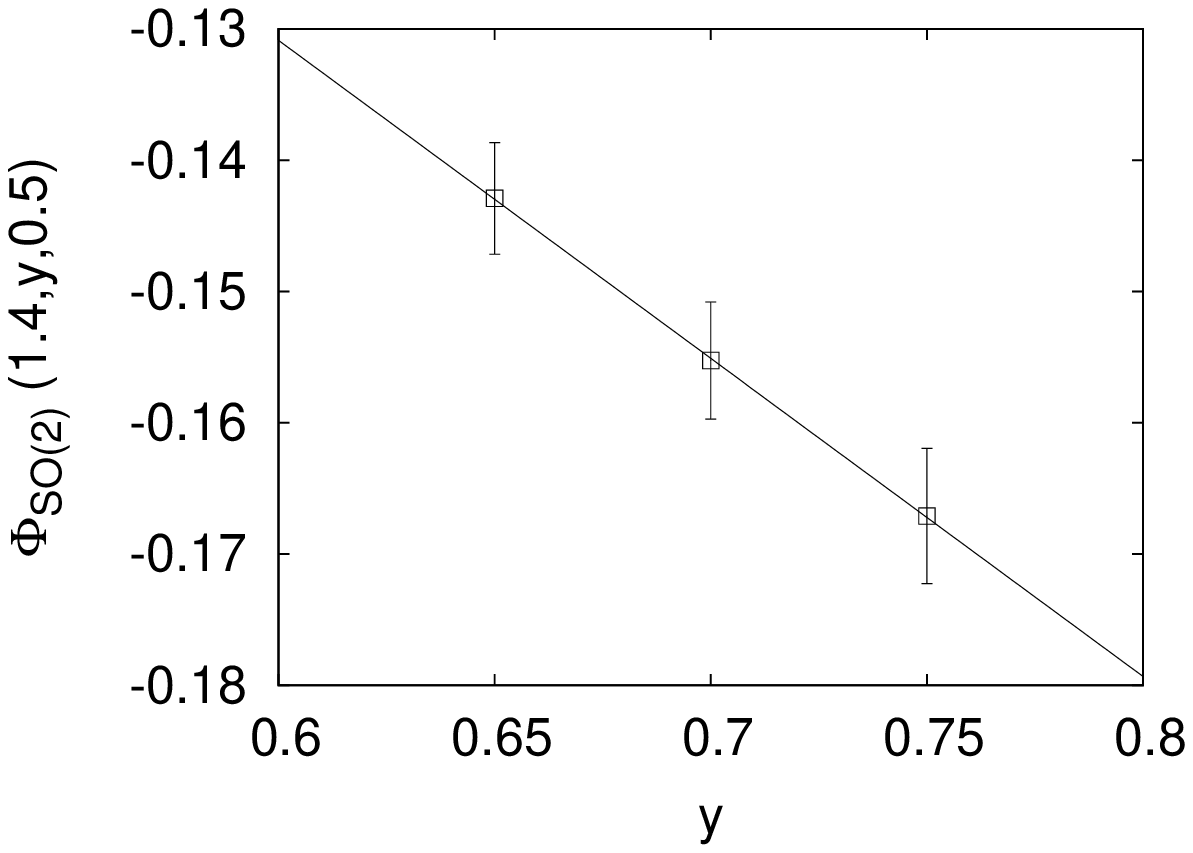,width=7.4cm}
\epsfig{file=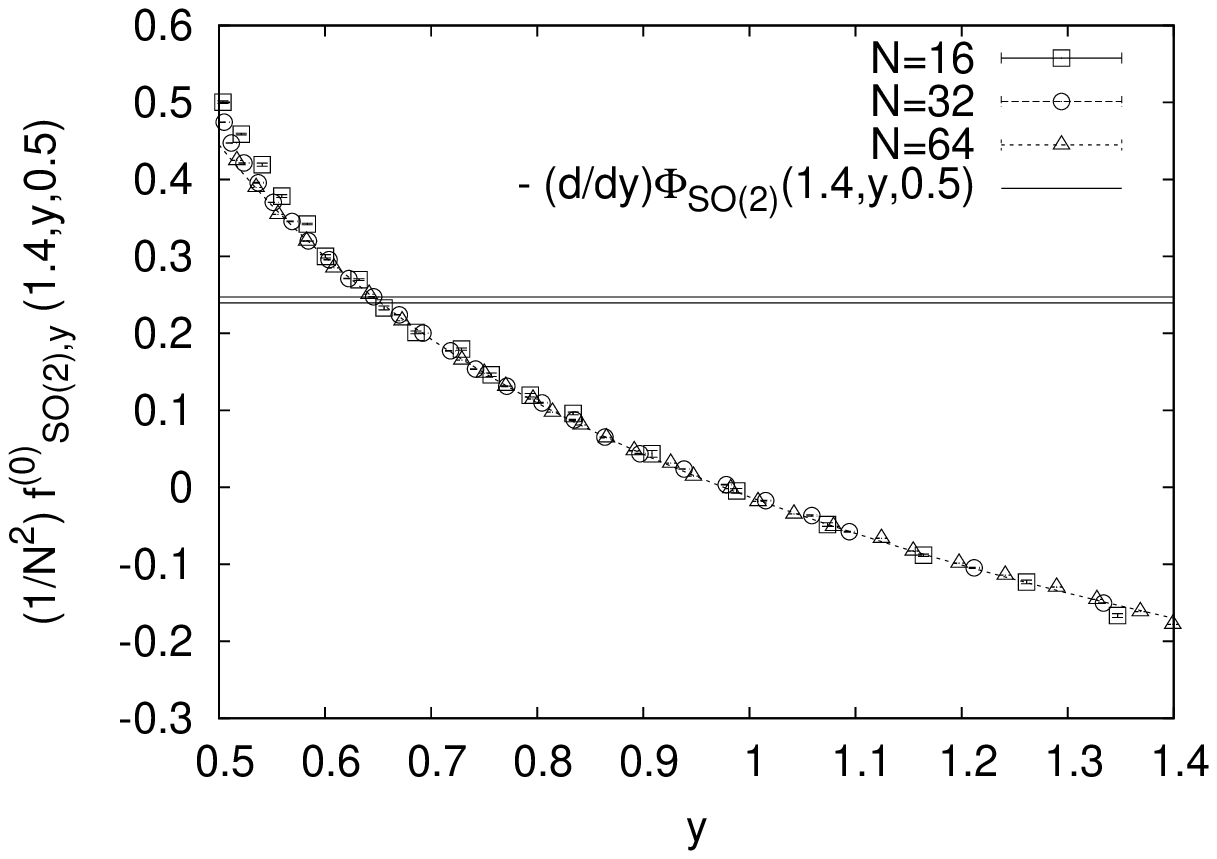,width=7.4cm}
   \caption{(Left) The values of the function
$\Phi_{\textrm{SO(2)}} (1.4,y,0.5)$ at $y=0.65,0.70,0.75$
are plotted against $y$.
%in the large-$N$ limit. 
(Right) The function
$\frac{1}{N^2} f^{(0)}_{\textrm{SO(2)},y} (1.4,y,0.5)$ 
is plotted against $y$
for $N=16,32,64$.
The solid lines represent
% the derivative
$- \frac{\partial}{\partial y} \Phi_{\textrm{SO(2)}} (1.4,y,0.5)$ 
at $y=0.7$ obtained from the plot on the left.}
   \label{result_so2_3}}

%\paragraph{Calculation of $\langle {\tilde \lambda}_{3} \rangle$} 
%\hspace{0mm} \\

Next 
let us consider 
%test that the GEM result solves 
the $\zeta=y$ component of eq.~(\ref{master_eq_so2}).
We set $x$ and $z$ to $1.4$ and $0.5$, respectively,
and solve the equation for $y$.
In order to obtain $\Phi_{\textrm{SO(2)}} (1.4,y,0.5)$,
we have to make a large-$N$ extrapolation.
The region of $y$ of the function 
$\frac{1}{N^2} \log w_{\textrm{SO(2)}} (1.4,y,0.5)$ 
is restricted to $0.5 \le y \le 1.4$, and
we do not have any asymptotic behavior with respect to $y$
that can be used for large-$N$ extrapolation.
Therefore we 
%do the same thing as
%repeat what we did in section
use the ``orthogonal'' method which was used in section
\ref{sec:res-SO3} for
studying the $\zeta=y$ component of eq.~(\ref{master_eq_so3}).
Since we were able to obtain
%$\frac{1}{N^2} \log w_{\textrm{SO(2)}} (x,y,z)$ for
$\Phi_{\textrm{SO(2)}} (x,y,z)$ for
$y=0.7$ and $z=0.5$, we do the same thing
for different $y$ such as $y=0.65,0.75$.
In fig.~\ref{result_so2_3} (Left)
we plot 
%$\frac{1}{N^2} \log w_{\textrm{SO(2)}} (1.4,y,0.5)$ 
$\Phi_{\textrm{SO(2)}} (1.4,y,0.5)$ 
for $y=0.65, 0.70, 0.75$.
By fitting the three points to a straight line, 
we obtain the derivative
%0.243347 \pm 0.003793$ 
$ \frac{\partial}{\partial y} \Phi_{\textrm{SO(2)}} (1.4,y,0.5) 
= - 0.243(4)$ at $y=0.7$. 
%0.329064 \pm 0.02356$
Figure \ref{result_so2_3} (Right) shows that 
the solution 
%of eq.~(\ref{master_eq_so2_3}) 
lies at $y=0.649(4)$,
which is consistent with the GEM value 0.7
given the uncertainties involved in the analysis.

  \FIGURE[t]{
\epsfig{file=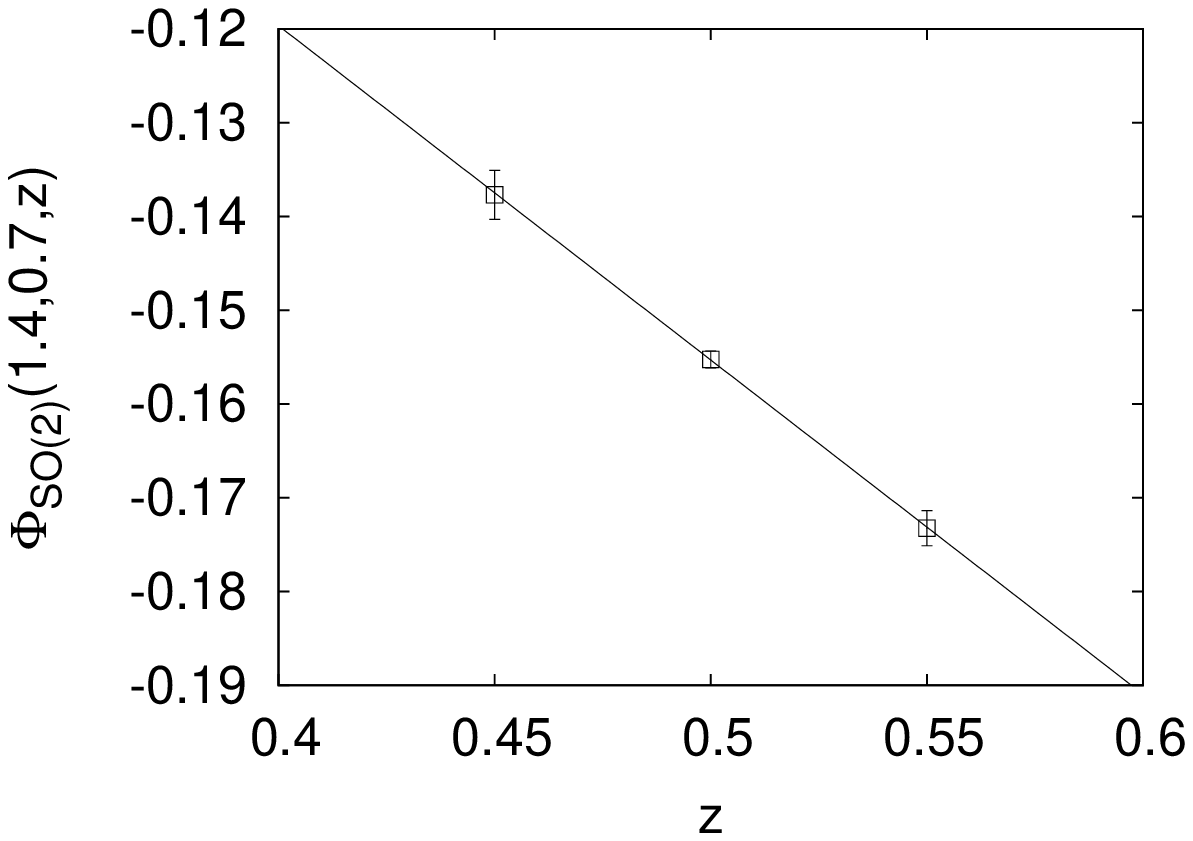,width=7.4cm}
\epsfig{file=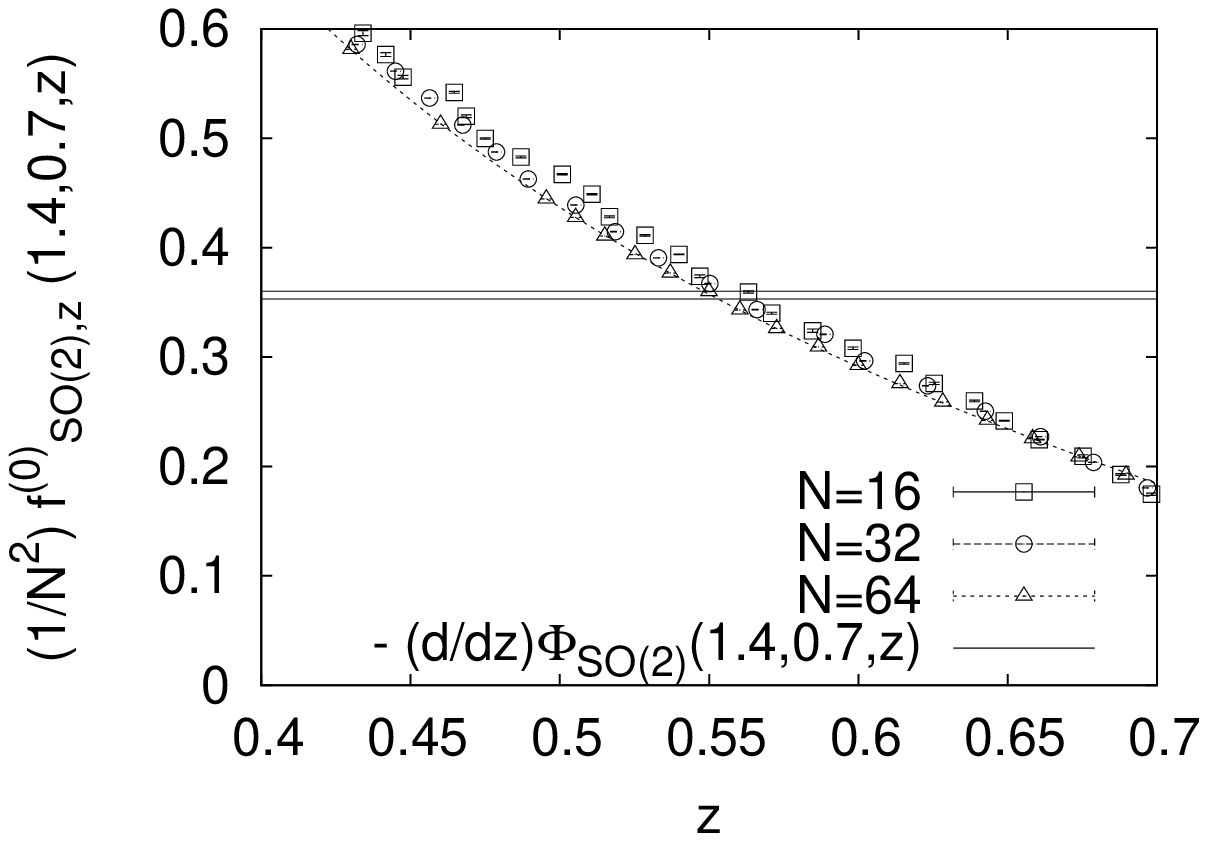,width=7.4cm}
   \caption{(Left) The values of the function
$\Phi_{\textrm{SO(2)}} (1.4,0.7,z)$ 
%in the large-$N$ limit 
at $z=0.45,0.5,0.55$
are plotted against $z$. 
(Right) The function 
$\frac{1}{N^2} f^{(0)}_{\textrm{SO(2)},z} (1.4,0.7,z)$ 
is plotted against $z$ for $N=16,32,64$.
The solid lines represent
% the derivative
$- \frac{\partial}{\partial z} \Phi_{\textrm{SO(2)}} (1.4,0.7,z)$
at $z=0.5$ obtained from the plot on the left.}
   \label{result_so2_4}}

%\paragraph{Calculation of $\langle {\tilde \lambda}_{4} \rangle$} 
%\hspace{0mm} \\

Finally let us consider
the $\zeta=z$ component of eq.~(\ref{master_eq_so2}).
We set $x$ and $y$ to $1.4$ and $0.7$, respectively,
and solve the equation for $z$.
In order to obtain $\Phi_{\textrm{SO(2)}} (1.4,0.7,z)$,
we 
%do the same thing as
use the ``orthogonal'' method.\footnote{We first tried to 
use the asymptotic behavior of
%$\log w_{\textrm{SO(3)}} (1.17,y)$ 
$\frac{1}{N^2}\log w_{\textrm{SO(2)}} (1.4,0.7,z)$ 
at small $z$.
However, it turned out that the finite-$N$ effects
in 
$\frac{1}{N^2}\log w_{\textrm{SO(2)}} (1.4,0.7,z)$ 
at small $z$
are much severer than in
$\frac{1}{N^2}\log w_{\textrm{SO(2)}} (x,0.7,0.5)$ 
at large $x$ investigated above.}
%% in section \ref{sec:res-SO3} used for
%% testing the $\zeta=y$ component of eq.~(\ref{master_eq_so3}).
Since we were able to obtain
%$\frac{1}{N^2} \log w_{\textrm{SO(2)}} (x,y,z)$ for
$\Phi_{\textrm{SO(2)}} (x,y,z)$ for
$y=0.7$ and $z=0.5$, we do the same thing
for different $z$ such as $z=0.45,0.55$.
In fig.~\ref{result_so2_4} (Left)
%we plot $\frac{1}{N^2} \log w_{\textrm{SO(2)}} (1.4,0.7,z)$ 
we plot $\Phi_{\textrm{SO(2)}} (1.4,0.7,z)$ 
at $z=0.45, 0.50, 0.55$.
By fitting the three points to a straight line, 
we obtain the derivative
%0.243347 \pm 0.003793$ 
$\frac{\partial}{\partial z} \Phi_{\textrm{SO(2)}} (1.4,0.7,z) 
= - 0.357(4)$ at $z=0.5$. 
%0.356575 \pm 0.003576$ 
Figure \ref{result_so2_4} (Right) shows that 
the solution 
%to the $\zeta=z$ component of eq.~(\ref{master_eq_so2})
%the solution of eq.~(\ref{master_eq_so2_4}) 
lies at $z=0.551(2)$,
%$x=0.54806667640638$ and $x=0.5530283387531921$ 
which is close to the GEM value 0.5.
%given the uncertainties involved in the analysis.

\subsection{estimates on the systematic error}
\label{sec:sys-error}

One thing we notice from the results in
sections \ref{sec:res-SO3} and \ref{sec:res-SO2}
(See Table \ref{tab:improvement} for a summary.)
is that the agreement with the GEM results is
better for the large eigenvalues $\vev{\tilde{\lambda}_n}>1$
than for the small eigenvalues $\vev{\tilde{\lambda}_n}<1$.
%Finally we would like to discuss the reason 
%In addition to the uncertainties due to 
Apart from the use of 
the ``orthogonal'' method in the latter case,
we consider that the systematic error may be
a possible reason.

Let us first consider the SO(3) symmetric vacuum.
The VEV $\vev{\tilde{\lambda}_n}$ is determined by solving
(\ref{master_eq_so3}).
The left-hand side is determined accurately since there is no
sign problem, and they behave as
%The question we would like to address is
\beqa
\frac{1}{N^2} f_{{\rm SO}(3),x}^{(0)}(x,0.5) 
%&\sim & -1.4(x-1.15)-0.12 \ .
&\sim & - a(x-1.17)-b \ , \quad a=1.43(2) \ , b=0.174(1)  \ ,
\label{fSO3-large} \\
\frac{1}{N^2} f_{{\rm SO}(3),y}^{(0)}(1.17,y) 
%&\sim & -1.4(x-1.15)-0.12 \ .
&\sim & - \tilde{a}(y-0.5)+\tilde{b} \ , 
%\quad \tilde{a}=1.7 \ , \tilde{b}=0.45  \ .
\quad \tilde{a}=1.70(4) \ , \tilde{b}=0.452(1)  \ ,
%= (-1.69722 +/- 0.04417) *(y-0.5) + (0.452404 +/- 0.00118)
\label{fSO3-small}
\eeqa
near the solution 
as one can obtain from the right panel of 
figs.~\ref{result_so3} and \ref{result_so3_2}.
%The main source of the error comes from 
%the determination of the right-hand side of (\ref{master_eq_so3}).
Let us denote the right-hand side of the $\zeta$ component
of (\ref{master_eq_so3}) by $D_{{\rm SO}(3),\zeta}$, 
and assume that it has an error $\Delta D_{{\rm SO}(3),\zeta}$, where
$\zeta=x,y$. This error includes the systematic error due to 
deviation from the asymptotic behavior as well as the one
due to the large-$N$ extrapolation. 
Then the error in $D_{{\rm SO}(3),x}$ 
propagates to the error of the solution $x$ as
$a \, \Delta x  =  \Delta D_{{\rm SO}(3),x}$, from which we obtain
\beq
\frac{\Delta x}{x} = \frac{|D_{{\rm SO}(3),x}|}{a x} 
\frac{\Delta D_{{\rm SO}(3),x}}{|D_{{\rm SO}(3),x}|} 
%\sim \frac{0.16}{1.4 \cdot 1.17} \frac{\Delta D_x}{D_x} 
\sim 0.1  \, 
\frac{\Delta D_{{\rm SO}(3),x}}{|D_{{\rm SO}(3),x}|} \ .
%\frac{\Delta D_x}{D_x}  \ .
\label{x-dx}
\eeq
%Similarly we denote the right-hand side of the $\zeta=y$ component
%of (\ref{master_eq_so3}) by $D_y$, 
%and assume that it has an error $\Delta D_y$.
On the other hand,  
the error in $D_{{\rm SO}(3),y}$ 
propagates to the error of the solution $y$ as
$\tilde{a} \, \Delta y  =  \Delta D_{{\rm SO}(3),y}$, 
from which we obtain
\beq
\frac{\Delta y}{y} = \frac{|D_{{\rm SO}(3),y}|}{\tilde{a} y} 
\frac{\Delta D_{{\rm SO}(3),y}}{|D_{{\rm SO}(3),y}|} 
%\sim \frac{0.32}{1.7 \cdot 0.5} \frac{\Delta D_y}{D_y} 
\sim 0.4 \, \frac{\Delta D_{{\rm SO}(3),y}}{|D_{{\rm SO}(3),y}|}  \ .
\label{y-dy}
\eeq
Thus we find that the coefficient is four times bigger for 
$y$ than for $x$.
A similar analysis can be made for the SO(2) symmetric vacuum,
and we obtain
\beqa
\frac{\Delta x}{x} 
&\sim& \frac{0.24}{0.82 \cdot 1.4} 
\frac{\Delta D_{{\rm SO}(2),x}}{|D_{{\rm SO}(2),x}|} 
\sim 0.21 \, \frac{\Delta D_{{\rm SO}(2),x}}{|D_{{\rm SO}(2),x}|}  \ , 
%\quad
\\
\frac{\Delta y}{y}
&\sim& \frac{0.25}{0.9 \cdot 0.7} 
\frac{\Delta D_{{\rm SO}(2),y}}{|D_{{\rm SO}(2),y}|} 
\sim 0.40 \, \frac{\Delta D_{{\rm SO}(2),y}}{|D_{{\rm SO}(2),y}|}  \ , 
%\quad
\\
\frac{\Delta z}{z}
&\sim& \frac{0.35}{1.6 \cdot 0.5} 
\frac{\Delta D_{{\rm SO}(2),z}}{|D_{{\rm SO}(2),z}|}  
\sim 0.43 \, \frac{\Delta D_{{\rm SO}(2),z}}{|D_{{\rm SO}(2),z}|}  \ ,
\eeqa
where $D_{{\rm SO}(2),\zeta}$ and 
$\Delta D_{{\rm SO}(2),\zeta}$ are defined analogously
to the SO(3) case. 
Thus we find that the relative errors for the
large eigenvalues tend to be smaller than those
for the small eigenvalues.

%% The essential point is that $D_y \sim 2 D_x$,
%% namely the solution
%% for the $\zeta$ component of (\ref{master_eq_so3}) 
%% lies in the region where each side of the equation becomes large,
%% and the second point is that $y \sim 0.4 x$, namely
%% the solution for $y$ is smaller than the solution for $x$,
%% and hence the relative error tends to become bigger.
%% The discrepancy for $\vev{\tilde{\lambda}_3}$ and 
%% $\vev{\tilde{\lambda}_4}$ is 2\% and 15\%, respectively,
%% which is not so surprising considering the above argument.

\subsection{comparison with the single-observable analysis}
\label{compare-single-multi}

In this section we compare the results obtained above by
the multi-observable analysis with those 
obtained by the single-observable analysis 
in section \ref{lambda-results-single}.
For that, we first need to reconsider the interpretation 
of the latter taking into account the artifacts
due to the remaining overlap problem.

%% From the viewpoint of the generalized formulation
%% in \ref{sec:generalized-form},
%% there are three important regions 
%% in the configuration space corresponding to 
%% each of the three cases (\ref{case_gen}).
Let us recall that the generalized distribution function
$\rho(x_1,x_2,x_3,x_4)$
is expected to have three local maxima
corresponding to each of the three cases (\ref{case_gen}).
%although we have studied only the first two cases
Therefore, the distribution function
$\rho_n(x)$ in the single-observable analysis
is also expected to have three local maxima
due to the relation (\ref{rho0-gen-full}).
The fact that we observe only two for $n=2,3$ and one for 
$n=1,4$ as we described in section \ref{lambda-results-single}
is due to the remaining overlap problem.

We reconsider the plots in the right column 
of fig.~\ref{res_r1-sol} from this point of view.
For instance, the intersection of the two curves
%% the curves $y=-\Phi_1 ' (x)$ 
%% and $y=\frac{1}{N^2} f^{(0)}_1(x)$
shown in the uppermost panel
can be naturally identified 
as the one corresponding to the case (c).
As $x$ decreases from the intersecting point,
the true curve $y=-\frac{d}{dx}\Phi_1  (x)$ is expected to 
have four more intersections with the curve
$y=\frac{1}{N^2} f^{(0)}_1(x)$ in the $x>1$ region.
Two of them represent the local maxima
of the distribution function $\rho_1(x)$
corresponding to the cases (a) and (b) in (\ref{case_gen}).
Such a structure is invisible in the results 
of the single-observable analysis due to the overlap problem.
%% The intersection that is seen 
%% in section \ref{lambda-results-single}
%% %figs.~\ref{res_r1-sol} and \ref{res_r2-sol}
%which are $x=1.17$ and $x=1.4$, respectively, according
%to the GEM results.
From this point of view,
it is misleading to compare the solution $x_{\rm l}$
for the $n=1$ case with the 
$\vev{\tilde{\lambda}_1}$ for the SO(2) symmetric vacuum.
Rather we should consider it as an estimate
for $\vev{\tilde{\lambda}_1}$ in the vacuum corresponding to 
the case (c) in (\ref{case_gen}).
This provides a prediction for the calculation using GEM,
which is not done yet. (See footnote \ref{footnote:gem-c}.)

Similar arguments apply to the $n=2,3,4$ cases.
The solutions $x_{\rm s}$ and $x_{\rm l}$ for the $n=2$ case 
should be interpreted as estimates
on $\vev{\tilde{\lambda}_2}$ 
in the vacuum corresponding to 
the case (c) in (\ref{case_gen})
and in the SO(2) symmetric vacuum, respectively.
The solutions $x_{\rm s}$ and $x_{\rm l}$ for the $n=3$ case 
should be interpreted as estimates
on $\vev{\tilde{\lambda}_3}$ for the 
SO(2) and SO(3) symmetric vacua, respectively.
The solution $x_{\rm s}$ for the $n=4$ case 
should be interpreted as an estimate
on $\vev{\tilde{\lambda}_4}$ for the SO(3) symmetric vacuum.

In Table \ref{tab:improvement} we summarize our results
for $\vev{\tilde{\lambda}_n}$ corresponding 
to the SO(2) and SO(3) symmetric vacua.
We also show the results of the single-observable analysis
with the new interpretation
except for an estimate on
$\vev{\tilde{\lambda}_4}$ for the SO(2) symmetric vacuum,
which is not available.
We do find that the results 
of the multi-observable analysis have better agreement 
with the GEM values. 
%The remaining discrepancies are due to the systematic errors

%the extrapolations based on the asymptotic behaviors.
%That said, we consider that the factorization method 

% %%%%%%%%%%%%%%%%%%%%%%%%%%%%%%%%%%%%%%%%%%%%%%%%%%%%%%%%%%%%%%%%%%%%%
\TABULAR[h]{|c|c|c|c|c|c|c|}{
\hline%---------------------------------------------------------
ansatz    & \multicolumn{3}{|c|}{SO(3)} &\multicolumn{3}{|c|}{SO(2)}\\
\hline%---------------------------------------------------------
method &  single-obs.  &  multi-obs.  & GEM &
%      $\vev{\tl_n}_{\mbox{\scriptsize{GEM}}}$  & 
      single-obs.  &  multi-obs. & GEM
% &  $\vev{\tl_n}_{\mbox{\scriptsize{GEM}}}$  
\\ 
\hline%---------------------------------------------------------
\vev{\tilde{\lambda}_1}   &  ---   & ---  & 1.17
%1.4&
&  ---  & ---  & 1.4
%& 1.7
\\
\vev{\tilde{\lambda}_2}    &  ---   & --- & 1.17
%1.4&
&  1.317(1)  & 1.373(2) & 1.4
%& 1.7
\\
% 3  & 0.67 & 1.13 & 0.7& 0.53 & 1.17 & 0.5\\
\vev{\tilde{\lambda}_3}      &  1.129(1)   & 1.151(2) & 1.17
%1.4&
&  0.62(2)  & 0.649(4) & 0.7
%& 0.5
\\
\vev{\tilde{\lambda}_4}   & 0.71(5) &  0.59(2)   & 0.5
%& 0.5
& not available &   0.551(2)    & 0.5
%& 0.1
%
%% \\
%% \hline%---------------------------------------------------------
%% free energy   & -1.5 &  -1.8  & 1.7
%% %& 0.5
%% & -1.9   &   -3.6  & 1.7
%% %& 0.1
\\
\hline%---------------------------------------------------------
}
%% % %%%%%%%%%%%%%%%%%%%%%%%%%%%%%%%%%%%%%%%%%%%%%%%%%%%%%%%%%%%%%%%%%%%%%
{The results for the normalized eigenvalues 
$\vev{\tilde{\lambda}_n}$ for $r=1$
% defined by (\ref{def_tilde})
obtained by the factorization method
with the single-observable and
%analysis and the 
multi-observable analyses
for the SO(3) and SO(2) symmetric vacua.
The dash implies that the result should be the same as the 
one below in the same column due to the imposed symmetry.
We also show the GEM results obtained at $N=\infty$
in ref.~\cite{0412194}.
\label{tab:improvement}
}  
% %%%%%%%%%%%%%%%%%%%%%%%%%%%%%%%%%%%%%%%%%%%%%%%%%%%%%%%%%%%%%%%%%%%%%

\subsection{calculation of the free energy}
\label{sec:calc-free}

%% One can also consider the case (c) in principle.
%% However, one has to have constraints on all 4 $\lambda$s.
%% Moreover, there is no corresponding analysis by GEM
%% due to a technical reason.
%% In fact the situation of the simulation will be similar
%% to the previous analyses
%% with $n=2$ at $x<1$ and with $n=1$ at $x>1$.
%% They are expected to give approximate values for 
%% $\vev{\lambda_2}$ and $\vev{\lambda_1}$ in the case (c),
%% respectively.
%% Constraining
%% $\lambda_1$, $\lambda_2$ and $\lambda_3$ to these values,
%% one can do an independent simulation to obtain approximate value
%% of $\vev{\lambda_4}$.
%% One can take these results as the first approximation, and 
%% try to improve them by iterating the procedure.

%In the previous sections, we have checked that the GEM results
In sections \ref{sec:res-SO3} and \ref{sec:res-SO2}, 
we have checked that the GEM results
with the SO(3) ansatz and the SO(2) ansatz
can indeed be obtained as solutions to (\ref{general-master-eq}).
In order to determine which solution dominates the path integral,
we need to compare the free energy.
This was done in the GEM and the result is shown 
in Table \ref{tab:GEMsummary},
from which we find that the SO(2) symmetric solution
dominates in the large-$N$ limit.
Here we try to obtain the difference of free energy 
for the two solutions by Monte Carlo simulation.
%Let us remind the readers that calculation of the free energy
%is generally a difficult problem even in a system without the sign problem.
The factorization method has the nice feature
that the difference of free energy is decomposed into 
the term coming from the phase-quenched model and the term
due to the effect of the phase.
The former can be calculated accurately, and the latter is the 
main source of the error.
%It is important that the latter can be calculated
%without 
%We will see that the uncertainties due to the extrapolations 
%have much larger effects on the determination of the free energy
%than the VEV $\vev{\tilde{\lambda}_n}$.

What we should calculate is
%This can be done by calculating 
the difference of
$\log \rho(x_1 , x_2 , x_3 , x_4)$ between the two solutions.
Let us denote the solution for cases (a) and (b) as
$\vec{x}_a = (X',X',X',Y')$ and $\vec{x}_b= (X,X,Y,Z)$, respectively.
From Table \ref{tab:GEMsummary} we find for $r=1$ that
\beqa
(a) &~& X' \simeq 1.17 \ , Y' \simeq 0.5 \ ,  \\
(b) &~& X \simeq 1.4 \ , Y \simeq 0.7 \ , Z \simeq 0.5 \ .
\label{XpYpZp}
\eeqa
%where $Y=Z'$ by accident. 
Using the factorization property (\ref{eq:rhodef_gen}), we obtain
\beqa
\Delta 
%&\equiv& 
&\stackrel{\textrm{def}}{=} &
\frac{1}{N^2} 
\Big\{ \log\rho(\vec{x}_a) - \log \rho(\vec{x}_b)  \Big\} 
\nonumber
\\
%% &=& \int_{\vec{x}_b}^{\vec{x}_a} d x_j 
%% \frac{1}{N^2} \frac{\del}{\del x_j} 
%% \log \rho(x_1 , x_2 , x_3 , x_4) \\
&=&
\Bigl\{ 
\Phi_{\rm SO(3)}(X',Y') - 
\Phi_{\rm SO(2)}(X,Y,Z)
%\frac{1}{N^2} \log w(\vec{x}_a) -  
%\frac{1}{N^2} \log w(\vec{x}_b) 
\Bigr\} 
+ \Xi \ ,
%% \int_{\vec{x}_b}^{\vec{x}_a} d x_j 
%% \frac{1}{N^2} \frac{\del}{\del x_j} 
%% \log \rho^{(0)}(x_1 , x_2 , x_3 , x_4)  \ .
\label{log-rho-diff} \\
\mbox{where~~~~~}\Xi  &\stackrel{\textrm{def}}{=}&
\int_{\vec{x}_b}^{\vec{x}_a} d x_j 
\frac{1}{N^2} \frac{\del}{\del x_j} 
\log \rho^{(0)}(x_1 , x_2 , x_3 , x_4)  \ .
\eeqa

The first term in (\ref{log-rho-diff}) can be estimated
% for $r=1$ 
as (See appendix \ref{largeN-multi}.)
%sections \ref{sec:res-SO3} and \ref{sec:res-SO2}.)
%from
%the function $\Phi(x_1 , x_2 , x_3 , x_4)$ at each solution.
%using the asymptotic behaviors (if necessary). 
\beqa
%\lim_{N\rightarrow \infty} 
%\frac{1}{N^2} \log w(\vec{x}_a) &=& 
\Phi_{\rm SO(3)}(1.17,0.5)
&=& -0.160(3) \ , 
\label{Phi-SO3}
\\
%\lim_{N\rightarrow \infty}  
%\frac{1}{N^2} \log w(\vec{x}_b) &=& 
\Phi_{\rm SO(2)}(1.4,0.7,0.5)
&=& - 0.155(1) \ .
\label{Phi-SO2}
\eeqa
%$$ \log w(a) = -0.1600409996 \pm 0.003077853185 $$
%$$ \log w(b) = - 0.1552658248 \pm 0.0009080238534$$
In order to calculate the second term $\Xi$ in (\ref{log-rho-diff}),
we first define a path in the  
$(x_1 , x_2 , x_3 , x_4)$-space connecting the two 
solutions $\vec{x}_a$ and $\vec{x}_b$, obtain 
the gradient of $\log \rho^{(0)}(x_1 , x_2 , x_3 , x_4)$
and integrate it along the path.
%the gradient 
As a path connecting the two solutions, we consider
\beqa
C_1&:& (X,X,Y,Z)  \mapsto (X,X,X,Z) \ , \nonumber \\
%\vec{x}_b \equiv 
C_2&:& (X,X,X,Z) \mapsto (X,X,X,Y')  \ , \nonumber \\
C_3&:& 
%\vec{x}_a\equiv 
(X,X,X,Y') \mapsto  (X',X',X',Y') \ . \nonumber
\eeqa
%% \beqa
%% C_1&:& 
%% %\vec{x}_a\equiv 
%% (X,X,X,Y) \mapsto (X',X',X',Y)  \ ,\\
%% C_2&:& (X',X',X',Y) \mapsto (X',X',X',Z')  \ , \\
%% C_3&:& (X',X',X',Z') \mapsto 
%% %\vec{x}_b \equiv 
%% (X',X',Y',Z')  \ .
%% \eeqa
The paths $C_1$ and $C_3$ are investigated already
in sections \ref{sec:res-SO2} and \ref{sec:res-SO3}, respectively,
while the path $C_2$ should be investigated newly.
%for determination of the free energy difference.
In the $r=1$ case, we have $Y \simeq Z'$ accidentally,
and therefore the study of the path $C_2$ can be totally omitted.
The second term in (\ref{log-rho-diff})
% integral $\int_{(a)}^{(b)} d x_j \frac{\del}{\del x_j} 
% \log \rho^{(0)}(x_1 , x_2 , x_3 , x_4) $ 
can then be evaluated as
\beqa
\Xi 
%&=&  \int_{\vec{x}_b}^{\vec{x}_a} d x_j \frac{\del}{\del x_j} 
%\log \rho^{(0)}(x_1 , x_2 , x_3 , x_4) \nonumber \\
&=& 
\int^{1.4}_{0.7} \frac{1}{N^2} f^{(0)}_{\textrm{SO(2)},y}(1.4,y,0.5) \, dy 
- \int^{1.4}_{1.17} \frac{1}{N^2} f^{(0)}_{\textrm{SO(3)},x}(x,0.5) \, dx  
\nonumber
\\
%&=& (-0.014 ) - (-0.083)  =  0.069 \ , 
&=& (-0.014 ) - (-0.079)  =  0.065 \ , 
\label{free_energy_f0}
\eeqa
%-0.083 --> - 0.078547558361082 (=-0.079)
%\Xi=0.069 --> 0.065
where the integrals are calculated by
fitting the data points for largest $N$
%$\frac{1}{N^2}f^{(0)}_{\textrm{SO(3)},x}(x,0.5)$ 
%and $\frac{1}{N^2}f^{(0)}_{\textrm{SO(2)},y}(1.4,y,0.5)$ 
in figs.~\ref{result_so3} and \ref{result_so2_3} (Right)
to some known functions with free parameters in the integration domain.
%% The functions 
%% $\frac{1}{N^2}f^{(0)}_{\textrm{SO(3)},x}(x,0.5)$ 
%% and $\frac{1}{N^2}f^{(0)}_{\textrm{SO(2)},y}(1.4,y,0.5)$ 
%% are plotted in figs.~\ref{result_so3} and \ref{result_so2_3}.
%% These integrals are calculated as
%% \beqa
%% \int^{1.4}_{0.7} \frac{1}{N^2} f^{(0)}_{\textrm{SO(2)},y}(1.4,y,0.5) \, dy
%% &=& -0.014 \ , \\
%% %= -0.01441393172326144, $$
%% \int^{1.4}_{1.17} \frac{1}{N^2} f^{(0)}_{\textrm{SO(3)},x}(x,0.5) \, dx 
%% &=& -0.083  \ ,
%% %= -0.08308256338057236, \hspace{3mm}  
%% \eeqa
%% and hence $\Xi = 0.069$,
The errors are negligible compared to 
those in the first term of (\ref{log-rho-diff}).
%of $\log w(\vec{x}_a)$ and $\log w(\vec{x}_b)$. 
Thus we obtain $\Delta = 0.060(4)$,
%% \begin{eqnarray}
%% % \log \rho(\vec{x}_b) - \log \rho(\vec{x}_a) 
%% \Delta = 0.064(4) \ . 
%% %= - 0.06389345686 \pm 0.003985877038
%% \label{Delta-result}
%% \end{eqnarray}
which should be compared with
the value $\Delta \simeq -0.3$ predicted by GEM
%the difference of free energy between the 2d solution and the 3d solution
%$as -0.5 for $r=1$ 
as one can see from Table \ref{tab:GEMsummary}.

%%   \FIGURE[t]{
%% \epsfig{file=f-naught-i24_3_r1.eps,width=7.4cm}
%%    \caption{$f^{(0)}_{\textrm{2d,3}}(x) 
%% = \frac{\partial}{\partial x} 
%% \log \rho^{(0)}_{\textrm{2d}} (y=1.4,x,z=0.5)$ for $N=16,32,64$.}
%%    \label{so2_f3}}

%% This may be rather peculiar to the present model, 
%% where the SO(2) vacuum and the SO(3) vacuum become
%% almost degenerate for small $r$ as we mentioned
%% in section \ref{section:model}.

According to the interpretation of the single-observable analysis
given in section \ref{compare-single-multi},
what we have done above may be regarded as 
a refined version of the calculation of $\Delta_3$
%(\ref{Delta-def})
in section \ref{lambda-results-single},
%if we interpret the single-observable analysis
%as explained in section (\ref{compare-single-multi}.
%in view of the interpretation of the single-observable analysis
%given in section (\ref{compare-single-multi}.
%what we did
%
%for the determinantion of $\Delta_3$
%defined by (\ref{Delta-def}).
where we obtained $\Delta_3= 0.11(4)$ for $r=1$.
%which is consistent with the present result.
%(\ref{Delta-result}).

%% First let us remark that 
%% %we note that while 
%% %While we consider that 
%% this value $-0.3$ should not be taken too serious.
%% %while we think that the sign is correct.
%% %$\Delta < 0$ is correct.
%% From figs.~1 and 2 in ref.~\cite{0412194},
%% we find that conservative estimates on the free energy at $r=1$ would be
%% would be $f_{{\rm SO}(3)}= -1.3 \sim - 1.6$ and 
%% $f_{{\rm SO}(2)}= -1.6 \sim - 1.8$,
%% so we consider that $\Delta = -0.3(2)$.

%% Next, note that the calculation
Let us recall that the calculation
of (\ref{free_energy_f0}) does not have much ambiguity.
Note also that $\Xi > 0$ implies that
the SO(3) symmetric configuration $(X' ,X' ,X' ,Y' )$
has lower free energy in the \emph{phase-quenched model}
than the SO(2) symmetric configuration $(X ,X ,Y  ,Z )$.
This is reasonable considering that
the former configuration 
is closer to the dominant SO(4) symmetric configuration
$\tilde{\lambda}_n = 1$ in the phase-quenched model.
% than the latter.

%% For the same reason, it is expected that
%% $\Phi_{\rm SO(3)}(X',Y') < \Phi_{\rm SO(2)}(X,Y,Z)$.
%% The values given in
%% (\ref{Phi-SO3}) and (\ref{Phi-SO2})
%% are consistent with this expectation.
On the other hand, the estimates (\ref{Phi-SO3}) and (\ref{Phi-SO2})
may be subject to systematic errors
due to the deviation 
of the functions
$\Phi_{\rm SO(3)}(x,0.5)$ and $\Phi_{\rm SO(2)}(x,0.7,0.5)$
from their asymptotic behaviors
and due also to large-$N$ extrapolations.
%In section \ref{compare-single-multi}
Let us recall that 
the error propagation in solving the $\zeta = x $ component
of eq.~(\ref{master_eq_so3}) is given by (\ref{x-dx}),
while the error in (\ref{Phi-SO3}) can be roughly
estimated to be
\beq
\frac{\Delta \Phi_{{\rm SO}(3)}(1.17,0.5)}
{|\Phi_{{\rm SO}(3)}(1.17,0.5)|}
\sim \frac{\Delta D_{{\rm SO}(3),x}}{|D_{{\rm SO}(3),x}|}    \ .
%% = \frac{\Delta \tilde{d}_1}{\tilde{d}_1}  
%% \left( 1 - 
%% \left( \frac{a x^3}{\tilde{d}_1} + 2 \right)^{-1} \right)   \ .
\eeq
Therefore, the relative error in $\Phi_{{\rm SO}(3)}(1.17,0.5)$
is ten times larger than that in $\vev{\tilde{\lambda}_3}$
in the SO(3) symmetric vacuum.
The crucial point here is that the function
$\Phi_{{\rm SO}(3)}(x,0.5)$ changes very rapidly in the
region where the solution to (\ref{master_eq_so3}) lies.
Therefore, a small systematic error in its estimation
affects the value at the solution much more than the solution itself.
%% The error in $\Phi_{{\rm SO}(3)}(x,0.5)$ 
%% propagates to 
%% the expectation values $\vev{\tilde{\lambda}_n}$,
%% but 
%% Let us discuss why this can be the case.
A similar argument applies also 
to the SO(2) symmetric vacuum.
It is therefore reasonable to conclude that
the first term in (\ref{log-rho-diff})
%$\Phi_{\rm SO(2)}(1.4,0.7,0.5)-\Phi_{\rm SO(3)}(1.17,0.5)$
is not obtained with the accuracy
that is sufficient 
for the determination of the sign of $\Delta$.
%for comparison with the second term given by (\ref{free_energy_f0}).
%the second term $\Xi = 0.065$.
%\gtrsim 0.07$, 

While this sounds a bit disappointing, let us recall that the
SO(2) and SO(3) symmetric vacua
%two vacua 
become degenerate as the parameter $r\equiv N_{\rm f}/N$
of the model goes to zero.
%approximate degeneracy of the two vacua
%SO(2) symmetric vacuum
%is 
%% Therefore, it is not so surprising that
%% the comparison of the free energy turned out to be difficult in this case.
%Therefore, the difficulty 
For instance, it would have been much easier to compare the free
energy for the solutions corresponding to the cases (b) and (c)
in (\ref{case_gen})
as the calculation of $\Delta_2$ in section \ref{lambda-results-single}
suggests.
Clearly the situation would be model dependent, and we consider that
the difficulty in the calculation of the free energy is not 
a generic feature of the factorization method.

%the first term in (\ref{log-rho-diff}) is 

%% While we do not have a clear understanding of this discrepancy,
%% we suspect 
%% that the extrapolations based on the asymptotic behavior
%% introduce a large systematic error in the determination of the free energy.
%compared with

%The problem there was that the calculation of $w_n(x)$ suffered
%from some overlap problem. That 
%From the $n=3$ analysis, 
%% There we were trying to evaluate the
%% difference of the free energy between the 2d solution and the 3d
%% solution. However, what we calculated was the difference of the
%% free energy between (1.07,1.02,0.7,0.7) and (1.13, 1.13, 1.13, 0.8?)
%% for $r=1$, which are \emph{not} minima of the generalized
%% distribution function $\rho(x_1 , x_2 , x_3 , x_4)$
%% due to the remaining overlap problem.
%% Of course we can view $\Delta_3$ as a rough estimate of the 
%% difference of the free energy between the actual solutions.
%% But obviously it cannot be taken as a precise one.
%% In fact the GEM predicts the difference of free energy 
%% between the 2d solution and the 3d solution
%% as -0.5 for $r=1$ as one can see from Table \ref{tab:GEMsummary}.
%% %and -1.7 for $r=2$, as one can read off from fig.\ 3 in ref.\ \cite{0412194}.

\subsection{including more observables}
\label{more-obs}

As we have seen above, 
the overlap problem can be reduced drastically by constraining
all the $\tilde{\lambda}_n$ ($n=1,2,3,4$) at the same time
instead of constraining just one of them.
A natural question that arises then is 
whether there is no more overlap problem we have to
worry about so that we actually do not need to include more
observables in our analysis.
Note that this is relevant not only to the calculation 
of the VEV of observables other than $\tilde{\lambda}_n$,
% ($n=1,2,3,4$),
but also to the calculation of
$\vev{\tilde{\lambda}_n}$ and the free energy
that we have discussed above.

%Let us consider a general observable ${\cal O}$.
Let us see directly
a possible overlap problem
associated with a general observable ${\cal O}$
in a simulation with constraints on $\tilde{\lambda}_n$ ($n=1,2,3,4$).
For that we rewrite the VEV $\vev{{\cal O}}$ as
\beqa
\langle {\cal O} \rangle 
&=& 
\int \prod_{k=1}^4 d x_k \, 
\Big\langle {\cal O} 
\prod_{k=1}^4 \delta (x_k -{\tilde \lambda}_{k}) \Big\rangle  \\
%% &=& 
%% \int \prod_{k=1}^4 d x_k \, \frac{1}{C}
%% \Big\langle {\cal O} \, \ee^{i \Gamma}
%% \prod_{k=1}^4 \delta (x_k -{\tilde \lambda}_{k}) \Big\rangle_0  \\
%% &=& 
%% \int \prod_{k=1}^4 d x_k \, \frac{1}{C} \, 
%% \rho^{(0)}(x_1 , x_2 , x_3 , x_4) 
%% \Big\langle {\cal O} \, \ee^{i \Gamma}
%%  \Big\rangle_{x_1 , x_2 , x_3 , x_4}  \\
&=& 
\label{rho-reweight}
\int \prod_{k=1}^4 d x_k \, 
\rho(x_1 , x_2 , x_3 , x_4) 
\frac{
\Big\langle {\cal O} \, \ee^{i \Gamma}
 \Big\rangle_{x_1 , x_2 , x_3 , x_4}
}{
\Big\langle \ee^{i \Gamma}
 \Big\rangle_{x_1 , x_2 , x_3 , x_4}
}  \ .
\eeqa
Assuming the usual equivalence between
the canonical ensemble and the microcanonical ensemble,
we may consider that the integral over $x_k$ in (\ref{rho-reweight})
is dominated by $(X,X,Y,Z)$ giving the absolute maximum of 
$\rho(x_1 , x_2 , x_3 , x_4)$. This leads to
\beq
\langle {\cal O} \rangle 
\simeq 
\frac{
\Big\langle {\cal O} \, \ee^{i \Gamma}
 \Big\rangle_{X,X,Y,Z}
}{
\Big\langle \ee^{i \Gamma}
 \Big\rangle_{X,X,Y,Z}
}  \ .
\label{ratioO}
\eeq
Thus we can calculate the VEV by using the standard reweighting
formula within the ``microcanonical ensemble''
characterized by $(X,X,Y,Z)$.
For ${\cal O}=\tilde{\lambda}_n$, the VEV will be
given by $X,X,Y,Z$ for $n=1,2,3,4$, respectively,
as it should.

For a general operator ${\cal O}$, however,
we can still have an overlap problem
due to the same reason that we have one
when we apply the reweighting method to the original model.
If it turns out that 
$\langle {\cal O} \rangle_{X,X,Y,Z}$ 
is actually close to (\ref{ratioO}), 
there is no more overlap problem as far as the operator ${\cal O}$
is concerned.
Note that this requires that
\beq
\Big\langle {\cal O} \, \ee^{i \Gamma}
 \Big\rangle_{X,X,Y,Z}
\sim \Big\langle {\cal O}  \Big\rangle_{X,X,Y,Z}
\Big\langle \ee^{i \Gamma} \Big\rangle_{X,X,Y,Z} \ ,
\label{no-corr}
\eeq
as we can see from (\ref{ratioO}).
Eq.~(\ref{no-corr}) only implies that
the correlation of
the operator
%a general operator 
${\cal O}$ and the phase factor $\ee^{i \Gamma}$
within the ``microcanonical ensemble'' is small.
If the same statement holds for arbitrary observable,
we may say that the sign problem is practically solved
\emph{even if} $\Big\langle \ee^{i \Gamma} \Big\rangle_{X,X,Y,Z}$
\emph{is not close to one at all}.

Thus we find that the success of the factorization method
in a general model relies on whether one can find the minimal
set of observables that is needed to ``solve'' the sign problem
in the above sense \cite{Anagnostopoulos:2010ux}.
We consider that the matrix model we are studying provides
an explicit example in which this can be done.

%% This can be achieved in a general model by constraining
%% %A general strategy of the factorization method should be to
%% %constrain 
%% all the observables that have strong correlation 
%% with the phase factor.
%% %Once this is done successfully, the 

%\subsection{addition of yet another observable}

In order to provide some evidence for this statement,
let us consider an observable 
\beq
{\cal O}= - \frac{1}{N} \sum_{\mu \neq \nu}\tr [A_\mu , A_\nu]^2 \ ,
\label{trf2}
\eeq
and the normalized one 
$\widetilde{\cal O}={\cal O}/\vev{{\cal O}}_0$.
When we constrain $\widetilde{\cal O}$ to a small value,
the dominant configurations have the property
$[A_\mu , A_\nu ] \approx 0$, meaning that $A_\mu$ are simultaneously 
diagonalizable, \emph{e.g.} as $A_\mu = \mbox{diag}(\alpha_\mu^{(1)},
\cdots , \alpha_\mu^{(N)})$.
For such configurations, the determinant becomes
\beq
\det {\cal D}= \prod_{i=1}^{N} 
\Bigl\{ \sum_\mu (\alpha_\mu^{(i)}) ^2 \Bigr\} \ge  0 \ .
\label{det-commute}
\eeq
%% Thus it is expected that
%% $\frac{1}{N^2} \log w_{\cal O}(x)$ approaches zero 
%% for $x\rightarrow 0$. 
Therefore, the observable (\ref{trf2})
%${\cal O}$ 
is considered as a ``dangerous'' one,
which can potentially
have strong correlation with the phase factor.

Clearly the formula (\ref{ratioO}) is not useful
for actually investigating the overlap problem 
associated with the operator ${\cal O}$.
For that purpose, we can use the factorization method
including not only $\tilde{\lambda}_n$ ($n=1,2,3,4$)
but also $\widetilde{\cal O}$.
Let us define the corresponding functions $\rho^{(0)}$ and $w$
with five arguments, and also define the reduced functions
\beqa
\rho^{(0)}_{\cal O}(x) &=& \rho^{(0)}(X,X,Y,Z,x)  \ ,  \\
w_{\cal O}(x) &=& w(X,X,Y,Z,x)  \ .
\label{w-O}
\eeqa
%% which can be obtained by simulating a phase quenched model
%% with constraints fixing 
%% $\tilde{\lambda}_k$ to the GEM result
%% and $\widetilde{\cal O}={\cal O}/\vev{{\cal O}}_0$ to some value $x$.
The VEV $\vev{\widetilde{\cal O}}$ can then be obtained by 
solving
\beq
\frac{1}{N^2} f^{(0)}_{\cal O}(x) 
= - \frac{d}{dx} \Phi_{\cal O}(x) \ ,
\eeq
where $f^{(0)}_{\cal O}(x)$ and $\Phi_{\cal O}(x)$ are defined as
\beqa
f^{(0)}_{\cal O}(x)  
&=&
%& \stackrel{\textrm{def}}{=} &
 \frac{d}{dx} \log  \rho^{(0)}_{\cal O}(x) \ ,   \\
\Phi_{\cal O}(x) &=& \lim_{N\rightarrow \infty}
\frac{1}{N^2} \log w_{\cal O}(x) \ .
\label{Phi-O}
\eeqa
%which is also shown in the same figure by the two solid lines.
%% The fact that we were able to reproduce the GEM results
%% with the factorization method specifying $\lambda_n$ at the same time
%% suggests that we do not need to specify more observables
%% to remove the main overlap problem.
%% In order to confirm this explicitly,

%% Under parity transformation $A_4\to -A_4$, $A_i\to A_i$ ($i
%% \ne D\equiv 4$) the fermion determinant transforms to its complex
%% conjugate. 

%% As we mentioned below (\ref{z_int_out}),
%% the fermion determinant transforms to its complex
%% conjugate under parity transformation.
%% Therefore for all parity even observables ${\cal O}$,
%% % we obtain 
%% eq.\ (\ref{eq:ovev}) reduces to
%% \begin{equation}
%%   \label{eq:covev}
%%   \vev{{\cal O}} = \frac
%%  {\vev{{\cal O}\,\cos{\Gamma}}_0}
%%  {\vev{          \cos{\Gamma}}_0}\, . 
%% \end{equation}
%% Thus the complex action problem becomes a sign problem,
%% which unfortunately is as hard to overcome as in \rf{eq:ovev}.

  \FIGURE[t]{
\epsfig{file=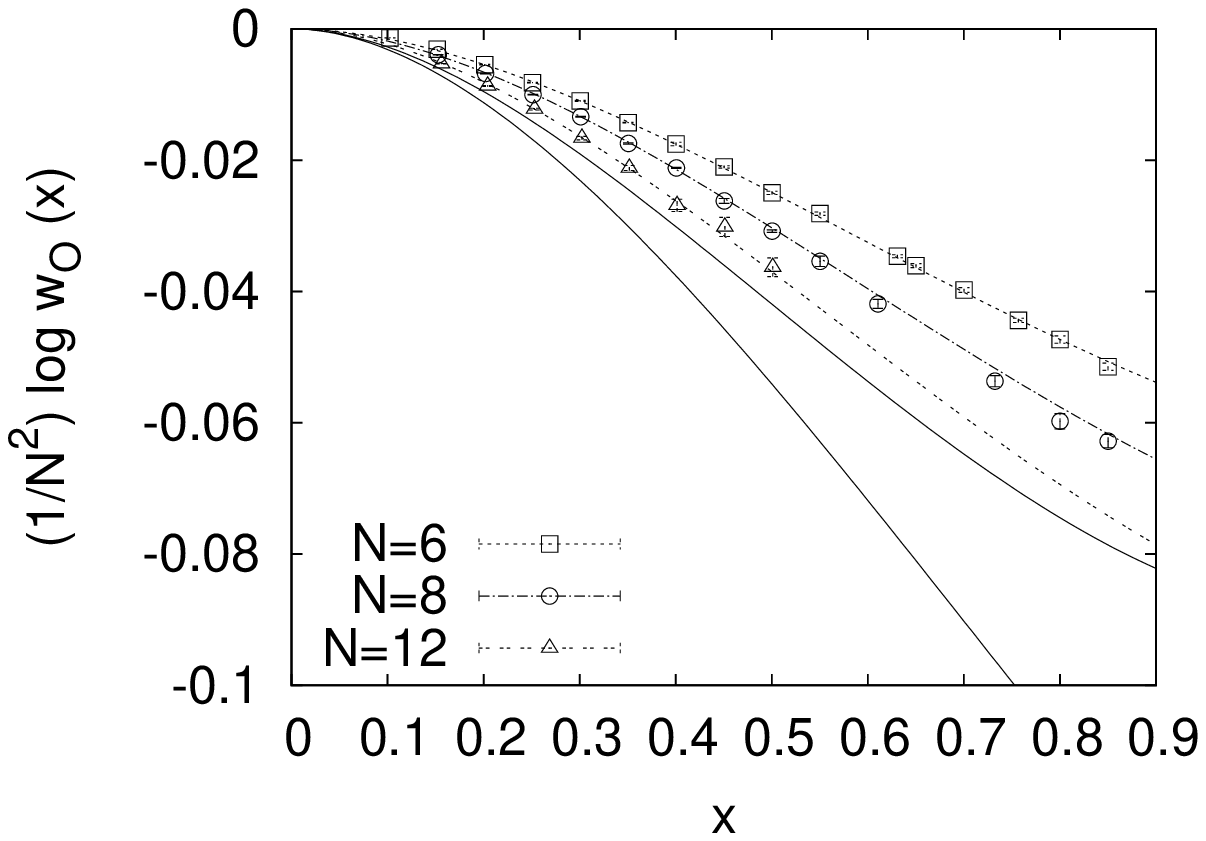,width=7.4cm}
\epsfig{file=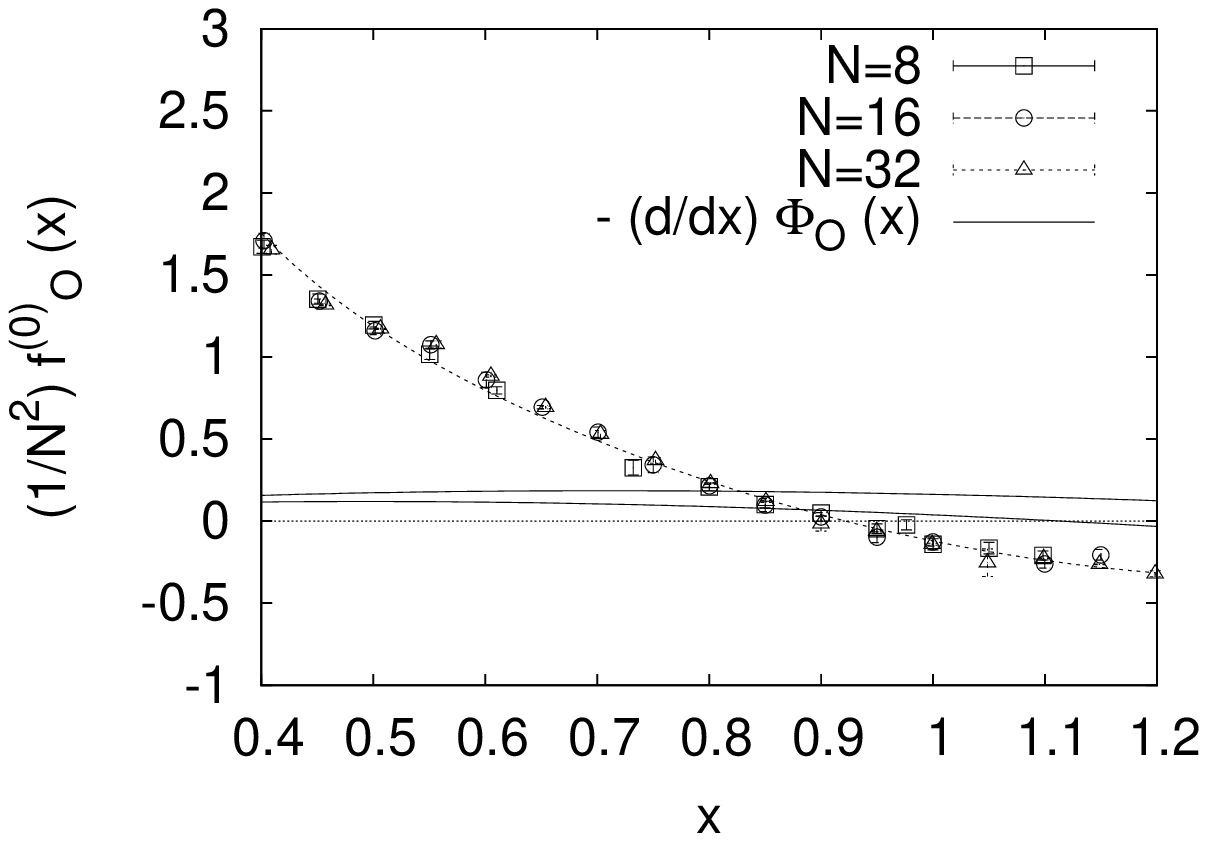,width=7.4cm}
   \caption{(Left) The function
$\frac{1}{N^2}  \log w_{{\cal O}}(x)$ is plotted against $x$
for $N=6,8,12$.
We also plot the function
$\Phi_{{\cal O}}(x)$ 
obtained by extrapolation to
$N=\infty$ as described in appendix \ref{SO2-1st}.
The two solid lines represent the margin of error.
(Right) The function 
$\frac{1}{N^2} \frac{d}{dx} \log \rho^{(0)}_{{\cal O}}(x)$ 
is plotted for $N=8,16,32$. 
%%% copied from PRD
%The dashed line is drawn to guide the eye.
% to the maximum of $\rho^{(0)}_{\cal O}(x)$.
% at the vanishing point of $f^{(0)}_{\cal O}(x)$.
We also plot $- \frac{d}{dx} \Phi_{{\cal O}}(x) $
obtained from the plot on the left.
%$- \frac{d}{dx} \lim_{N\rightarrow \infty} 
%\{ \frac{1}{N^2} \log w_{{\cal O}}(x) \} $, 
%where the two solid lines show the margin of error. 
}
\label{result_so2_trF2}}
%%%%%%%%%%%%%%%%%%%
%As a further check,

%lnw-fsq-r1_b3.eps    raw data of logw
%lnw-fsq-r1_c3.eps    subleading
%f-naught-fsq-r1_2.eps   solution
% Azuma-kun's e-mail on 09/12

From fig.~\ref{result_so2_trF2} (Left) 
we find that
$\frac{1}{N^2} \log w_{\cal O}(x)$ approaches zero 
for $x\rightarrow 0$ as expected from (\ref{det-commute}).
We make a large-$N$ extrapolation as described in
appendix \ref{SO2-1st} to obtain 
$\Phi_{\cal O}(x)$,
which is also shown in the same figure by the two solid lines
showing the margin of error.
From fig.~\ref{result_so2_trF2} (Right) 
we find that the effect of the phase 
%represented by (\ref{sys-error})
is to shift the estimate of 
$\langle \widetilde{\cal O} \rangle$
by $\Delta x = 0.07(3)$.
On the other hand, the standard deviation of 
the distribution $\rho^{(0)}_{\cal O}(x)$ is estimated
as $\sigma \sim 0.7/N$ from the slope of the function
plotted in fig.~\ref{result_so2_trF2} (Right) around $x\sim 0.92$.
This means that the deviation $\Delta x$ is 
$\lesssim 2 \, \sigma$ for $N \le 16$.
Thus, the remaining overlap problem 
associated with this observable (\ref{trf2})
%in obtaining the data 
%in fig.~\ref{result_so2_2} 
is practically small.
%as far as this observable (\ref{trf2}) is concerned.
This is consistent with the fact that we were able to reproduce
the GEM result by constraining only the four observables 
$\lambda_n$ ($n=1,2,3,4$). 
\section{Summary and discussions}
\label{letzte_ep}
% %%%%%%%%%%%%%%%%%%%%%%%%%%%%%%%%%%%%%%%%%%%%%%%%%%%%%%%%%%%%%%%%%%%%%
% %%%%%%%%%%%%%%%%%%%%%%%%%%%%%%%%%%%%%%%%%%%%%%%%%%%%%%%%%%%%%%%%%%%%%

In this paper we tested the proposed scenario for dynamical
compactification of space-time in the IIB matrix model,
in which the phase of the fermion determinant induces
the spontaneous breaking of rotational symmetry.
We have shown that this can indeed occur
in a simplified model \cite{0108070} by performing Monte Carlo
simulations.
The model exhibits SSB of $\textrm{SO}(4)$ rotational symmetry 
in agreement with the results of GEM \cite{0412194},
and the phase of
the fermion determinant plays a crucial role in the mechanism of SSB.

%% This model was chosen because it is
%% expected from analytical arguments \cite{0108070,0412194} to
%% dynamically break its $\textrm{SO}(4)$ rotational symmetry 
%% in a similar manner to the IIB matrix model. 
%% It is also easier to simulate than the IIB matrix model 
%% since it is defined in $D=4<10$ and the fermions are
%% in the fundamental representation.  Monte Carlo simulation can provide
%% a non-perturbative confirmation of the results in
%% ref.\ \cite{0108070,0412194} which gives confidence in the use of their
%% methods in the IIB matrix model and vice-versa, reproducing numerically
%% the analytical results can help developing efficient numerical methods
%% to be used in the study of the IIB matrix model. This toy model has a
%% strong complex action problem which is expected to play a crucial role
%% in the mechanism of SSB similar to the case of the IIB matrix model.
%% This makes the analogy to the IIB matrix model even
%% stronger. Overcoming the notorious technical problems associated with
%% the Monte Carlo simulations of such systems can have a great impact on
%% a wide range of physical problems.  In this work we applied the
%% ``factorization method'' proposed in \cite{0108041}, which has also been
%% proven useful in models with complex action in statistical physics and
%% finite density QCD \cite{Ambjorn:2003rr,Azcoiti:2002vk,Fodor:2007vv}.

First, in the absence of the phase, we have 
confirmed that the model has no SSB
by calculating the VEV of the eigenvalues of $T_{\mu\nu}$,
which is analogous to the moment of inertia tensor. 
% in complete agreement with \cite{0108070}. 
%% We showed that the density of eigenvalues
%% $\rho^{(0)}_n(x)$ has asymptotic behaviors when some of the
%% dimensions become very large or small.
%% We were able to understand these behaviors quantitatively
%% from simple theoretical arguments.
%% This is nice since it gives us analytical handle on
%% how the collapsed configurations are suppressed in the phase quenched model.
%
%``thin'' or ``thick''. 
%% This is possible only
%% with the factorization method, since one needs to sample regions of
%% configuration space that are impossible to sample with the usual
%% importance sampling methods\footnote{We conducted tests with a
%%   multicanonical \cite{berg-mc} algorithm but we could not reach very
%%   small values of $x$.}. These scaling properties are crucial in the
%% study of the full model because they can be used to extrapolate the
%% results to the region suppressed by the complex phase fluctuations.
The effect of the phase has been studied by using the 
factorization method originally proposed in ref.~\cite{0108041}.
From the single-observable analysis, we find
that the phase fluctuations strongly
suppress the region of the configuration space favored in the phase
quenched model and result in different length scales for each
dimension of space-time. 
%We cleanly verify SSB of SO(4) in the $N\rightarrow \infty$ limit.
%as in ref.~\cite{0412194}, 
While our results are in partial agreement
%While the results are partially consistent 
with the GEM results in ref.\ \cite{0412194},
we also observe some discrepancies and puzzles, which we attributed to 
the remaining overlap problem.
This motivated us to generalize the method
%apply the factorization method
to multiple observables.
%perform the analysis with multiple variables.
%specifying all the eigenvalues at the same time.
In particular, we find that controlling the four eigenvalues of the
``moment of inertia tensor'' is good enough to reduce the overlap
problem to a sufficient level.
Restricting ourselves to the SO(2) and SO(3) symmetric vacua,
we were able to reproduce the GEM results for the
VEV of the eigenvalues consistently.

%% The numerical calculation turned out to be quite tedious. 
%% The computational effort to simulate the model is smaller than the
%% corresponding IIB matrix model but the fluctuations of the phase are
%% quite strong and increase considerably with the value of $r$. The use
%% of the factorization method is important for calculating accurately
%% the scaling properties of the density of states and the phase $w_n(x)$
%% in highly suppressed regions of the configuration space. For this the
%% choice of $\tl_n$ as the order parameters is crucial because of the
%% property of the collapsed configurations
%% that yield a real fermion determinant.
%% Similar choices should be made when one attempts to use this method on
%% a different system.

Our results are an encouragement for pursuing similar studies
on the IIB matrix model. Although it is computationally more
demanding, the supersymmetry of the IIB matrix model
could make the SSB of rotational symmetry easier to see. 
In particular, we have observed in ref.\ \cite{0108041} that
the $\frac{1}{N^2}f^{(0)}_n(x)$ that appears
in 
%(\ref{extrema})
(\ref{master-eq})
is actually very close to zero,
which can be understood as a result of 
cancellations by fermionic and bosonic
contributions to the interactions among space-time points.
This implies that the solutions to eq.\ (\ref{master-eq})
%\rf{extrema} 
appear
in the region where the sign problem is not severe.
%Also the extrapolations using the asymptotic behaviors become
%much more reliable in that case.
We are currently studying 
the $6$ dimensional version of the IIB matrix model
by Monte Carlo simulation \cite{6dIKKT} and trying to compare
the results against the predictions 
obtained by the GEM recently \cite{Aoyama:2010ry}. 
%Results will be reported elsewhere.

We would like to emphasize that the model studied in this paper
has a severe sign problem despite its simpleness.
It is encouraging that such a system can be studied
by Monte Carlo simulation using 
%a rather direct approach like 
the factorization method,
which gives us a lot of useful insights into the effect of the phase.
The method itself is quite general and it can be applied
to any interesting system which suffers from the sign problem.
The crucial step is
%How well it does depends crucially on 
to find
%whether one can find
an appropriate set of observables that one can control
in order to determine and sample efficiently 
the region of the configuration space 
favored by the relatively small fluctuation 
of the phase.
We expect that this is possible in many interesting 
systems.

% %%%%%%%%%%%%%%%%%%%%%%%%%%%%%%%%%%%%%%%%%%%%%%%%%%%%%%%%%%%%%%%%%%%%%
% %%%%%%%%%%%%%%%%%%%%%%%%%%%%%%%%%%%%%%%%%%%%%%%%%%%%%%%%%%%%%%%%%%%%%
\acknowledgments
% %%%%%%%%%%%%%%%%%%%%%%%%%%%%%%%%%%%%%%%%%%%%%%%%%%%%%%%%%%%%%%%%%%%%%
% %%%%%%%%%%%%%%%%%%%%%%%%%%%%%%%%%%%%%%%%%%%%%%%%%%%%%%%%%%%%%%%%%%%%%
The authors would like to thank Shinji Ejiri, Sourendu Gupta and Yuko
Okamoto for valuable discussions. This work was partially funded by
the National Technical University of Athens through the ``Basic
Research Support Programme 2009 and 2010''.
The work of T.A.\ and J.N.\ was
supported in part by Grant-in-Aid for Scientific Research (No.\ 03740,
23740211 for T.A.\ and 20540286 for J.N.)
%, respectively)
from Japan Society for the Promotion of Science.
%from the Ministry of Education, Culture, Sports, Science and Technology. 

% %%%%%%%%%%%%%%%%%%%%%%%%%%%%%%%%%%%%%%%%%%%%%%%%%%%%%%%%%%%%%%%%%%%%%
% %%%%%%%%%%%%%%%%%%%%%%%%%%%%%%%%%%%%%%%%%%%%%%%%%%%%%%%%%%%%%%%%%%%%%
 \appendix
% %%%%%%%%%%%%%%%%%%%%%%%%%%%%%%%%%%%%%%%%%%%%%%%%%%%%%%%%%%%%%%%%%%%%%

% %%%%%%%%%%%%%%%%%%%%%%%%%%%%%%%%%%%%%%%%%%%%%%%%%%%%%%%%%%%%%%%%%%%%%
 \section{Details of the Monte Carlo simulation}
 \label{algorithm}
% %%%%%%%%%%%%%%%%%%%%%%%%%%%%%%%%%%%%%%%%%%%%%%%%%%%%%%%%%%%%%%%%%%%%%
% %%%%%%%%%%%%%%%%%%%%%%%%%%%%%%%%%%%%%%%%%%%%%%%%%%%%%%%%%%%%%%%%%%%%%

 In this section we present the details of the algorithm used
 for our Monte Carlo simulation. 
 It is essentially the hybrid Monte Carlo (HMC) algorithm, 
 which has been applied in similar models in refs.\ 
 \cite{Ambjorn:2000bf,Ambjorn:2000dx,0108041,Ambjorn:2003rr,0506062}.
 The computational effort grows as O$(N^{3})$ in the present model.
%%  which is much reduced
%%  compared with O$(N^{6})$ for the IIB matrix model.
%%  This is because the fermions are in the fundamental representation
%%  instead of the adjoint representation as in the IIB matrix model.
%
% In the IIB matrix model the CPU time required is O$(N^{6})$ since the
% fermions are the adjoint representation. In our case 

We first discuss the HMC algorithm for the phase-quenched 
model (\ref{action-z0}). We introduce
auxiliary bosonic Hermitian matrices $P_{\mu}$ and consider the action
\begin{equation}
S_{\rm HMC}[P,A] = \frac{1}{2} \, \tr (P_{\mu})^{2} + S_{0}[A] \ .
\label{S-HMC}
\end{equation}
The original model is obtained by integrating out $P_{\mu}$. We regard
the action $S_{\rm HMC}[P,A]$ as the Hamiltonian of a classical system
described by $A_{\mu}(\tau)$ and its conjugate momentum $P_{\mu}
(\tau)$, where $\tau$ denotes the fictitious time of the classical
system. 
The Hamiltonian equation of motion is obtained as
\begin{eqnarray}
\frac{d (A_{\mu})_{kl}}{d \tau} 
&=& \frac{\partial S_{\rm HMC}}{\partial (P_{\mu})_{kl}} 
= (P_{\mu})_{lk} \ , 
\label{a_hm_eq} \\
\frac{d (P_{\mu})_{kl}}{d \tau} &=& 
- \frac{\partial S_{\rm HMC}}{\partial (A_{\mu})_{kl}} 
\nonumber \\  &=& 
%=
- N (A_{\mu})_{lk} + \frac{N_{f}}{2} 
\left\{ \Tr \left( 
{\cal D}^{-1} \frac{\partial {\cal D}}{\partial (A_{\mu})_{kl}} \right) 
+ \Tr \left( {\cal D}^{-1} 
\frac{\partial {\cal D}}{\partial (A_{\mu})_{lk}} \right)^{*} \right\} \ , 
\label{p_hm_eq}
\end{eqnarray}
where $\Tr$ denotes a trace with respect to the $2N$-dimensional index.
% \times 2N$ matrix $\cal D$. 
The indices $k,l$ run over $k,l=1,2,\cdots,N$. 
The updating procedure consists of 
(i) refreshing the momentum variables $P_\mu$ by Gaussian random numbers,
which obey the distribution $e^{-S_{\rm HMC}}$
and (ii) solving the Hamiltonian equation of motion for 
a fixed time interval $\tau$.
In actual
calculations we have to discretize the Hamiltonian equation
(\ref{a_hm_eq}) and (\ref{p_hm_eq}). The reversibility of the time
evolution is preserved by using the so-called leap-frog
discretization, which gives
  \begin{eqnarray}
  (P_\mu^{(1/2)})_{kl}  &=& (P_{\mu}^{(0)})_{kl}
   - \frac{\Delta \tau}{2} \, 
  \frac{\partial S_{\rm HMC}}{\partial (A_\mu)_{kl}}(A_\mu^{(0)}) \ , 
   \nonumber \\
    (A_\mu^{(1)})_{kl}  &=& (A_\mu^{(0)})_{kl} + \Delta \tau \, 
    (P_\mu^{(1/2)})_{lk}  \ ,
   \label{lf-step1}  \\
   (P_\mu^{(r+1/2)})_{kl} 
 &=& (P_\mu^{(r-1/2)})_{kl}
   - \Delta \tau \, 
   \frac{\partial S_{\rm HMC}}{\partial (A_\mu)_{kl}} (A^{(r)}_\mu )
   \ , \nonumber \\
    (A_\mu^{(r+1)})_{kl} &=&
   (A_\mu^{(r)})_{kl} + \Delta \tau \, 
    (P_\mu^{(r+1/2)})_{lk}  \ ,  \label{lf-step2} \\
  (P_\mu^{(\nu)})_{kl}  &=& (P_\mu^{(\nu - 1/2)})_{kl} 
   - \frac{\Delta \tau}{2}  \, 
  \frac{\partial S_{\rm HMC}}{\partial (A_\mu)_{kl}}
(A_\mu^{(\nu)})  \ ,
  \label{lf-step3}
  \end{eqnarray}
where $r=1,2, \cdots , \nu-1$ and $T = \nu \, \Delta \tau$, 
and we have introduced the 
short-hand notation $P_\mu ^{(r)} = P_\mu(r \, \Delta \tau)$ 
and $A_\mu ^{(r)} = A_\mu(r \, \Delta \tau)$.
The conservation of the Hamiltonian 
($S_{\rm HMC}$) is violated by the discretization. 
The detailed balance can be preserved, however, by adding
a Metropolis accept/reject procedure at the
end of each trajectory.
Namely we accept the trial configuration with the probability
${\rm min}(1,e^{-\Delta S_{\rm HMC}})$, where 
$\Delta S_{\rm HMC}$ represents the difference of
$S_{\rm HMC}$ between the trial and original configurations.
% we can preserve the detailed balance condition.
The step size $\Delta \tau$ for the time evolution should be small
enough to keep the acceptance rate reasonably high. 
%The discretized Hamiltonian equation is given by

Next we discuss how to simulate the constrained system 
defined by the partition function $Z_{x_1,x_2,x_3,x_4}$ 
in eq.~(\ref{ix_pf_gen}).
As we discussed in section \ref{prac-implement}
for the single-observable analysis, 
we approximate the delta function by the Gaussian function as
\beqa
\label{ix_pf_gen_gauss} 
 Z_{x_1,x_2,x_3,x_4,V} &=& 
\int dA \, 
\exp \left[- \left(
S_0+
\sum_{n=1}^4 V_n (\lambda_n)
\right)
\right] 
 \ , \\
%\quad
V_n(\lambda_n)
&=& \frac{1}{2}  \gamma_{n} (\lambda_{n} - \xi_{n})^2 \ . 
\label{ix_pf_action_gen}
\eeqa
The system 
%(\ref{ix_pf})
(\ref{iV_partition})
used for the single-observable analysis
in section \ref{section:mc}
%\ref{prac-implement}
%\ref{single_factorization}
can be obtained by setting $\gamma_k = 0$ for $k\neq n$.

Taking into account the potential term $V_n(\lambda_n)$ in 
%(\ref{ix_pf_action_gen}), 
(\ref{ix_pf_gen_gauss}),
we need to subtract the term
$ \frac{\partial V_{n} (\lambda_{n})}{\partial
  (A_{\mu})_{kl}}$ 
from the right-hand side
of eq.~(\ref{p_hm_eq}).
This term can be calculated explicitly as follows.
%% we replace the action (\ref{S-HMC}) by
%% \begin{eqnarray}
%% S_{\textrm{HMC},V} [P,A] = S_{\textrm{HMC}}[P,A] 
%% + \sum_{n=1}^{4} V_{n} (\lambda_{n}) \ .
%% \end{eqnarray}
%% The only difference is that we need to add the contribution of
%% $ \frac{\partial V_{n} (\lambda_{n})}{\partial
%%   (A_{\mu})_{kl}}$ to the Hamiltonian equation (\ref{p_hm_eq}). 
Let us note first that the eigenvalues $\lambda_{n}$ satisfy
\begin{eqnarray}
 \sum_{\rho=1}^{4} T_{\nu \rho} v_{\rho}^{(n)} 
= \lambda_{n} v_{\nu}^{(n)}, \label{app_eigenv}
\end{eqnarray}
where $v_{\mu}^{(n)}$ is the eigenvector of the $4 \times 4$ matrix
$T_{\mu \nu}$ normalized as $\sum_{\mu=1}^{4}
v^{(n)}_{\mu} v^{(n)}_{\mu} = 1$. (No summation over the index $n$.)
%We do not take a summation with
%respect to the index $n$. 
Taking the derivative of
eq.\ (\ref{app_eigenv}) with respect to $(A_{\mu})_{kl}$, we obtain
\begin{eqnarray}
 \sum_{\rho=1}^{4} 
\left(  \frac{\partial T_{\nu \rho}}{\partial (A_{\mu})_{kl}}
v_{\rho}^{(n)} + 
T_{\nu \rho} \frac{\partial v_{\rho}^{(n)}}{\partial (A_{\mu})_{kl}} 
\right)
 =  \frac{\partial \lambda_{n}}{\partial (A_{\mu})_{kl} } 
v_{\nu}^{(n)} + \lambda_{n} 
\frac{\partial v^{(n)}_{\nu}}{\partial (A_{\mu})_{kl} }  \ . 
\label{app_eigenv2}
\end{eqnarray}
Multiplying both sides 
of eq.\ (\ref{app_eigenv2}) 
by $v_{\nu}^{(n)}$ and taking a sum over $\nu$, we obtain
\begin{eqnarray}
\sum_{\nu,\rho=1}^{4} v_{\nu}^{(n)} 
 \frac{\partial T_{\nu \rho}}{\partial (A_{\mu})_{kl}} 
 v_{\rho}^{(n)}  = 
\sum_{\nu=1}^{4} v_{\nu}^{(n)} 
 \frac{\partial \lambda_{n}}{\partial (A_{\mu})_{kl} } 
 v_{\nu}^{(n)} = 
 \frac{\partial \lambda_{n}}{\partial (A_{\mu})_{kl} }  \ ,
\end{eqnarray}
where the second terms of each side of eq.\ (\ref{app_eigenv2}) cancel. 
Therefore we obtain
\begin{eqnarray}
\frac{\partial V_{n}(\lambda_{n})}{\partial (A_{\mu})_{kl}} 
= \gamma_{n} (\lambda_{n} - \xi_{n}) 
\frac{\partial \lambda_{n}}{\partial (A_{\mu})_{kl}} 
= \frac{2 \gamma_{n}}{N} (\lambda_{n} - \xi_{n}) 
\sum_{\nu=1}^{4} v^{(n)}_{\mu} v^{(n)}_{\nu} (A_{\nu})_{lk} \ .
\end{eqnarray}
%We add this term (summed over $n$ in the multi-observable analysis)
%to the right-hand side of eq.~(\ref{p_hm_eq}).
%and simulate this system in a similar fashion. 

%\section{Power law behaviors of $\Phi_i (x)$}

Let us comment on actual values we have chosen for the parameters in
the potential (\ref{ix_pf_action_gen}).
For the single-observable analysis in 
section \ref{section:mc},
%\ref{single_factorization},
we typically\footnote{
In fact we have used $\gamma_n$ in the
range $10^3$---$10^7$. 
We have checked that our results are independent of
$\gamma_n$ within a wide range.} used $\gamma_n = 10^4$.
For the multi-observable analysis 
in section \ref{generalized_factorization}, 
we studied the cases (a) and (b) in (\ref{case_gen})
corresponding to the SO(3) and SO(2) symmetric vacua, respectively.
In the case (a), we set $\gamma_1 = \gamma_2 = 0$
since $\lambda_1$ and $\lambda_2$ are automatically close to 
$\lambda_3$. 
When we fix $\lambda_{4}$ to the value obtained by GEM
and vary $\lambda_{3}$ as we do in section \ref{sec:res-SO3},
we set $\gamma_4$ much larger than $\gamma_3$ 
(typically $\gamma_4 = 10^5$ and $\gamma_3=10^2$), 
and set $\xi_{4}$ to the GEM value for $\langle \lambda_{4} \rangle$
obtained with the SO(3) ansatz.
In the case (b),
we set $\gamma_1 = 0$ 
since $\lambda_1$ is automatically close to $\lambda_2$. 
When we fix $\lambda_{3}$ and $\lambda_{4}$ to the values obtained by GEM
and vary $\lambda_{2}$
as we do in section \ref{sec:res-SO2},
we set $\gamma_3$ and $\gamma_4$ 
much larger than $\gamma_2$ 
(typically $\gamma_3 = \gamma_4 = 10^5$ and $\gamma_2 = 10^2$), 
and set $\xi_{3}$ and $\xi_4$ to the GEM values for $\vev{\lambda_{3}}$ 
and $\vev{\lambda_{4}}$, respectively, obtained with the SO(2) ansatz.

\section{Asymptotic behaviors of the functions 
$f^{(0)}_n(x)$}
\label{sec:pqm}

%% We simulate the system for $r=1, 2$ and compute the eigenvalues
%% $\vev{\lambda_n}_0$.
%% %, which are predicted as eq.\ \rf{eq:pqmlam} in the large-$N$ limit.
%% Let us compute the left-hand side of 
%% (\ref{master-eq}), which is defined by
%% eq.\ 
%% %\rf{f_naught_def} 
%% \rf{extrema}
%% in terms of 
%% the density of states $\rho^{(0)}_n(x)$.
%% %% Next we proceed to compute the density of states $\rho^{(0)}_n(x)$ of
%% %% eq.\ \rf{rho_naught}. 
%% %% According to the discussion in Section \rf{section:mc} 
%% %% we need to evaluate $f_n^{(0)}(x)$ in eq.\ \rf{f_naught_def}. 
%% We simulate the system $Z_{n,V}$ in (\ref{iV_partition})
%% for $r= 1$ and $r=2$ for matrix size up to $N= 64$. 
%% We measure the value of $\vev{\tl_n}_{i,V}$, which is the estimator of $x_p$
%% according to eq.\ \rf{vev_nv}. 
%% We use eq.\ \rf{fi_vev} to compute $f_n^{(0)}(x)$ at $x=x_p$.

In this section we discuss
the asymptotic behaviors of the functions 
$f^{(0)}_n(x)$ that appears in (\ref{master-eq}).
%at $x \ll 1$ and $x\gg 1$ 
While these behaviors are not explicitly used in our analysis,
they help us understand the dynamical properties of 
the eigenvalues $\tilde{\lambda}_n$ in the matrix model.
In fact the functions $f^{(0)}_n(x)$ in the IIB matrix model
behaves quite differently as we mention in section \ref{letzte_ep}, 
and we hope that our results given below would be useful
for comparison.

Let us recall that the function $f_n^{(0)}(x)$ is defined
by eq.~(\ref{extrema}) in terms of $\rho^{(0)}_n(x)$, which is 
the distribution function of $\tilde{\lambda}_n$ 
in the phase-quenched model.
It is therefore expected to have simple scaling behavior 
when $x\ll 1$ or $x\gg 1$. 
%% This is crucial to verify since we need to extrapolate, 
%% using possible corrections to this scaling, to
%% intermediate values of $x$. 
At small $x$ we expect
\begin{equation}
 \frac{1}{N^{2}} f^{(0)}_{n}(x) \simeq 
\left\{   \frac{1}{2} (5-n)  + r \delta_{n1}  \right\} \frac{1}{x}
+ a_{n} \ .
%\qquad x\ll 1\, ,
\label{power-law-f}
\end{equation}
%% where the constants $a_n$ are expected to take
%% values close to $-1$ as explained below.
This behavior is due to the phase space suppression 
since $(5-n)$ directions shrink as we decrease $x$.
Considering that the shrunken directions have the extent proportional
to $\sqrt{x}$, we obtain
$\rho_n^{(0)}(x)\sim (\sqrt{x})^{(5-n) N^2}$.
%For small $n$ this
%suppression will be stronger because $(5-n)$ directions shrink. Then
The $n=1$ case differs since all the eigenvalues of
$A_\mu$ collapse to $0$ at $x\ll 1$ and the suppression factor comes
also from the fermion determinant, which is a homogeneous polynomial of
$A_\mu$ of degree $2 r N^2$. 
This gives an extra suppression factor of $(\sqrt{x})^{2r N^2}$. 
From the definition 
\rf{extrema}, 
%\rf{f_naught_def}
we obtain the leading behavior of \rf{power-law-f}. 
%% The density of
%% states $\rho^{(0)}_n(x)$ is peaked near $x=1$, therefore
%% $f^{(0)}_n(x)$ becomes zero in the neighborhood of $1$ 
%% and $a_n\approx -1$. 
Assuming that we have some analytic function of $x$ multiplied
to the leading power-law behavior of $\rho_n^{(0)}(x)$,
we obtain some constant $a_n$ in \rf{power-law-f}
as the subleading term. 

% %%%%%%%%%%%%%%%%%%%%%%%%%%%%%%%%%%%%%%%%%%%%%%%%%%%%%%%%%%%%%%%%%%%%%
\FIGURE{
    \epsfig{file=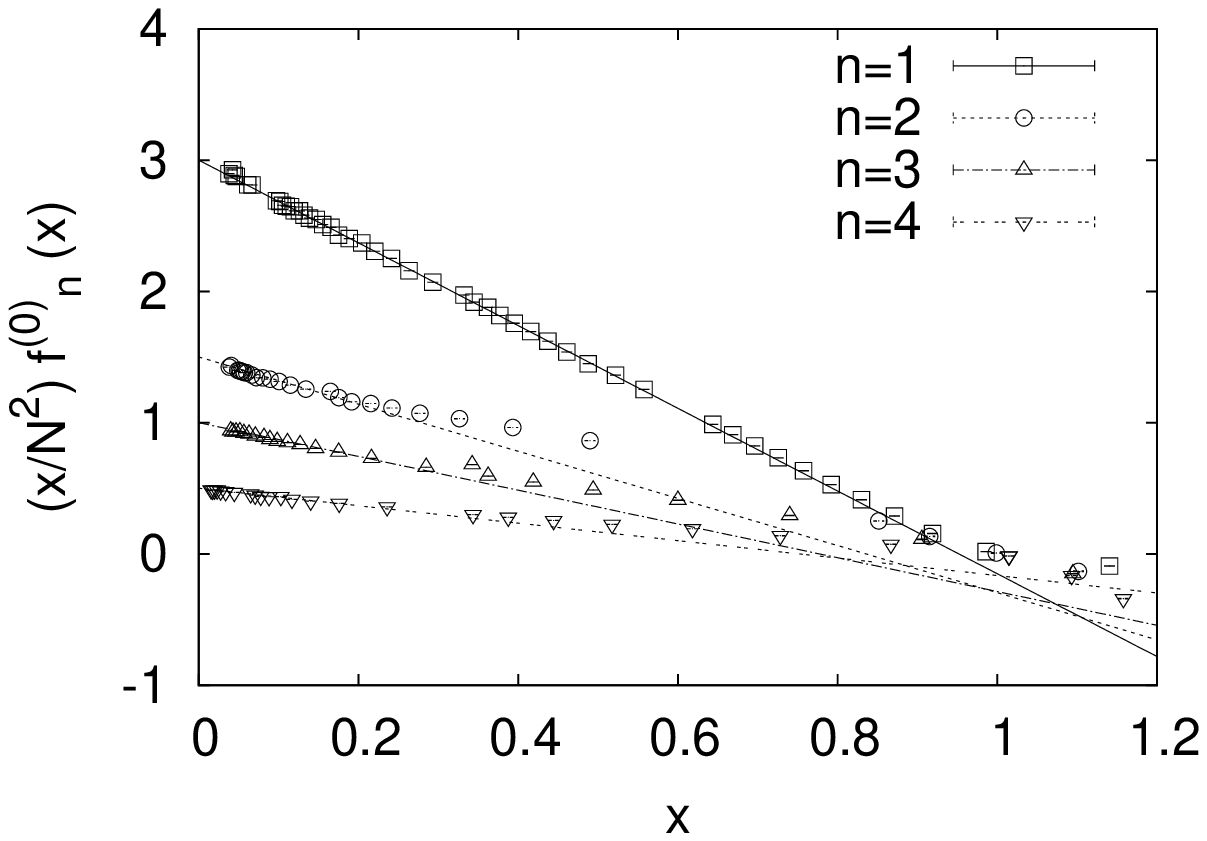,width=7.4cm}
    \epsfig{file=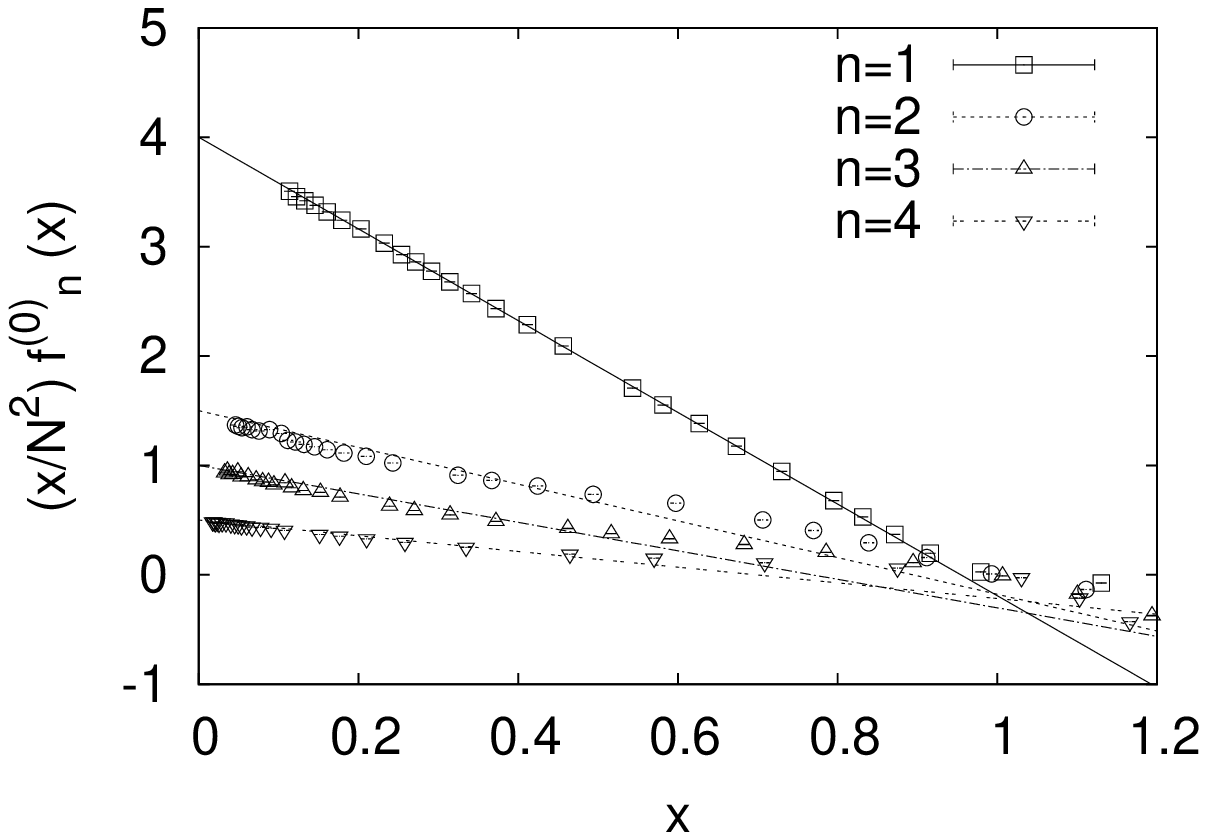,width=7.4cm}

    \caption{The function
%      $\displaystyle  \frac{x}{N^{2}} f^{(0)}_{n}(x)$
      $\frac{x}{N^{2}} f^{(0)}_{n}(x)$
      at $N=64$ is plotted against $x$ for $r=1$ 
      (Left) and $r=2$ (Right). The straight lines are fits to the
    behavior \rf{power-law-f}.}
   \label{f-loglog}
}
% %%%%%%%%%%%%%%%%%%%%%%%%%%%%%%%%%%%%%%%%%%%%%%%%%%%%%%%%%%%%%%%%%%%%%
\FIGURE{
    \epsfig{file=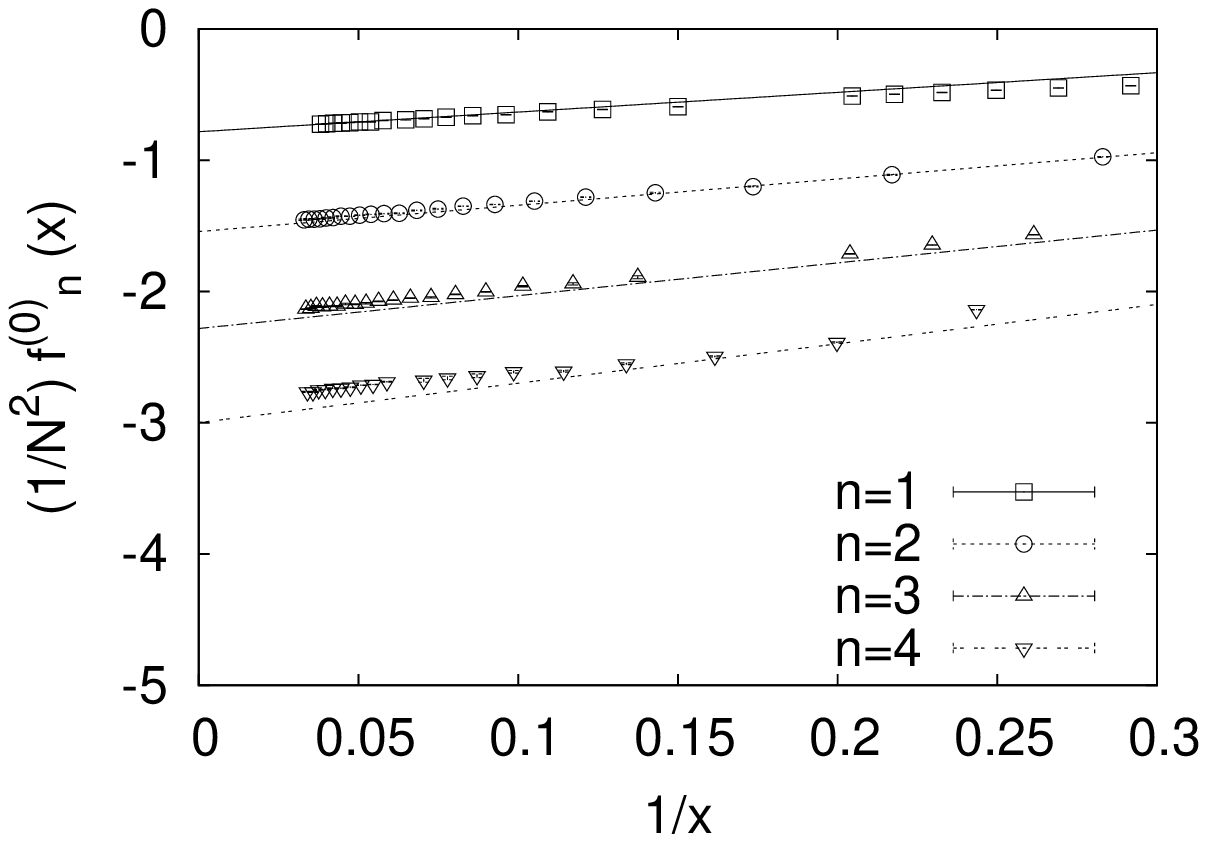,width=7.4cm} 
    \epsfig{file=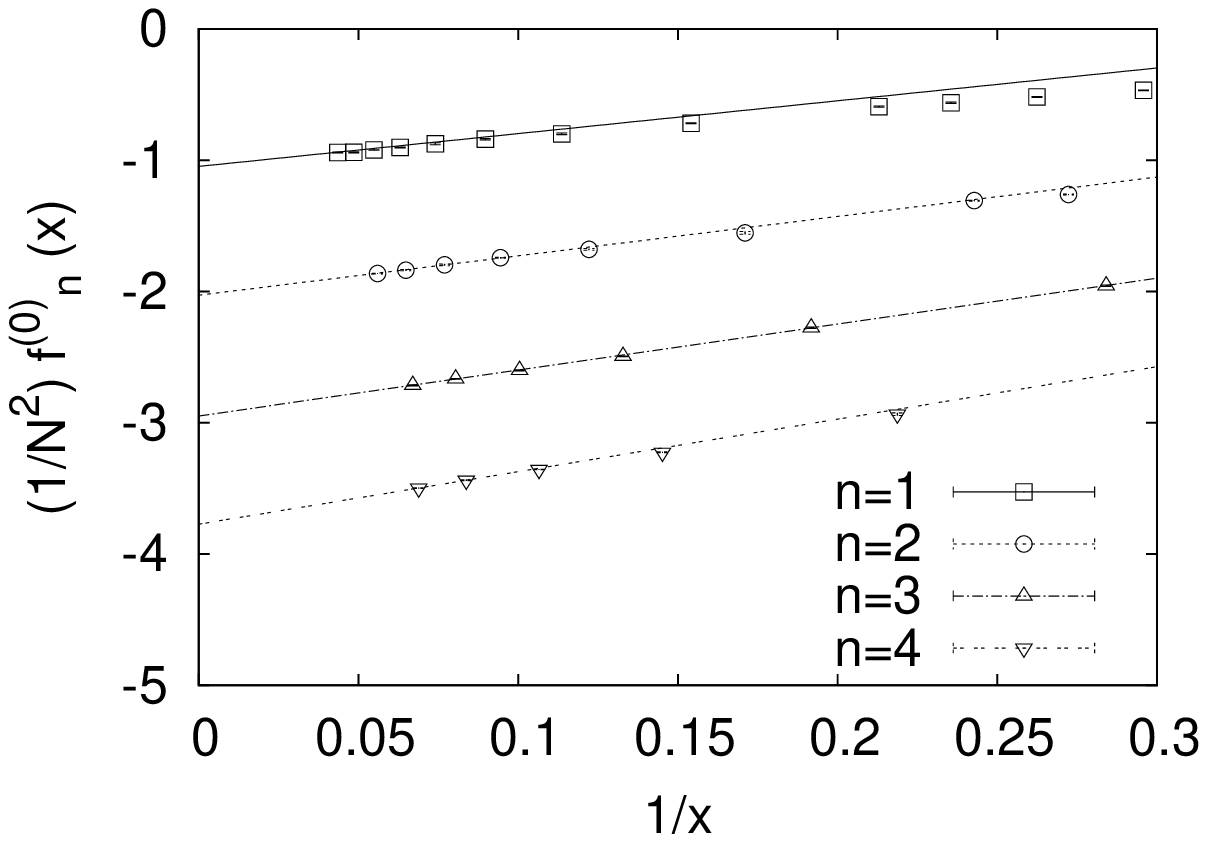,width=7.4cm}
    \caption{The function
      $ \frac{1}{N^{2}} f^{(0)}_{n}(x)$
      at $N=64$ is plotted against $1/x$ for $r=1$ 
      (Left) and $r=2$ (Right). The straight lines 
%are fits to the
represent the behavior \rf{power-law-f-largex}.
}
   \label{f-loglog_r2}
}
% %%%%%%%%%%%%%%%%%%%%%%%%%%%%%%%%%%%%%%%%%%%%%%%%%%%%%%%%%%%%%%%%%%%%%

At large $x$ we expect
\begin{equation}
 \frac{1}{N^2} f^{(0)}_{n} (x) 
  \simeq 
  - \frac{1}{2} n \vev{\lambda_n}_0 
%\left( 1 + \frac{r}{2} \right) 
+ \left( \frac{n}{2} + r  \right) \frac{1}{x}  \ .
%+ \frac{b_{n}}{x}  \ .
%  \hspace{2mm} (x\gg 1,n=1,2,3,4)\, . 
  \label{power-law-f-largex}
\end{equation}
As we increase $x$, the eigenvalues 
$\lambda_1,\cdots,\lambda_n$ are
forced to be large and the $(n+1)$-th through the $4$-th directions
remain relatively small. In that case it is the bosonic action that
dominates the suppression of those configurations and we expect that 
\beq
 \rho^{(0)}_n(x) 
\sim \exp{\left( -\Bigl\langle \frac{1}{2} N \tr \, (A_\mu)^2 
\Bigr\rangle_{n,V}
\right)} 
\sim \exp{\left( -\frac{1}{2} N^2  
      \sum_{k=1}^n \vev{\lambda_k}_{n,V} \right)} \ .
\eeq
Since
$\vev{\lambda_k}_{n,V}\sim x \vev{\lambda_n}_0$ at large $x$ for
$k=1,\cdots, n$,
we obtain the first term of \rf{power-law-f-largex},
which becomes $-\frac{1}{2}n(1+\frac{r}{2})$ at large $N$.
%% $ \frac{1}{N^2} f^{(0)}_n(x) \sim
%% \frac{1}{2} n x \vev{\lambda_n}_0  $.
%$ \frac{1}{N^2}\log \rho^{(0)}_n(x) \sim  $
%From the definition \rf{extrema}, 
We also expect a power-law correction 
$(\sqrt{x})^{(n+2 r) N^2} $
%(\sqrt{x})^{2 r N^2}$
to $\rho^{(0)}_n(x)$
that comes from the measure and the fermion determinant as we discussed
in the case of $x \ll 1$.
Thus we obtain the second term in \rf{power-law-f-largex}
as the subleading term.
Figures \ref{f-loglog} and \ref{f-loglog_r2} confirm 
the above asymptotic behaviors at small $x$ and large $x$, respectively.
%for $r=1$ and $r=2$.
% The $N=64$ case is displayed as an example.

%\section{Large-$N$ extrapolation of $w_n(x)$}
\section{Large-$N$ extrapolations in the single-observable analysis}
\label{sec:largeNextrapol}

In this section we explain the large-$N$ extrapolations
made
%for the function $w_n(x)$ 
in the single-observable analysis 
based on the asymptotic behavior \rf{ii-fit}.
%discussed in section \ref{section:mc}.

% %%%%%%%%%%%%%%%%%%%%%%%%%%%%%%%%%%%%%%%%%%%%%%%%%%%%%%%%%%%%%%%%%%%%%
\TABULAR[h]{|c|c|c|c|c|}{
\hline%----------------------------------------------------
    & \multicolumn{2}{|c|}{$r=1$} &\multicolumn{2}{|c|}{$r=2$}\\
\hline%----------------------------------------------------
$n$ & $c_{n}$      & $d_{n}$      & $c_{n}$      & $d_{n}$      \\
\hline%----------------------------------------------------
 1  &            & $2.26(6)$  &            &  $2.02(5)$  \\
 2  & $2.3(1)$  & $0.378(2) $ &  $2.407(2)$  &  $0.53(2)$ \\
% 3  & 0.4055(80) & 0.2426(25) & 0.6760(45) &  0.3852(61) \\
% 3  & $0.52(4)$ & $0.246(2)$ &  <--- this is wrong (2011.7.29)
 3  & $0.52(4)$ & $0.20(3)$ & 
$0.96(6)$ & $0.353(3) $ \\
 4  & $0.26(4)$ &            & $0.46(8)$ &             \\
\hline%----------------------------------------------------
%% \hline%----------------------------------------------------
%%     & \multicolumn{2}{|c|}{$r=1$} &\multicolumn{2}{|c|}{$r=2$}\\
%% \hline%----------------------------------------------------
%% $n$ & $c_{n}$      & $d_{n}$      & $c_{n}$      & $d_{n}$      \\
%% \hline%----------------------------------------------------
%%  1  &            & $2.2563 \pm 0.05679$  &            &  $2.02256 \pm 0.051$  \\
%%  2  & $2.30499 \pm 0.121$  & $0.3777 \pm 0.001723$ &  $2.40698 \pm 0.00233$  &  $0.526483 \pm 0.02246$ \\
%% % 3  & 0.4055(80) & 0.2426(25) & 0.6760(45) &  0.3852(61) \\
%%  3  & $0.515564 \pm 0.03909$ & $0.24619 \pm 0.00182$ & $0.962089 \pm 0.05777$ & $0.353374 \pm 0.002736$ \\
%%  4  & $0.257238 \pm 0.03668$ &            & $0.461022 \pm 0.08255$ &             \\
%% \hline%----------------------------------------------------
}
{Large-$N$ extrapolated values of
the coefficients $c_{n}$ and $d_{n}$ in the 
asymptotic behaviors \rf{ii-fit}. 
%scaling relations
\label{tab:cd}
}
% %%%%%%%%%%%%%%%%%%%%%%%%%%%%%%%%%%%%%%%%%%%%%%%%%%%%%%%%%%%%%%%%%%%%%

\begin{figure}[H]
\begin{center}

\begin{minipage}{.50\linewidth}
\includegraphics[width=0.9 \linewidth]{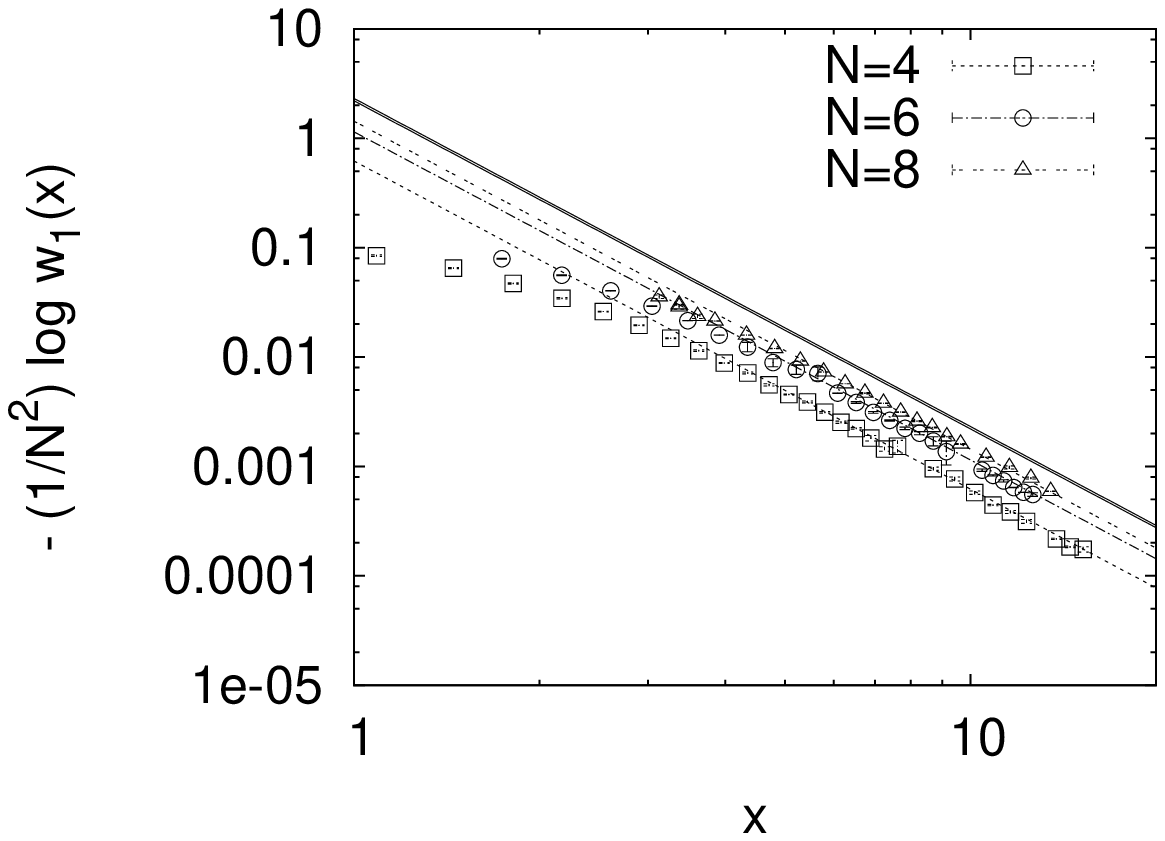}
  \end{minipage}
%  \end{center}
%  \hspace{1.0pc}
%\begin{center}
%\label{fig1}

\vspace{1.0pc}

\begin{tabular}{cc}

\begin{minipage}{.50\linewidth}
\includegraphics[width=0.9 \linewidth]{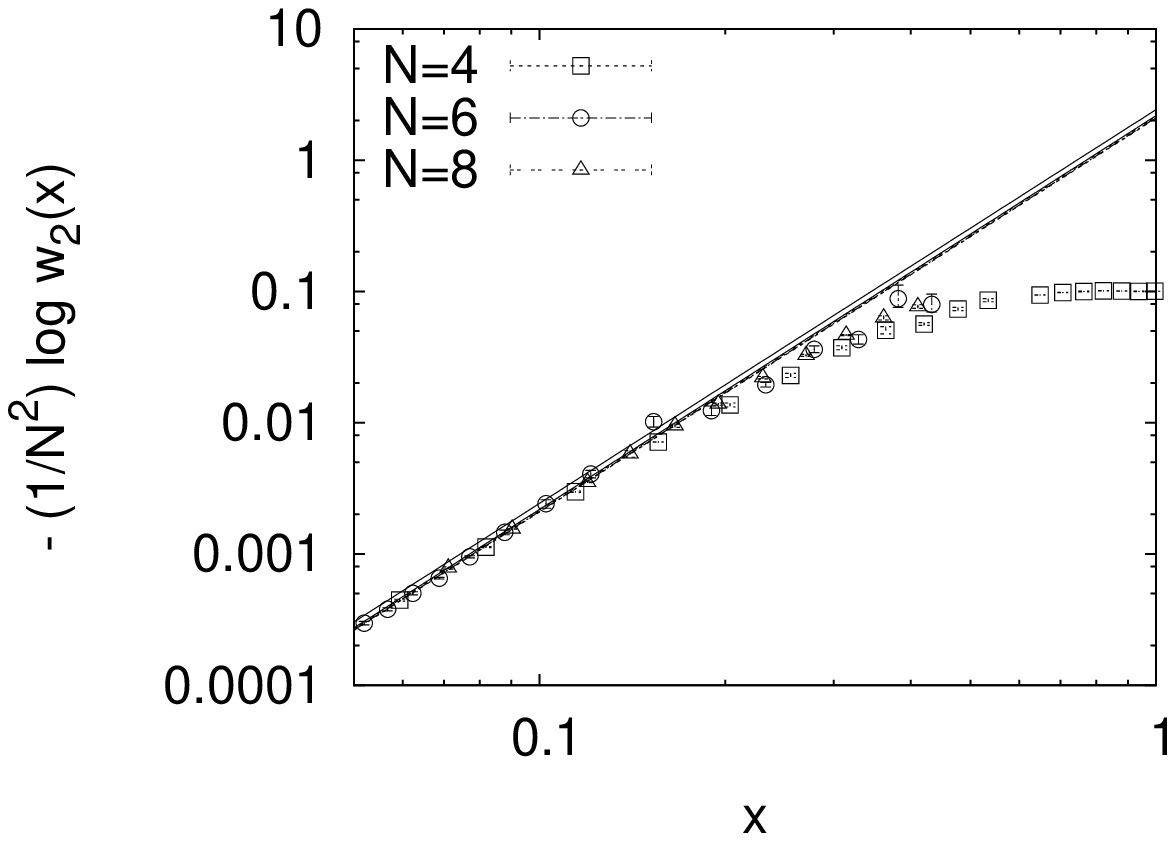}
  \end{minipage}
%  \end{center}
%  \hspace{1.0pc}
%\begin{center}
%\label{fig1}

\begin{minipage}{.50\linewidth}
\includegraphics[width=0.9 \linewidth]{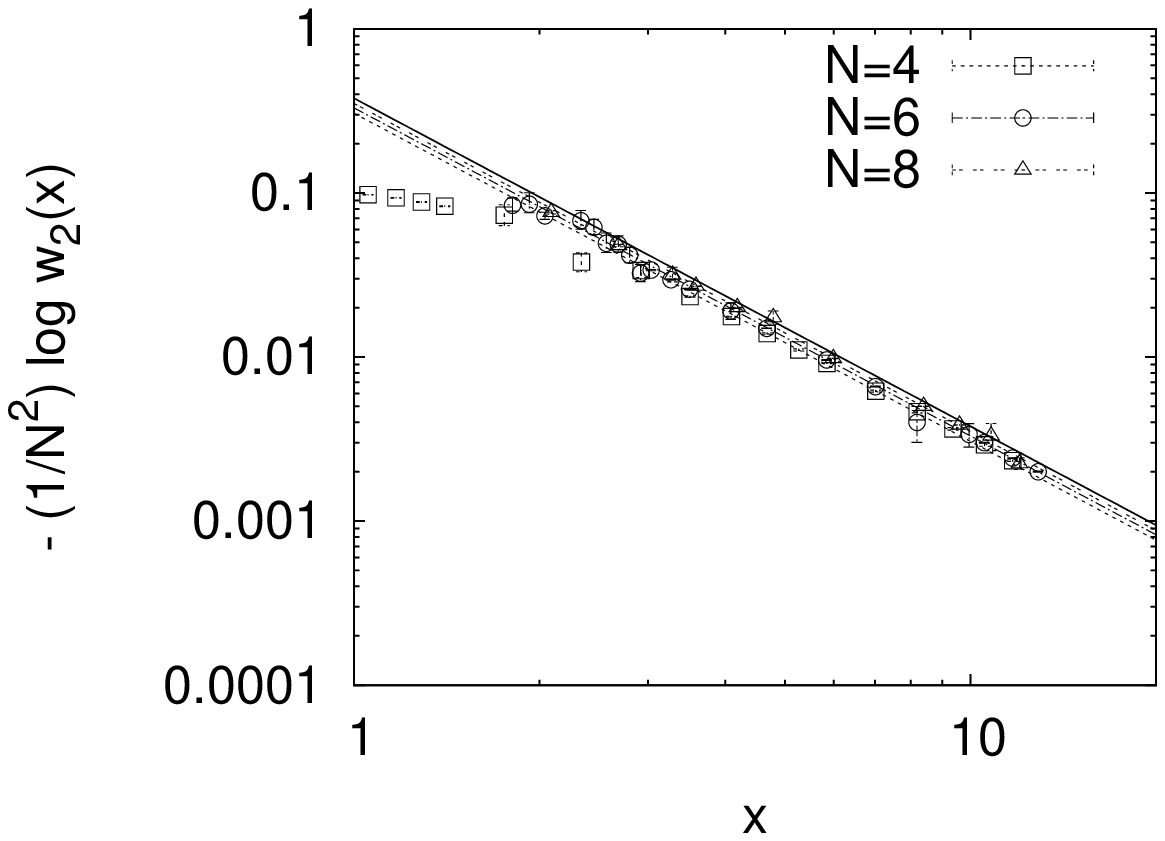}
  \end{minipage}

\end{tabular}

\vspace{1.0pc}

\begin{tabular}{cc}

\begin{minipage}{0.5\linewidth}
\includegraphics[width=0.9 \linewidth]{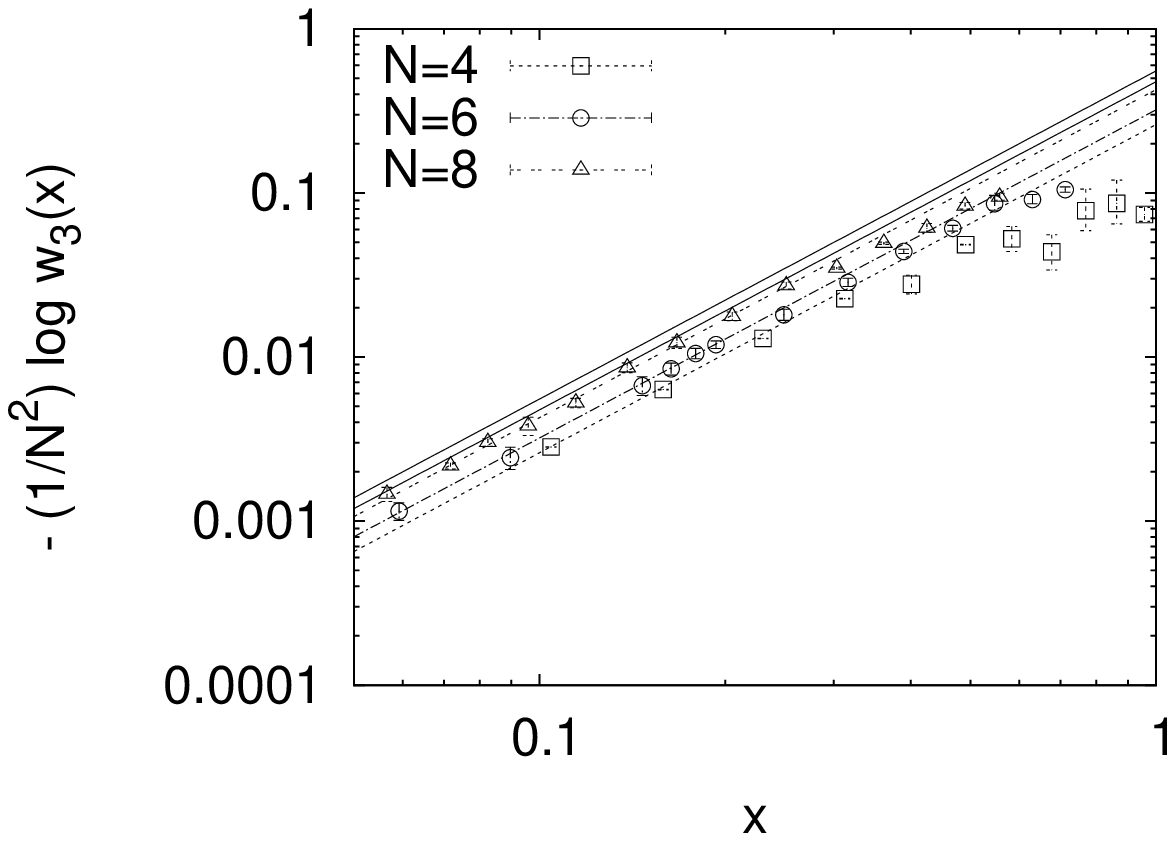}
  \end{minipage}

\begin{minipage}{0.5\linewidth}
\includegraphics[width=0.9 \linewidth]{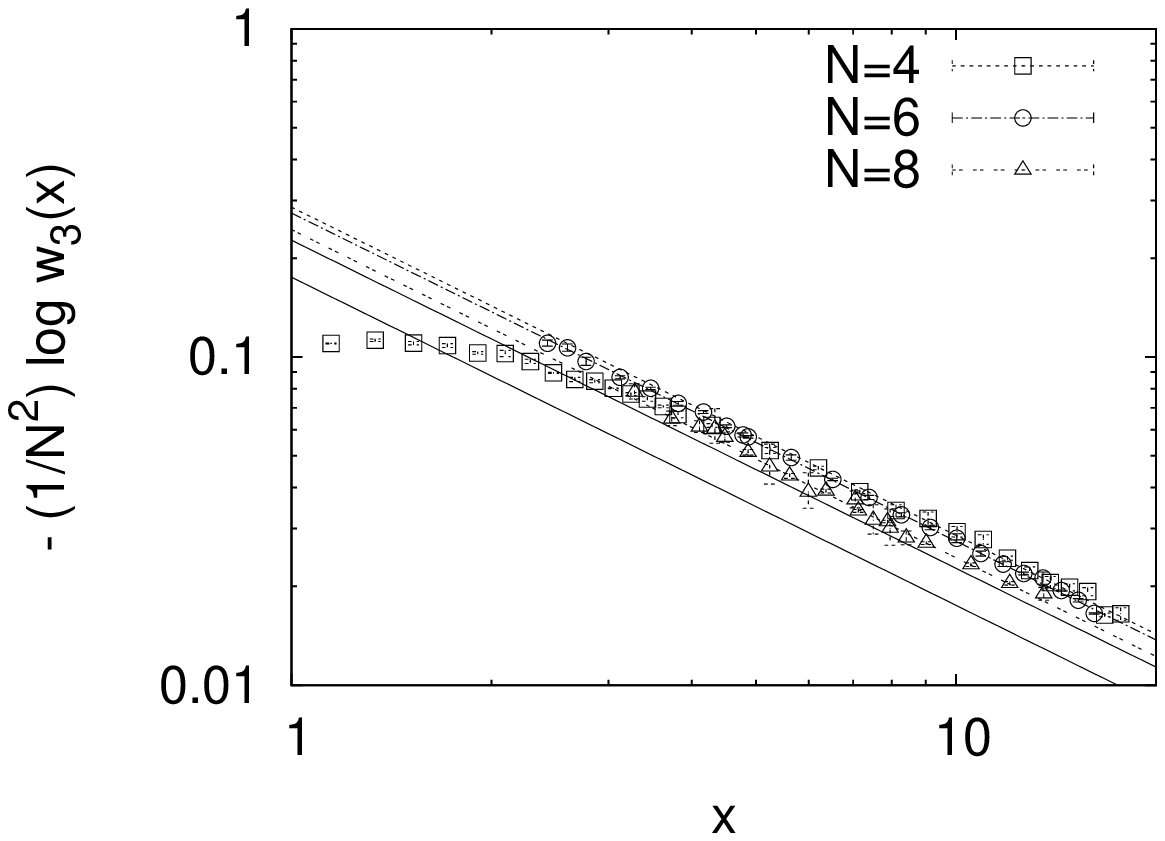}
  \end{minipage}

\end{tabular}

\vspace{1.0pc}

\begin{minipage}{0.5\linewidth}
\includegraphics[width=0.9 \linewidth]{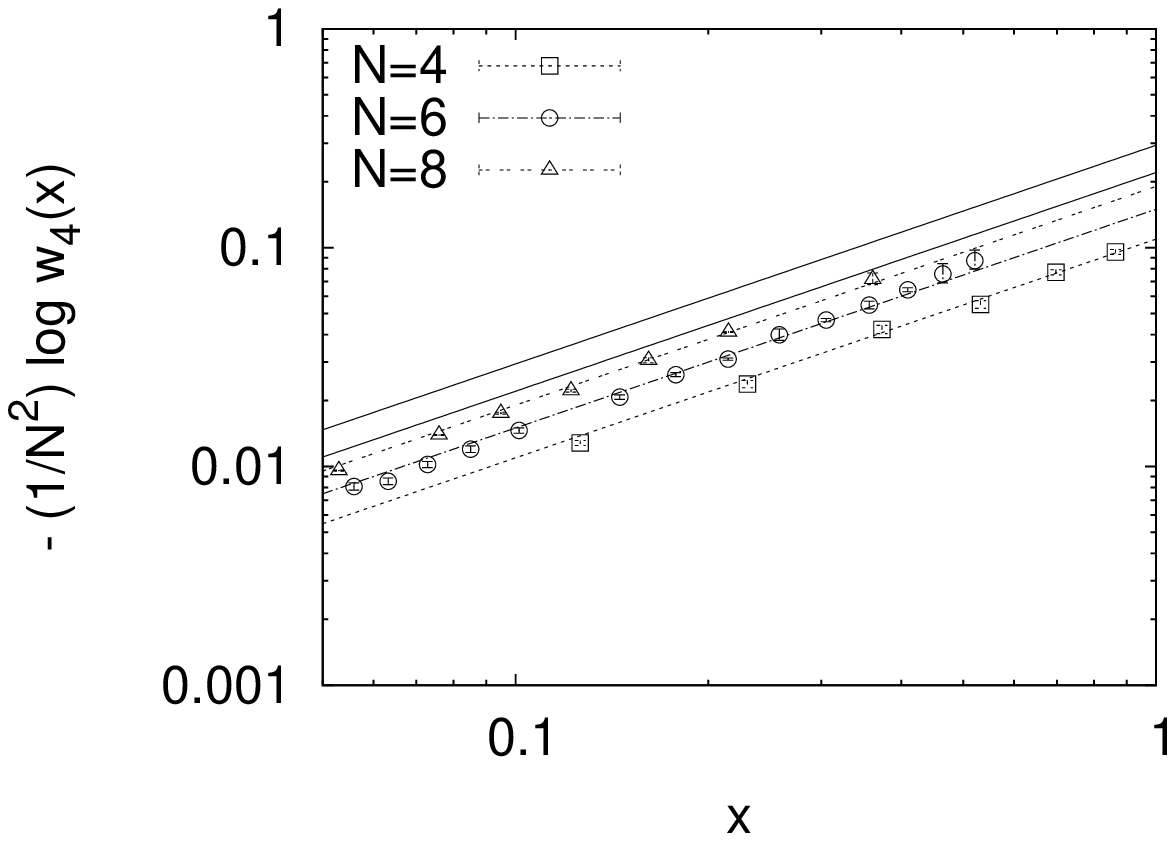}
  \end{minipage}

\end{center}
\caption{The log-log plot of $- \frac{1}{N^2} \log w_n (x)$ for $r=1$.
%which confirms the asymptotic behavior (\ref{ii-fit}).
The straight solid lines represent the 
%asymptotic predicted 
power-law behavior 
(\ref{ii-fit}) with the coefficients presented 
in Table \ref{tab:cd}, which are 
obtained by the large-$N$ extrapolation from the $N=4,6,8$ data
as described in fig.~\ref{cdn_ext_r1}.
A clear trend towards large-$N$ scaling is observed,
in particular, for the $n=2,3$ cases.
}
   \label{res_r1}
  \end{figure}

\begin{figure}[H]
\begin{center}

\begin{minipage}{.50\linewidth}
\includegraphics[width=0.9 \linewidth]{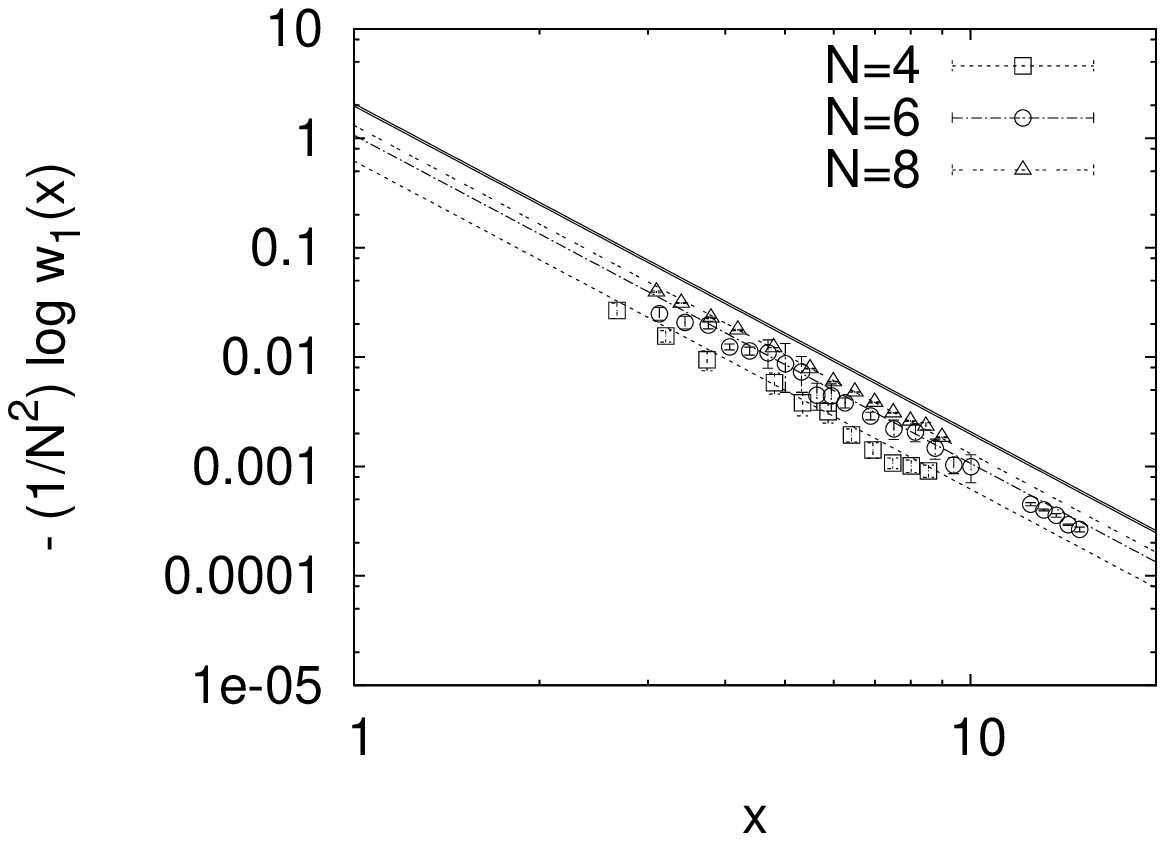}
  \end{minipage}
%  \end{center}
%  \hspace{1.0pc}
%\begin{center}
%\label{fig1}

\vspace{1.0pc}

\begin{tabular}{cc}

\begin{minipage}{.50\linewidth}
\includegraphics[width=0.9 \linewidth]{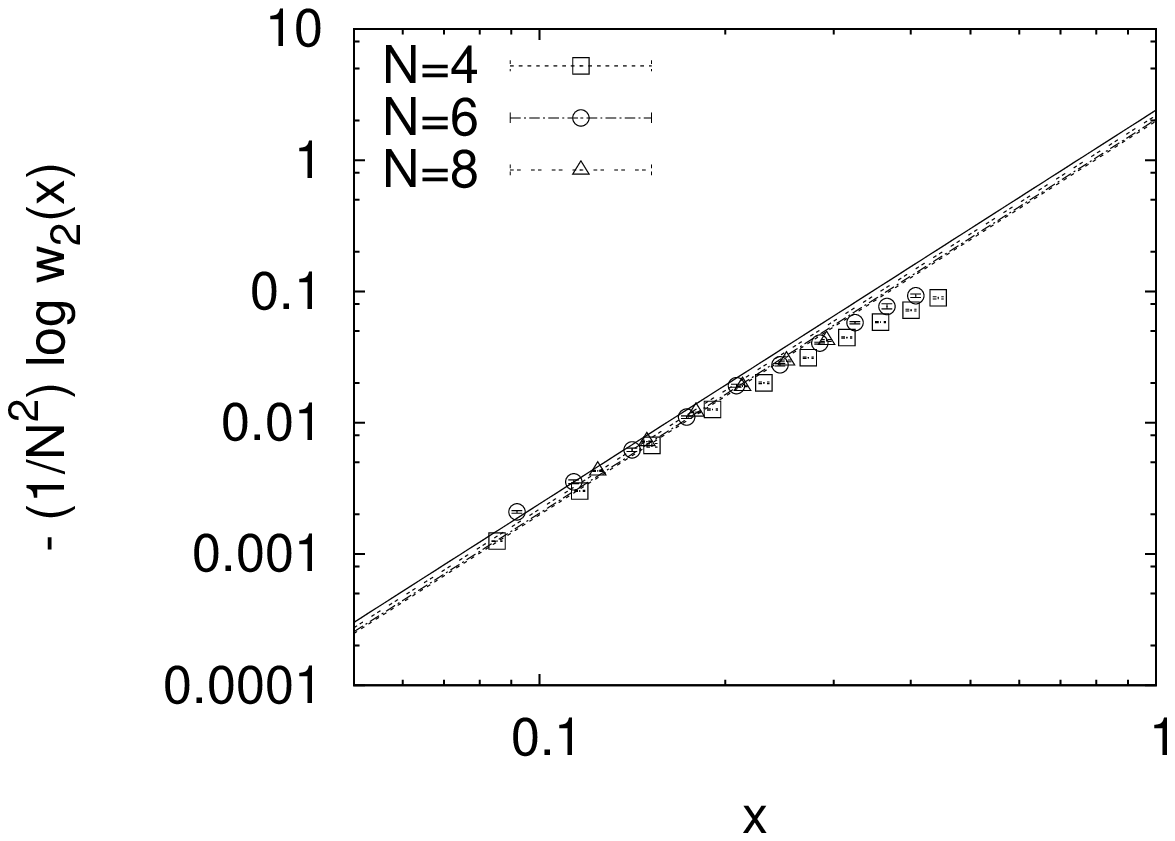}
  \end{minipage}
%  \end{center}
%  \hspace{1.0pc}
%\begin{center}
%\label{fig1}

\begin{minipage}{.50\linewidth}
\includegraphics[width=0.9 \linewidth]{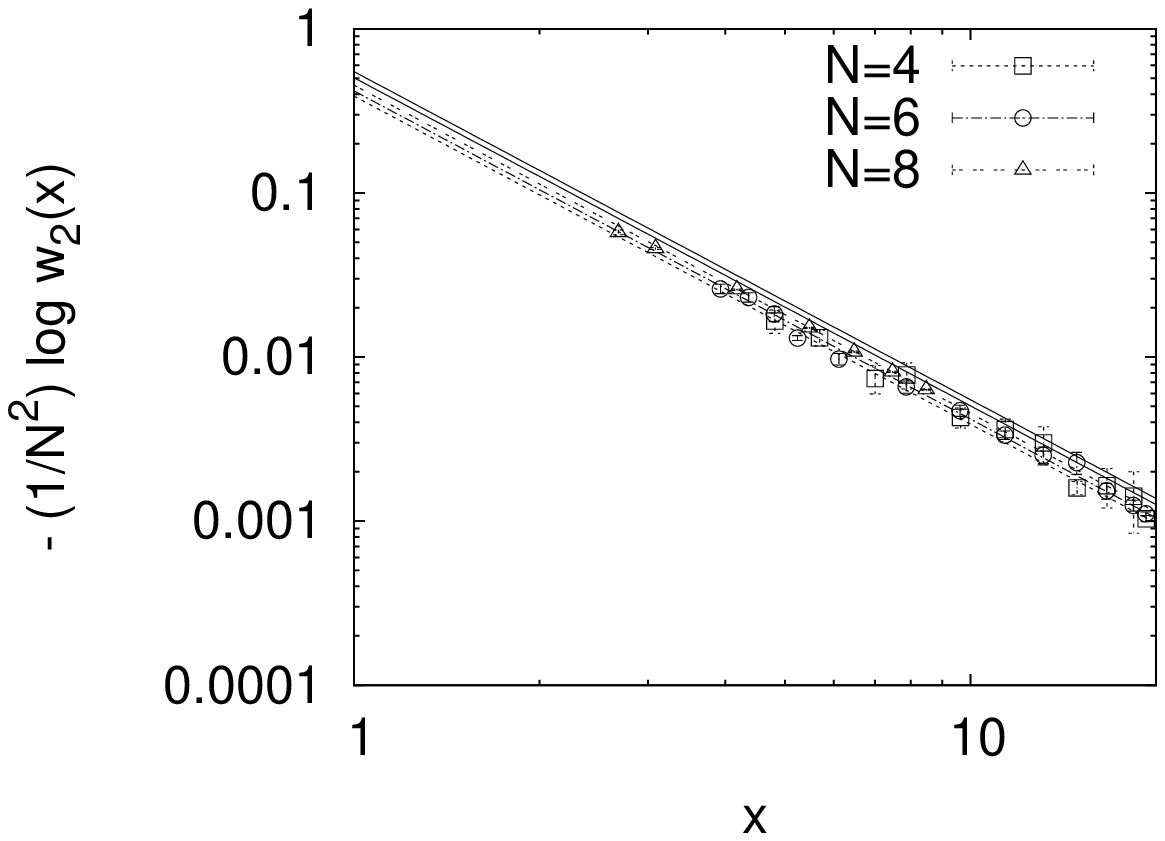}
  \end{minipage}

\end{tabular}

\vspace{1.0pc}

\begin{tabular}{cc}

\begin{minipage}{0.5\linewidth}
\includegraphics[width=0.9 \linewidth]{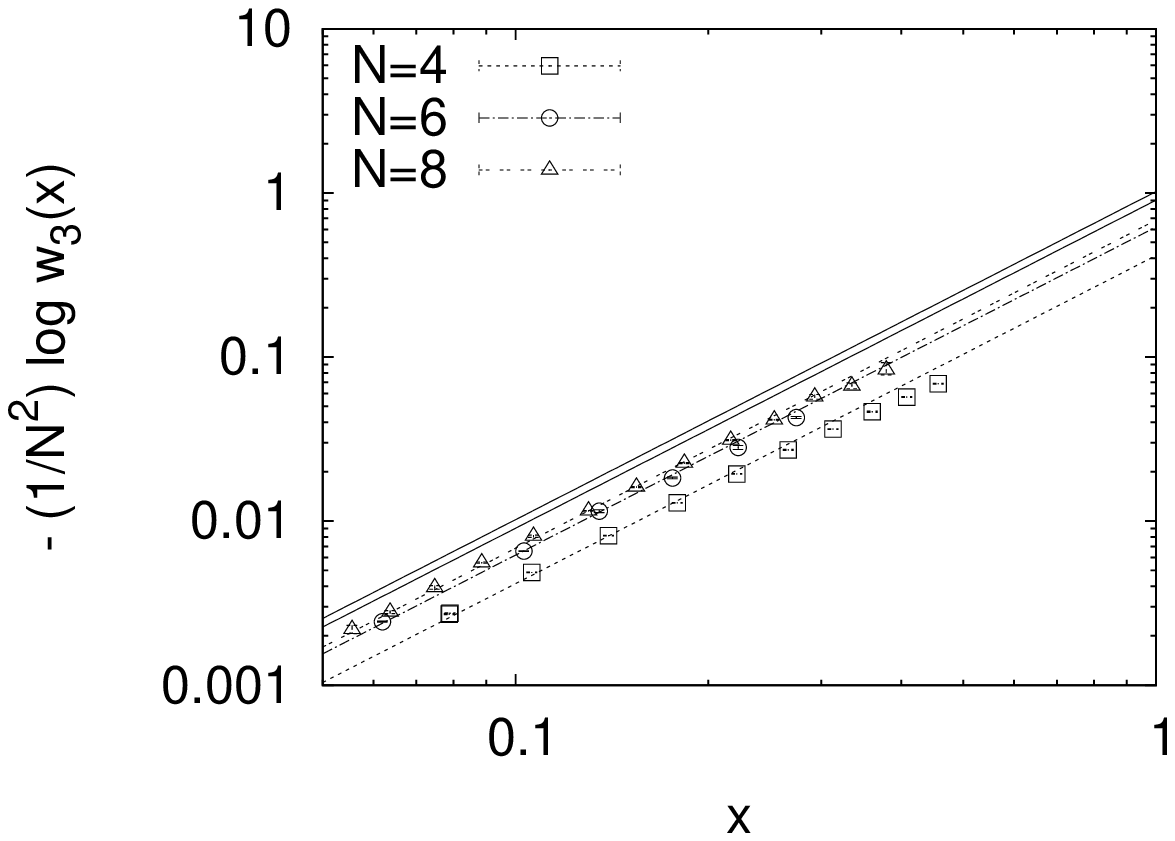}
  \end{minipage}

\begin{minipage}{0.5\linewidth}
\includegraphics[width=0.9 \linewidth]{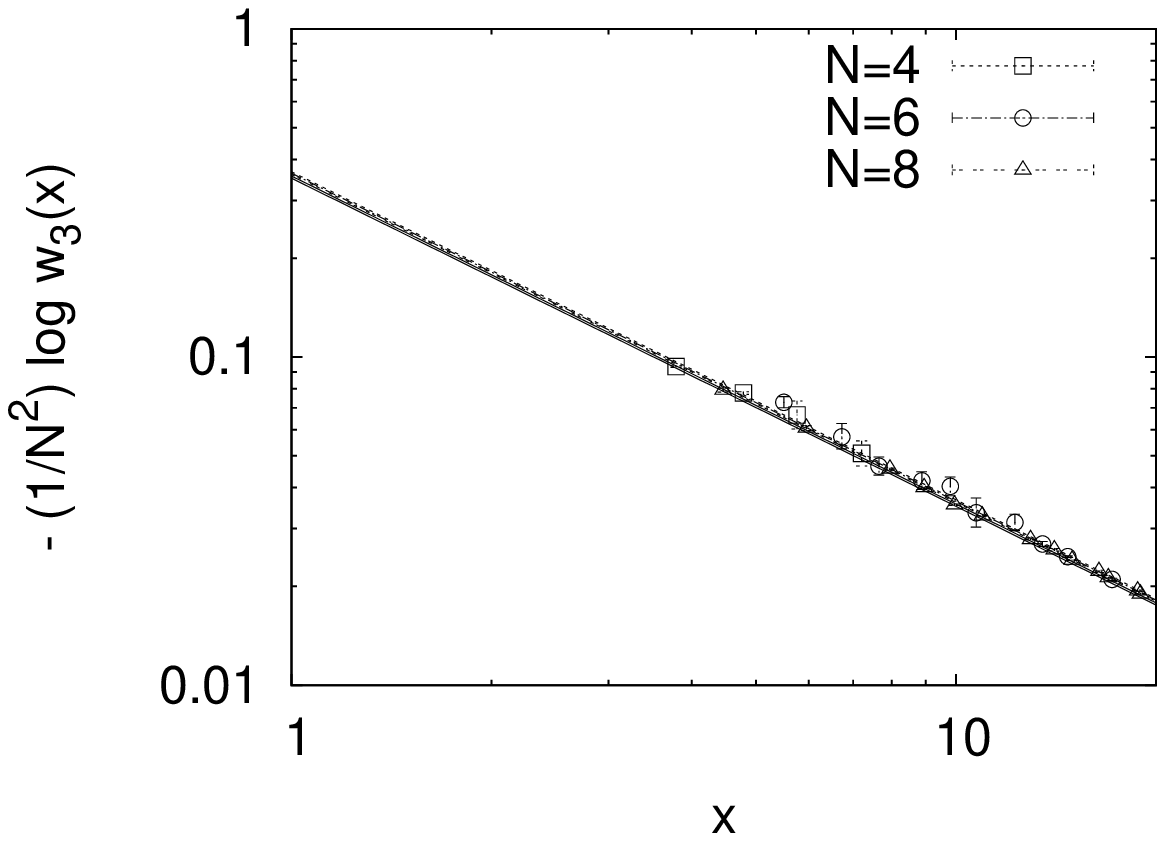}
  \end{minipage}

\end{tabular}

\vspace{1.0pc}

\begin{minipage}{0.5\linewidth}
\includegraphics[width=0.9 \linewidth]{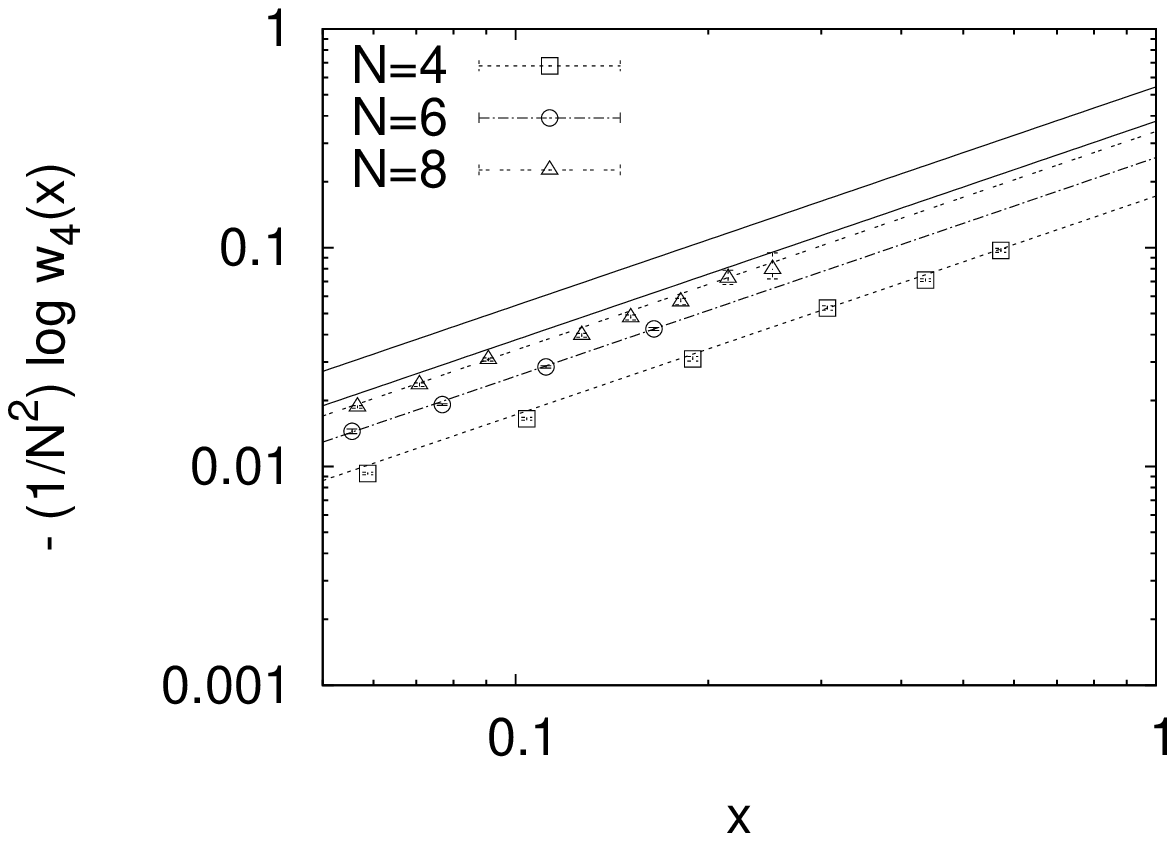}
  \end{minipage}

\end{center}
\caption{
The log-log plot of $- \frac{1}{N^2} \log w_n (x)$ for $r=2$.
%which confirms the asymptotic behavior (\ref{ii-fit}).
The straight solid lines represent the 
%asymptotic predicted 
power-law behavior 
(\ref{ii-fit}) with the coefficients presented 
in Table \ref{tab:cd}, which are 
obtained by the large-$N$ extrapolation from the $N=4,6,8$ data
as described in fig.~\ref{cdn_ext_r2}.
A clear trend towards large-$N$ scaling is observed,
in particular, for the $n=2,3$ cases.
%The same as fig.~\ref{res_r1} but for $r=2$.
%
%except for $n=3$ and $x>1$, where we used $N=6$.
%% The effect of the phase
%% $- \frac{1}{N^2} \log w_n (x)$ is plotted against $x$
%% in the log-log scale for $r=2$.
}
   \label{res_r2}
  \end{figure}

\begin{figure}[H]
\begin{center}

\begin{minipage}{.50\linewidth}
\includegraphics[width=0.9 \linewidth]{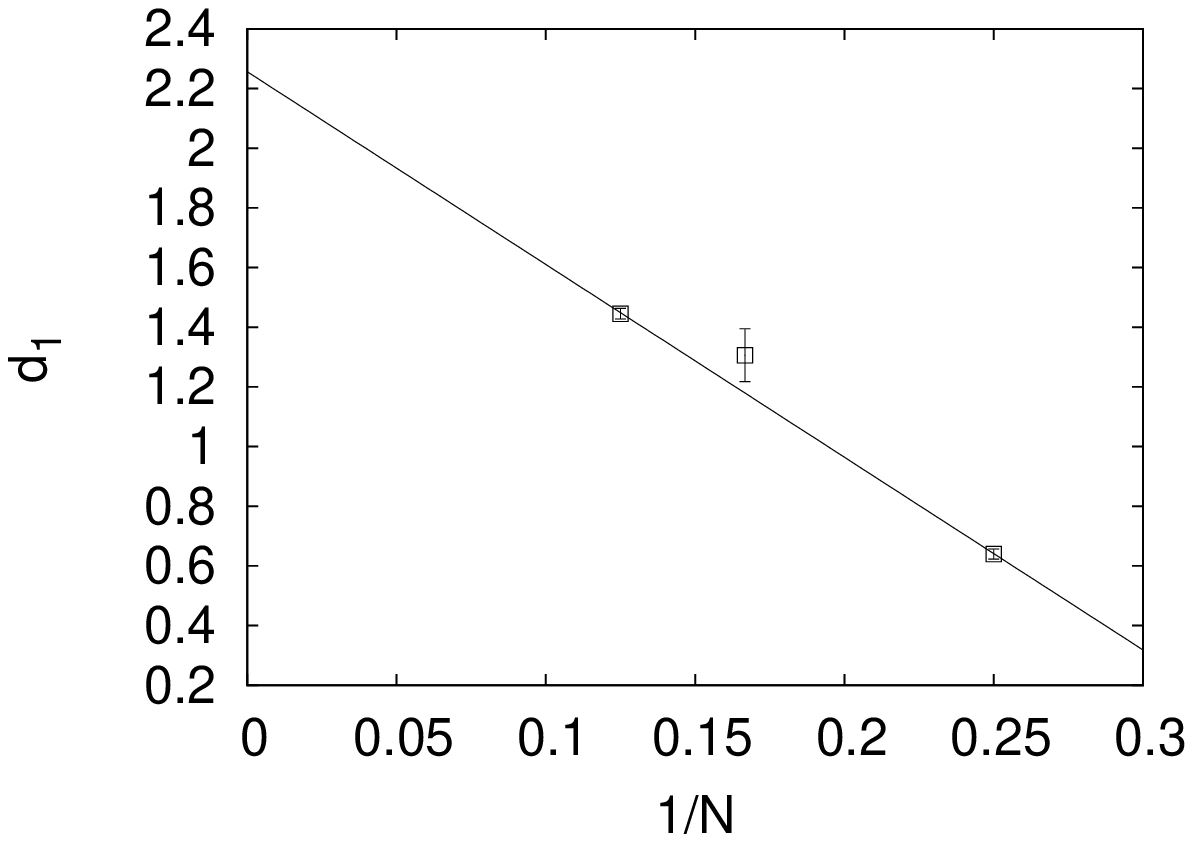}
  \end{minipage}
%  \end{center}
%  \hspace{1.0pc}
%\begin{center}
%\label{fig1}

\vspace{1.0pc}

\begin{tabular}{cc}

\begin{minipage}{.50\linewidth}
\includegraphics[width=0.9 \linewidth]{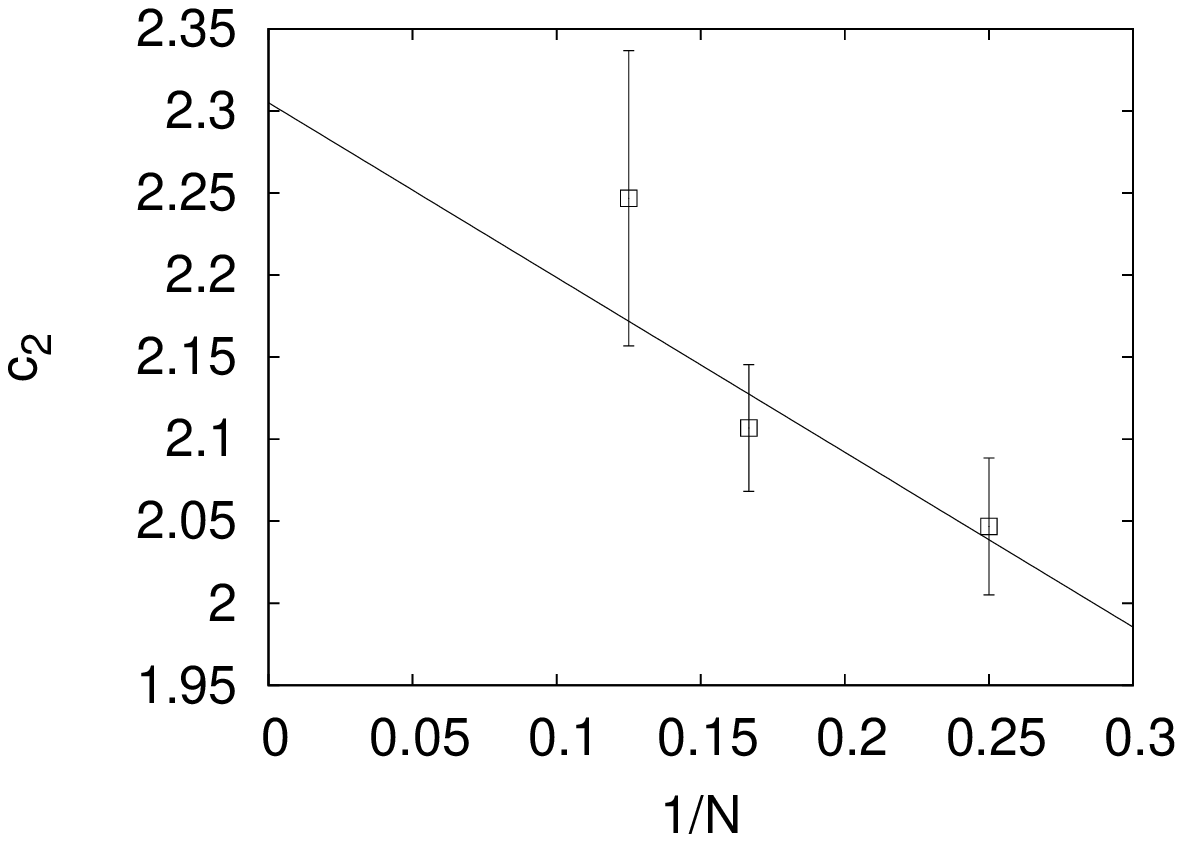}
  \end{minipage}
%  \end{center}
%  \hspace{1.0pc}
%\begin{center}
%\label{fig1}

\begin{minipage}{.50\linewidth}
\includegraphics[width=0.9 \linewidth]{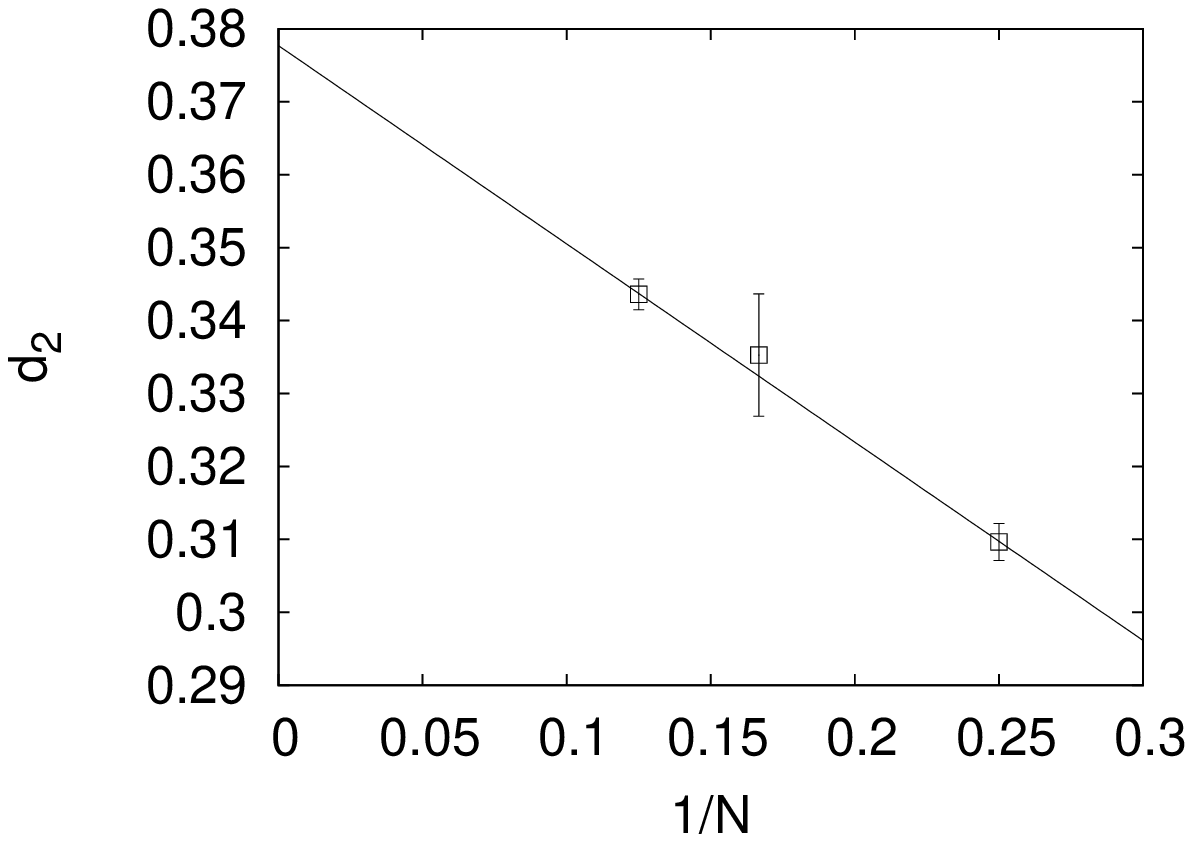}
  \end{minipage}

\end{tabular}

\vspace{1.0pc}

\begin{tabular}{cc}

\begin{minipage}{0.5\linewidth}
\includegraphics[width=0.9 \linewidth]{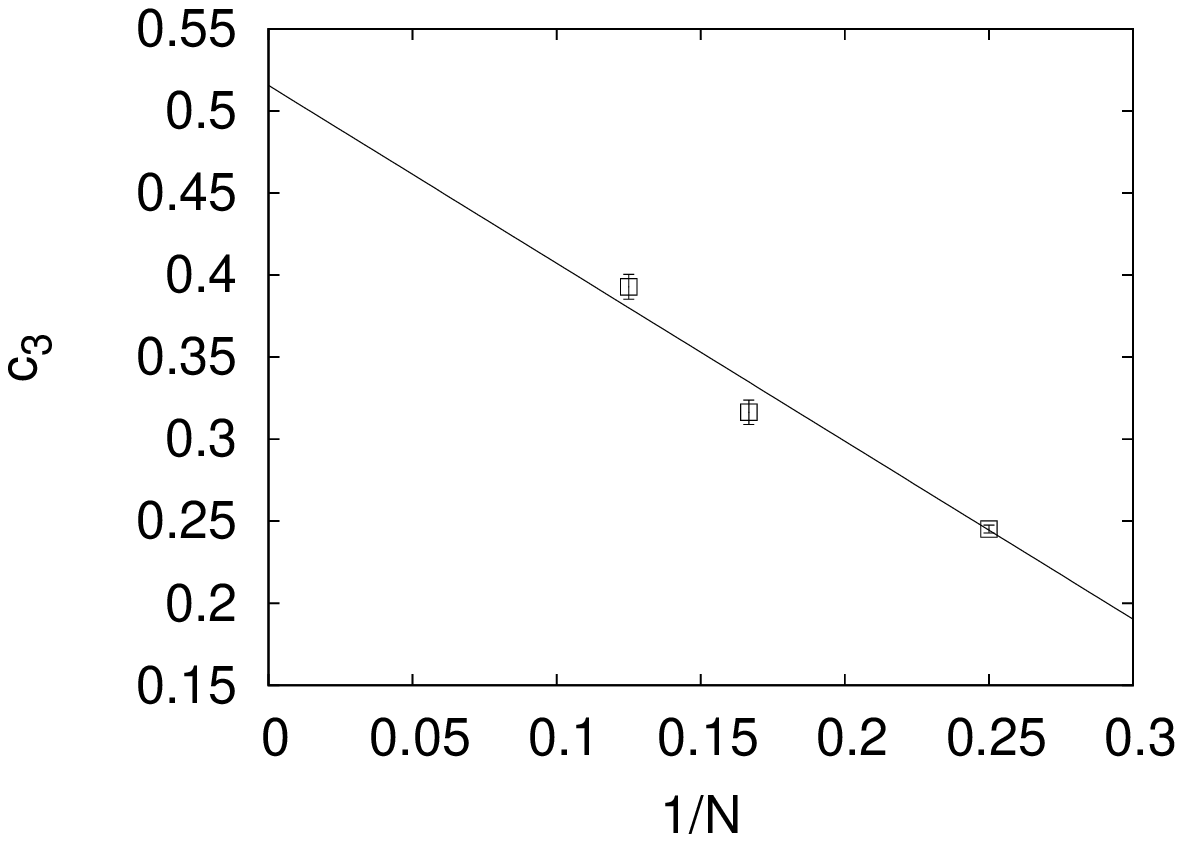}
  \end{minipage}

\begin{minipage}{0.5\linewidth}
\includegraphics[width=0.9 \linewidth]{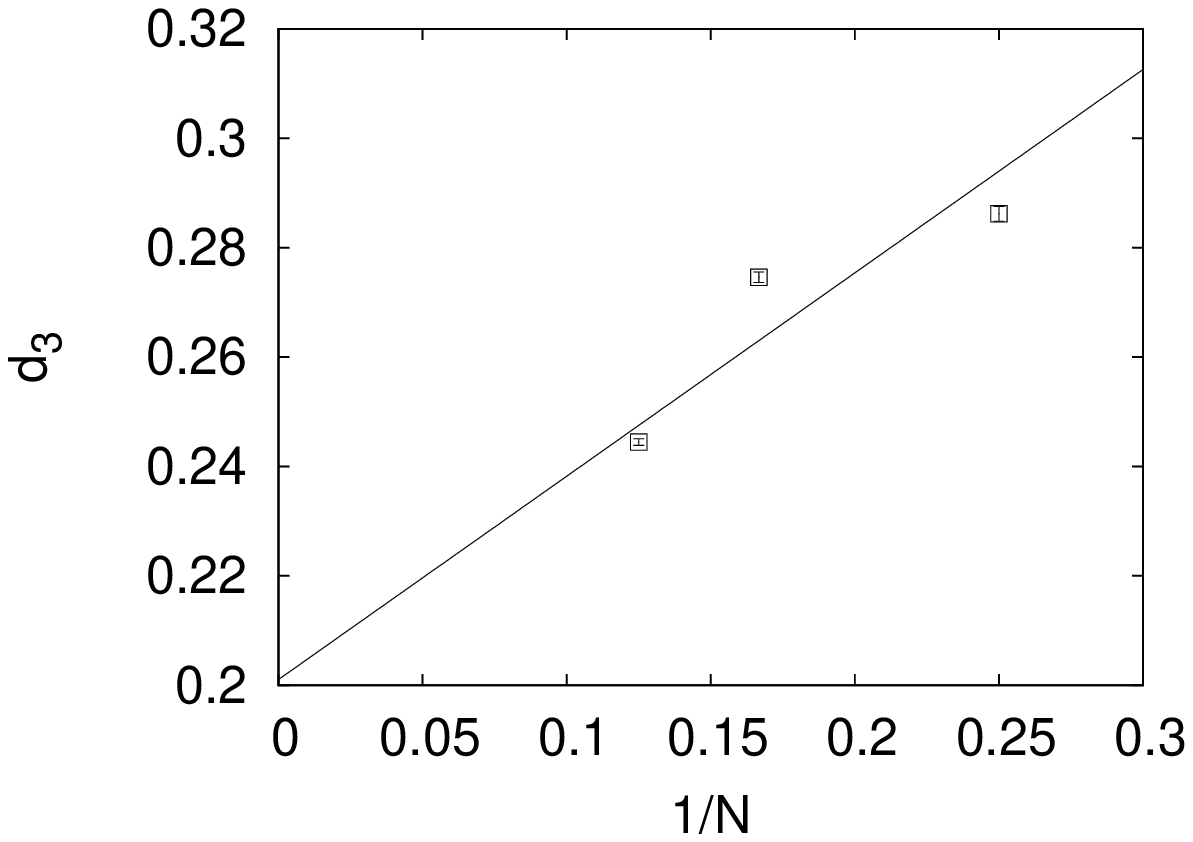}
  \end{minipage}

\end{tabular}

\vspace{1.0pc}

\begin{minipage}{0.5\linewidth}
\includegraphics[width=0.9 \linewidth]{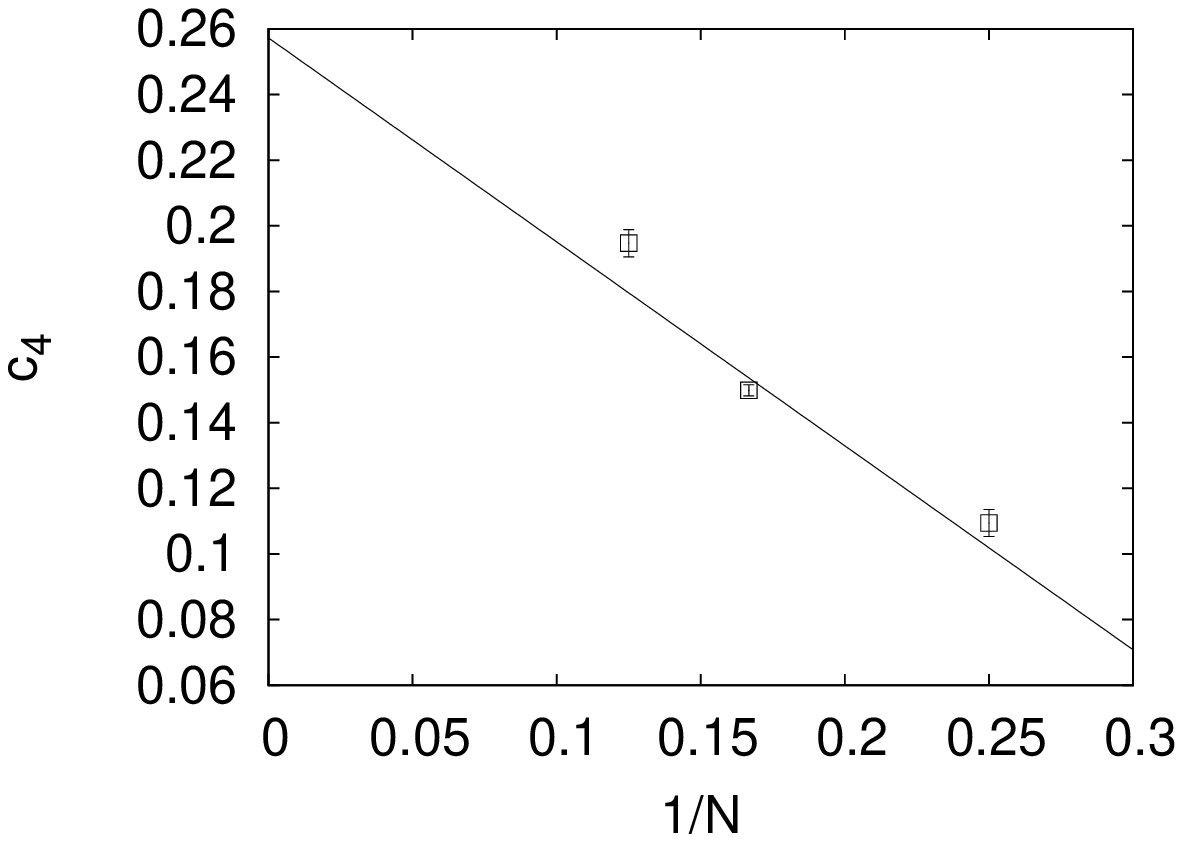}
  \end{minipage}

\end{center}
\caption{The coefficients $c_n$ and $d_n$
in the asymptotic formula (\ref{ii-fit}) for $r=1$
extracted from fig.~\ref{res_r1} are plotted against $\frac{1}{N}$.
The data points are fitted to a straight line.
The large-$N$ extrapolated values obtained in this way are presented
in Table \ref{tab:cd}.
}
   \label{cdn_ext_r1}
  \end{figure}

\begin{figure}[H]
\begin{center}

\begin{minipage}{.50\linewidth}
\includegraphics[width=0.9 \linewidth]{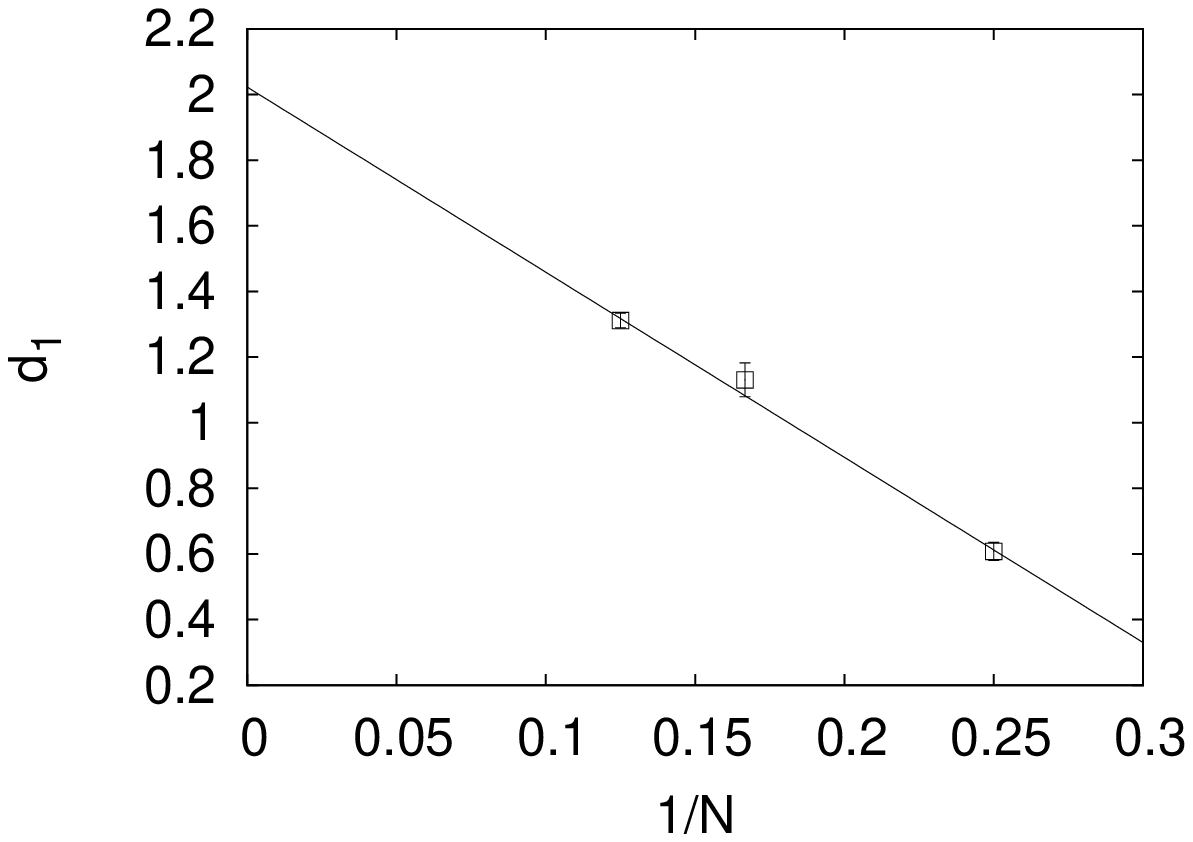}
  \end{minipage}
%  \end{center}
%  \hspace{1.0pc}
%\begin{center}
%\label{fig1}

\vspace{1.0pc}

\begin{tabular}{cc}

\begin{minipage}{.50\linewidth}
\includegraphics[width=0.9 \linewidth]{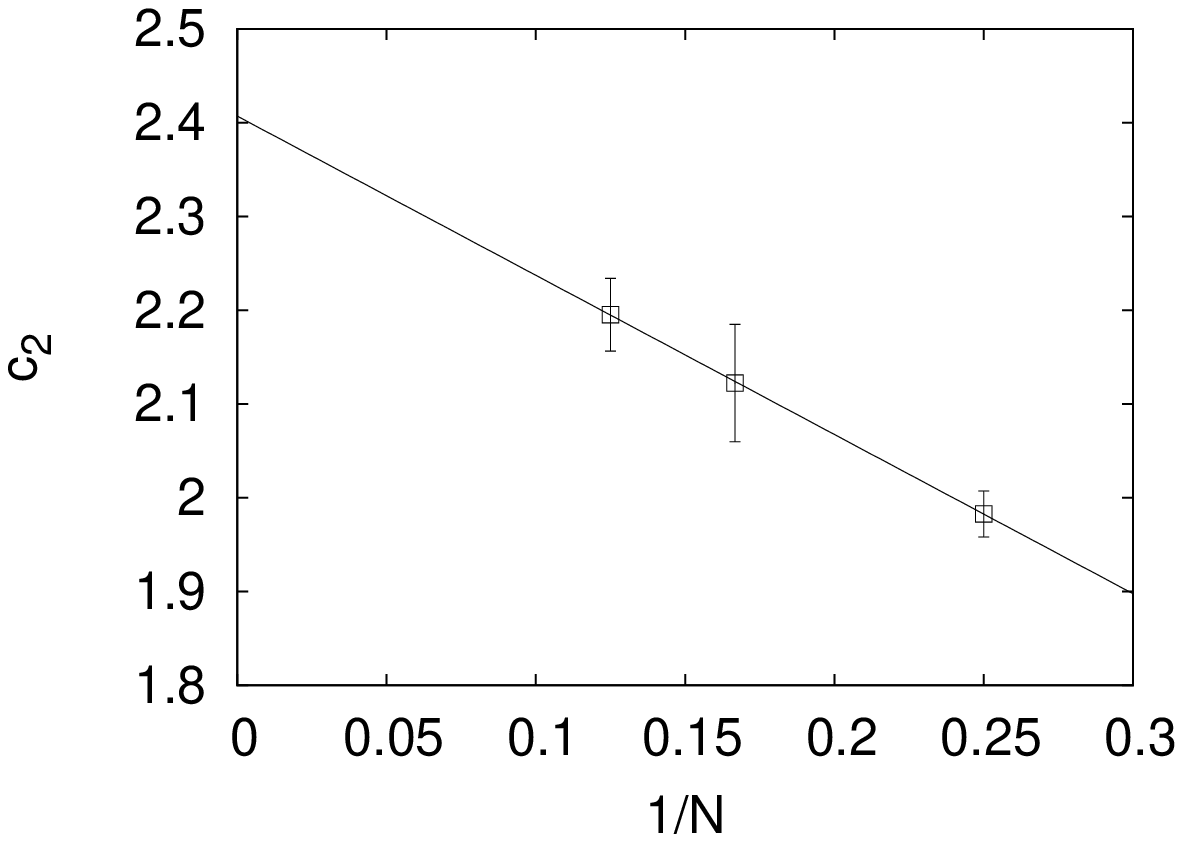}
  \end{minipage}
%  \end{center}
%  \hspace{1.0pc}
%\begin{center}
%\label{fig1}

\begin{minipage}{.50\linewidth}
\includegraphics[width=0.9 \linewidth]{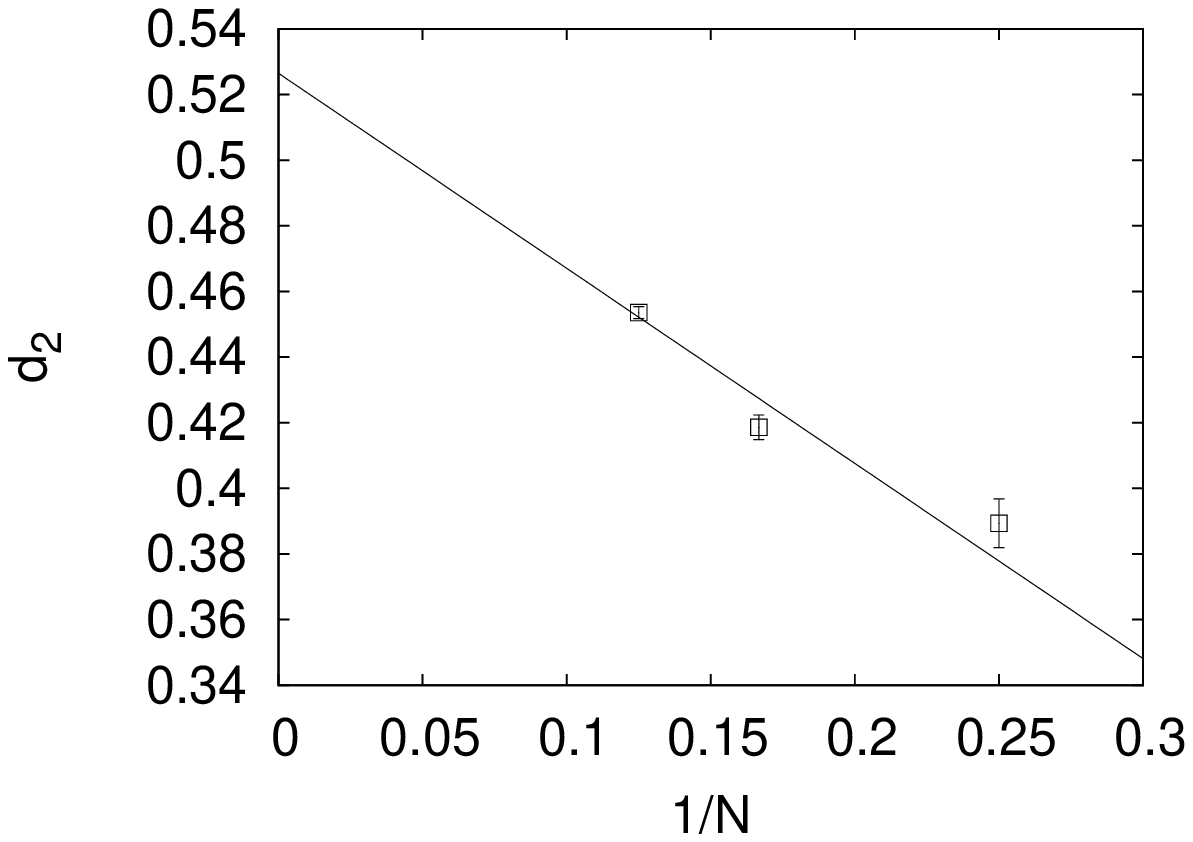}
  \end{minipage}

\end{tabular}

\vspace{1.0pc}

\begin{tabular}{cc}

\begin{minipage}{0.5\linewidth}
\includegraphics[width=0.9 \linewidth]{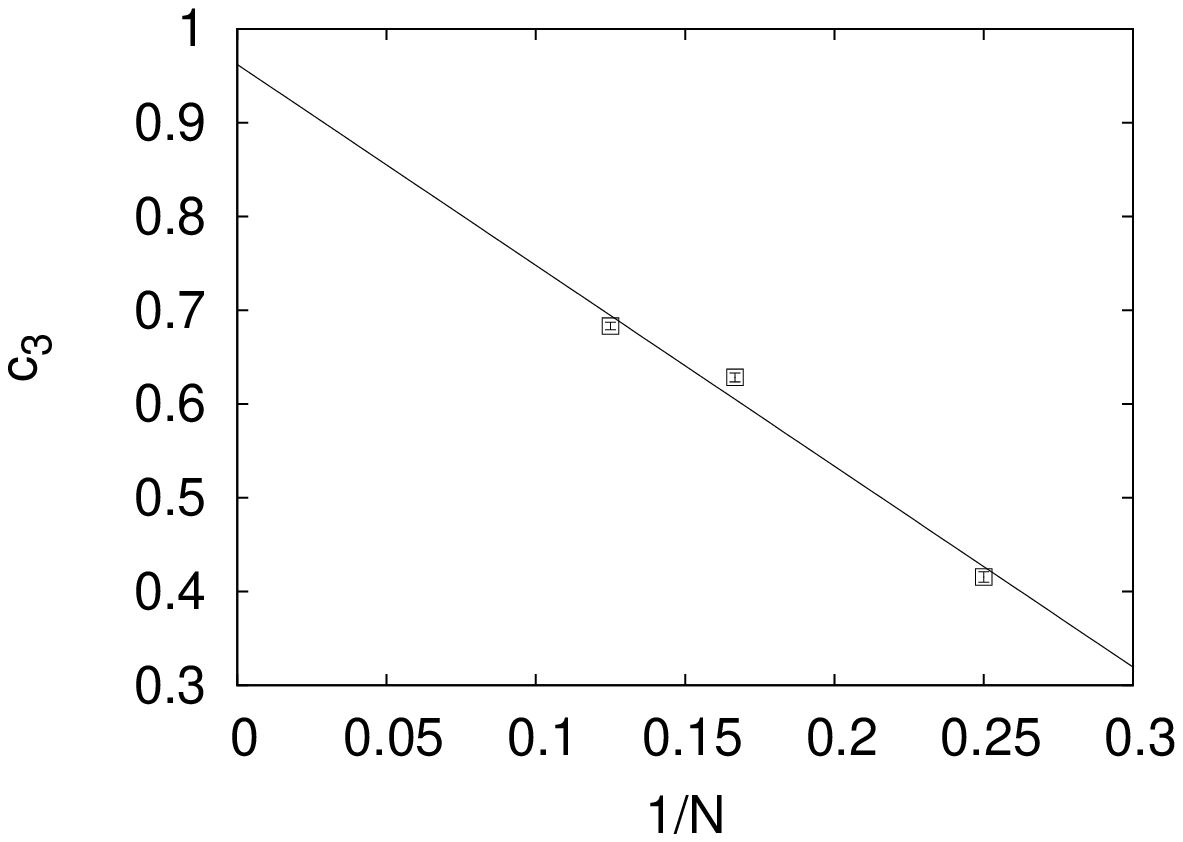}
  \end{minipage}

\begin{minipage}{0.5\linewidth}
\includegraphics[width=0.9 \linewidth]{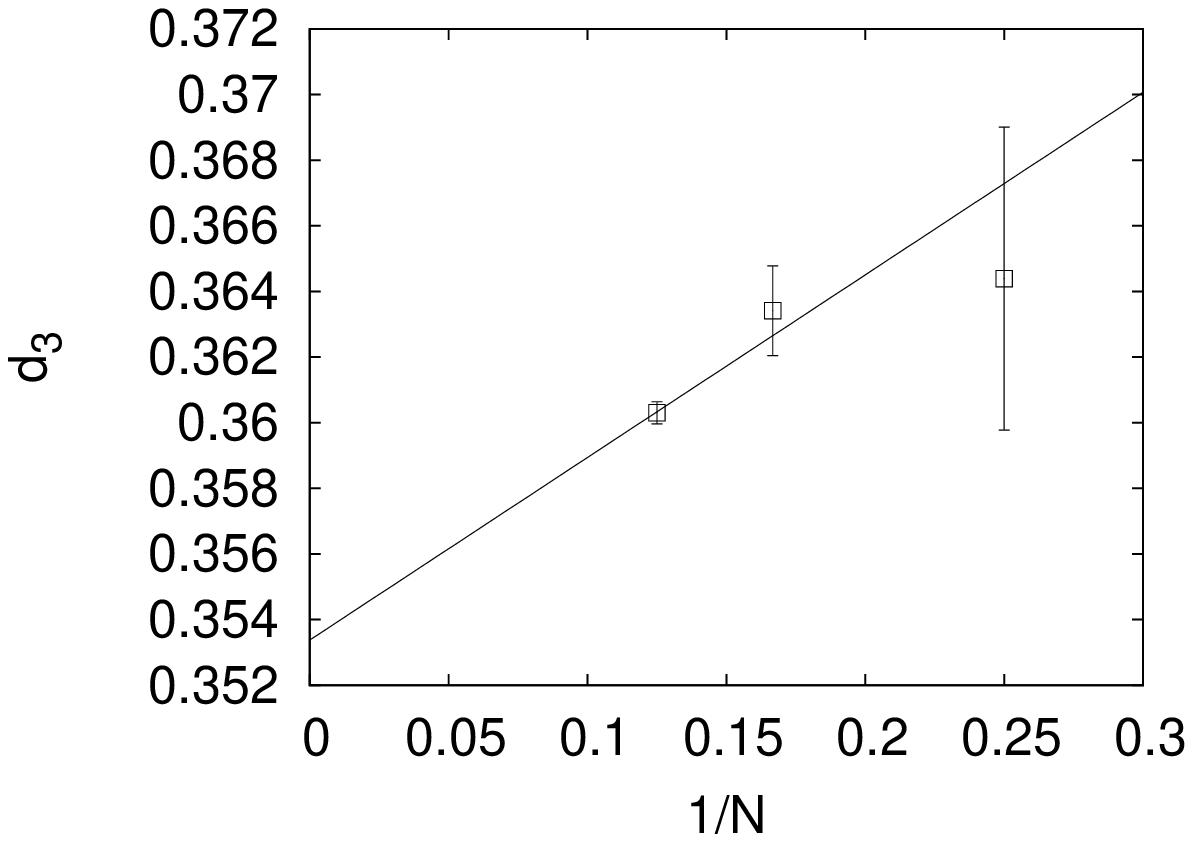}
  \end{minipage}

\end{tabular}

\vspace{1.0pc}

\begin{minipage}{0.5\linewidth}
\includegraphics[width=0.9 \linewidth]{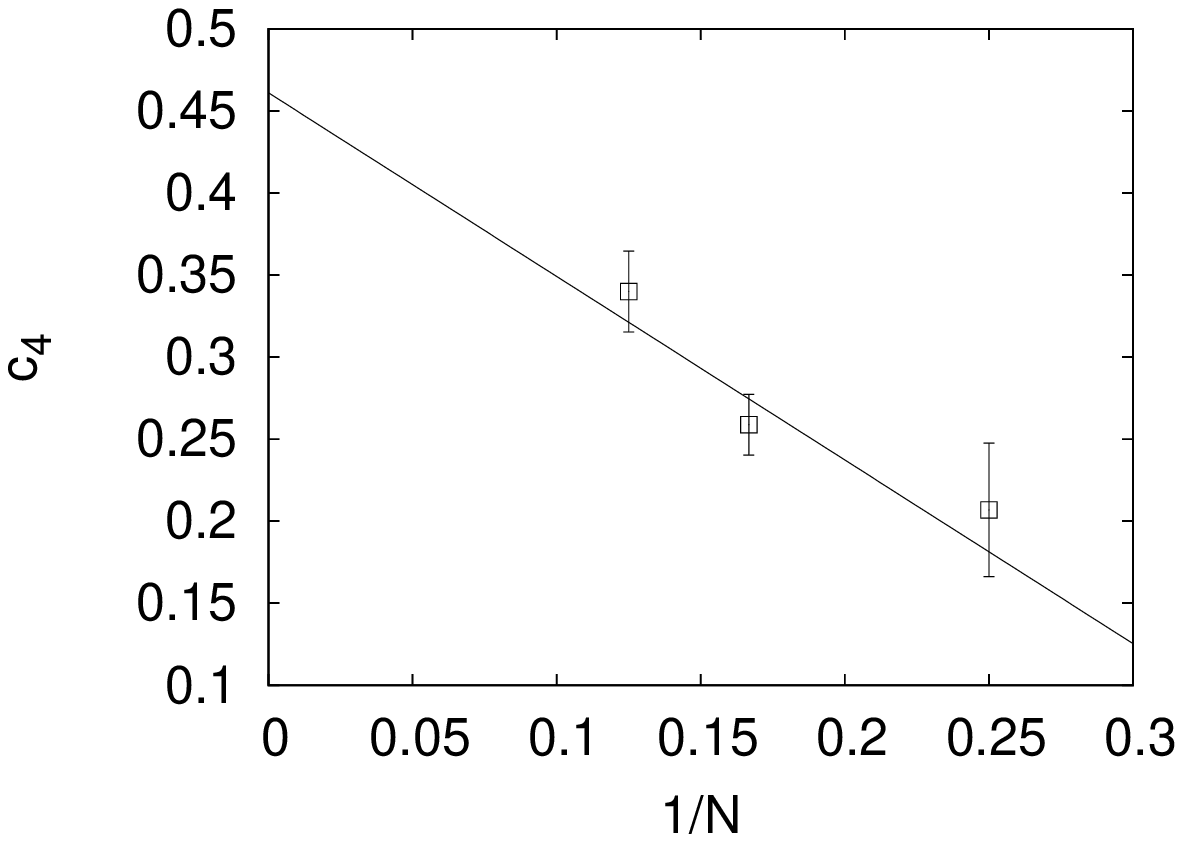}
  \end{minipage}

\end{center}
\caption{
The coefficients $c_n$ and $d_n$
in the asymptotic formula (\ref{ii-fit}) for $r=2$
extracted from fig.~\ref{res_r2} are plotted against $\frac{1}{N}$.
The data points are fitted to a straight line.
The large-$N$ extrapolated values obtained in this way are presented
in Table \ref{tab:cd}.
%The same as fig.~\ref{cdn_ext_r1}, but for $r=2$.
}
   \label{cdn_ext_r2}
  \end{figure}

%, which represent the effect of the phase.
In figs.~\ref{res_r1} and \ref{res_r2}
we show the log-log plots of $-\frac{1}{N^2}\log w_n(x)$
for $r=1$ and $r=2$, respectively.
We fit the data to 
the behavior \rf{ii-fit} 
at $x\ll 1$ and at $x \gg 1$,
and extract the coefficients 
$c_{n}$ and $d_{n}$ in \rf{ii-fit} for each $N$.
The results are plotted against $\frac{1}{N}$ 
in figs.~\ref{cdn_ext_r1} and \ref{cdn_ext_r2} for $r=1$ and $r=2$, 
respectively. 
Assuming that the finite-$N$ effects are ${\rm O}(1/N)$,
%which seems reasonable from the data,
we make large-$N$ extrapolations, which give the values
shown in Table \ref{tab:cd}.
The corresponding functions \rf{ii-fit},
which we define as the scaling functions $\Phi_n(x)$ in each region,
are also plotted in figs.~\ref{res_r1} and \ref{res_r2}.
%\ref{loglog_i1},\ref{loglog_i2},\ref{loglog_i3} and \ref{loglog_i4}
(The two solid lines show the margin of error.)
%Note that there is a clear trend towards large-$N$ scaling,
%in particular, for $n=2$ and $n=3$ cases.
The scaling functions thus obtained are plotted
in the left column of
figs.~\ref{res_r1-sol} and \ref{res_r2-sol}.

%\begin{figure}[htbp]

%%%%%%%

%\begin{figure}[htbp]

%%%%%%%%%%%%%%%%%%%%%%%%%%%%%%%%%%%%%%%%%%%%%%%%%%%%%%%%%%%%%%%

%\begin{figure}[htbp]
  
%%%%%%%%%%%%%%%%%%%%%%%%%%%%

\section{Large-$N$ extrapolations in the multi-observable analysis}
\label{largeN-multi}

In this section we summarize the 
large-$N$ extrapolations we made in the multi-observable case.
We describe them for the SO(3) 
and SO(2) symmetric vacua separately.

\subsection{extrapolations for the SO(3) symmetric vacuum}
\label{SO3-lambda3}

Similarly to the asymptotic behavior (\ref{ii-fit})
used in the single-observable analysis,
it is expected that 
\begin{eqnarray}
 \frac{1}{N^2} \log w_{\textrm{SO(3)}} (x,0.5)  
\sim  - {\tilde d}_{1} x^{-1} + {\tilde d}_{2} x^{-3/2}
\label{w_3d_x3}
\end{eqnarray}
at $x \gg 1$. Here we have added the subleading term motivated
from the fact that $\delta A/|A| \propto 1/\sqrt{x}$
as described below (\ref{ii-fit}).

  \FIGURE[b]{
\epsfig{file=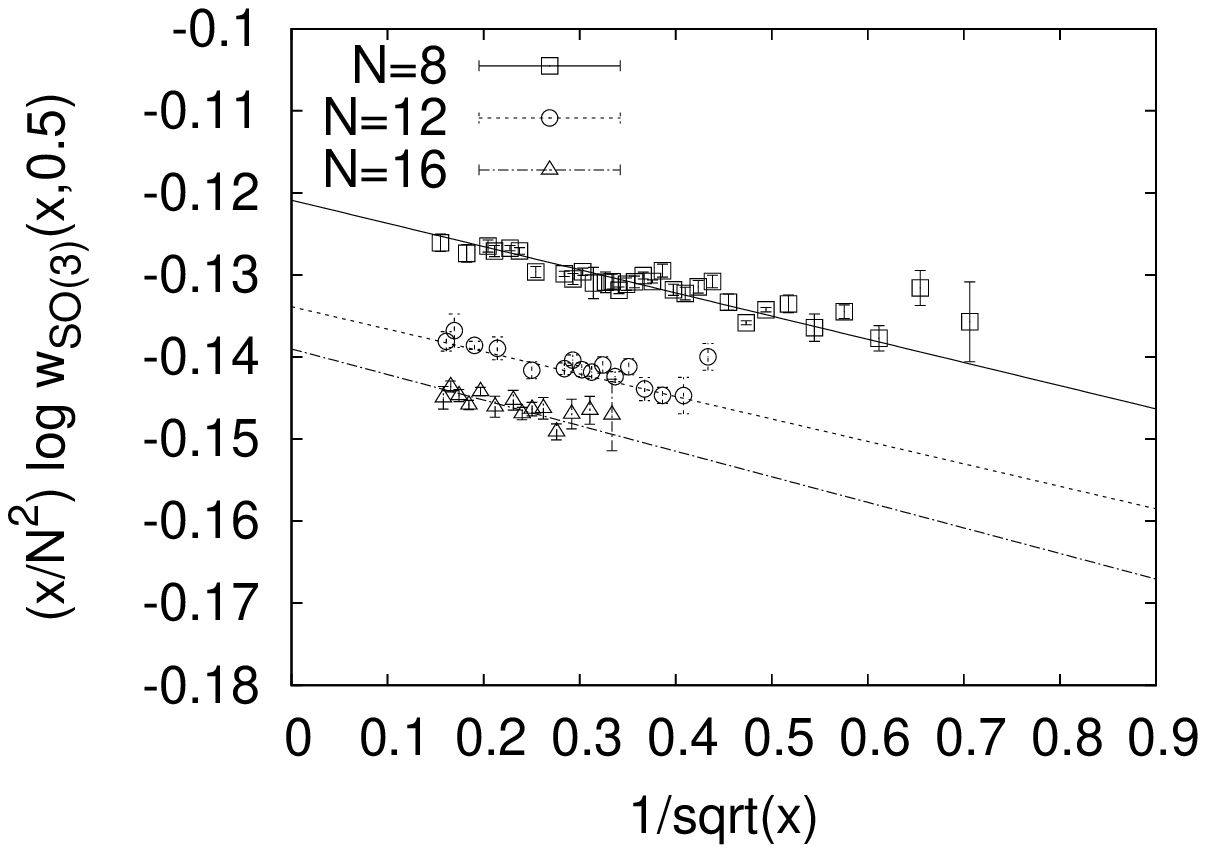,width=7.4cm} \\
\epsfig{file=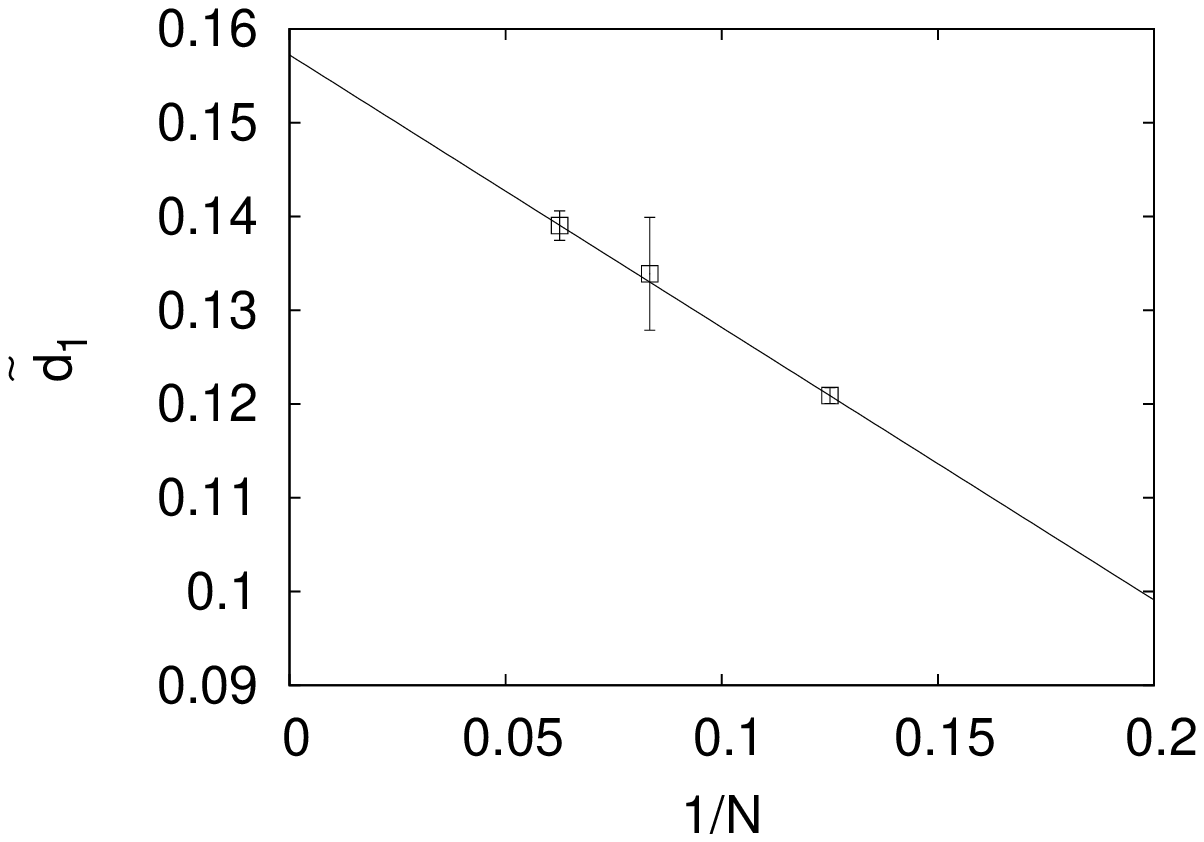,width=7.4cm}
\epsfig{file=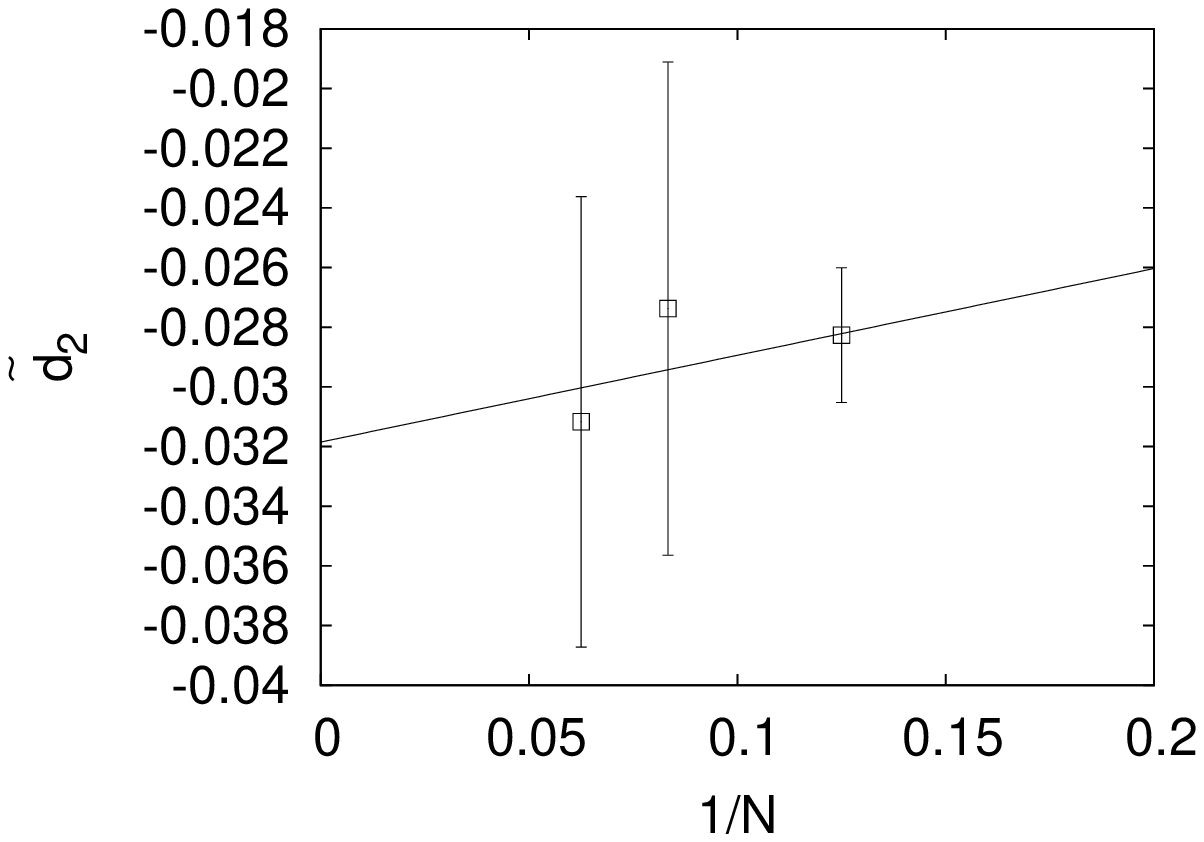,width=7.4cm}
   \caption{(Top) The function
$\frac{x}{N^2} \log w_{\textrm{SO(3)}} (x,0.5)$ 
is plotted against $\frac{1}{\sqrt{x}}$ for $N=8,12,16$. 
The straight lines represent the fits to the asymptotic behavior
(\ref{w_3d_x3}).
(Bottom) The coefficients ${\tilde d}_{1}$ and ${\tilde d}_{2}$ 
in the asymptotic behavior (\ref{w_3d_x3}) 
extracted from the top figure are
plotted against $\frac{1}{N}$.
The data points can be fitted nicely to a straight line.
}
   \label{d12_extrapolation}}
%% \begin{eqnarray}
%%  {\tilde d}_{1} &=& 0.157225 \pm 0.0004712  \hspace{3mm} 
%% {\tilde d}_{2} = -0.0318523 \pm 0.003845  \label{d12_extrapolation2}
%% \end{eqnarray}

In fig.~\ref{d12_extrapolation} (Top), 
we plot $\frac{x}{N^2} \log w_{\textrm{SO(3)}} (x,0.5)$ 
against $\frac{1}{\sqrt{x}}$, which
confirms (\ref{w_3d_x3}) including the subleading term.
We plot the coefficients ${\tilde d}_{1}$ and ${\tilde d}_{2}$ 
against $\frac{1}{N}$ in fig.~\ref{d12_extrapolation} (Bottom),
which shows that the finite $N$ effects are of the order of $1/N$.
Based on this observation, we obtain the large-$N$ extrapolated values
${\tilde d}_{1} = 0.1572(5)$ and ${\tilde d}_{2} = - 0.032(4)$.
The scaling function (\ref{def-Phi-so3}) obtained in this way
is plotted in fig.~\ref{result_so3} (Left).
By plugging $x=1.17$ into (\ref{w_3d_x3}), we obtain
$\Phi_{\rm SO(3)}(1.17,0.5)= -0.160(3)$ as in eq.~(\ref{Phi-SO3}).

%% While $\frac{1}{N^2} \log w_{\textrm{SO(3)}} (1.17,y)$ behaves as 
%% %(\ref{5_20}) 
%% at $y<1$, 
%% this behavior is subject to finite-$N$ effect, 
%% which makes simulation difficult. 
We redo the calculation for 
$y=0.45, 0.55$ and obtain the large-$N$ extrapolated values
%% From this result, 
%% we read off the coefficients 
%% ${\tilde d}_{1}$ and ${\tilde d}_2$ 
%% in figure \ref{d12_045_055}. 
%% We thus obtain the coefficients as
\begin{eqnarray}
 {\tilde d}_{1} &=& 0.1433(2)  \ , \quad
{\tilde d}_{2} = -0.023(1) \quad
\textrm{for } y=0.45 \ , \label{d12_extrapolation_045} \\
 {\tilde d}_{1} &=& 0.1749(2) \ , \quad
{\tilde d}_{2} = -0.029(3) \quad 
\textrm{for } y=0.55  \ ,
\label{d12_extrapolation_055}
\end{eqnarray}
from which 
%By plugging in $x=1.17$ in (\ref{w_3d_x3}),
we obtain $\Phi_{\textrm{SO(3)}} (1.17,y)$ for 
$y=0.45$ and $0.55$.
These results, together with the value at $y=0.5$,
are plotted in fig.~\ref{result_so3_2} (Left).
%used to give an estimate on the derivative 
%$\frac{\partial}{\partial y} \Phi_{\textrm{SO(3)}} (1.17,y)$
%at $y = 0.5$ as described in section \ref{sec:res-SO3}.
%
%% \begin{eqnarray}
%%  {\tilde d}_{1} &=& 0.143289 \pm 0.0001532, \hspace{3mm} 
%% {\tilde d}_{2} = - 0.022898 \pm 0.001222 \hspace{3mm} 
%% (\textrm{for } x=0.45) \label{d12_extrapolation_045} \\
%%  {\tilde d}_{1} &=& 0.174892 \pm 0.0002147, \hspace{3mm} 
%% {\tilde d}_{2} = - 0.0292455 \pm 0.003312  \hspace{3mm} 
%% (\textrm{for } x=0.55) \label{d12_extrapolation_055}
%% \end{eqnarray}

  \FIGURE[b]{
\epsfig{file=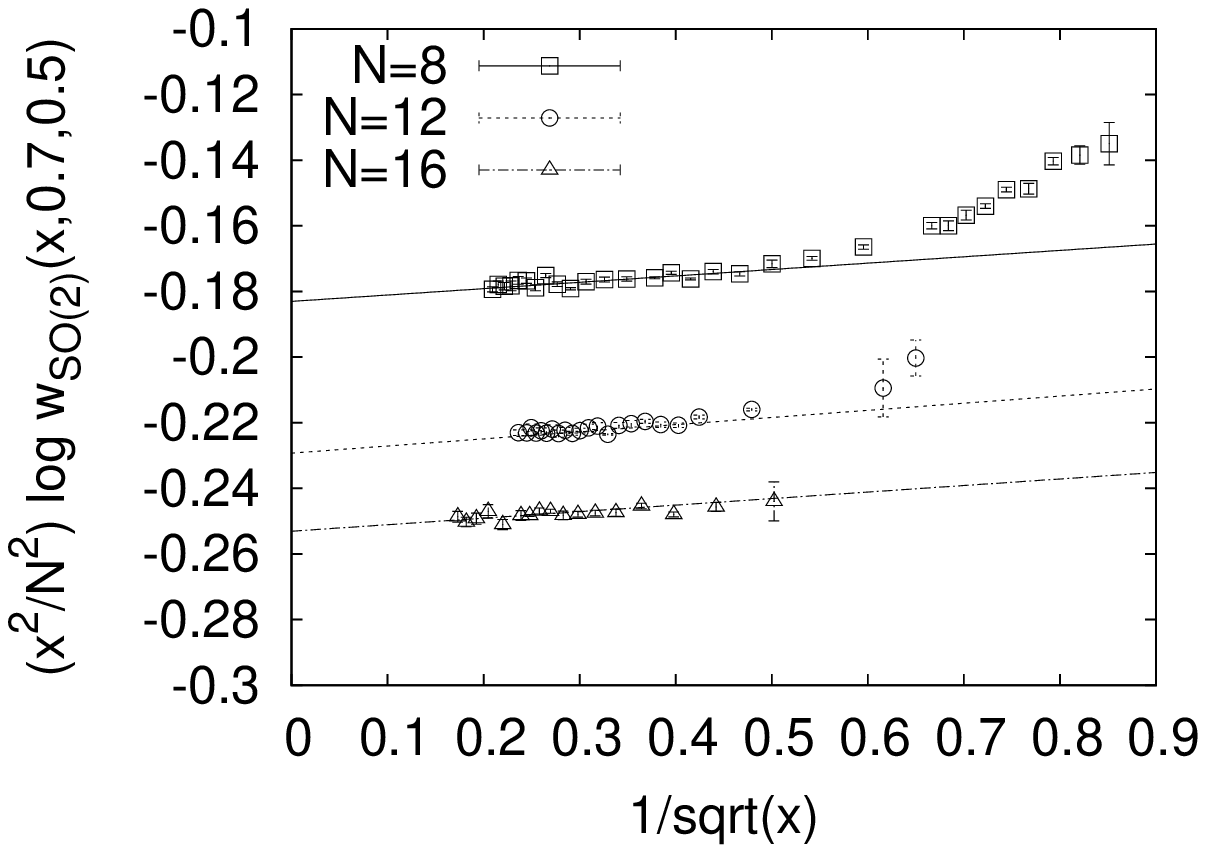,width=7.4cm} \\
\epsfig{file=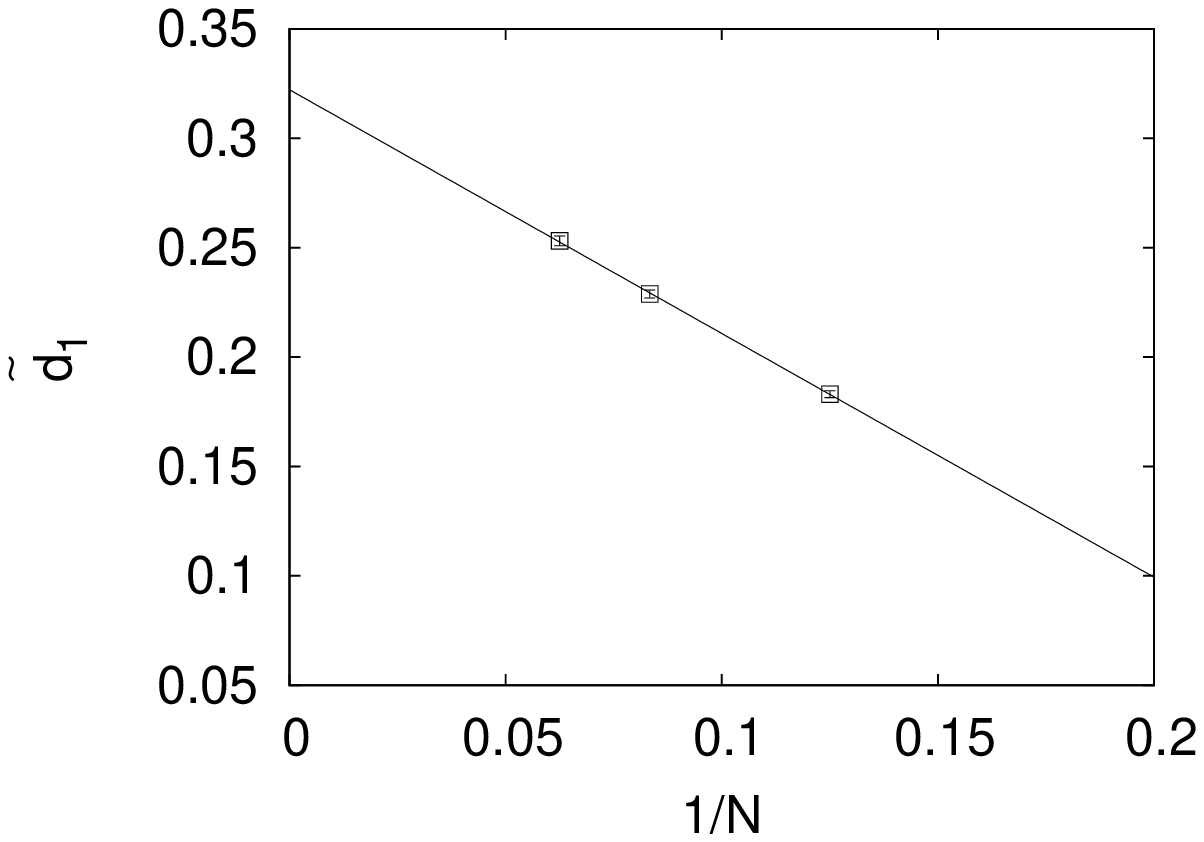,width=7.4cm}
\epsfig{file=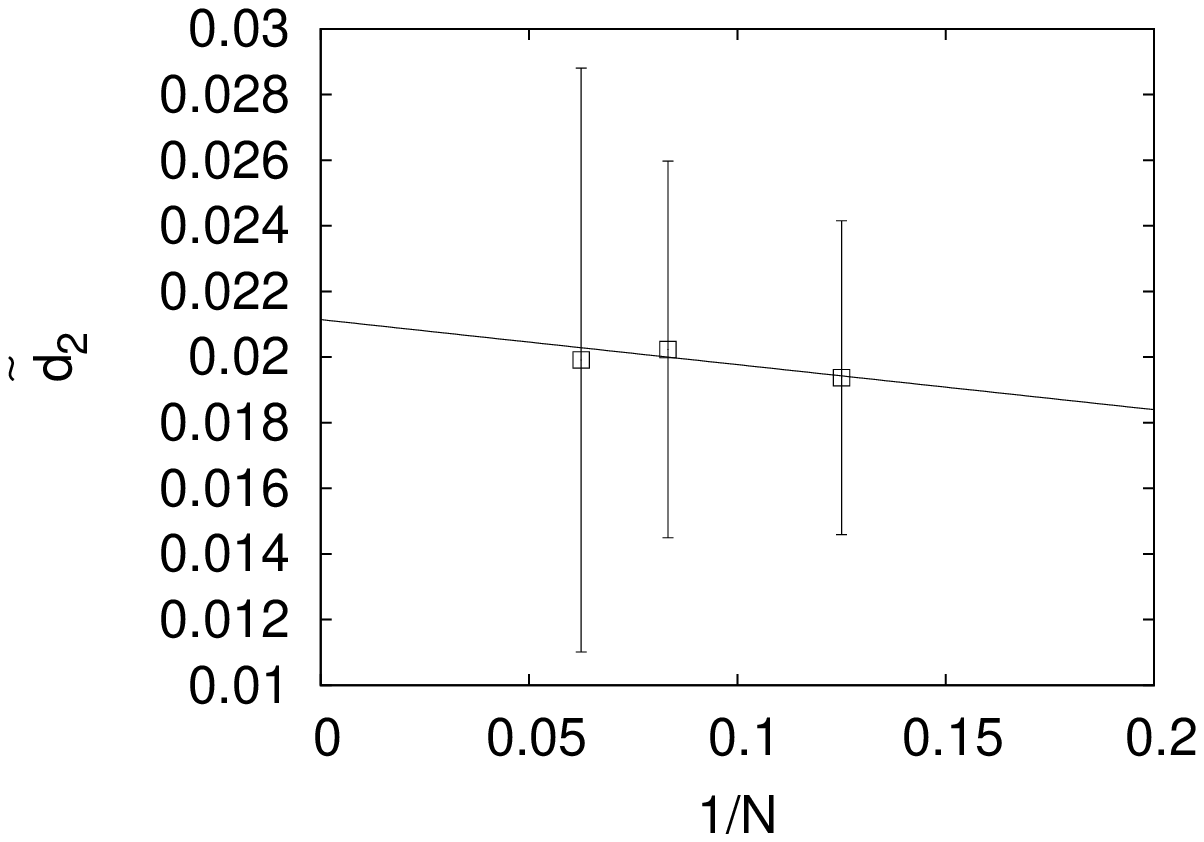,width=7.4cm}
   \caption{(Top) The function 
$\frac{x^2}{N^2} \log w_{\textrm{SO(2)}} (x,0.7,0.5)$ 
is plotted against $\frac{1}{\sqrt{x}}$ for $N=8,12,16$. 
The straight lines represent the fits to the asymptotic behavior
(\ref{w_2d_x2}).
(Bottom) The coefficients ${\tilde d}_{1}$ and ${\tilde d}_{2}$ 
in the asymptotic behavior (\ref{w_2d_x2}) 
extracted from the top figure
are plotted 
against $\frac{1}{N}$.
The data points can be fitted nicely to a straight line.
% for $N=8,12,16$
}
   \label{d12_extrapolation_34_2}}

\subsection{extrapolations for the SO(2) symmetric vacuum}
\label{SO2-1st}

Similarly to (\ref{w_3d_x3})
for the SO(3) symmetric vacuum,
it is expected that 
\begin{eqnarray}
 \frac{1}{N^2} \log w_{\textrm{SO(2)}} (x,0.7,0.5) 
\simeq - {\tilde d}_{1} x^{-2} + {\tilde d}_{2} x^{-5/2} 
\label{w_2d_x2}
\end{eqnarray}
at $x \gg 1$. 
In fig.~\ref{d12_extrapolation_34_2} (Top),
we plot $\frac{x^2}{N^2} \log w_{\textrm{SO(2)}} (x,0.7,0.5)$ 
against $\frac{1}{\sqrt{x}}$,
% for $N=8,12,16$,
which confirms (\ref{w_2d_x2}) including the subleading term.
We extract the coefficients ${\tilde d}_{1}$ 
and ${\tilde d}_{2}$ for $N=8,12,16$, and 
make a large-$N$ extrapolation for 
${\tilde d}_{1}$ and ${\tilde d}_{2}$ as 
in fig.~\ref{d12_extrapolation_34_2} (Bottom),
from which we get the large-$N$ extrapolated values
$ {\tilde d}_{1} = 0.322(2)$
and ${\tilde d}_{2} = 0.021(1)$.
The scaling function (\ref{def-Phi-so2}) obtained in this way
is plotted in fig.~\ref{result_so2_2} (Left).
By plugging $x=1.4$ into (\ref{w_2d_x2}), we obtain
$\Phi_{\rm SO(2)}(1.4,0.7,0.5)= - 0.155(1)$
as in eq.~(\ref{Phi-SO2}).
%% We thus obtain 
%% \begin{eqnarray}
%%  {\tilde d}_{1} = 0.322188 \pm 0.001628, 
%% \hspace{3mm} {\tilde d}_{2} = 
%% 0.0211405 \pm 0.0008508 \label{d12_extrapolation_34_2_2}
%% \end{eqnarray}

We redo the calculation for $y=0.65,0.75$, and obtain
the large-$N$ extrapolated values
%% by calculating $\frac{1}{N^2} \log w_{\textrm{SO(2)}} (x,y,0.5) 
%% \sim  - {\tilde d}_{1} y^{-2} + {\tilde d}_{2} y^{-2.5}$ 
%% with fixed $y=0.65, 0.70, 0.75$ at large $N$ and 
%% then plugging $x=1.4$. 
%
%% In figure \ref{orthogonal_so2_x_0_65_75}, 
%% we plot $\frac{y^2}{N^2} \log w_{\textrm{SO(2)}} (x,y,0.5)$ 
%% against $\frac{1}{\sqrt{x}}$ for $N=8,12,16$ at $y=0.65$ and $y=0.75$.
%%   \FIGURE{
%% \epsfig{file=ln-1x-w-r1td24_3_log_x065_2.eps,width=7.4cm}
%% \epsfig{file=ln-1x-w-r1td24_3_log_x075_2.eps,width=7.4cm}
%%    \caption{$\frac{x^2}{N^2} \log w_{\textrm{SO(2)}} (x,y,0.5)$ 
%% against $\frac{1}{\sqrt{x}}$ for $N=8,12,16$ 
%% at $y=0.65$ (Left) and $y=0.75$ (Right)}
%%    \label{orthogonal_so2_x_0_65_75}}
%
%% From this result, we read off the coefficients 
%% ${\tilde d}_{1}$ and ${\tilde d}_2$ in figure \ref{d12_so2_065_075}. 
%% We thus obtain the coefficients as
\begin{eqnarray}
 {\tilde d}_{1} &=& 0.295(3) \ , \quad
{\tilde d}_{2} = 0.018(1) \quad \textrm{for } y=0.65  \ , \\
 {\tilde d}_{1} &=& 0.344(3)  \ , \quad
{\tilde d}_{2} = 0.019(3)   \quad
\textrm{for } y=0.75 \ ,
\end{eqnarray}
from which we obtain
$\Phi_{\textrm{SO(2)}}(1.4,y,0.5) $
at $y=0.65, 0.75$.
These results, together with the value at $y=0.7$,
are plotted in fig.~\ref{result_so2_3} (Left).
%These results are used to give an estimate on
%the derivative 
%$\frac{\partial}{\partial y} \Phi_{\textrm{SO(2)}} (1.4,y,0.5)$
%at $y = 0.7$ as described in section \ref{sec:res-SO2}.

  \FIGURE[b]{
\epsfig{file=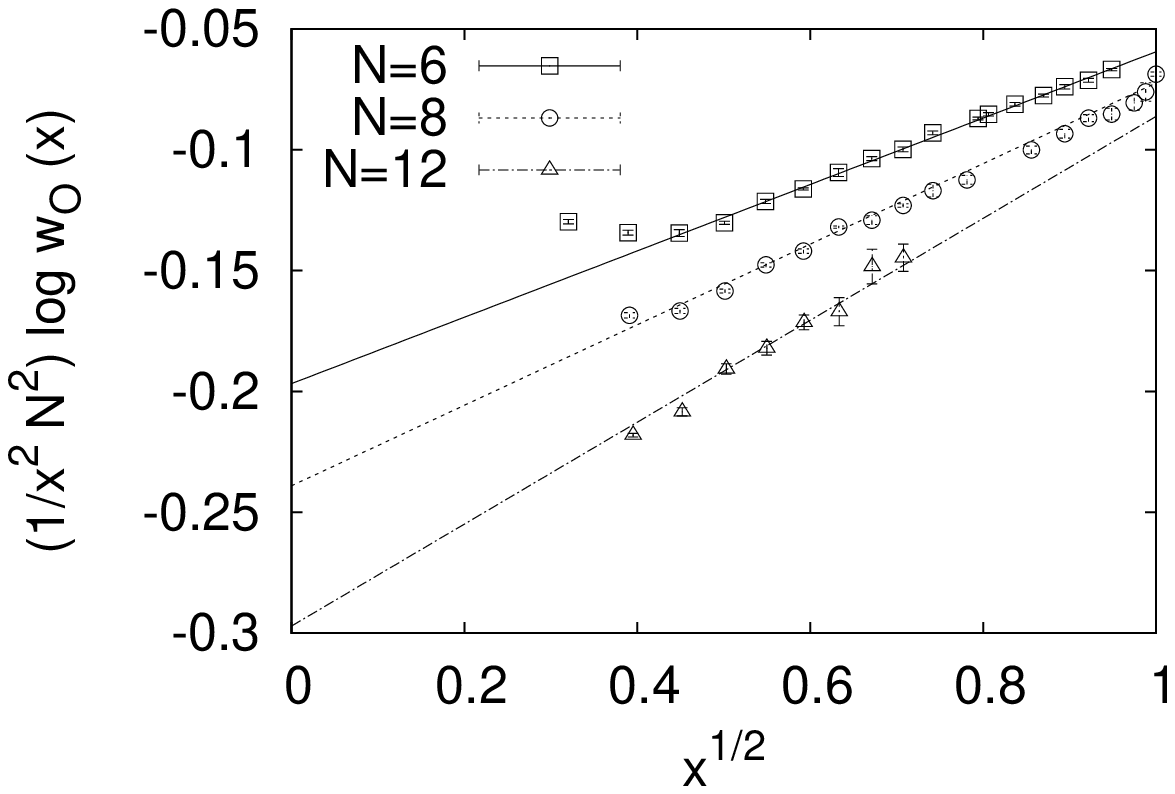,width=7.4cm} \\
\epsfig{file=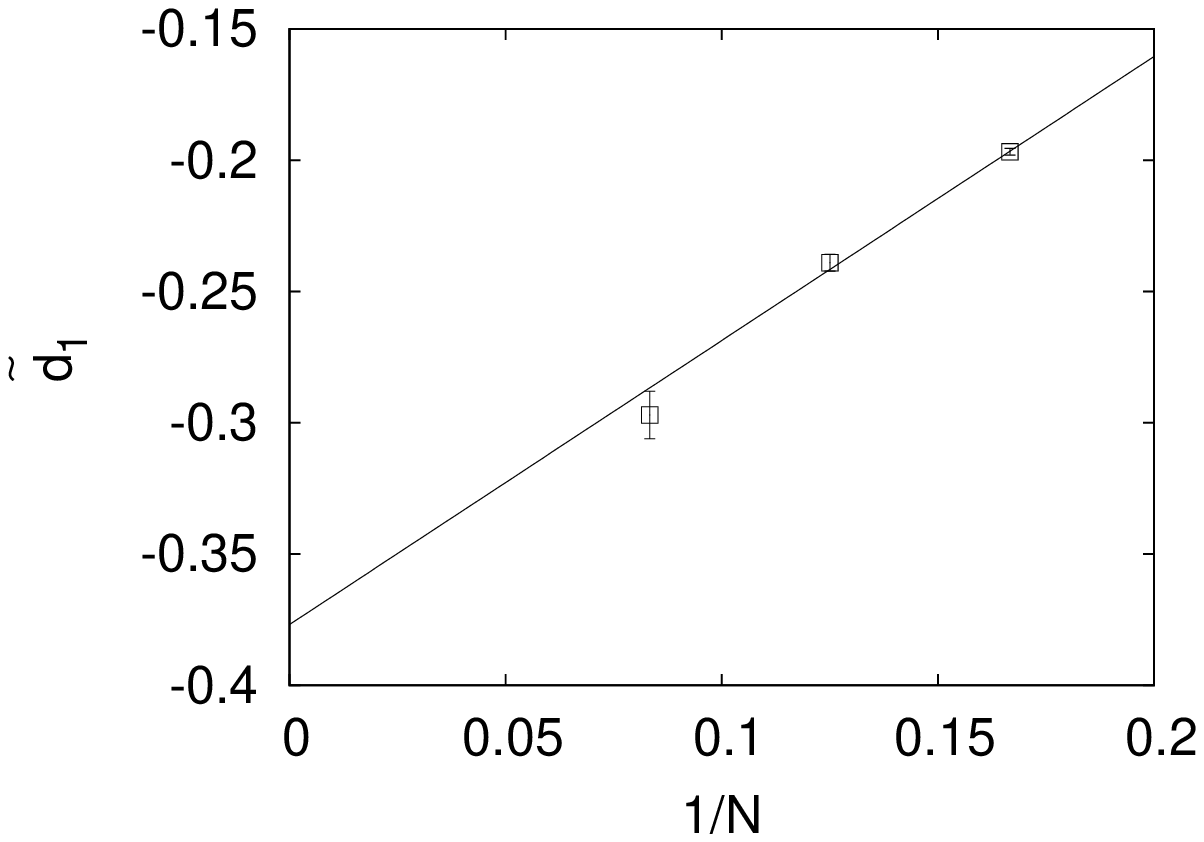,width=7.4cm}
\epsfig{file=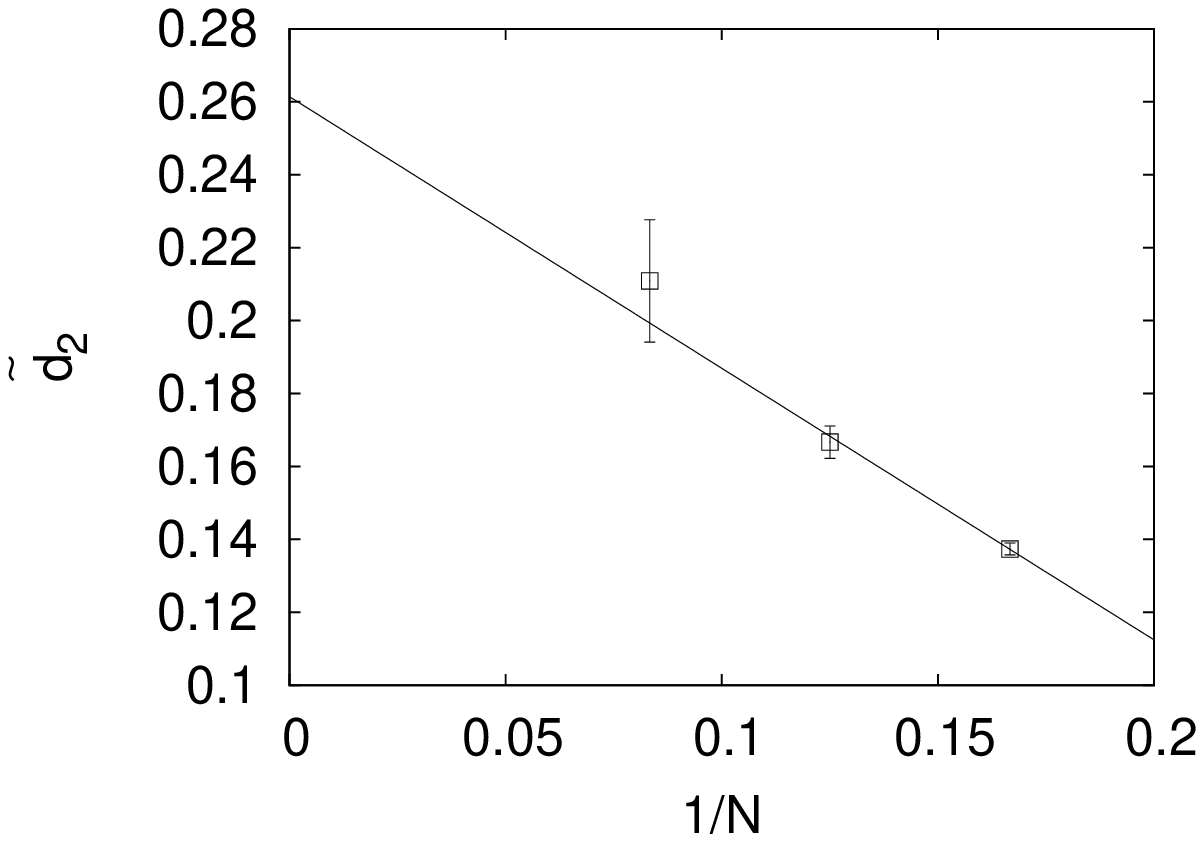,width=7.4cm}
   \caption{(Top) The function
$\frac{1}{x^2 N^2} \log w_{{\cal O}} (x)$ 
is plotted against 
$\sqrt{x}$ for $N=8,12,16$. 
(Bottom) The coefficients ${\tilde d}_{1}$ and ${\tilde d}_{2}$ 
in the asymptotic behavior (\ref{otildescal})
are plotted against $\frac{1}{N}$.
The data points can be fitted nicely to a straight line.
%The solid lines represent the linear fitting.
}
\label{d12_fsq_extrapolation_34_2}
}

%% \begin{eqnarray}
%%  {\tilde d}_{1} &=& 0.29532 \pm 0.002506, \hspace{3mm} 
%% {\tilde d}_{2} = 0.0179898 \pm 0.0008176 \hspace{3mm} 
%% (\textrm{for } x=0.65) \label{d12_so2_extrapolation_045-2} \\
%%  {\tilde d}_{1} &=& 0.343889 \pm 0.002832, \hspace{3mm} 
%% {\tilde d}_{2} = 0.0193388 \pm 0.003456 \hspace{3mm} 
%% (\textrm{for } x=0.75) \label{d12_so2_extrapolation_055-2}
%% \end{eqnarray}

%%   \FIGURE{
%% \epsfig{file=d1_24_3_x065_r1_extrapolate_3.eps,width=7.4cm}
%% \epsfig{file=d2_24_3_x065_r1_extrapolate_3.eps,width=7.4cm}
%% \epsfig{file=d1_24_3_x075_r1_extrapolate_3.eps,width=7.4cm}
%% \epsfig{file=d2_24_3_x075_r1_extrapolate_3.eps,width=7.4cm}
%%    \caption{Coefficients ${\tilde d}_{1}$ and ${\tilde d}_{2}$ 
%% in the asymptotic behavior (\ref{w_2d_x2}) against 
%% $\frac{1}{N}$ for $N=8,12,16$ at $x=0.65$ (top) and $x=0.75$ (bottom).}
%%    \label{d12_so2_065_075}}

We redo the calculation for $z=0.45,0.55$, and obtain
the large-$N$ extrapolated values
%% We evaluate $\langle {\tilde \lambda}_{4} \rangle$ 
%% by calculating $\frac{1}{N^2} \log w_{\textrm{SO(2)}} (y,z=0.7,x) 
%% \sim  - {\tilde d}_{1} y^{-2} + {\tilde d}_{2} y^{-2.5}$ 
%% with fixed $x=0.45, 0.50, 0.55$ at large $N$ and then plugging $y=1.4$. 
%
%% In figure \ref{orthogonal_so2_x_0_45_55}, 
%% we plot $\frac{y^2}{N^2} \log w_{\textrm{2d}} (y,z=0.7,x)$ 
%% against $\frac{1}{\sqrt{y}}$ for $N=8,12,16$ at $x=0.45$ and $x=0.55$.
%%   \FIGURE{
%% \epsfig{file=ln-1x-w-r1td23_4_log_x045_2_2.eps,width=7.4cm}
%% \epsfig{file=ln-1x-w-r1td23_4_log_x055_2_2.eps,width=7.4cm}
%%    \caption{$\frac{y^2}{N^2} \log w_{\textrm{2d}} (y,z=0.7,x)$ against $\frac{1}{\sqrt{y}}$ for $N=8,12,16$ at $x=0.45$ (Left) and $x=0.55$ (Right)}
%%    \label{orthogonal_so2_x_0_45_55}}
%
%% From this result, we read off the coefficients 
%% ${\tilde d}_{1}$ and ${\tilde d}_2$ in figure \ref{d12_so2_045_055}. 
%% We thus obtain the coefficients as
\begin{eqnarray}
 {\tilde d}_{1} &=& 0.292(5) \ ,   \quad 
 {\tilde d}_{2} = 0.026(2)  \quad
\textrm{for } z=0.45 \ , 
\label{d12_so2_extrapolation_045-3} \\
 {\tilde d}_{1} &=& 0.352(2) \ , \quad
{\tilde d}_{2} = 0.015(3) \quad
\textrm{for } z=0.55 \ ,
\label{d12_so2_extrapolation_055-3}
\end{eqnarray}
from which we obtain
$\Phi_{\textrm{SO(2)}}(1.4,0.7,z) $
at $z=0.45, 0.55$.
These results, together with the value at $z=0.5$,
are plotted in fig.~\ref{result_so2_4} (Left).
%These results are used to give an estimate on
%the derivative 
%$\frac{\partial}{\partial z} \Phi_{\textrm{SO(2)}} (1.4,0.7,z)$
%at $z = 0.5$ as described in section \ref{sec:res-SO2}.

%% \begin{eqnarray}
%%  {\tilde d}_{1} &=& 0.292087 \pm 0.004701, 
%% \hspace{3mm} {\tilde d}_{2} = 0.0262557 \pm 0.002352 \hspace{3mm} 
%% (\textrm{for } x=0.45) \label{d12_so2_extrapolation_045-3} \\
%%  {\tilde d}_{1} &=& 0.351839 \pm 0.002383, \hspace{3mm} 
%% {\tilde d}_{2} = 0.0145501 \pm 0.003302 \hspace{3mm} 
%% (\textrm{for } x=0.55) \label{d12_so2_extrapolation_055-3}
%% \end{eqnarray}

%%   \FIGURE{
%% \epsfig{file=d1_23_4_x045_r1_extrapolate_3.eps,width=7.4cm}
%% \epsfig{file=d2_23_4_x045_r1_extrapolate_3.eps,width=7.4cm}
%% \epsfig{file=d1_23_4_x055_r1_extrapolate_3.eps,width=7.4cm}
%% \epsfig{file=d2_23_4_x055_r1_extrapolate_3.eps,width=7.4cm}
%%    \caption{Coefficients ${\tilde d}_{1}$ and ${\tilde d}_{2}$ 
%% in the asymptotic behavior (\ref{w_2d_x2}) against $\frac{1}{N}$ 
%% for $N=8,12,16$ at $x=0.45$ (top) and $x=0.55$ (bottom).}
%%    \label{d12_so2_045_055}}

%\paragraph{Calculation of 
%$\langle {\tilde \lambda}_{4} \rangle$} \hspace{0mm} \\

%%% add figure

Finally let us describe the large-$N$ extrapolations 
used in section \ref{more-obs} in the analysis for the 
observable (\ref{trf2}).
First the asymptotic behavior of $\frac{1}{N^2} \log w_{\cal O}(x)$
is expected to be
\beq
\label{otildescal}
\frac{1}{N^2} \log w_{\cal O}(x)=
  -\tilde{d}_1 x^{2} + \tilde{d}_2 x^{5/2} \  .
\eeq
The leading term is obtained as follows.
Let us consider a small perturbation around 
a diagonal configuration.
The phase of the fermion determinant appears at the second
order of this perturbation.
On the other hand, the observable ${\cal O}$ becomes nonzero
also at the second order.
Therefore the distribution of the phase is expected
to have a width $\sigma \propto x$.
Applying the formula (\ref{ejiri}), we obtain
the leading term.
%The subleading term is suggested from the argument
%below (\ref{w_3d_x3}).
The power of the subleading term can be deduced by expanding 
the fermion determinant around a diagonal configuration.

In fig.~\ref{d12_fsq_extrapolation_34_2} (Top) we plot
$\frac{1}{x^2 N^2} \log w_{{\cal O}} (x)$ against 
$\sqrt{x}$.
The data points can be fitted nicely by straight lines,
which confirms the asymptotic behavior (\ref{otildescal}).
We extract the coefficients ${\tilde d}_{1}$ 
and ${\tilde d}_{2}$ for $N=6,8,12$, and 
make a large-$N$ extrapolation for 
${\tilde d}_{1}$ and ${\tilde d}_{2}$ as 
in fig.~\ref{d12_fsq_extrapolation_34_2} (Bottom).
We obtain $ {\tilde d}_{1} = 0.38(2)$
and ${\tilde d}_{2} = 0.26(1)$.
%d_{1}=0.376915 +/- 0.01535 and d_{2}=0.261401 +/- 0.01272.
The scaling function (\ref{Phi-O}) obtained in this way
is plotted in fig.~\ref{result_so2_trF2} (Left).

%------------------------------------------------------------

% %%%%%%%%%%%%%%%%%%%%%%%%%%%%%%%%%%%%%%%%%%%%%%%%%%%%%%%%%%
% %%%%%%%%%%%%%%%%%%%%%%%%%%%%%%%%%%%%%%%%%%%%%%%%%%%%%%%%%%
% %%%%%%%%%%%%%%%%%%%%%%%%%%%%%%%%%%%%%%%%%%%%%%%%%%%%%%%%%%

% %%%%%%%%%%%%%%%%%%%%%%%%%%%%%%%%%%%%%%%%%%%%%%%%%%%%%%%%%%
% %%%%%%%%%%%%%%%%%%%%%%%%%%%%%%%%%%%%%%%%%%%%%%%%%%%%%%%%%%


\begin{thebibliography}{99}
% %%%%%%%%%%%%%%%%%%%%%%%%%%%%%%%%%%%%%%%%%%%%%%%%%%%%%%%%%%
% %%%%%%%%%%%%%%%%%%%%%%%%%%%%%%%%%%%%%%%%%%%%%%%%%%%%%%%%%%
% %%%%%%%%%%%%%%%%%%%%%%%%%%%%%%%%%%%%%%%%%%%%%%%%%%%%%%%%%%

% %%%%%%%%%%%%%%%%%%%%%%%%%%%%%%%%%%%%%%%%%%%%%%%%%%%%%%%%%%
\bibitem{Maldacena:1997re}
  J.~M.~Maldacena,
\emph{The large N limit of superconformal field theories and supergravity},
\emph{Adv.\ Theor.\ Math.\ Phys.} {\bf 2} (1998) 231 
[\emph{Int.\ J.\ Theor.\ Phys.} {\bf 38} (1999) 1113]
[{\tt hep-th/9711200}].
  %%CITATION = IJTPB,38,1113;%%


\bibitem{Hanada:2007ti}
  M.~Hanada, J.~Nishimura and S.~Takeuchi,
\emph{Non-lattice simulation for supersymmetric gauge theories in one
dimension},
\emph{Phys.\ Rev.\ Lett.} {\bf 99} (2007) 161602 
[{\tt arXiv:0706.1647}].
%[{\tt arXiv:0706.1647 [hep-lat]}].
  %%CITATION = PRLTA,99,161602;%%

\bibitem{Catterall:2007fp}
  S.~Catterall and T.~Wiseman,
\emph{Towards lattice simulation of the gauge theory duals to 
black holes and hot strings},
\emph{JHEP} {\bf 0712} (2007) 104 
[{\tt arXiv:0706.3518}].
%[arXiv:0706.3518 [hep-lat]].
  %%CITATION = JHEPA,0712,104;%%
%


% %%%%%%%%%%%%%%%%%%%%%%%%%%%%%%%%%%%%%%%%%%%%%%%%%%%%%%%%%%
\bibitem{Anagnostopoulos:2007fw}
  K.~N.~Anagnostopoulos, M.~Hanada, J.~Nishimura and S.~Takeuchi,
\emph{Monte Carlo studies of supersymmetric matrix quantum mechanics 
with sixteen supercharges at finite temperature},
\emph{Phys.\ Rev.\ Lett.} {\bf 100} (2008) 021601 
[{\tt arXiv:0707.4454}].
%[arXiv:0707.4454 [hep-th]].
  %%CITATION = PRLTA,100,021601;%%

\bibitem{Catterall:2008yz}
  S.~Catterall and T.~Wiseman,
\emph{Black hole thermodynamics from simulations of 
lattice Yang-Mills theory},
\emph{Phys.\ Rev.\  D} {\bf 78} (2008) 041502 
[{\tt arXiv:0803.4273}].
%[arXiv:0803.4273 [hep-th]].
  %%CITATION = PHRVA,D78,041502;%%


\bibitem{Hanada:2008ez}
  M.~Hanada, Y.~Hyakutake, J.~Nishimura and S.~Takeuchi,
\emph{Higher derivative corrections to black hole thermodynamics 
from supersymmetric matrix quantum mechanics},
\emph{Phys.\ Rev.\ Lett.} {\bf 102} (2009) 191602 
[{\tt arXiv:0811.3102}].
%  [arXiv:0811.3102 [hep-th]].
  %%CITATION = PRLTA,102,191602;%%
%

%\cite{Catterall:2009xn}
%\cite{Catterall:2009xn}
\bibitem{Catterall:2009xn}
  S.~Catterall and T.~Wiseman,
\emph{Extracting black hole physics from the lattice},
\emph{JHEP} {\bf 1004} (2010) 077
[{\tt arXiv:0909.4947}].
%[arXiv:0909.4947 [hep-th]].
  %%CITATION = JHEPA,1004,077;%%


\bibitem{Hanada:2008gy}
  M.~Hanada, A.~Miwa, J.~Nishimura and S.~Takeuchi,
\emph{Schwarzschild radius from Monte Carlo calculation 
of the Wilson loop in supersymmetric matrix quantum mechanics},
\emph{Phys.\ Rev.\ Lett.} {\bf 102} (2009) 181602 
[{\tt arXiv:0811.2081}].
%  [arXiv:0811.2081 [hep-th]].\\
  %%CITATION = PRLTA,102,181602;%%

%
\bibitem{HNSY}
%\cite{Hanada:2009ne}
%\bibitem{Hanada:2009ne}
  M.~Hanada, J.~Nishimura, Y.~Sekino and T.~Yoneya,
\emph{Monte Carlo studies of Matrix theory correlation functions},
\emph{Phys.\ Rev.\ Lett.} {\bf 104} (2010) 151601
[{\tt arXiv:0911.1623}];\\
%[{\tt arXiv:0911.1623 [hep-th]}].
  %%CITATION = PRLTA,104,151601;%%
%
%\cite{Hanada:2011fq}
%\bibitem{Hanada:2011fq}
%  M.~Hanada, J.~Nishimura, Y.~Sekino and T.~Yoneya,
\emph{Direct test of the gauge-gravity correspondence for Matrix theory
correlation functions},
{\tt arXiv:1108.5153}.
%  arXiv:1108.5153 [hep-th].
  %%CITATION = ARXIV:1108.5153;%%


%\cite{Nishimura:2009xm}
\bibitem{Nishimura:2009xm}
  J.~Nishimura,
\emph{Non-lattice simulation of supersymmetric gauge theories as a probe 
to quantum black holes and strings},
\emph{PoS} {\bf LAT2009} (2009) 016
[{\tt arXiv:0912.0327}].
%  [arXiv:0912.0327 [hep-lat]].
  %%CITATION = POSCI,LAT2009,016;%%


\bibitem{Itzhaki:1998dd}
  N.~Itzhaki, J.~M.~Maldacena, J.~Sonnenschein and S.~Yankielowicz,
\emph{Supergravity and the large N limit of theories with sixteen
supercharges},
\emph{Phys.\ Rev.\  D} {\bf 58} (1998) 046004 
[{\tt hep-th/9802042}].
%[arXiv:hep-th/9802042].
  %%CITATION = PHRVA,D58,046004;%%

%%%%%

%% \bibitem{Banks:1996vh}
%%   T.~Banks, W.~Fischler, S.~H.~Shenker and L.~Susskind,
%%   ``M theory as a matrix model: A conjecture,''
%%   Phys.\ Rev.\  D {\bf 55} (1997) 5112
%%   [arXiv:hep-th/9610043];\\ 
%%   %%CITATION = PHRVA,D55,5112;%%
%% %%%%%%%
%% %%%%%%%%%%%%%%%%%%%
%% %\cite{Taylor:2001vb}
%% %\bibitem{Taylor:2001vb}
%%   for a comprehensive review see e.g. 
%%   W.~Taylor,
%%   ``M(atrix) theory: Matrix quantum mechanics as a fundamental theory,''
%%   Rev.\ Mod.\ Phys.\  {\bf 73}, 419 (2001)
%%   [arXiv:hep-th/0101126].
%%   %%CITATION = RMPHA,73,419;%%

%% %\cite{Witten:1995ex}
%% \bibitem{Witten:1995ex}
%%   E.~Witten,
%%   ``String theory dynamics in various dimensions,''
%%   Nucl.\ Phys.\  B {\bf 443}, 85 (1995)
%%   [arXiv:hep-th/9503124].
%%   %%CITATION = NUPHA,B443,85;%%
% %%%%%%%%%%%%%%%%%%%%%%%%%%%%%%%%%%%%%%%%%%%%%%%%%%%%%%%%%%

\bibitem{Ishibashi:1996xs}
  N.~Ishibashi, H.~Kawai, Y.~Kitazawa and A.~Tsuchiya,
\emph{A large-N reduced model as superstring},
\emph{Nucl.\ Phys.\  B} {\bf 498} (1997) 467 
[{\tt hep-th/9612115}];\\
  %%CITATION = NUPHA,B498,467;%%
for a comprehensive review see e.g.
%\bibitem{Aoki:1998bq}
  H.~Aoki, S.~Iso, H.~Kawai, Y.~Kitazawa, A.~Tsuchiya and T.~Tada,
  \emph{IIB matrix model},
\emph{Prog.\ Theor.\ Phys.\ Suppl.} {\bf 134} (1999) 47
[{\tt hep-th/9908038}];\\
  %%CITATION = PTPSA,134,47;%%
%\bibitem{Azuma:2004vs}
  T.~Azuma,
\emph{Matrix models and the gravitational interaction},
{\tt hep-th/0401120}.
  %%CITATION = HEP-TH/0401120;%%
% %%%%%%%%%%%%%%%%%%%%%%%%%%%%%%%%%%%%%%%%%%%%%%%%%%%%%%%%%%

\bibitem{9802085}
H.~Aoki, S.~Iso, H.~Kawai, Y.~Kitazawa and T.~Tada,
\emph{Space-time structures from IIB matrix model,}
\ptp{99}{1998}{713} [\hepth{9802085}].
%%CITATION = HEP-TH 9802085;%%


%\cite{Kostelecky:1988zi}
\bibitem{Kostelecky:1988zi}
  V.~A.~Kostelecky and S.~Samuel,
\emph{Spontaneous breaking of Lorentz symmetry in string theory},
\emph{Phys.\ Rev.\ D} {\bf 39} (1989) 683.
  %%CITATION = PHRVA,D39,683;%%



% %%%%%%%%%%%%%%%%%%%%%%%%%%%%%%%%%%%%%%%%%%%%%%%%%%%%%%%%%%
\bibitem{Nishimura:2001sx}
  J.~Nishimura and F.~Sugino,
\emph{Dynamical generation of four-dimensional space-time 
in the IIB matrix model},
\emph{JHEP} {\bf 0205} (2002) 001
[{\tt hep-th/0111102}].
  %%CITATION = JHEPA,0205,001;%%

\bibitem{higher}
%\bibitem{Kawai:2002jk}
  H.~Kawai, S.~Kawamoto, T.~Kuroki, T.~Matsuo and S.~Shinohara,
\emph{Mean field approximation of IIB matrix model and emergence of 
four dimensional space-time},
\emph{Nucl.\ Phys.\  B} {\bf 647} (2002) 153 
[{\tt hep-th/0204240}];\\
  %%CITATION = NUPHA,B647,153;%%
%\bibitem{Kawai:2002ub}
  H.~Kawai, S.~Kawamoto, T.~Kuroki and S.~Shinohara,
\emph{Improved perturbation theory and four-dimensional space-time 
in IIB matrix model},
\emph{Prog.\ Theor.\ Phys.} {\bf 109} (2003) 115 
[{\tt hep-th/0211272}].\\
  %%CITATION = PTPKA,109,115;%%
% %%%%%%%%%%%%%%%%%%%%%%%%%%%%%%%%%%%%%%%%%%%%%%%%%%%%%%%%%%
%\cite{Aoyama:2006rk}
%\bibitem{Aoyama:2006rk}
  T.~Aoyama and H.~Kawai,
\emph{Higher order terms of improved mean field approximation 
for IIB matrix model and emergence of four-dimensional space-time},
\emph{Prog.\ Theor.\ Phys.} {\bf 116} (2006) 405
[{\tt hep-th/0603146}].
  %%CITATION = PTPKA,116,405;%%


%\cite{Nishimura:2011xy}
\bibitem{Nishimura:2011xy}
  J.~Nishimura, T.~Okubo and F.~Sugino,
\emph{Systematic study of the SO(10) symmetry breaking vacua in the matrix model
for type IIB superstrings},
{\tt arXiv:1108.1293}.
%  arXiv:1108.1293 [hep-th].
  %%CITATION = ARXIV:1108.1293;%%

%\cite{Kim:2011cr}
\bibitem{Kim:2011cr}
  S.-W.~Kim, J.~Nishimura and A.~Tsuchiya,
\emph{Expanding (3+1)-dimensional universe from a Lorentzian matrix model for
superstring theory in (9+1)-dimensions},
{\tt arXiv:1108.1540}.
%arXiv:1108.1540 [hep-th].
  %%CITATION = ARXIV:1108.1540;%%


% %%%%%%%%%%%%%%%%%%%%%%%%%%%%%%%%%%%%%%%%%%%%%%%%%%%%%%%%%%
\bibitem{Krauth:1998xh}
  W.~Krauth, H.~Nicolai and M.~Staudacher,
\emph{Monte Carlo approach to M-theory},
\emph{Phys.\ Lett.\  B} {\bf 431} (1998) 31 
[{\tt hep-th/9803117}];\\
  %%CITATION = PHLTA,B431,31;%%
%\bibitem{Krauth:1998yu}
  W.~Krauth and M.~Staudacher,
\emph{Finite Yang-Mills integrals},
\emph{Phys.\ Lett.\  B} {\bf 435} (1998) 350 
[{\tt hep-th/9804199}];\\
  %%CITATION = PHLTA,B435,350;%%
%\bibitem{Krauth:1999qw}
  W.~Krauth and M.~Staudacher,
\emph{Eigenvalue distributions in Yang-Mills integrals},
\emph{Phys.\ Lett.\  B} {\bf 453} (1999) 253 
[{\tt hep-th/9902113}].
  %%CITATION = PHLTA,B453,253;%%
% %%%%%%%%%%%%%%%%%%%%%%%%%%%%%%%%%%%%%%%%%%%%%%%%%%%%%%%%%%
%\bibitem{Hotta:1998en}
\bibitem{9811220}
  T.~Hotta, J.~Nishimura and A.~Tsuchiya,
\emph{Dynamical aspects of large N reduced models},
\emph{Nucl.\ Phys.\  B} {\bf 545} (1999) 543 
[{\tt hep-th/9811220}].
  %%CITATION = NUPHA,B545,543;%%
% %%%%%%%%%%%%%%%%%%%%%%%%%%%%%%%%%%%%%%%%%%%%%%%%%%%%%%%%%%

\bibitem{Ambjorn:2000dx}
  J.~Ambjorn, K.~N.~Anagnostopoulos, W.~Bietenholz, T.~Hotta and J.~Nishimura,
\emph{Monte Carlo studies of the IIB matrix model at large N},
\emph{JHEP} {\bf 0007} (2000) 011 
[{\tt hep-th/0005147}].

\bibitem{Ambjorn:2000bf}
  J.~Ambjorn, K.~N.~Anagnostopoulos, W.~Bietenholz, T.~Hotta and J.~Nishimura,
\emph{Large N dynamics of dimensionally reduced 4D SU(N) super Yang-Mills
theory},
\emph{JHEP} {\bf 0007} (2000) 013 
[{\tt hep-th/0003208}];\\
  %%CITATION = JHEPA,0007,013;%%
%
  %%CITATION = JHEPA,0007,011;%%
%\bibitem{Ambjorn:2001xs}
  J.~Ambjorn, K.~N.~Anagnostopoulos, W.~Bietenholz, F.~Hofheinz and
  J.~Nishimura, 
\emph{On the spontaneous breakdown of Lorentz symmetry in matrix models of
superstrings},
\emph{Phys.\ Rev.\  D} {\bf 65} (2002) 086001 
[{\tt hep-th/0104260}].
  %%CITATION = PHRVA,D65,086001;%%

% %%%%%%%%%%%%%%%%%%%%%%%%%%%%%%%%%%%%%%%%%%%%%%%%%%%%%%%%%%
\bibitem{Bialas:2000gf}
  P.~Bialas, Z.~Burda, B.~Petersson and J.~Tabaczek,
\emph{Large N limit of the IKKT matrix model},
\emph{Nucl.\ Phys.\  B} {\bf 592} (2001) 391 
[{\tt hep-lat/0007013}];\\
  %%CITATION = NUPHA,B592,391;%%
%\bibitem{Burda:2000mn}
  Z.~Burda, B.~Petersson and J.~Tabaczek,
\emph{Geometry of reduced supersymmetric 4D Yang-Mills integrals},
\emph{Nucl.\ Phys.\  B} {\bf 602} (2001) 399 
[{\tt hep-lat/0012001}];\\
  %%CITATION = NUPHA,B602,399;%%
%\bibitem{Burda:2005we}
  Z.~Burda, B.~Petersson and M.~Wattenberg,
\emph{Semiclassical geometry of 4D reduced supersymmetric Yang-Mills
integrals},
\emph{JHEP} {\bf 0503} (2005) 058 
[{\tt hep-th/0503032}].
  %%CITATION = JHEPA,0503,058;%%
% %%%%%%%%%%%%%%%%%%%%%%%%%%%%%%%%%%%%%%%%%%%%%%%%%%%%%%%%%%
% %%%%%%%%%%%%%%%%%%%%%%%%%%%%%%%%%%%%%%%%%%%%%%%%%%%%%%%%%%
% %%%%%%%%%%%%%%%%%%%%%%%%%%%%%%%%%%%%%%%%%%%%%%%%%%%%%%%%%%
\bibitem{0108041}
  K.~N.~Anagnostopoulos and J.~Nishimura,
\emph{New approach to the complex-action problem and its application to a
nonperturbative study of superstring theory},
\emph{Phys.\ Rev.\  D} {\bf 66} (2002) 106008 
[{\tt hep-th/0108041}].
  %%CITATION = PHRVA,D66,106008;%%
% %%%%%%%%%%%%%%%%%%%%%%%%%%%%%%%%%%%%%%%%%%%%%%%%%%%%%%%%%%
\bibitem{Ambjorn:2003rr}
%\bibitem{Ambjorn:2002pz}
  J.~Ambjorn, K.~N.~Anagnostopoulos, J.~Nishimura and J.~J.~M.~Verbaarschot,
\emph{The factorization method for systems with a complex action: a test in
Random Matrix Theory for finite density QCD},
\emph{JHEP} {\bf 0210} (2002) 062
[{\tt hep-lat/0208025}];\\
  %%CITATION = JHEPA,0210,062;%%
  J.~Ambjorn, K.~N.~Anagnostopoulos, J.~Nishimura and J.~J.~M.~Verbaarschot,
\emph{Non-commutativity of the zero chemical potential limit and the
thermodynamic limit in finite density systems},
\emph{Phys.\ Rev.\  D} {\bf 70} (2004) 035010
[{\tt hep-lat/0402031}].
  %%CITATION = PHRVA,D70,035010;%%
% %%%%%%%%%%%%%%%%%%%%%%%%%%%%%%%%%%%%%%%%%%%%%%%%%%%%%%%%%%
\bibitem{Azcoiti:2002vk}
  V.~Azcoiti, G.~Di Carlo, A.~Galante and V.~Laliena,
\emph{New proposal for numerical simulations of $\theta$-vacuum 
like systems},
\emph{Phys.\ Rev.\ Lett.} {\bf 89} (2002) 141601
[{\tt hep-lat/0203017}].
  %%CITATION = PRLTA,89,141601;%%
% %%%%%%%%%%%%%%%%%%%%%%%%%%%%%%%%%%%%%%%%%%%%%%%%%%%%%%%%%%
\bibitem{Fodor:2007vv}
  Z.~Fodor, S.~D.~Katz and C.~Schmidt,
\emph{The density of states method at non-zero chemical potential},
\emph{JHEP} {\bf 0703} (2007) 121
[{\tt hep-lat/0701022}].\\
  %%CITATION = JHEPA,0703,121;%%
  Z.~Fodor and S.~D.~Katz,
\emph{A new method to study lattice QCD at finite temperature and chemical
potential},
\emph{Phys.\ Lett.\  B} {\bf 534} (2002) 87 
[{\tt hep-lat/0104001}].\\
   %%CITATION = PHLTA,B534,87;%%
  Z.~Fodor and S.~D.~Katz,
\emph{Critical point of QCD at finite T and mu, lattice results for physical
quark masses},
\emph{JHEP} {\bf 0404} (2004) 050 
[{\tt hep-lat/0402006}].
  %%CITATION = JHEPA,0404,050;%%


\bibitem{07063549}
  %\cite{Ejiri:2007ga}
  %\bibitem{Ejiri:2007ga}
  S.~Ejiri,
\emph{On the existence of the critical point 
in finite density lattice QCD},
\emph{Phys.\ Rev.\ D} {\bf 77} (2008) 014508 
[{\tt arXiv:0706.3549}].
%[arXiv:0706.3549 [hep-lat]].
  %%CITATION = PHRVA,D77,014508;%%
% %%%%%%%%%%%%%%%%%%%%%%%%%%%%%%%%%%%%%%%%%%%%%%%%%%%%%%%%%%
\bibitem{fdQCD}
%\cite{Lombardo:2009aw}
%\bibitem{Lombardo:2009aw}
  M.~P.~Lombardo, K.~Splittorff and J.~J.~M.~Verbaarschot,
\emph{Distributions of the phase angle of the fermion determinant in QCD},
\emph{Phys.\ Rev.\ D} {\bf 80} (2009) 054509
[{\tt arXiv:0904.2122}];\\
%[arXiv:0904.2122 [hep-lat]];\\
  %%CITATION = PHRVA,D80,054509;%%
%\cite{Lombardo:2009rt}
%\bibitem{Lombardo:2009rt}
  M.~P.~Lombardo, K.~Splittorff and J.~J.~M.~Verbaarschot,
\emph{The fluctuations of the quark number and of the chiral condensate},
\emph{Phys.\ Rev.\  D} {\bf 81} (2010) 045012
[{\tt arXiv:0910.5482}].
%  arXiv:0910.5482 [hep-lat].
  %%CITATION = ARXIV:0910.5482;%%
% %%%%%%%%%%%%%%%%%%%%%%%%%%%%%%%%%%%%%%%%%%%%%%%%%%%%%%%%%%



\bibitem{complex-action-prob}
%
%\bibitem{Unger:2011it}
  W.~Unger, P.~de Forcrand,
  \emph{Continuous time Monte Carlo for lattice QCD in the strong coupling limit,}''
  {\tt arXiv:1107.1553};\\
%\bibitem{deForcrand:2010he}
  P.~de Forcrand, O.~Philipsen,
  \emph{Constraining the QCD phase diagram by tricritical lines at imaginary
chemical potential},
  \emph{Phys.\ Rev.\ Lett.}  {\bf 105 } (2010)  152001
  [{\tt arXiv:1004.3144}].

%\bibitem{Bloch:2011jx}
  J.~Bloch,
\emph{Evading the sign problem in random matrix simulations},
{\tt arXiv:1103.3467};\\
%[arXiv:1103.3467 [hep-lat]]; \\
 J.~Bloch, T.~Wettig,
\emph{The QCD sign problem and dynamical simulations of random matrices},
\emph{JHEP} {\bf 05} (2011) 048 
[{\tt arXiv:1102.3715}]; \\
%  [arXiv:1102.3715 [hep-lat]]; \\
%\cite{Aarts:2011ax}
%\bibitem{Aarts:2011ax}
G.~Aarts, F.~A.~James, E.~Seiler, I.~-O.~Stamatescu,
\emph{Complex Langevin: Etiology and diagnostics of its main problem},
{\tt arXiv:1101.3270}; \\
%  [arXiv:1101.3270 [hep-lat]]; \\
  G.~Aarts, F.~A.~James,
\emph{On the convergence of complex Langevin dynamics: The
three-dimensional XY model at finite chemical potential},
\emph{JHEP} {\bf 08} (2010) 020
[{\tt arXiv:1005.3468}]; \\
%  [arXiv:1005.3468 [hep-lat]]; \\
%\cite{Chandrasekharan:2010iy}
%\bibitem{Chandrasekharan:2010iy}
  S.~Chandrasekharan, A.~Li,
\emph{Fermion bag approach to the sign problem in strongly coupled
lattice QED with Wilson fermions},
\emph{JHEP} {\bf 01} (2011) 018
[{\tt arXiv:1008.5146}]; \\
%[arXiv:1008.5146 [hep-lat]]; \\
% \bibitem{Allton:2002zi}
  C.~R.~Allton, S.~Ejiri, S.~J.~Hands, O.~Kaczmarek, F.~Karsch,
E.~Laermann, C.~Schmidt, L.~Scorzato,
\emph{The QCD thermal phase transition in the presence of a small
chemical potential},
\emph{Phys.\ Rev.\ D} {\bf 66} (2002) 074507
[{\tt hep-lat/0204010}]; \\
%  [hep-lat/0204010]; \\
%\bibitem{Gavai:2003mf}
  R.~V.~Gavai, S.~Gupta,
\emph{Pressure and nonlinear susceptibilities in QCD at finite chemical
potentials},
\emph{Phys.\ Rev.\ D} {\bf 68} (2003) 034506
[{\tt hep-lat/0303013}]; \\
%  [hep-lat/0303013]; \\
%\bibitem{D'Elia:2002gd}
  M.~D'Elia, M.~-P.~Lombardo,
\emph{Finite density QCD via imaginary chemical potential},
\emph{Phys.\ Rev.\ D} {\bf 67} (2003) 014505
[{\tt hep-lat/0209146}]; \\
%[hep-lat/0209146]; \\
%\bibitem{de Forcrand:2002ci}
  P.~de Forcrand, O.~Philipsen,
\emph{The QCD phase diagram for small densities from imaginary chemical
potential},
\emph{Nucl.\ Phys.\ B} {\bf 642} (2002) 290
[{\tt hep-lat/0205016}]; \\
%[hep-lat/0205016]; \\
%\cite{Bietenholz:1995zk}
%\bibitem{Bietenholz:1995zk}
  W.~Bietenholz, A.~Pochinsky, U.~J.~Wiese,
\emph{Meron cluster simulation of the theta vacuum in the 2-d O(3) model},
\emph{Phys.\ Rev.\ Lett.} {\bf 75} (1995) 4524
[{\tt hep-lat/9505019}].
%  [hep-lat/9505019].


% %%%%%%%%%%%%%%%%%%%%%%%%%%%%%%%%%%%%%%%%%%%%%%%%%%%%%%%%%%
\bibitem{dos}
%\cite{Creutz:1983ra}
%\bibitem{Creutz:1983ra}
  M.~Creutz,
\emph{Microcanonical Monte Carlo simulation},
\emph{Phys.\ Rev.\ Lett.} {\bf 50} (1983) 1411;\\
  %%CITATION = PRLTA,50,1411;%%
%\cite{Bhanot:1986hi}
%\bibitem{Bhanot:1986hi}
%%   G.~Bhanot, S.~Black, P.~Carter and R.~Salvador,
%% \emph{A new method for the partition function of discrete systems with
%% application to the 3-D Ising model},
%% \emph{Phys.\ Lett.\  B} {\bf 183} (1987) 331;\\
  %%CITATION = PHLTA,B183,331;%%
%\cite{Bhanot:1986ku}
%\bibitem{Bhanot:1986ku}
%%   G.~Bhanot, K.~Bitar, S.~Black, P.~Carter and R.~Salvador,
%% \emph{The partition function of Z(2) and Z(8) lattice gauge theory 
%% in four-dimensions: A novel approach to simulations of lattice systems},
%% \emph{Phys.\ Lett.\  B} {\bf 187} (1987) 381;\\
  %%CITATION = PHLTA,B187,381;%%
%\cite{Bhanot:1986kv}
%\bibitem{Bhanot:1986kv}
  G.~Bhanot, K.~Bitar and R.~Salvador,
\emph{On solving four-dimensional SU(2) gauge theory 
by numerically finding its partition function},
\emph{Phys.\ Lett.\  B} {\bf 188} (1987) 246;\\
  %%CITATION = PHLTA,B188,246;%%
%\cite{Bhanot:1987nv}
%\bibitem{Bhanot:1987nv}
  G.~Bhanot, A.~Gocksch and P.~Rossi,
\emph{On simulating complex actions},
\emph{Phys.\ Lett.\  B} {\bf 199} (1987) 101;\\
  %%CITATION = PHLTA,B199,101;%%
%\cite{Gocksch:1988iz}
%\bibitem{Gocksch:1988iz}
  A.~Gocksch,
\emph{Simulating lattice QCD at finite density},
\emph{Phys.\ Rev.\ Lett.} {\bf 61} (1988) 2054;\\
  %%CITATION = PRLTA,61,2054;%%
%\cite{Karliner:1987cu}
%\bibitem{Karliner:1987cu}
  M.~Karliner, S.~R.~Sharpe and Y.~F.~Chang,
\emph{Zeroing in on SU(3)},
\emph{Nucl.\ Phys.\  B} {\bf 302} (1988) 204;\\
  %%CITATION = NUPHA,B302,204;%%
%\cite{Gocksch:1988ma}
%\bibitem{Gocksch:1988ma}
  A.~Gocksch,
\emph{The Riemann walk: A method for simulating complex actions},
\emph{Phys.\ Lett.\ B} {\bf 206} (1988) 290.
  %%CITATION = PHLTA,B206,290;%%
% %%%%%%%%%%%%%%%%%%%%%%%%%%%%%%%%%%%%%%%%%%%%%%%%%%%%%%%%%%


\bibitem{Nishimura:2000ds}
  J.~Nishimura and G.~Vernizzi,
\emph{Spontaneous breakdown of Lorentz invariance in IIB matrix model},
\emph{JHEP} {\bf 0004} (2000) 015 
[{\tt hep-th/0003223}];\\
  %%CITATION = JHEPA,0004,015;%%
%\bibitem{Nishimura:2000wf}
  J.~Nishimura and G.~Vernizzi,
\emph{Brane world from IIB matrices},
\emph{Phys.\ Rev.\ Lett.} {\bf 85} (2000) 4664
[{\tt hep-th/0007022}].
  %%CITATION = PRLTA,85,4664;%%
% %%%%%%%%%%%%%%%%%%%%%%%%%%%%%%%%%%%%%%%%%%%%%%%%%%%%%%%%%%

\bibitem{Vernizzi:2002mu}
  G.~Vernizzi and J.~F.~Wheater,
\emph{Rotational symmetry breaking in multi-matrix models},
\emph{Phys.\ Rev.\  D} {\bf 66} (2002) 085024 
[\emph{Erratum-ibid.\  D} {\bf 67} (2003) 029904]
[{\tt hep-th/0206226}].
  %%CITATION = PHRVA,D66,085024;%%

% %%%%%%%%%%%%%%%%%%%%%%%%%%%%%%%%%%%%%%%%%%%%%%%%%%%%%%%%%%
\bibitem{0108070}
  J.~Nishimura,
\emph{Exactly solvable matrix models 
for the dynamical generation of space-time in superstring theory},
\emph{Phys.\ Rev.\  D} {\bf 65} (2002) 105012 
[{\tt hep-th/0108070}].
  %%CITATION = PHRVA,D65,105012;%%
% %%%%%%%%%%%%%%%%%%%%%%%%%%%%%%%%%%%%%%%%%%%%%%%%%%%%%%%%%%
\bibitem{0412194}
  J.~Nishimura, T.~Okubo and F.~Sugino,
\emph{Gaussian expansion analysis of a matrix model with the spontaneous
breakdown of rotational symmetry},
\emph{Prog.\ Theor.\ Phys.} {\bf 114} (2005) 487 
[{\tt hep-th/0412194}].
  %%CITATION = PTPKA,114,487;%%
% %%%%%%%%%%%%%%%%%%%%%%%%%%%%%%%%%%%%%%%%%%%%%%%%%%%%%%%%%%


%\cite{Anagnostopoulos:2010ux}
\bibitem{Anagnostopoulos:2010ux}
  K.~N.~Anagnostopoulos, T.~Azuma and J.~Nishimura,
\emph{A general approach to the sign problem: 
the factorization method with multiple observables},
\emph{Phys.\ Rev.\  D} {\bf 83} (2011) 054504
[{\tt arXiv:1009.4504}].
%[arXiv:1009.4504 [cond-mat.stat-mech]].
  %%CITATION = PHRVA,D83,054504;%%


%% % %%%%%%%%%%%%%%%%%%%%%%%%%%%%%%%%%%%%%%%%%%%%%%%%%%%%%%%%%%
%% \bibitem{lat0412035}
%% %\cite{Gavai:2004sd}
%% %\bibitem{Gavai:2004sd}
%%   R.~V.~Gavai and S.~Gupta,
%%   ``The critical end point of QCD,''
%%   Phys.\ Rev.\  D {\bf 71} (2005) 114014
%%   [arXiv:hep-lat/0412035].
%%   %%CITATION = PHRVA,D71,114014;%%
%% % %%%%%%%%%%%%%%%%%%%%%%%%%%%%%%%%%%%%%%%%%%%%%%%%%%%%%%%%%%



% %%%%%%%%%%%%%%%%%%%%%%%%%%%%%%%%%%%%%%%%%%%%%%%%%%%%%%%%%%
\bibitem{6dIKKT} K.~N.~Anagnostopoulos, 
T.~Azuma and J.~Nishimura, work in progress.


%\cite{Aoyama:2010ry}
\bibitem{Aoyama:2010ry}
  T.~Aoyama, J.~Nishimura and T.~Okubo,
\emph{Spontaneous breaking of the rotational symmetry 
in dimensionally reduced super Yang-Mills models},
\emph{Prog.\ Theor.\ Phys.} {\bf 125} (2011) 537
[{\tt arXiv:1007.0883}].
%[arXiv:1007.0883 [hep-th]].
  %%CITATION = PTPKA,125,537;%%


% %%%%%%%%%%%%%%%%%%%%%%%%%%%%%%%%%%%%%%%%%%%%%%%%%%%%%%%%%%

% %%%%%%%%%%%%%%%%%%%%%%%%%%%%%%%%%%%%%%%%%%%%%%%%%%%%%%%%%%
%% \bibitem{berg-mc}
%%   %%CITATION = PHLTA,B267,249;%%
%% %\cite{Berg:1991cf}
%% %\bibitem{Berg:1991cf}
%%   B.~A.~Berg and T.~Neuhaus,
%%   ``Multicanonical Algorithms For First Order Phase Transitions,''
%%   Phys.\ Lett.\  B {\bf 267}, 249 (1991);\\
%%   %%CITATION = PHLTA,B267,249;%%
%% %\cite{Berg:2002we}
%% %\bibitem{Berg:2002we}
%%   B.~A.~Berg,
%%   ``Multicanonical Simulations Step by Step,''
%%   Comput.\ Phys.\ Commun.\  {\bf 153}, 397 (2003)
%%   [arXiv:cond-mat/0206333].
%%   %%CITATION = CPHCB,153,397;%%
%% % %%%%%%%%%%%%%%%%%%%%%%%%%%%%%%%%%%%%%%%%%%%%%%%%%%%%%%%%%%


% %%%%%%%%%%%%%%%%%%%%%%%%%%%%%%%%%%%%%%%%%%%%%%%%%%%%%%%%%%
\bibitem{0506062}
%\cite{Anagnostopoulos:2005cy}
%\bibitem{Anagnostopoulos:2005cy}
  K.~N.~Anagnostopoulos, T.~Azuma, K.~Nagao and J.~Nishimura,
\emph{Impact of supersymmetry on the nonperturbative dynamics 
of fuzzy spheres},
\emph{JHEP} {\bf 0509} (2005) 046 
[{\tt hep-th/0506062}].
  %%CITATION = JHEPA,0509,046;%%
% %%%%%%%%%%%%%%%%%%%%%%%%%%%%%%%%%%%%%%%%%%%%%%%%%%%%%%%%%%



% %%%%%%%%%%%%%%%%%%%%%%%%%%%%%%%%%%%%%%%%%%%%%%%%%%%%%%%%%%
% %%%%%%%%%%%%%%%%%%%%%%%%%%%%%%%%%%%%%%%%%%%%%%%%%%%%%%%%%%
\end{thebibliography}
\end{document}